\documentclass[reprint,amsmath,amssymb]{revtex4-2}
\usepackage[utf8]{inputenc}
\usepackage[T1]{fontenc}
\usepackage{graphicx}
\usepackage[svgnames]{xcolor}
\usepackage{dcolumn}
\usepackage{bm}
\usepackage{enumerate}
\usepackage{hyperref}
\usepackage{mathcomp}
\hypersetup{
 colorlinks=true,%
 linkcolor=DarkSlateBlue,
 citecolor=DarkSlateBlue,
 urlcolor=DarkSlateBlue,
}
\everymath{\displaystyle}
\pdfoutput=1
\begin{document}

\preprint{}

\title{Cotton gravity and 84 galaxy rotation curves}

\author{Junpei Harada}
 \email{jharada@hoku-iryo-u.ac.jp}
\affiliation{Health Sciences University of Hokkaido, 1757 Kanazawa, Tobetsu-cho, Ishikari-gun, Hokkaido 061-0293, Japan}

\date{September 23, 2022}

\begin{abstract}
Recently, as a generalization of general relativity, a gravity theory has been proposed in which gravitational field equations are described by the Cotton tensor. That theory allows an additional contribution to the gravitational potential of a point mass that rises linearly with radius as $\Phi = -GM/r + \gamma r/2$, where $G$ is the Newton constant. The coefficients $M$ and $\gamma$ are the constants of integration and should be determined individually for each physical system. When applied to galaxies, the coefficient $\gamma$, which has the dimension of acceleration, should be determined for each galaxy. This is the same as having to determine the mass $M$ for each galaxy. If $\gamma$ is small enough, the linear potential term is negligible at short distances, but can become significant at large distances. In fact, it may contribute to the extragalactic systems. In this paper, we derive the effective field equation for Cotton gravity applicable to extragalactic systems. We then use the effective field equation to numerically compute the gravitational potential of a sample of 84~rotating galaxies. The 84~galaxies span a wide range, from stellar disk-dominated spirals to gas-dominated dwarf galaxies. We do not assume the radial density profile of the stellar disk, bulge, or gas; we use only the observed data. We find that the rotation curves of 84 galaxies can be explained by the observed distribution of baryons. This is due to the flexibility of Cotton gravity to allow the integration constant $\gamma$ for each galaxy. In the context of Cotton gravity, “dark matter” is in some sense automatically included as a curvature of spacetime. Consequently, even galaxies that have been assumed to be dominated by dark matter do not need dark matter. 
\end{abstract}

\maketitle

\section{Introduction}
The missing gravity problem in extragalactic systems appears when gravity is extremely weak. 
Empirically, the typical magnitude of gravitational acceleration is $1 \ {\rm km^2 \ s^{-2} \ pc^{-1}} \sim 10^{-11} \ {\rm m \ s^{-2}}$ or less. 
If gravity is much stronger than that scale, no acceleration discrepancy has been observed. 
For understanding the missing gravity problem, it is necessary to test the laws of gravity at the extremely weak regimes.
To date, no gravity theory has been experimentally established applicable to such extremely weak regimes.

Recently, a new gravity theory has been proposed~\cite{Harada:2021bte}, in which gravitational field equations are described by the Cotton tensor (named after mathematician, \'Emile Cotton)~\cite{Cotton:1899}.
Here, the theory is called {\it Cotton gravity}.
Cotton gravity is a generalization of general relativity. 
The field equations of Cotton gravity have solutions that are not solutions of Einstein equations. In particular, Cotton gravity allows an extra contribution to the gravitational potential of a point mass that rises linearly with radius, $\Phi = -GM/r + \gamma r/2$. 
Thus, Cotton gravity may give the extra contributions at large distances, which cannot be explained by general relativity.

The purpose of this paper is to report that in Cotton gravity, the galaxy rotation curves can be explained by the observed distribution of baryons. We investigate 84 rotating galaxies with very different properties; galaxies extend a wide range from largest galaxies known to the smallest, low mass to high mass, low surface brightness to high surface brightness, and low gas fraction to high.

In this paper, we do not assume a radial density profile for the stellar disk, bulge, or gas---only the data is used.
We also do not assume any profiles for the dark matter halo. 
As we will see, the observed baryonic mass and the gravitational field equation of Cotton gravity alone can explain galaxy rotation curves. 

Before beginning this work, it should be mentioned about Milgromian dynamics (MOND)~\cite{Milgrom:1983pn}. 
MOND is one of the most extensively studied modified gravity approaches to the missing gravity problem. 
The very detailed recent review of MOND (and its applications) is given in~\cite{Banik:2021woo}, and the detailed fits to the galaxy rotation curves without an extra free parameter per galaxy was given in~\cite{Li:2018tdo}. The comparison between MOND and dark matter approach was presented in~\cite{McGaugh:2020ppt}.

It may be also mentioned about conformal gravity. Conformal gravity also allows the linear term in the potential of a point mass~\cite{Mannheim:1988dj}. 
Applications of conformal gravity to galaxy rotation curves were reviewed in~\cite{Mannheim:2012qw},
but some severe problems have been reported in~\cite{Horne:2016ajh,Campigotto:2017ytw,Hobson:2022ahe}.
Although a spherically symmetric exact solution of conformal gravity is approximately equivalent to that of Cotton gravity, 
it should be noted that Cotton gravity is a different theory from conformal gravity~\cite{Harada:2021bte}. 

This paper is organized as follows. Section~\ref{sec:Cotton} briefly summarizes Cotton gravity. In Sec.~\ref{sec:EFE}, the effective field equation of Cotton gravity is derived. Section~\ref{sec:RC} plots the rotation velocities of 84 galaxies. Section~\ref{sec:discussion} is devoted to discussion and conclusions. 
The appendix shows the details of the calculations.
The signature of metric is $(-,+,+,+)$ throughout the paper. 

\newpage

\section{\label{sec:Cotton}Cotton gravity}
In Cotton gravity, the field equations are given by~\cite{Harada:2021bte}
\begin{equation}
	C_{\nu\rho\sigma} = 16\pi G \nabla_\mu T^\mu{}_{\nu\rho\sigma},
	\label{eq:FE}
\end{equation}
where $C_{\nu\rho\sigma}$ is the Cotton tensor, $G$ is the Newton constant, $\nabla_\mu$ is a covariant derivative associated with the Levi-Civita connection, and $T_{\mu\nu\rho\sigma}$ is defined by Eq.~\eqref{eq:T4_tensor}.

Equation~\eqref{eq:FE} is a generalization of Einstein equations of general relativity. 
Indeed, Eq.~\eqref{eq:FE} has all solutions of Einstein equations---with or without the nonzero cosmological constant---and other solutions that are not solutions of Einstein equations~\cite{Harada:2021bte}. 
This means the following; if the Einstein equations ($G_{\mu\nu}=8\pi G T_{\mu\nu}$) are satisfied, then Eq.~\eqref{eq:FE} is also satisfied. 
Furthermore, if the Einstein equations with the nonzero cosmological constant ($G_{\mu\nu}+\Lambda g_{\mu\nu}=8\pi G T_{\mu\nu}$) are satisfied, then Eq.~\eqref{eq:FE} is still satisfied (note that Eq.~\eqref{eq:FE} does not include the cosmological constant $\Lambda$). Most importantly, even if Einstein equations are {\it not} satisfied, Eq.~\eqref{eq:FE} can be satisfied.

Cotton gravity is more general than general relativity in that sense. 
Since Eq.~\eqref{eq:FE} has solutions that are not solutions of Einstein equations, it may describe physics that cannot be explained by general relativity. 
This paper shows that galaxies may be typical systems, in which deviations from general relativity are significant.

In Eq.~\eqref{eq:FE}, the Cotton tensor $C_{\nu\rho\sigma}$ is defined by~\cite{Cotton:1899}
\begin{equation}
	C_{\nu\rho\sigma} = \nabla_\rho R_{\nu\sigma} - \nabla_\sigma R_{\nu\rho} - \frac{1}{6}(g_{\nu\sigma}\nabla_\rho {\cal R} - g_{\nu\rho}\nabla_\sigma {\cal R}),
	\label{eq:Cotton_tensor}
\end{equation}
where $R_{\mu\nu}$ is the Ricci tensor and ${\cal R}$ is the Ricci scalar. Here and hereafter, we denote the Ricci scalar with ${\cal R}$---following the convention in astronomy, we will denote the radial distance in a cylindrical coordinate system with $R$.
The tensor $T_{\mu\nu\rho\sigma}$ in Eq.~\eqref{eq:FE} is defined by~\cite{Harada:2021bte}
\begin{eqnarray}
	T_{\mu\nu\rho\sigma}
	&=&
	\frac{1}{2}(g_{\mu\rho}T_{\nu\sigma} - g_{\nu\rho}T_{\mu\sigma} - g_{\mu\sigma}T_{\nu\rho} + g_{\nu\sigma}T_{\mu\rho})\nonumber\\
	&&-\frac{1}{6} (g_{\mu\rho}g_{\nu\sigma} - g_{\nu\rho}g_{\mu\sigma})T,
	\label{eq:T4_tensor}
\end{eqnarray}
where $T:=g^{\mu\nu}T_{\mu\nu}$, and $g^{\nu\sigma} T_{\mu\nu\rho\sigma} = T_{\mu\rho}$ holds. 

Multiplying Eq.~\eqref{eq:FE} by $g^{\nu\sigma}$, we find that the conservation law $\nabla_\mu T^\mu{}_\nu=0$ is a consequence of Eq.~\eqref{eq:FE} as
\begin{equation}
	g^{\nu\sigma}C_{\nu\rho\sigma} = 16 \pi G \nabla_\mu T^\mu{}_{\rho} = 0,
\end{equation}
where the Bianchi identity $g^{\nu\sigma}C_{\nu\rho\sigma}=0$ has been used. 
With $\nabla_\mu T^\mu{}_\nu=0$, the right-hand side of Eq.~\eqref{eq:FE} is 
\begin{eqnarray}
	\nabla_\mu T^\mu{}_{\nu\rho\sigma}
	=\frac{1}{2}(\nabla_\rho T_{\nu\sigma} - \nabla_\sigma T_{\nu\rho}) - \frac{1}{6}(g_{\nu\sigma}\nabla_\rho - g_{\nu\rho}\nabla_\sigma)T.
	\nonumber \\
\end{eqnarray}
In the next section, we will derive the effective field equation of Eq.~\eqref{eq:FE} applicable to galaxies.

\section{\label{sec:EFE}Effective field equation}
When gravity is weak, a metric can be written as
\begin{equation}
	g_{\mu\nu} = \eta_{\mu\nu} + h_{\mu\nu},
\end{equation}
where $\eta_{\mu\nu}$ is the Minkowski metric, and $h_{\mu\nu}$ is small. 
For galaxies studied in this work, the energy-momentum tensor $T_{\mu\nu}$ is approximately given by
\begin{equation}
	T_{00} = \rho, \quad
	T_{ij} = 0, \quad
	T_{0i} = 0, \quad
	T = -\rho,
\end{equation}
where $\rho$ is a mass density. We assume that $h_{\mu\nu}$ and $\rho$ are time independent (or assume that their time dependence sufficiently small so that time derivative can be ignored).

For $\nu=\sigma=0$ and $\rho=1,2,3$ in Eq.~\eqref{eq:FE}, we have
\begin{equation}
	\nabla \left(R_{00} + \frac{1}{6}{\cal R}\right) = \frac{16\pi G}{3} \nabla \rho,
\end{equation}
or equivalently
\begin{equation}
	R_{00} + \frac{1}{6}{\cal R} = \frac{16\pi G}{3} \rho + {\rm const}.
	\label{eq:FE2}
\end{equation}
The constant in Eq.~\eqref{eq:FE2} is the cosmological constant---recall that in Cotton gravity, the cosmological constant is a constant of integration~\cite{Harada:2021bte}. We assume that the contributions of the cosmological constant are negligible at galactic scales. In that case, Eq.~\eqref{eq:FE2} reduces to 
\begin{equation}
	R_{00} + \frac{1}{6}{\cal R} = \frac{16\pi G}{3} \rho.
	\label{eq:FE3}
\end{equation}

Let us calculate $R_{00}$ and ${\cal R}$ in Eq.~\eqref{eq:FE3}.
At far from a center, a metric can be approximately written by 
\begin{equation}
	ds^2 = -(1+2\Phi)dt^2 +  (1-2\Phi)dr^2 + r^2 d\Omega^2,
\end{equation}
where $d\Omega^2=d\theta^2+\sin^2\theta d\phi^2$, and $\Phi$ is the potential. Transforming from spherical to Cartesian coordinates, we find that 
\begin{eqnarray}
	h_{00}= -2\Phi,\quad
	h_{ij} = -2\Phi n_i n_j,\quad
	h_{0i} = 0,
	\label{eq:metric_h}
\end{eqnarray}
where ${\bm n}:={\bm r}/|\bm{r}|$, and $h:=\eta^{\mu\nu}h_{\mu\nu}=2\Phi - 2\Phi \bm{n} \cdot \bm{n}=0$.
At a linear order in $h_{\mu\nu}$, the Ricci tensor is given by
\begin{equation}
	R_{\mu\nu} = \frac{1}{2}\left( - \partial_\mu\partial_\nu h + \partial_\mu \partial^\rho h_{\nu\rho} + \partial_\nu \partial^\rho h_{\mu\rho} - \Box h_{\mu\nu}\right).
	\label{eq:Ricci_linear}
\end{equation}
Plugging Eq.~\eqref{eq:metric_h} into Eq.~\eqref{eq:Ricci_linear}, we obtain
\begin{equation}
	R_{00} = -\frac{1}{2}\nabla^2 h_{00} = \nabla^2 \Phi.
\end{equation}

If Einstein equations are satisfied, then the field equation~\eqref{eq:FE3} just reduces to the usual Poisson equation as
\begin{equation}
	\nabla^2 \Phi = \frac{16\pi G}{3} \rho - \frac{8\pi G}{6}\rho = 4\pi G \rho,
	\label{eq:Poisson}
\end{equation}
where the Ricci scalar ${\cal R} = 8\pi G \rho$ has been used.
In Cotton gravity, however, Einstein equations are {\it not} necessarily satisfied~\cite{Harada:2021bte}. In that case, the field equation~\eqref{eq:FE3} does {\it not} reduce to the Poisson equation as follows. 

At a linear order in $h_{\mu\nu}$, the Ricci scalar is given by
\begin{equation}
		{\cal R} = \partial^\mu \partial^\nu h_{\mu\nu} - \Box h.
		\label{eq:Ricciscalar_linear}
\end{equation}
Plugging Eq.~\eqref{eq:metric_h} into Eq.~\eqref{eq:Ricciscalar_linear}, we find that 
\begin{equation}
	{\cal R} 
	=- \frac{2}{|\bm{r}|^2} \left( (\bm{r}\cdot \nabla)^2 + 4(\bm{r}\cdot\nabla)+2\right) \Phi,
	\label{eq:Ricciscalar_linear2}
\end{equation}
and the field equation~\eqref{eq:FE3} reduces to
\begin{equation}
	\left[\nabla^2 - \frac{1}{3|\bm{r}|^2} \left( (\bm{r}\cdot \nabla)^2 + 4(\bm{r}\cdot\nabla)+2\right) \right]\Phi
	= \frac{16\pi G}{3}\rho.
	\label{eq:EFE}
\end{equation}
This is the effective field equation of Eq.~\eqref{eq:FE}. 
Thus, operators other than $\nabla^2$ are necessary to determine $\Phi$. 

Equation~\eqref{eq:EFE} yields a crucial difference from the Poisson equation~\eqref{eq:Poisson} at large distances.
To see it, in the spherically symmetric system for example, we have
\begin{equation}
	R_{00}=\nabla^2 \Phi 
	= \frac{1}{r^2} \frac{\partial}{\partial r}\left(r^2 \Phi^\prime \right) 
	= \Phi^{\prime\prime} + \frac{2\Phi^\prime}{r},
\end{equation}
where the primes denote the derivative with respect to the radial coordinates $r$. 
From Eq.~\eqref{eq:Ricciscalar_linear2}, we also have
\begin{equation}
	{\cal R} = -2\left(\Phi^{\prime\prime}+\frac{4\Phi^\prime}{r}+\frac{2\Phi}{r^2}\right).
\end{equation}
Therefore, the left-hand side of Eq.~\eqref{eq:EFE} is given by
\begin{eqnarray}
	&&R_{00} + \frac{1}{6} {\cal R}
	=\left[\nabla^2 - \frac{1}{3|\bm{r}|^2} \left( (\bm{r}\cdot \nabla)^2 + 4(\bm{r}\cdot\nabla)+2\right) \right]\Phi \nonumber\\
	&&=\frac{2}{3}\left(\Phi^{\prime\prime}+\frac{\Phi^\prime}{r}-\frac{\Phi}{r^2}\right)
	=\frac{2}{3r^2}\frac{\partial}{\partial r}\left[r^3\frac{\partial}{\partial r}\left(\frac{\Phi}{r}\right)\right].
\end{eqnarray}
Then, Eq.~\eqref{eq:EFE} is given in a simple form as,
\begin{eqnarray}
	\frac{\partial}{\partial r}\left[r^3\frac{\partial}{\partial r}\left(\frac{\Phi}{r}\right)\right]
	= 8\pi G r^2 \rho(r).
	\label{eq:EFE_spherical}
\end{eqnarray}

For the mass density $\rho(r)=0 \ (r\geq a)$ and $\rho (r) \not = 0 \ (r \leq a)$, 
the solution of Eq.~\eqref{eq:EFE_spherical} is given by
\begin{eqnarray}
	\Phi (r \geq a) &=& - \frac{GM}{r} + \frac{\gamma}{2} r, \quad M=\int_0^a 4\pi \xi^2 \rho(\xi)d\xi, \nonumber\\
	\Phi (r \leq a) &=& - \frac{G}{r}\int_0^r 4\pi\xi^2\rho(\xi)d\xi - Gr\int_r^a 4\pi \rho(\xi)d\xi + \frac{\gamma}{2}r,
	\nonumber\\
	\label{eq:sol_Cotton}
\end{eqnarray}
where $\gamma$ is an integration constant($1/2$ is a convention). 
While the mass $M$ is determined by $\rho(r)$, $\gamma$ is arbitrary unless the boundary condition will be determined.
The linear potential term $\gamma r/2$---which cannot be obtained from general relativity---is significant at large $r$.

In Sec.~\ref{sec:RC}, it is shown that Eq.~\eqref{eq:EFE} can explain galaxy rotation curves from the distribution of baryons. In the context of Cotton gravity, the missing gravity problem is originated from the Ricci curvature~\eqref{eq:Ricciscalar_linear2}, rather than dark matter.

\begin{figure*}[t]
\includegraphics[width=60mm]{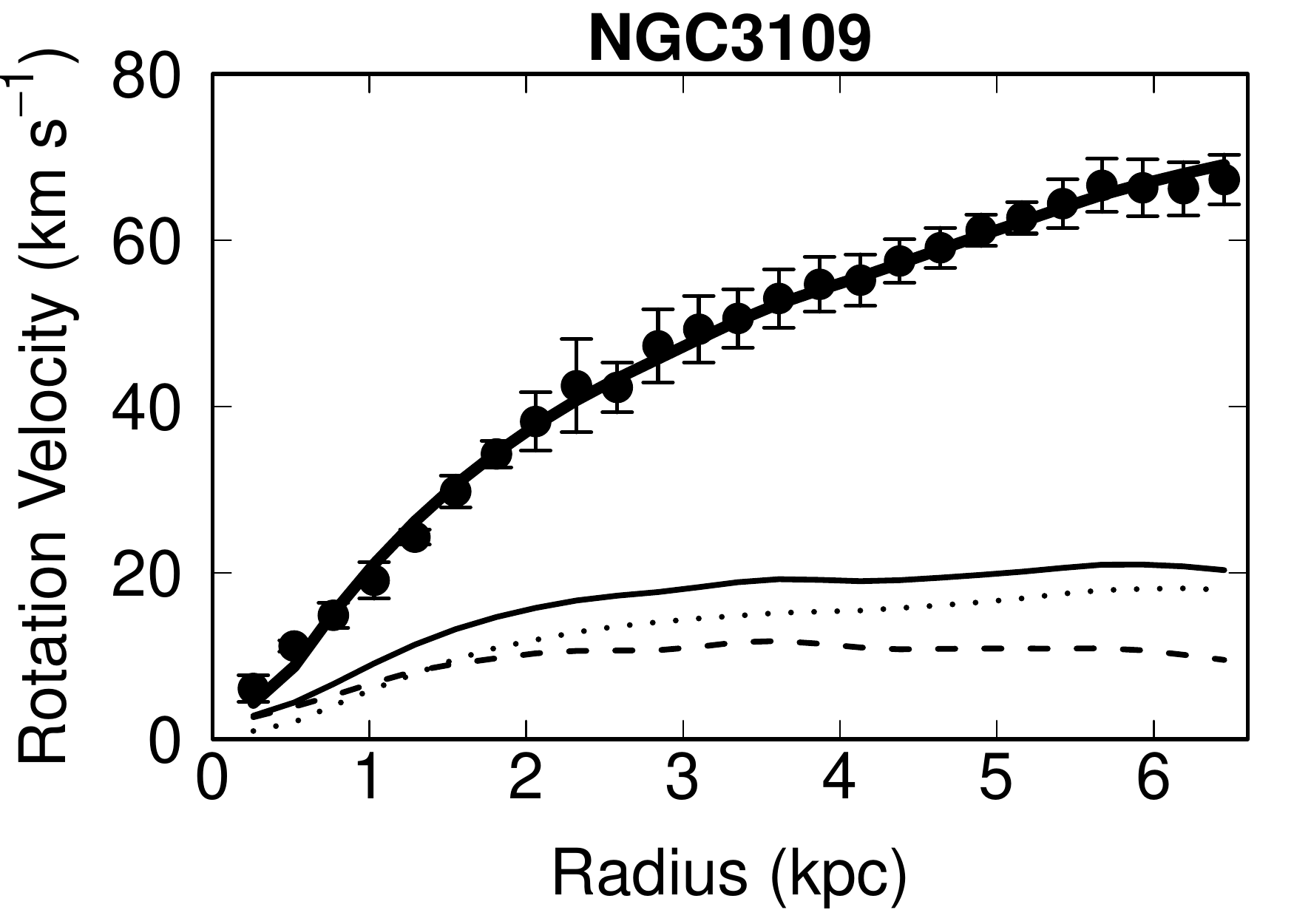}%
\includegraphics[width=60mm]{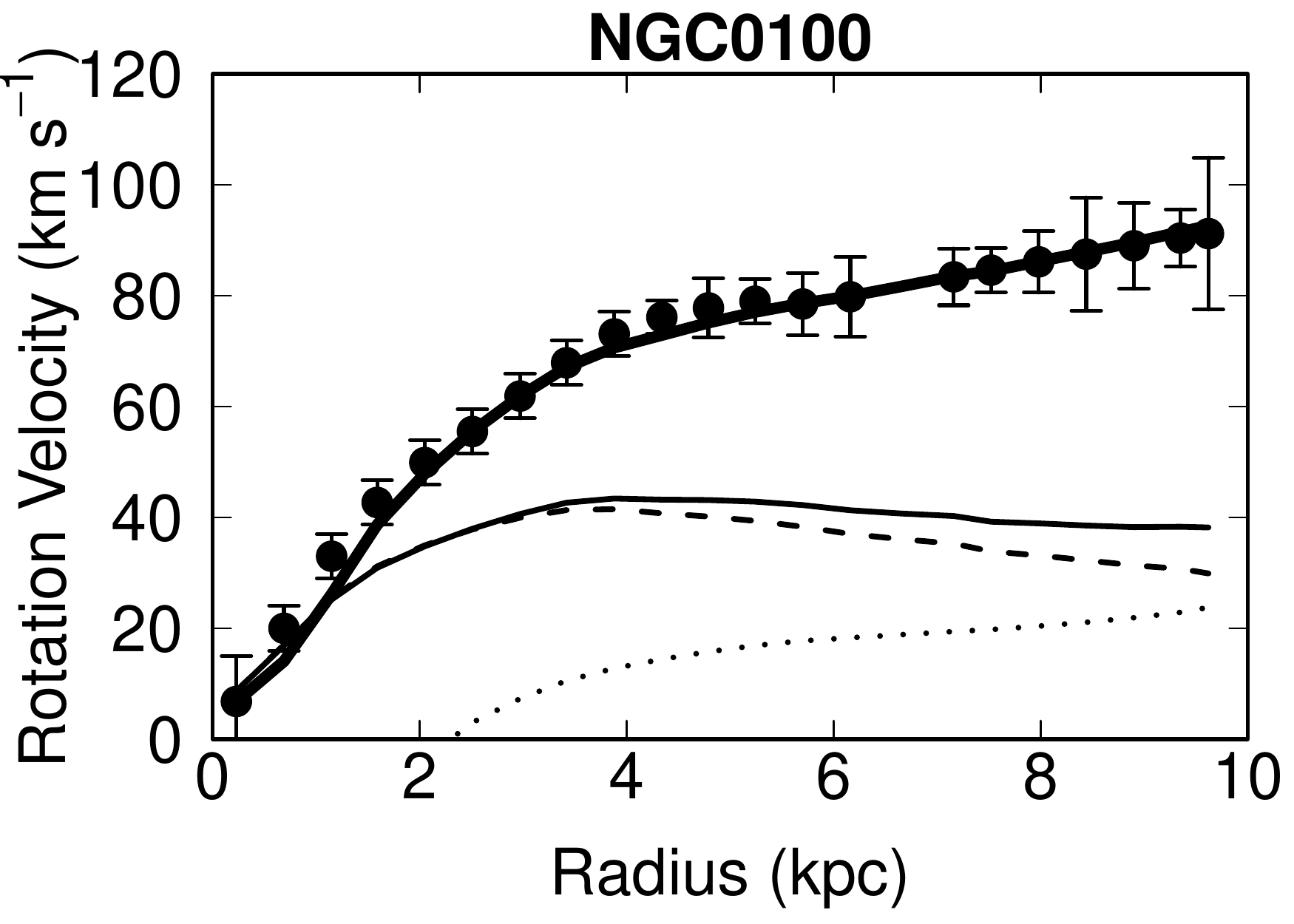}%
\includegraphics[width=60mm]{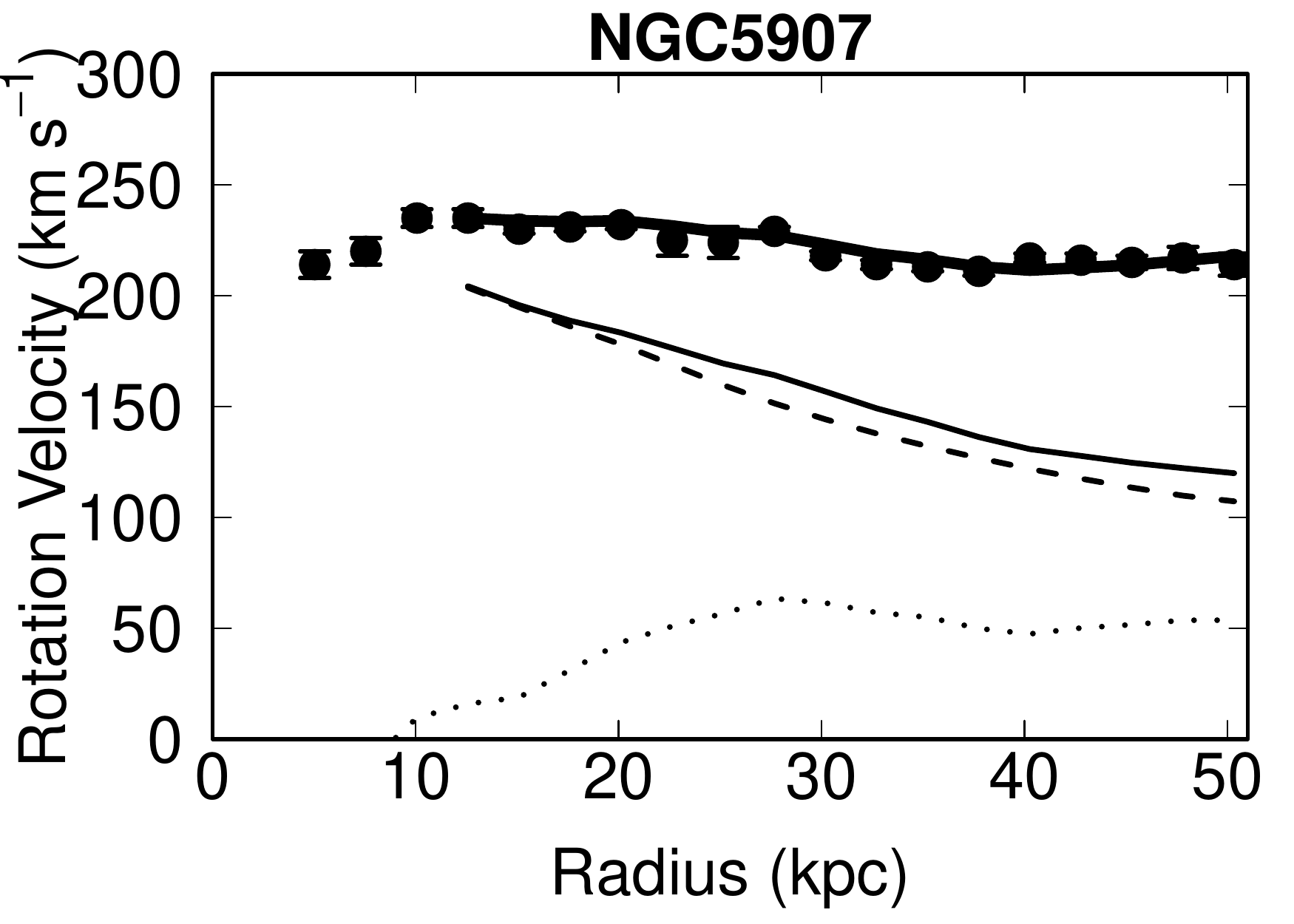}%
\\ \ \\
\includegraphics[width=60mm]{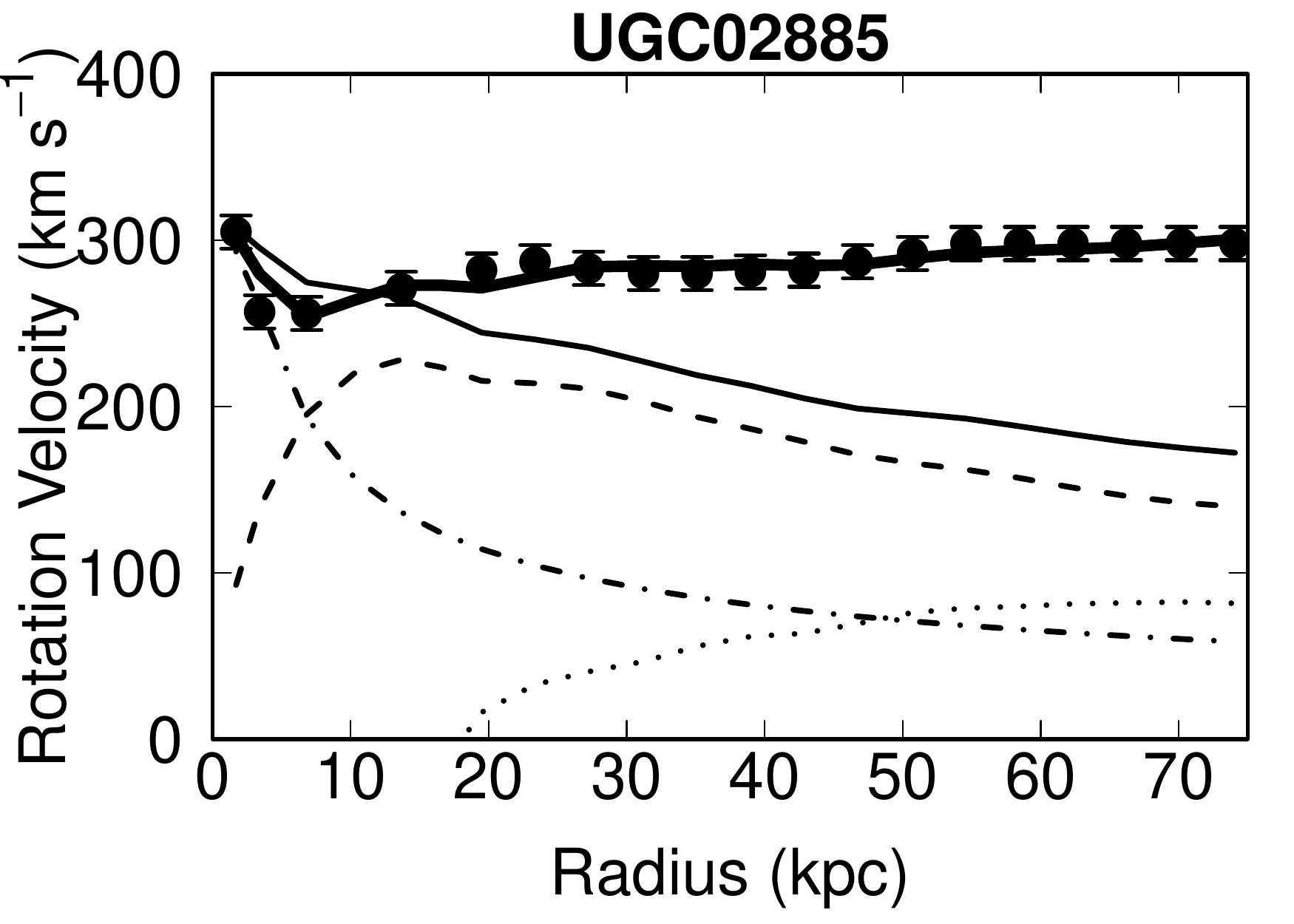}%
\includegraphics[width=60mm]{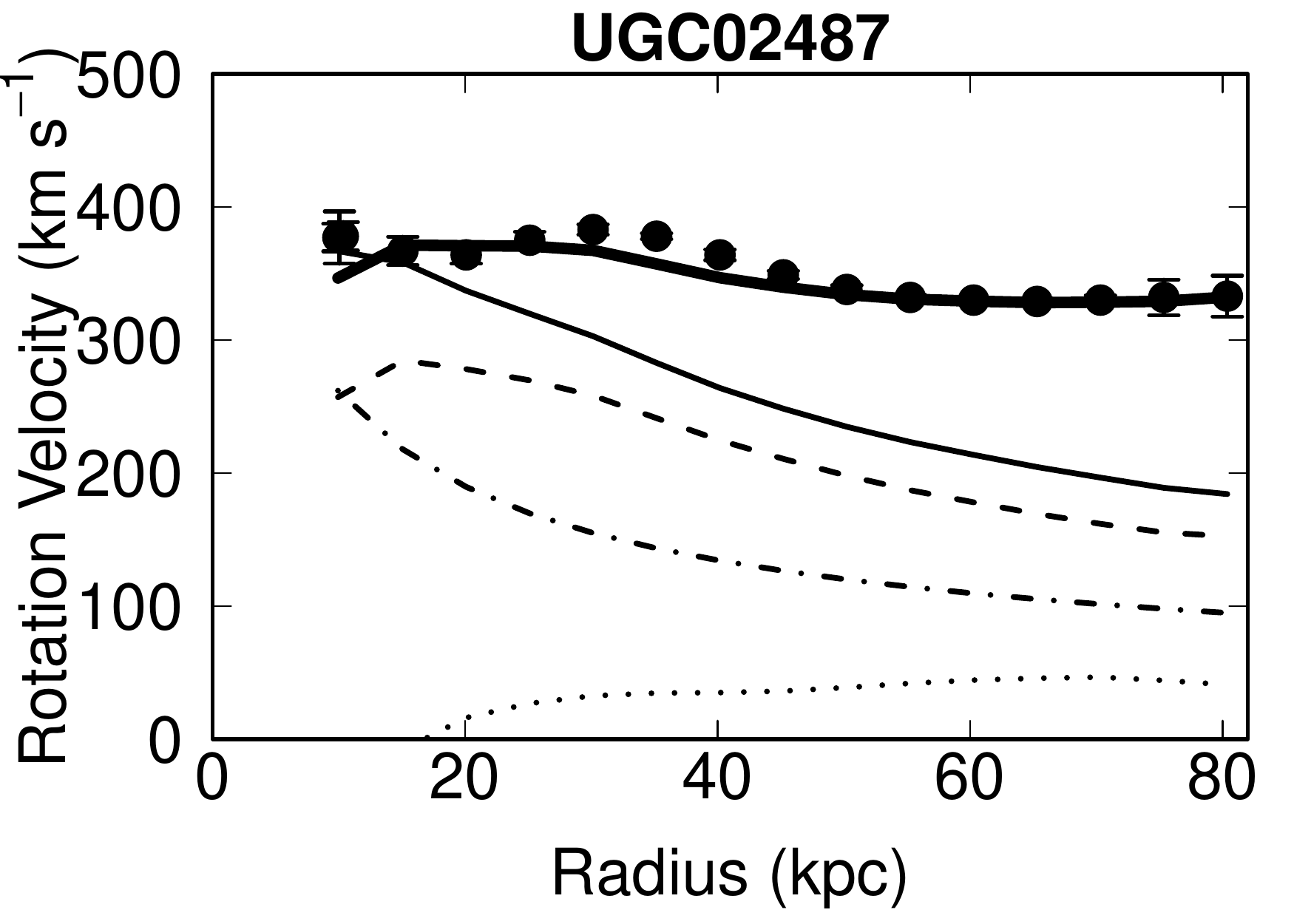}%
\includegraphics[width=60mm]{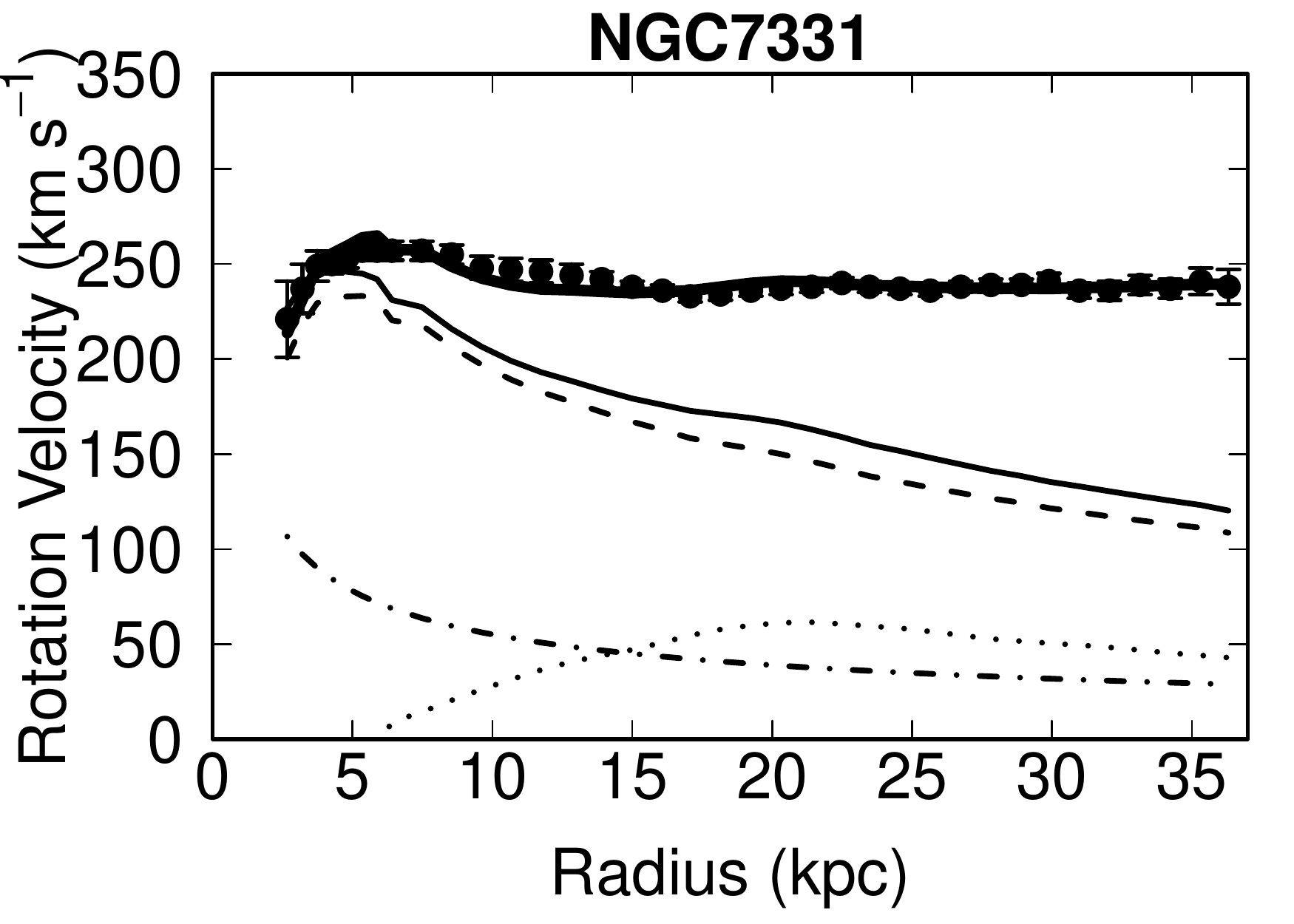}%
\caption{\label{fig:RCex}Examples of rotation curves for galaxies. The points with error bars are the observed rotation velocity $V_{\rm obs}(R)$, which are available from the SPARC database~\cite{Lelli:2016zqa}. Four thin lines (solid, dashed, dotted, dash-dotted) represent the calculated rotation velocity $V(R)=\sqrt{R \partial \Phi/\partial R}$, where $\Phi$ is determined by solving the Poisson equation~\eqref{eq:Poisson}. Each baryonic component is represented: dotted lines for the gas, dashed lines for the stellar disk, dash-dotted lines for the bulge when present, and thin solid lines for the sum of these components. In all galaxies, the observed data exceed the thin solid lines, indicating the need for dark matter. Thick solid lines represent the calculated rotation velocity $V(R)=\sqrt{R \partial \Phi/\partial R}$, where $\Phi$ is determined by solving the effective field equation~\eqref{eq:EFE}, rather than the Poisson equation. In all galaxies, the observed data are consistent with the thick solid lines. NGC~3109 has been assumed that dark matter dominates. Nevertheless, in the context of Cotton gravity, the rotation velocities can be explained without the need for dark matter. UGC~2885 (Rubin's galaxy, named after Vera Rubin) is one of the largest known. Even for such largest galaxies, the calculated velocities are in agreement with the observed data.
Some galaxies have large mass density at small radii. In those cases, the surface brightness profiles are very sensitive to the results at small radii, and lines are not shown near the center (NGC~5907). Further investigations are required for those cases.}
\end{figure*}

\section{Galaxy rotation curve\label{sec:RC}}

\subsection{Data and galaxy sample}
We use the SPARC database~\cite{Lelli:2016zqa} (Spitzer Photometry and Accurate Rotation Curves).
SPARC includes near-infrared (3.6 ${\rm \mu m}$) observations that trace the distribution of stellar mass. 
The surface density profiles of the stellar disk and bulge are available from the database. 
SPARC also includes the observed rotation velocity $V_{\rm obs}(R)$. 

In this work, we investigate 84 galaxies listed in Table~\ref{tab:table1}. 
The list includes galaxies with the morphologies from S0 to Im, baryonic masses $ 4 \times 10^7 < M_{\rm bar}/M_\odot < 5 \times 10^{11}$, gas fraction $ 0.03 < f_{\rm gas} = M_{\rm gas}/M_{\rm bar} < 0.94$, and rotation velocities $30 < V_{\rm flat}/({\rm km \ s^{-1}})<330$. This range extends from the largest galaxies known to the smallest. 
Of 84 galaxies, 53 galaxies are disk-dominated ($f_{\rm disk}=M_{\rm disk}/M_{\rm bar}>0.5$), 30 galaxies are gas-dominated ($f_{\rm gas}>0.5$), and only 1 galaxy---NGC~7814---is bulge-dominated ($f_{\rm bulge}=M_{\rm bulge}/M_{\rm bar}>0.5$). Only 11 of 84 galaxies have the central bulge.
 The radial H {\scriptsize I} (atomic hydrogen) surface density profiles are not available from the SPARC database. 
We collect them from the references listed in column~(11) of Table~\ref{tab:table1}.

\renewcommand{\arraystretch}{1.3}
\begin{table*}[htb]
\caption{\label{tab:table1}%
Parameters for galaxies. Column~(1) gives the galaxy name. Column~(2) gives the numerical Hubble type adopting the following scheme: $0={\rm S}0$, $1={\rm Sa}$, $2={\rm Sab}$, $3={\rm Sb}$, $4={\rm Sbc}$, $5={\rm Sc}$, $6={\rm Scd}$, $7={\rm Sd}$, $8={\rm Sdm}$, $9={\rm Sm}$, $10={\rm Im}$.  Column~(3) gives the assumed distance. Column~(4) gives the assumed inclination angle ($i$). The parenthesis shows the error. Columns~(3) and (4) are obtained from the SPARC~\cite{Lelli:2016zqa}. Column~(5) gives the constant in Eq.~\eqref{eq:bc}. Column~(6) gives the mass-to-light ratio at 3.6 ${\rm \mu m}$ band ($\Upsilon_\star$) of the stellar disk. Column~(7) gives the baryonic mass ($M_{\rm bar}$). The baryonic mass is a sum of the stellar mass $M_\star$ and the gas mass $M_{\rm gas}$. The stellar mass is a sum of the masses of the stellar disk $M_{\rm disk}$ and the central bulge $M_{\rm bulge}$. Column~(8) gives the disk fraction ($f_{\rm disk}=M_{\rm disk}/M_{\rm bar}$). Column~(9) gives the bulge fraction ($f_{\rm bulge}=M_{\rm bulge}/M_{\rm bar}$). The mass-to-light ratio for the bulge is assumed to be $1.4\Upsilon_\star$ for all galaxies with the bulge. Column~(10) gives the gas fraction ($f_{\rm gas}=M_{\rm gas}/M_{\rm bar}$). Gas mass is estimated by $M_{\rm gas}=1.33 M_{\rm HI}$, where $M_{\rm HI}$ is the H {\scriptsize I} mass and $1.33$ represents an enhancement factor to account for the cosmic abundance of helium. Columns~(5), (6), (7), (8), (9), and (10) are obtained in this work. Column~(11) gives the references for the radial H {\scriptsize I} surface density profiles ($\Sigma_{\rm H I}$) used in this work: Al15~\cite{Allaert:2015}, An22~\cite{Annibali:2022}, Ba05~\cite{Barbieri:2005di}, Ba06~\cite{Battaglia:2005ky}, BC04~\cite{Begum:2004sj}, BW94~\cite{Broeils:1994}, Ca90~\cite{Carignan_1990}, CB89~\cite{Carignan_1989}, Co91~\cite{Cote_1991}, Co00~\cite{Cote:2000}, CP90~\cite{Carignan:1990b}, Fr02~\cite{Fraternali:2002hr}, Fr11~\cite{Fraternali:2011vm}, Ga02~\cite{Garcia-Ruiz:2002vmq}, Ge04~\cite{Gentile:2004tb}, Ha14~\cite{Hallenbeck:2014}, Ho01~\cite{Hoekstra:2000iz}, JC90~\cite{Jobin:1990}, Ke07~\cite{Kepley:2007gm}, Le14~\cite{Lelli:2014sxa}, MC94~\cite{Martimbeau_1994}, No05~\cite{Noordermeer:2005ey}, Rh96~\cite{Rhee:1996}, SG06~\cite{Spekkens:2006vd}, Sw02~\cite{Swaters:2002fp}, VH93~\cite{van_der_Hulst_1993}, VS01~\cite{Verheijen:2001ym}.}
\begin{ruledtabular}
\begin{tabular}{lcdcccdcccc}
\textrm{Name} & \textrm{Type} & \multicolumn{1}{c}{$D$} & $i$ & $\gamma_{\rm galaxy}/2$ & $\Upsilon_\star$ 
& \multicolumn{1}{c}{$M_{\rm bar}$} & $f_{\rm disk}$ & $f_{\rm bulge}$ & $f_{\rm gas}$ & \textrm{Ref.} \\
& & \multicolumn{1}{c}{(Mpc)} & \multicolumn{1}{c}{($\tcdegree$)} & (${\rm km^2 \ s^{-2} \ pc^{-1}}$) & $(M_\odot/L_\odot)$ & \multicolumn{1}{c}{$(10^9 M_\odot)$} & & &\\
 (1) & (2) & \multicolumn{1}{c}{(3)} & (4) & (5) & (6) & \multicolumn{1}{c}{(7)} & \multicolumn{1}{c}{(8)} & (9) & (10) & (11) \\
\hline
DDO064 		& 10 & 6.80 	& 60(5) 		& 0.75	& 0.50 	& 0.391		& 0.208	& 0.000	& 0.792 	& Sw02 \\
DDO154 		& 10 & 4.04 	& 64(3) 		& 0.39	& 0.65 	& 0.385		& 0.084	& 0.000	& 0.916	& CB89 \\
DDO161 		& 10 & 7.50 	& 70(10)		& 0.28 	& 0.34 	& 2.778		& 0.066	& 0.000	& 0.934 	& Co00 \\
DDO168 		& 10 & 4.25 	& 63(6)		& 0.67  	& 0.50 	& 0.432		& 0.212	& 0.000	& 0.788 	& Ho01 \\
DDO170 		& 10 & 15.40 	& 66(7) 		& 0.31 	& 0.50 	& 1.033		& 0.244	& 0.000	& 0.756 	& Ho01 \\
ESO079-G014 	& 4 	& 28.70	& 79(5)		& 1.38 	& 0.67 	& 37.932		& 0.886	& 0.000	& 0.114 	& Ge04 \\
ESO116-G012 	& 7 	& 13.00 	& 74(3)		& 1.15 	& 0.77 	& 4.837		& 0.683	& 0.000	& 0.317 	& Ge04 \\
ESO444-G084 	& 10 & 4.83	& 32(6)		& 1.20 	& 0.50 	& 0.209		& 0.168	& 0.000	& 0.832 	& Co00 \\
IC2574 		& 9 	& 3.91	& 75(7)		& 0.37 	& 0.20 	& 1.595		& 0.142	& 0.000	& 0.858 	& MC94 \\
KK98-251 	& 10 & 6.80 	& 59(5) 		& 0.26	& 0.55 	& 0.197		& 0.222	& 0.000	& 0.778 	& BC04 \\
NGC0055 	& 9 	& 2.11	& 77(3)		& 0.51 	& 0.55 	& 4.108		& 0.571	& 0.000	& 0.429	& Ho01 \\
NGC0100 	& 6 	& 13.50	& 89(1)		& 0.83 	& 0.57 	& 4.574		& 0.405	& 0.000	& 0.595 	& Rh96 \\
NGC0247 	& 7 	& 3.70	& 74(3)		& 0.53 	& 1.10 	& 10.065		& 0.772	& 0.000	& 0.228 	& CP90 \\
NGC0300 	& 7 	& 2.08	& 42(10)		& 0.75 	& 1.20 	& 4.386		& 0.773	& 0.000	& 0.227 	& Ho01 \\
NGC0801 	& 5 	& 80.70	& 80(1)		& 0.39 	& 0.57 	& 212.118		&0.882	& 0.000	& 0.118 	& Ho01 \\
NGC0891         & 3	& 9.91	& 90(1)		& 1.50	& 0.38	& 61.876		& 0.728 	& 0.179 	& 0.093	& Fr11\\
NGC1003 	& 6 	& 11.40	& 67(5)		& 0.38 	& 0.75 	& 25.349		& 0.191	& 0.000	& 0.809 	& BW94 \\
NGC1090 	& 4 	& 37.00	& 64(3)		& 0.60 	& 0.60 	& 72.626		& 0.791	& 0.000	& 0.209 	& Ge04 \\
NGC2366 	& 10 & 3.27	& 68(5)		& 0.43 	& 0.90 	& 1.030		& 0.204	& 0.000	& 0.796 	& Le14 \\
NGC2403 	& 6 	& 3.16	& 63(3)		& 0.82 	& 1.00 	& 14.418		& 0.696	& 0.000	& 0.304 	& Fr02 \\
NGC2841		& 3	& 14.10	& 76(10)		& 1.20	& 1.00	& 234.347		& 0.608 	& 0.296	& 0.096	& Ho01\\
NGC2903 	& 4 	& 6.60	& 66(3)		& 1.40 	& 0.45 	& 37.783		& 0.933	& 0.000	& 0.067 	& Ho01 \\
NGC2998 	& 5 	& 68.10	& 58(2)		& 0.89 	& 0.60 	& 118.241		& 0.788	& 0.000	& 0.212 	& Ho01 \\
NGC3109 	& 9 	& 1.33	& 70(5)		& 0.68 	& 0.60 	& 0.651		& 0.174	& 0.000	& 0.826 	& JC90 \\
NGC3198 	& 5 	& 13.80	& 73(3)		& 0.42 	& 1.00 	& 49.745		& 0.763	& 0.000	& 0.237 	& Ho01 \\
NGC3726 	& 5 	& 18.00	& 53(2)		& 0.72 	& 0.50 	& 42.923		& 0.770	& 0.000	& 0.230 	& VS01 \\
NGC3741 	&10	& 3.21	& 70(4)		& 0.42 	& 1.00 	& 0.238		& 0.102	& 0.000	& 0.898 	& An22 \\
NGC3769 	& 3	& 18.00	& 70(2)		& 0.35 	& 0.75 	& 17.697		& 0.635	& 0.000	& 0.365 	& VS01 \\
NGC3893 	& 5 	& 18.00	& 49(2)		& 1.07 	& 0.62 	& 39.662		& 0.808	& 0.000	& 0.192	& VS01 \\
NGC3972 	& 4 	& 18.00	& 77(1)		& 1.64 	& 0.59 	& 9.851		& 0.840	& 0.000	& 0.160 	& VS01 \\
NGC4010 	& 7 	& 18.00	& 89(1)		& 1.10 	& 0.45 	& 12.257		& 0.623	& 0.000	& 0.377 	& VS01 \\
NGC4068 	&10 	& 4.37	& 44(6)		& 0.43 	& 0.50 	& 0.362		& 0.324	& 0.000	& 0.676	& Le14 \\
NGC4085 	& 5 	& 18.00	& 82(2)		& 1.75 	& 0.30 	& 7.948		& 0.781	& 0.000	& 0.219 	& VS01 \\
NGC4088 	& 4 	& 18.00	& 69(2)		& 1.00 	& 0.32 	& 43.415		& 0.757	& 0.000	& 0.243 	& VS01 \\
NGC4100 	& 4 	& 18.00 	& 73(2)		& 0.90 	& 0.48 	& 29.897		& 0.864	& 0.000	& 0.136 	& VS01 \\
NGC4157		& 3	& 18.00	& 82(3)		& 0.80	& 0.45	& 57.581		& 0.785	& 0.024	& 0.191	& VS01\\
NGC4214 	&10	& 2.87 	& 15(10)		& 1.05 	& 1.05  	& 1.746		& 0.664	& 0.000	& 0.336 	& Le14 \\
\end{tabular}
\end{ruledtabular}
\end{table*}

\renewcommand{\arraystretch}{1.3}
\addtocounter{table}{-1}
\begin{table*}[htb]
\caption{\label{tab:table2}%
 \textit{(continued)}.}
\begin{ruledtabular}
\begin{tabular}{lcdcccdcccc}
\textrm{Name} & \textrm{Type} & \multicolumn{1}{c}{$D$} & $i$ & $\gamma_{\rm galaxy}/2$ & $\Upsilon_\star$ 
& \multicolumn{1}{c}{$M_{\rm bar}$} & $f_{\rm disk}$ & $f_{\rm bulge}$ & $f_{\rm gas}$ & \textrm{Ref.} \\
& & \multicolumn{1}{c}{(Mpc)} & \multicolumn{1}{c}{($\tcdegree$)} & (${\rm km^2 \ s^{-2} \ pc^{-1}}$) & $(M_\odot/L_\odot)$ & \multicolumn{1}{c}{$(10^9 M_\odot)$} & & &\\
 (1) & (2) & \multicolumn{1}{c}{(3)} & (4) & (5) & (6) & \multicolumn{1}{c}{(7)} & \multicolumn{1}{c}{(8)} & (9) & (10) & (11) \\
\hline
NGC4559 	& 6 	& 9.00	& 67(1)		& 0.41 	& 0.70  	& 22.080		&0.605	& 0.000	& 0.395 	& Ba05 \\
NGC5055 	& 4 	& 9.90 	& 55(6)		& 0.55 	& 0.50  	& 82.099		& 0.894	& 0.000	& 0.106	& Ba06 \\
NGC5585 	& 7 	& 7.06	& 51(2)		& 0.65 	& 0.82  	& 4.253		& 0.563 	& 0.000	& 0.437 	& Co91 \\
NGC5907 	& 5 	& 17.30	& 88(2)		& 0.70 	& 0.59  	& 125.032		& 0.838	& 0.000	& 0.162 	& Al15 \\
NGC6195		& 3	& 127.80	& 62(5)		& 0.50	& 0.55	& 278.553		& 0.568	& 0.309	& 0.123	& SG06 \\
NGC6503 	& 6 	& 6.26	& 74(2) 		& 0.60 	& 0.75  	& 11.367		& 0.822	& 0.000	& 0.178 	& Ho01 \\
NGC6674		& 3	& 51.20	& 54(6)		& 0.55	& 1.00	& 255.953 	& 0.775	& 0.117	& 0.108	& Ho01 \\
NGC7331		& 3	& 14.70	& 75(2)		& 1.10	& 0.40	& 114.914		& 0.827	& 0.061	& 0.112	& Ho01 \\
NGC7793 	& 7 	& 3.61	& 47(9)		& 0.35 	& 0.90  	& 7.658		& 0.821	& 0.000	& 0.179	& Ca90 \\
NGC7814		& 2	& 14.40	& 90(1)		& 1.90	& 0.55	& 56.548		& 0.233	& 0.741	& 0.026	& Fr11 \\
UGC00128	& 8	& 64.50	& 57(10)		& 0.25	& 2.30	& 38.381		& 0.716	& 0.000	& 0.284	& VH93 \\
UGC01281 	& 8 	& 5.27	& 90(1) 		& 0.75  	& 0.38  	& 0.533		& 0.264	& 0.000	& 0.736 	& Sw02 \\
UGC02487	& 0	& 69.10	& 36(5)		& 0.95	& 1.00	& 564.712		& 0.667	& 0.298	& 0.035	& No05 \\
UGC02885	& 5	& 80.60	& 64(4)		& 0.75	& 0.80	& 404.295		& 0.698	& 0.146	& 0.155	& Ho01 \\
UGC03546	& 1	& 28.70	& 55(5)		& 1.05	& 0.43	& 52.426		& 0.607	& 0.342	& 0.051	& No05 \\
UGC04278 	& 7 	& 9.51	& 90(3)		& 1.05  	& 0.45  	& 2.142		& 0.275	& 0.000	& 0.725 	& Sw02 \\
UGC04483 	& 10 & 3.34	& 58(3)		& 0.42  	& 0.65  	& 0.045		& 0.184	& 0.000	& 0.816 	& Le14 \\
UGC04499 	& 8 	& 12.50	& 50(3)		& 0.60  	& 0.60  	& 2.511		& 0.339	& 0.000	& 0.661 	& Sw02 \\
UGC05414 	& 10 & 9.40	& 55(3)		& 0.72  	& 0.50  	& 1.539		& 0.402	& 0.000	& 0.598 	& Sw02 \\
UGC05829 	& 10 & 8.64	& 34(10)		& 0.52  	& 0.65  	& 1.813		& 0.209	& 0.000	& 0.791 	& Sw02 \\
UGC05986 	& 9 	& 8.63	& 90(3)		& 1.15  	& 1.00  	& 7.904		& 0.583	& 0.000	& 0.417 	& Sw02 \\
UGC06399 	& 9 	& 18.00 	& 75(2)		& 1.00 	& 0.50  	& 2.187		& 0.571	& 0.000	& 0.429 	& VS01 \\
UGC06446 	& 7 	& 12.00	& 51(3)		& 0.55  	& 3.00  	& 4.666		& 0.612	& 0.000	& 0.388 	& VS01 \\
UGC06917 	& 9 	& 18.00 	& 56(2)		& 0.57  	& 1.37  	& 11.685		& 0.782	& 0.000	& 0.218 	& VS01 \\
UGC06923 	& 10	& 18.00	& 65(2)		& 0.75  	& 0.70  	& 2.880		& 0.651	& 0.000	& 0.349 	& VS01 \\
UGC06973	& 2	& 18.00	& 71(3)		& 3.00	& 0.20	& 12.632		& 0.668	& 0.175	& 0.157	& VS01\\
UGC06983 	& 6 	& 18.00 	& 49(1) 		& 0.37  	& 2.20  	& 14.421		& 0.731	& 0.000	& 0.269 	& VS01 \\
UGC07089 	& 8 	& 18.00 	& 80(3) 		& 0.58  	& 0.32  	& 2.979		& 0.393	& 0.000	& 0.607 	& Ga02 \\
UGC07151 	& 6 	& 6.87	& 90(3)		& 0.72  	& 0.80  	& 2.804		& 0.707	& 0.000	& 0.293 	& Sw02 \\
UGC07261 	& 8 	& 13.10	& 30(10)		& 0.70  	& 0.75  	& 3.256		& 0.377	& 0.000	& 0.623 	& Sw02 \\
UGC07323 	& 8 	& 8.00 	& 47(3) 		& 0.96  	& 0.35  	& 2.467		& 0.604	& 0.000	& 0.396 	& Sw02  \\
UGC07399 	& 8 	& 8.43 	& 55(3) 		& 1.55  	& 2.00  	& 3.379		& 0.700	& 0.000	& 0.300 	& Sw02 \\
UGC07559 	& 10 & 4.97	& 61(3)		& 0.29  	& 0.60  	& 0.309		& 0.207	& 0.000	& 0.793 	& Sw02 \\
UGC07603 	& 7 	& 4.70 	& 78(3)		& 0.78  	& 1.20  	& 0.761		& 0.576	& 0.000	& 0.424 	& Sw02 \\
UGC07608 	& 10 & 8.21 	& 25(10) 		& 0.94  	& 0.70  	& 0.949		& 0.235	& 0.000	& 0.765 	& Sw02 \\
UGC07690 	& 10 & 8.11 	& 41(5) 		& 0.26  	& 1.20  	& 1.509		& 0.613	& 0.000	& 0.387	& Sw02 \\
UGC07866 	& 10 & 4.57	& 44(5) 		& 0.35  	& 0.50  	& 0.243		& 0.265	& 0.000	& 0.735 	& Sw02 \\
UGC08490 	& 9 	& 4.65 	& 50(3) 		& 0.45  	& 2.20  	& 3.111		& 0.696	& 0.000	& 0.304 	& Sw02 \\
UGC08550 	& 7 	& 6.70 	& 90(3)		& 0.55  	& 0.70  	& 0.613		& 0.325	& 0.000	& 0.675 	& Sw02 \\
UGC08837 	& 10 & 7.21 	& 80(5) 		& 0.38  	& 0.30  	& 0.603		& 0.249	& 0.000	& 0.751 	& Sw02 \\
UGC09037 	& 6 	& 83.60	& 65(5) 		& 0.55  	& 0.50  	& 58.087		& 0.520	& 0.000	& 0.480 	& Ha14  \\
UGC11455 	& 6 	& 78.60 	& 90(1)		& 1.00 	& 0.65  	& 373.377		& 0.918	& 0.000	& 0.082 	& SG06 \\
UGC11557 	& 8 	& 24.20	& 30(10) 		& 0.28 	& 0.40  	& 8.554		& 0.571	& 0.000	& 0.429	& Sw02 \\
UGC12506 	& 6 	& 100.60	& 86(4) 		& 0.48 	& 1.50  	& 250.494		& 0.811	& 0.000	& 0.189 	& Ha14 \\
UGC12732 	& 9 	& 13.20 	& 39(6) 		& 0.55  	& 2.20  	& 8.785		& 0.410	& 0.000	& 0.590 	& Sw02 \\
UGCA442 	& 9 	& 4.35	& 64(7)		& 0.55  	& 0.50  	& 0.489		& 0.137	& 0.000	& 0.863 	& Co00 \\
UGCA444 	& 10 & 0.98 	& 78(4) 		& 0.51  	& 0.50  	& 0.107		& 0.056	& 0.000	& 0.944	& Ke07 \\
\end{tabular}
\end{ruledtabular}
\end{table*}

\subsection{Gravitational potential}

We solve the effective field equation~\eqref{eq:EFE} numerically to determine the gravitational potential $\Phi$---we also solve the Poisson equation~\eqref{eq:Poisson} for comparison. The appendix presents the formulas and the details of calculations. The mass density is totally from the stellar disk, bulge, and gas. In particular, we do not assume a radial profile for the stellar disk, bulge, or gas---we use only the data. 
For the vertical distributions from the galactic planes, we assume the finite thickness for the stellar disk and gas. We assume that bulge is spherical as an approximation. These details are shown in the appendix.

When we solve Eq.~\eqref{eq:EFE}, following the solution~\eqref{eq:sol_Cotton}, we adopt the boundary condition at far from the center,
\begin{equation}
	\Phi = \gamma_{\rm galaxy}\sqrt{R^2+z^2}/2,
	\label{eq:bc}
\end{equation}
where $R$ is the radius in a cylindrical coordinate system, $z$ is the vertical  coordinate, and $\gamma_{\rm galaxy}$ is a constant.
The values of $\gamma_{\rm galaxy}$ should be determined for each galaxy. This is the same as the mass $M$ should be determined for each galaxy. 
The constant $\gamma_{\rm galaxy}$ is just an integration constant rather than a fundamental parameter of the theory---the field equations~\eqref{eq:FE} does not include $\gamma_{\rm galaxy}$. 
We determine the value of $\gamma_{\rm galaxy}$ as a fitting parameter for each galaxy. 
We find that many galaxies have typically $\gamma_{\rm galaxy} \sim 1 \  {\rm km^2 \ s^{-2} \ pc^{-1}}$ (column~(5) in Table~\ref{tab:table1}). The flexibility to allow the individual values $\gamma_{\rm galaxy}$ for each galaxy is a significant advantage of Eq.~\eqref{eq:FE}.

\subsection{Rotation curves}
In Cotton gravity, the observed rotation curves can be explained by the distributions of baryons (Fig.~\ref{fig:RCex}). 
The flat shape of rotation curves (NGC~5907, UGC 2885, UGC 2487, and NGC7331 in Fig.~\ref{fig:RCex}) and the nonflat/rising shape (NGC~3109 and NGC~100 in Fig.~\ref{fig:RCex}) can be obtained in the same context without assumptions. 
In the context of Cotton gravity, the calculated shape of rotation curves are essentially determined by the distributions of baryons. 
This is in agreement with one of the most important observations: “For any feature in the luminosity profile there is a corresponding feature in the rotation curve and vise verse”~\cite{Sancisi:2004}.

\begin{figure*}[b]
\includegraphics[width=60mm]{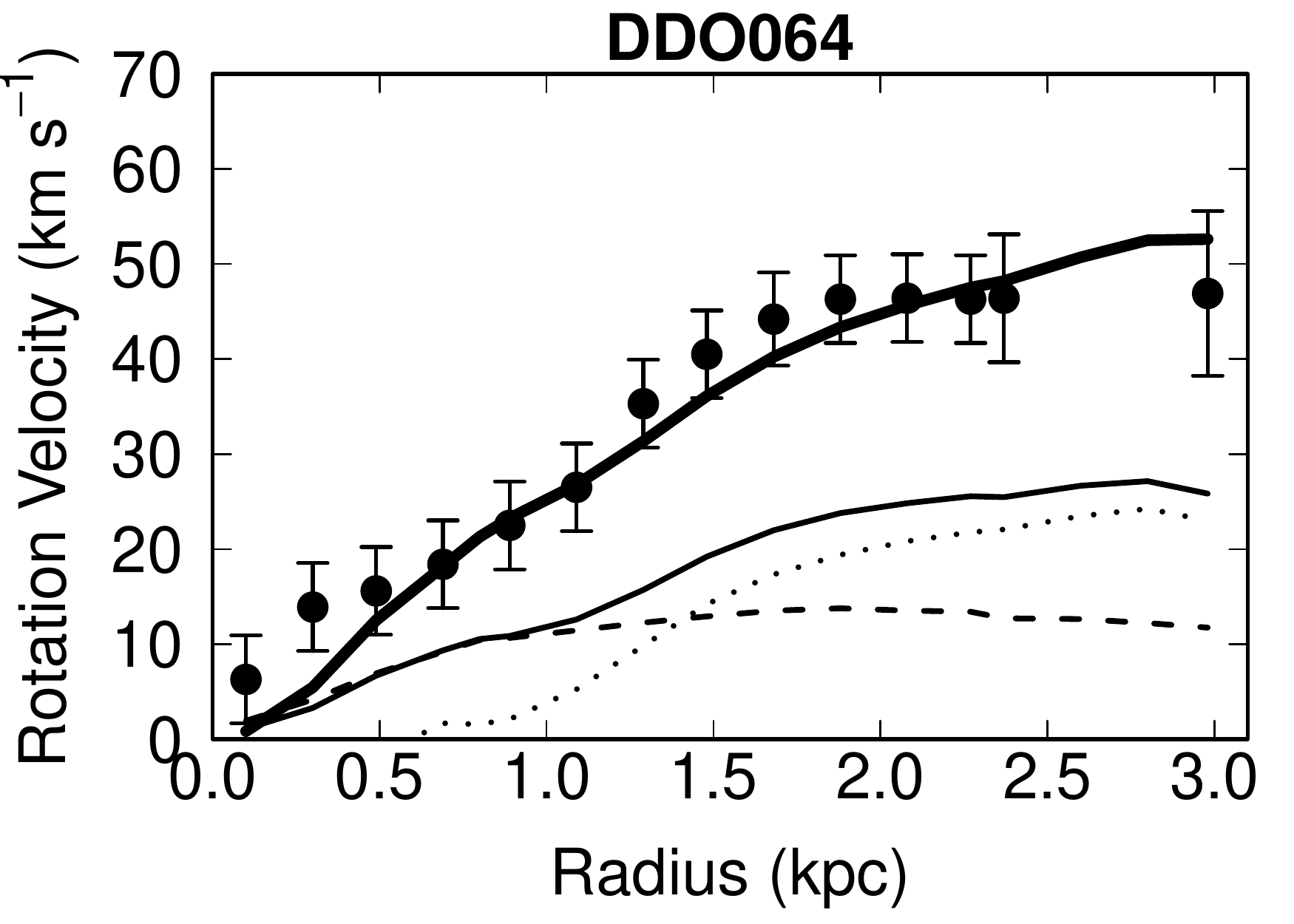}%
\includegraphics[width=60mm]{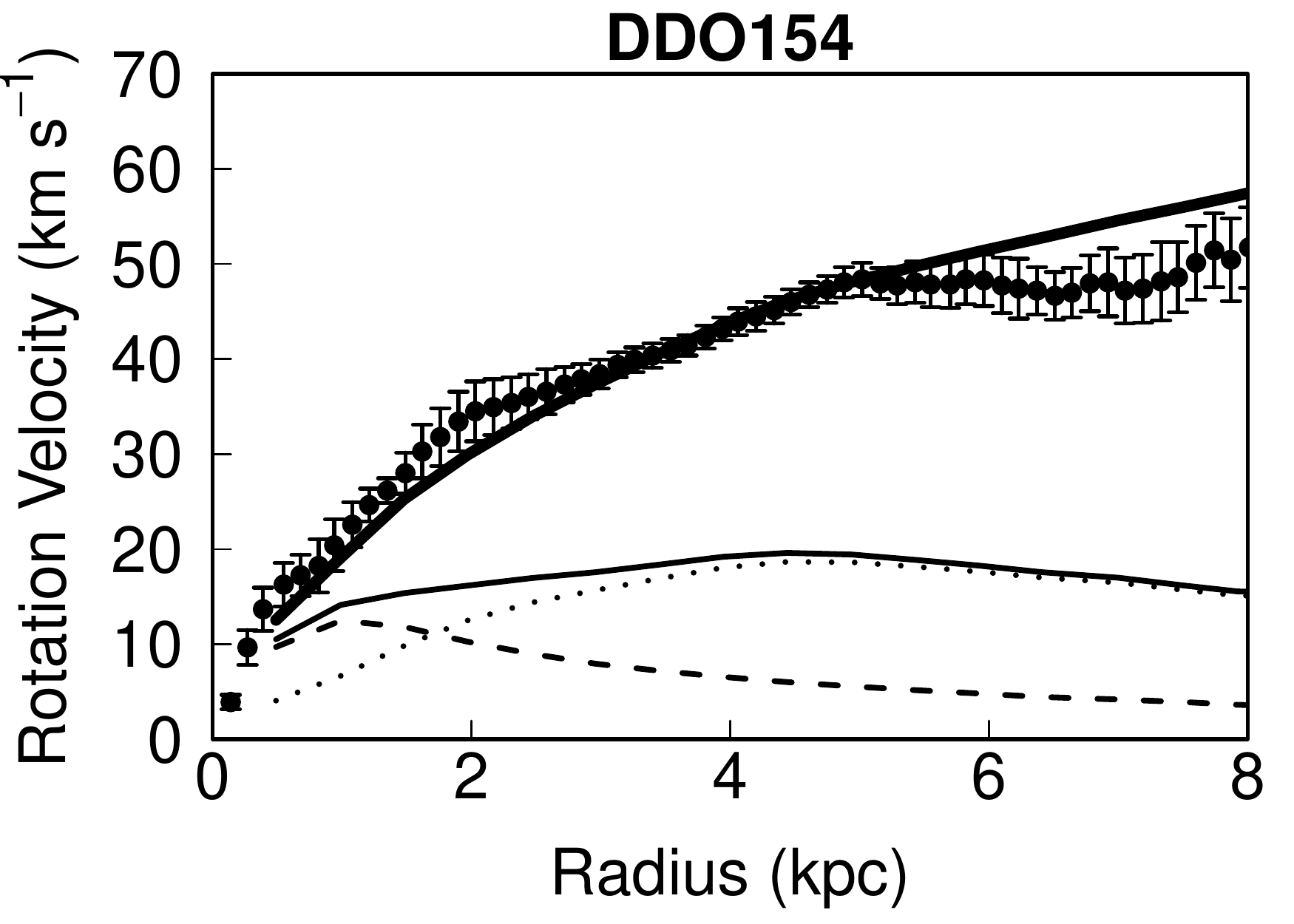}%
\includegraphics[width=60mm]{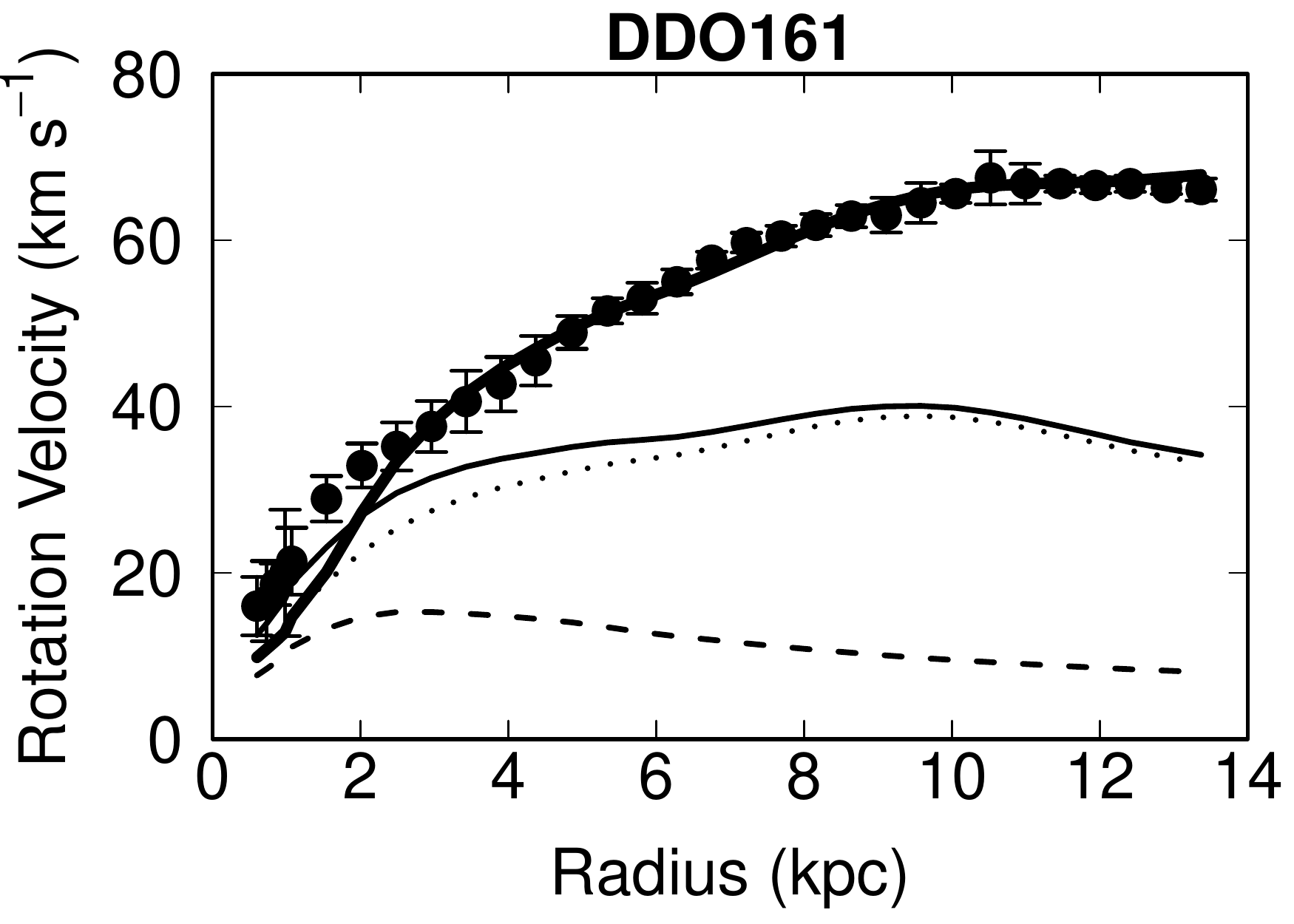}%
\\ \ \\
\includegraphics[width=60mm]{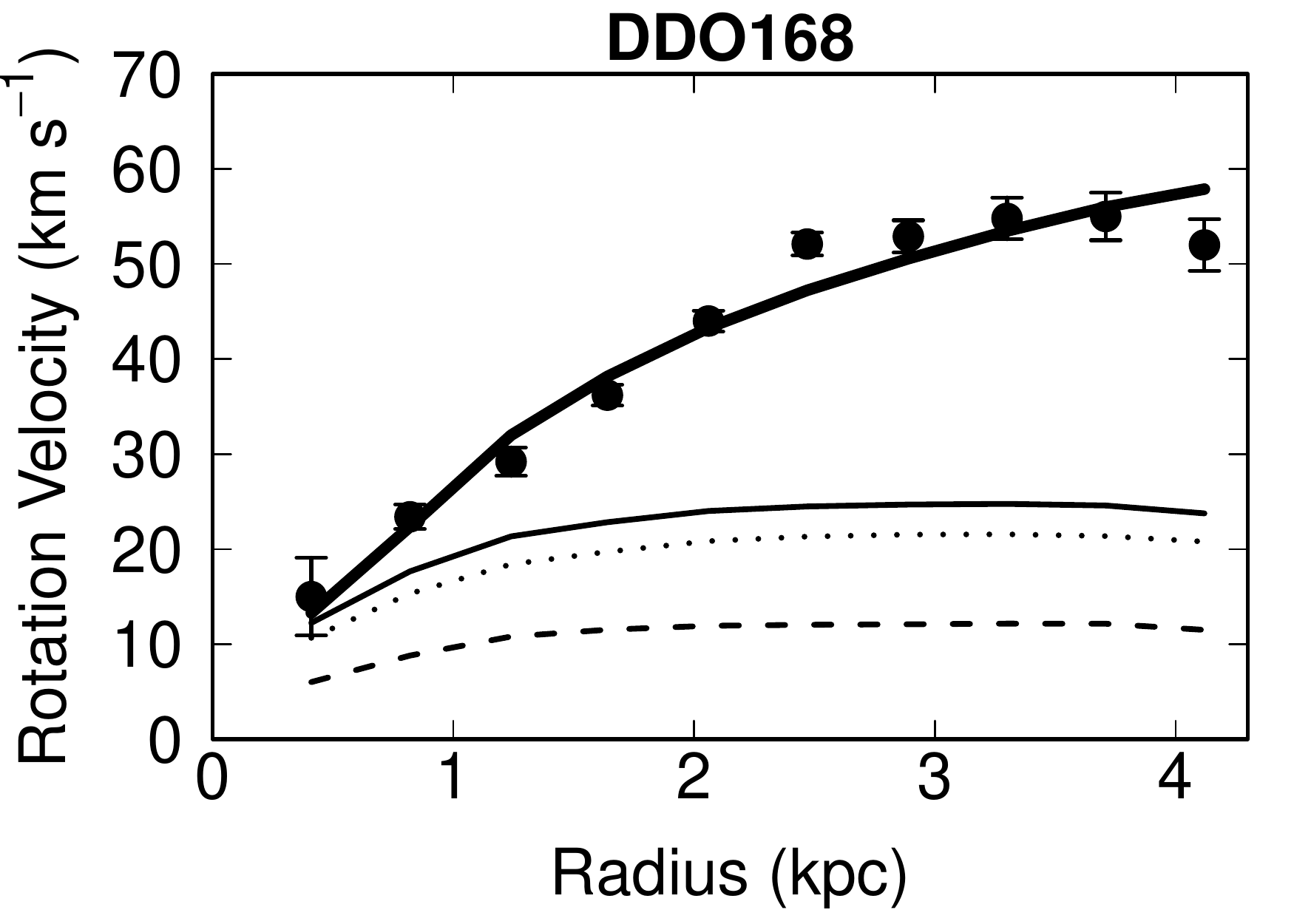}%
\includegraphics[width=60mm]{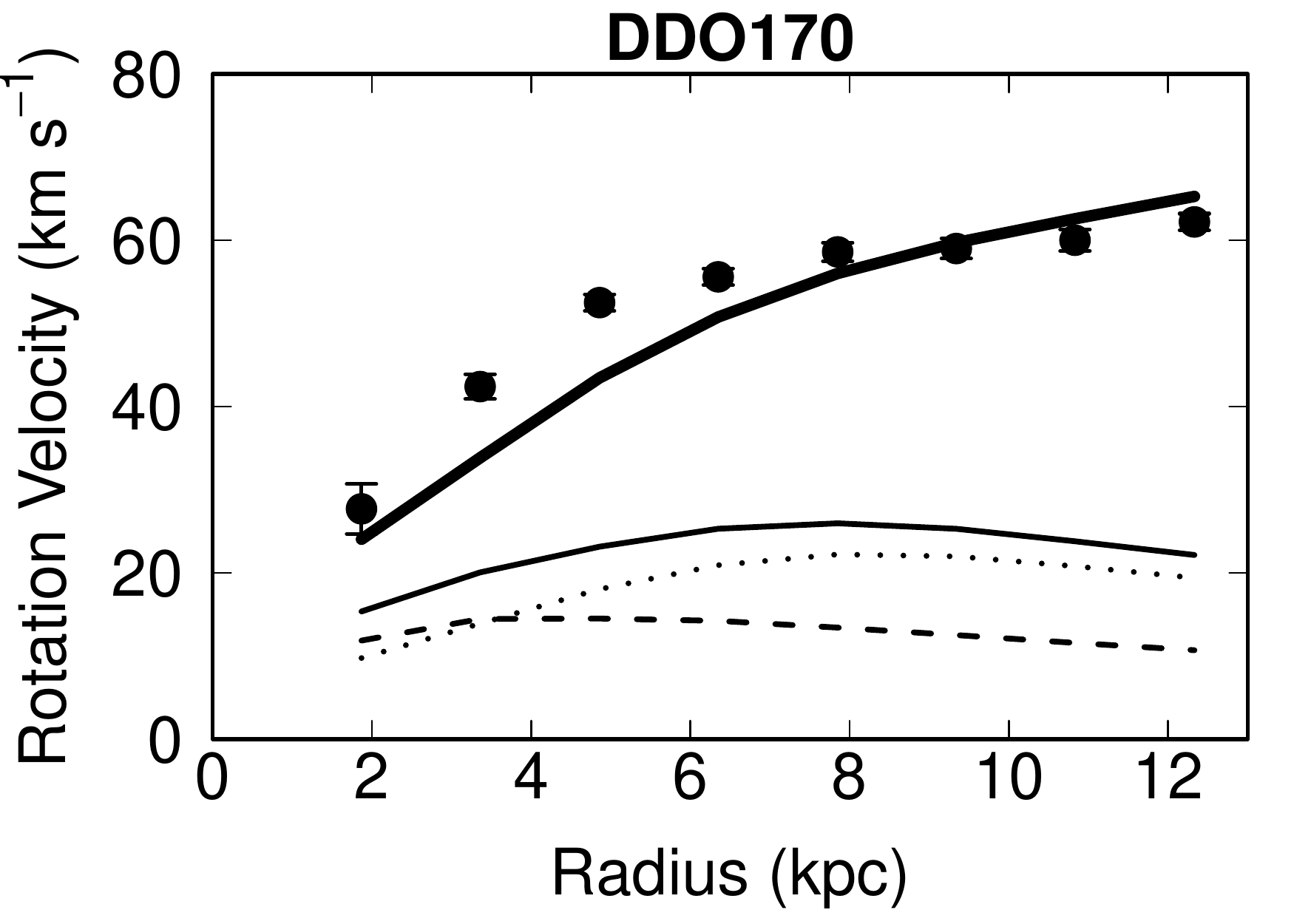}%
\includegraphics[width=60mm]{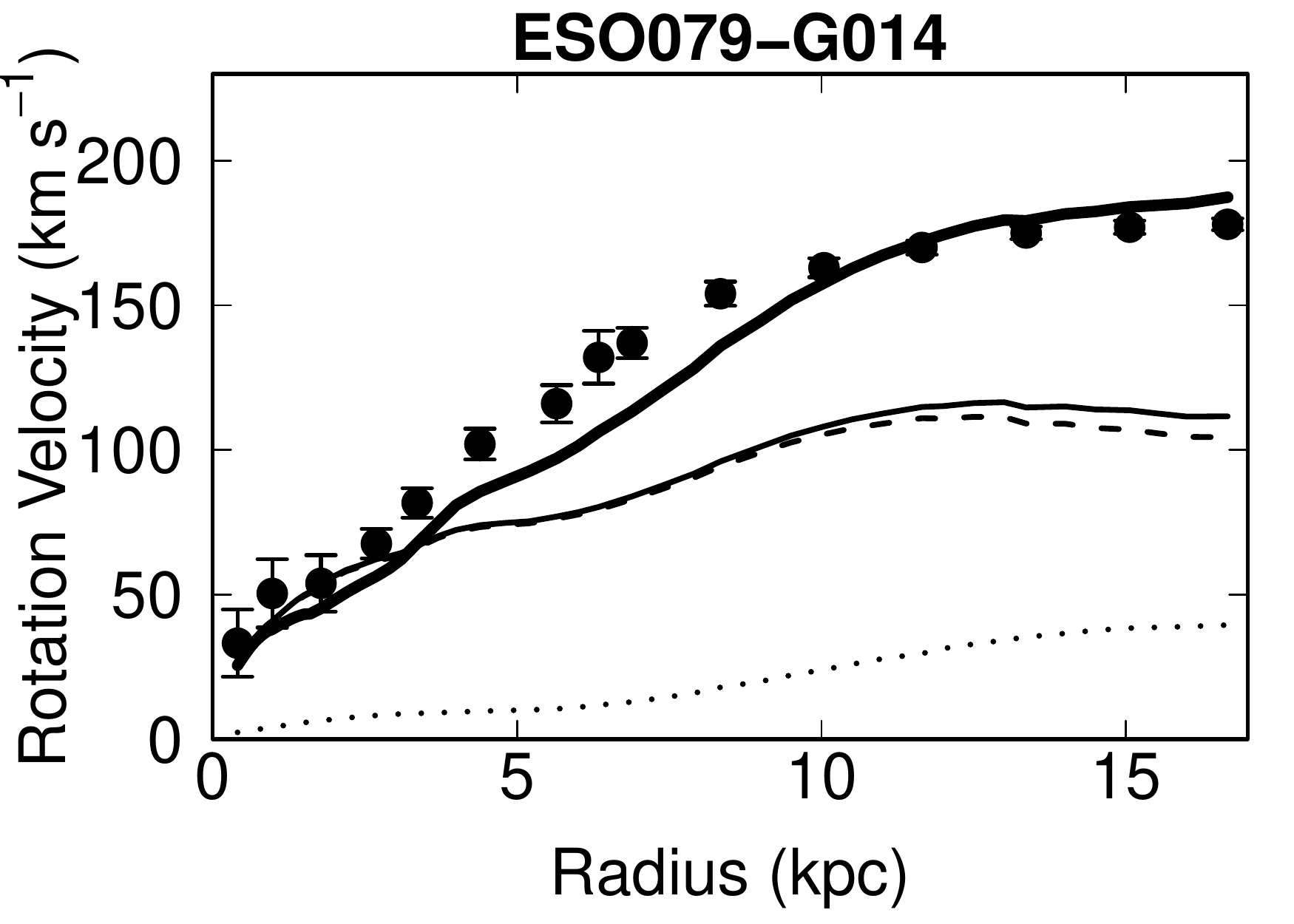}%
\\ \ \\
\includegraphics[width=60mm]{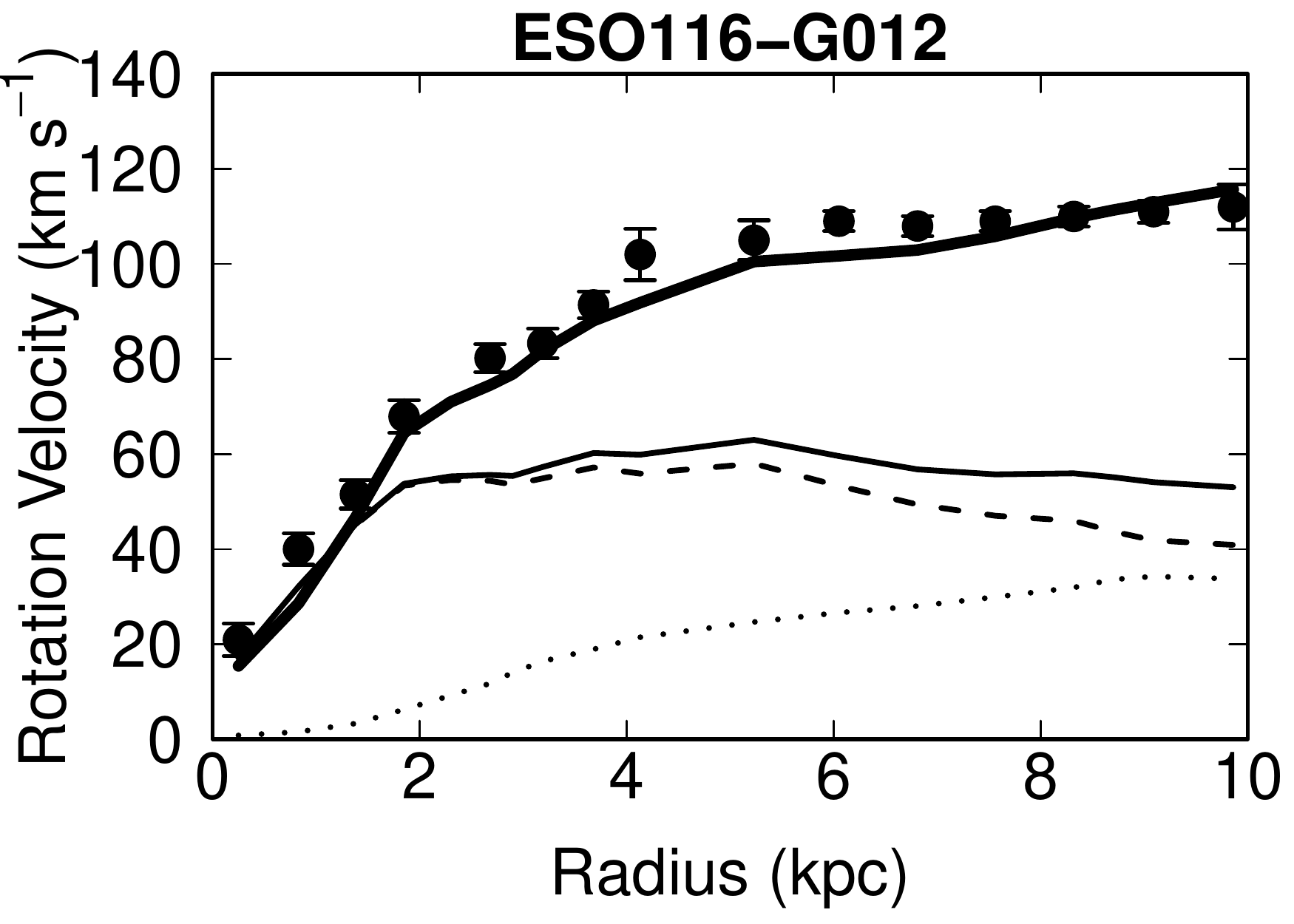}%
\includegraphics[width=60mm]{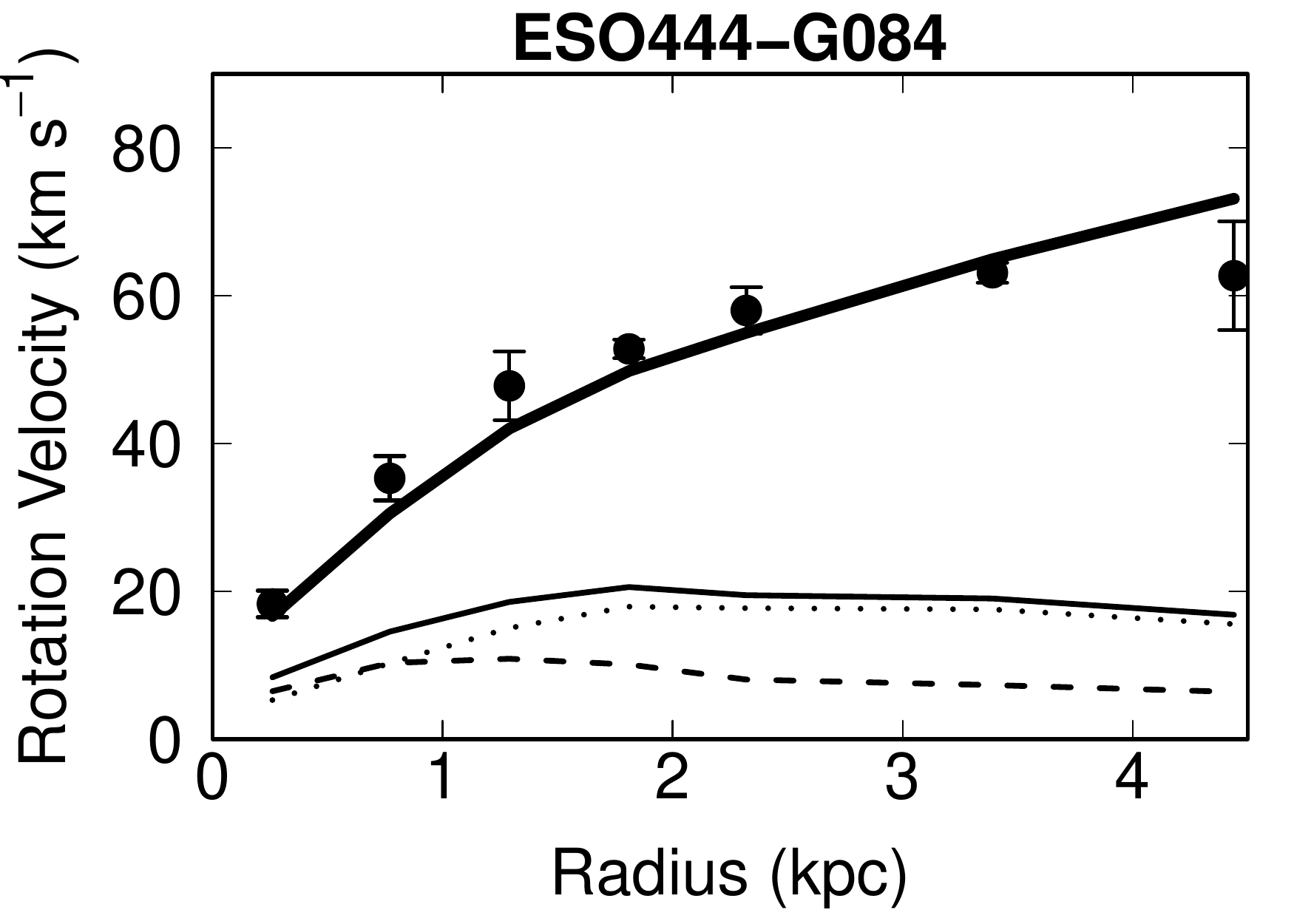}%
\includegraphics[width=60mm]{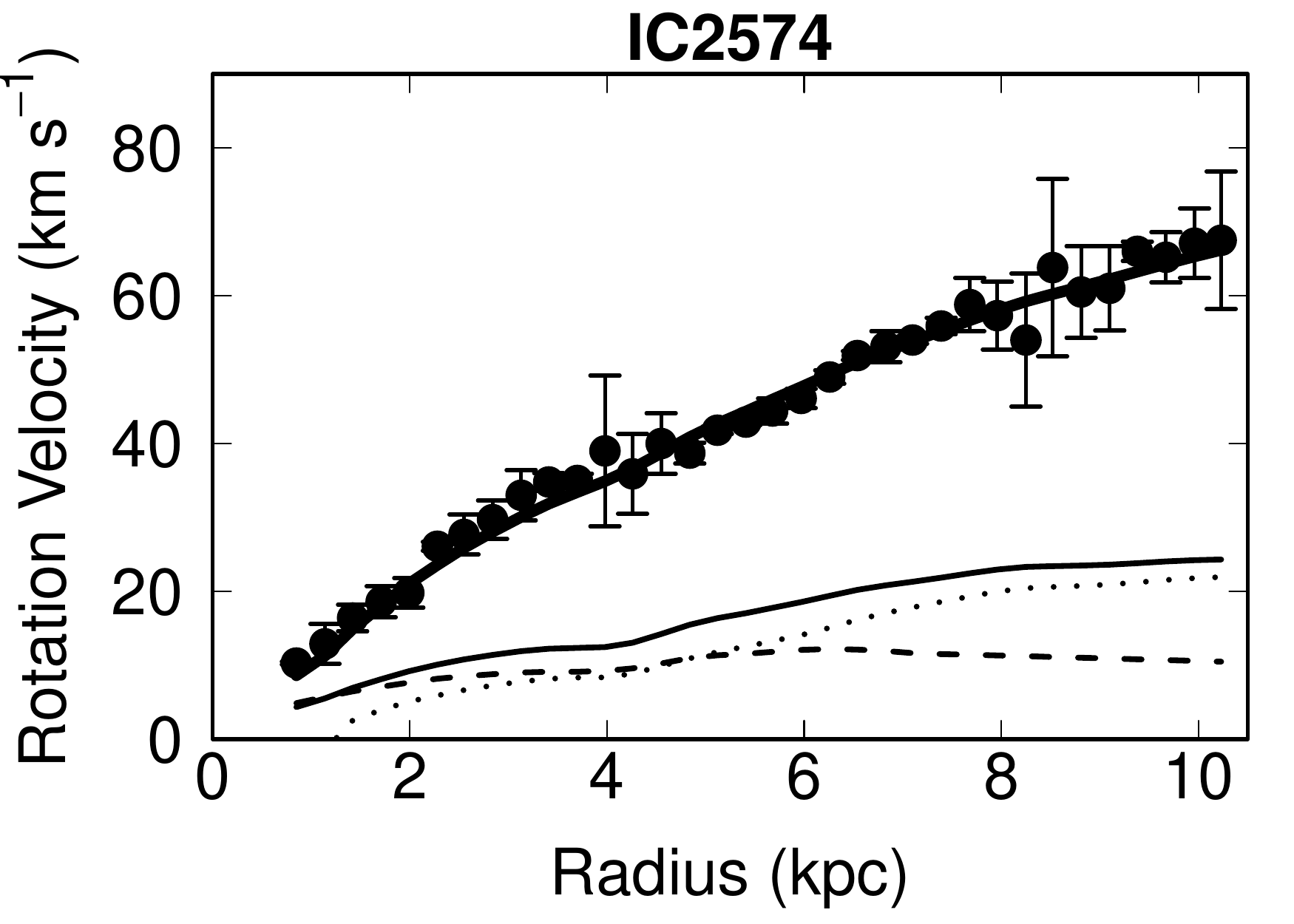}%
\\ \ \\
\includegraphics[width=60mm]{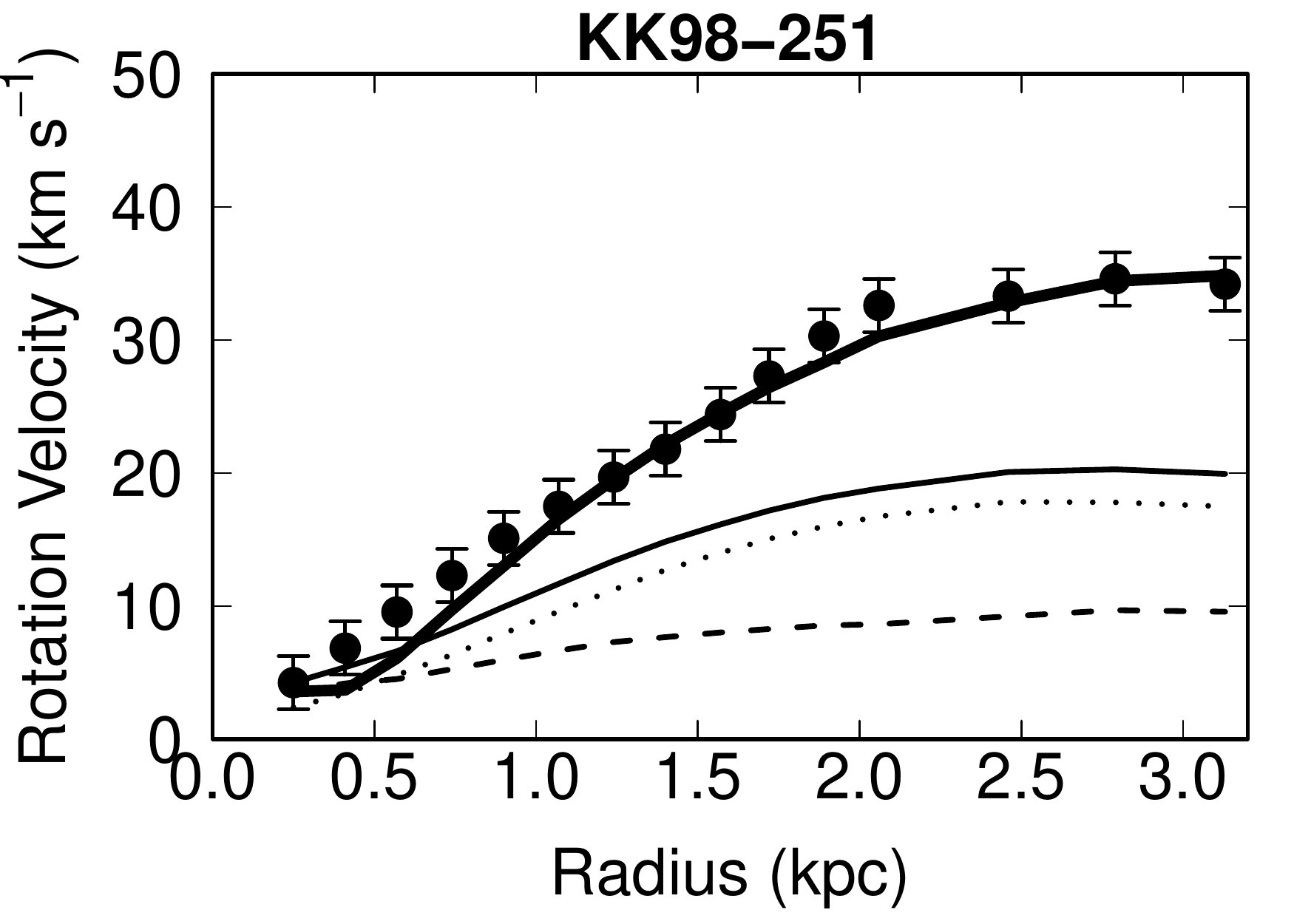}%
\includegraphics[width=60mm]{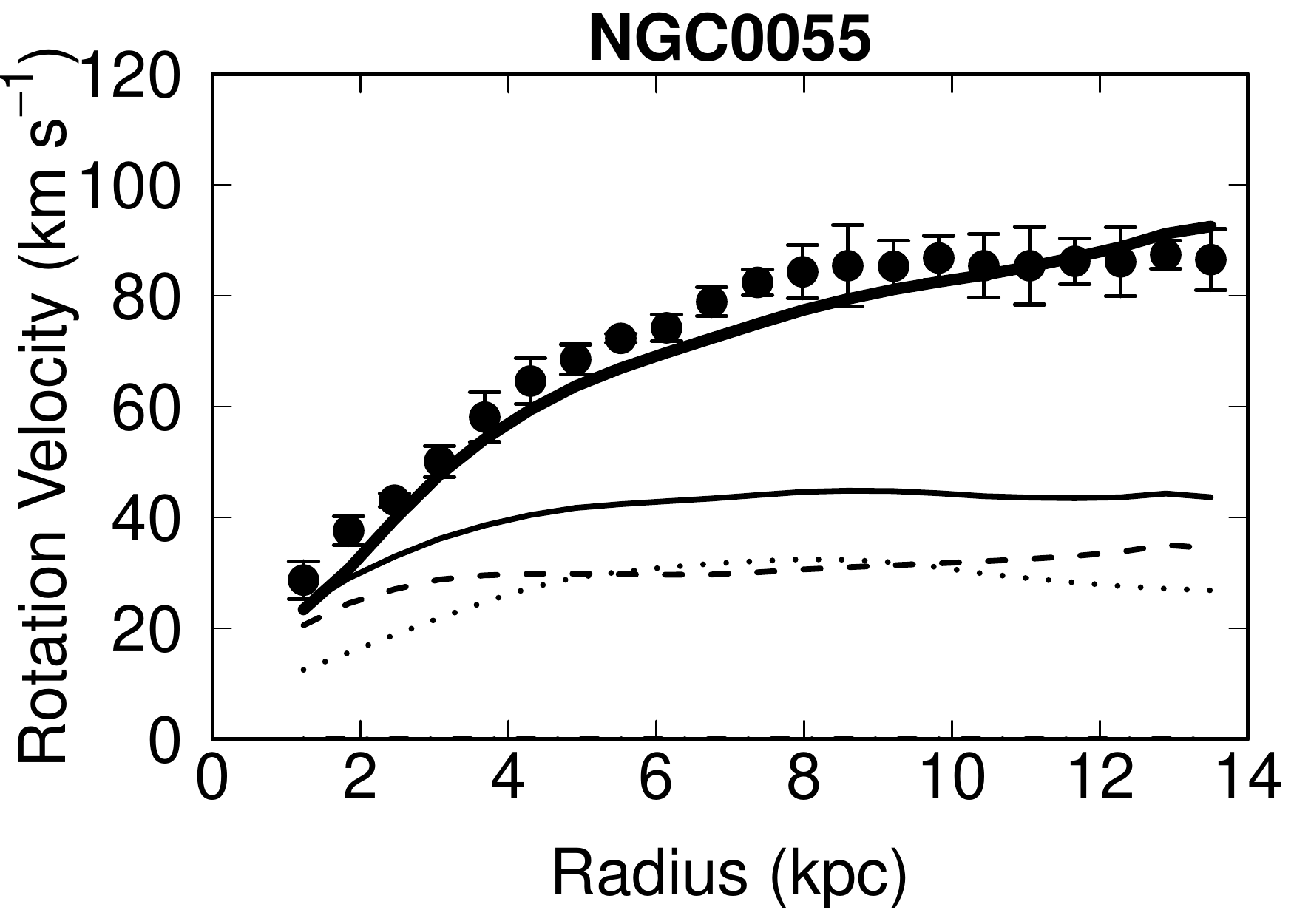}%
\includegraphics[width=60mm]{NGC0100_RC_V5.pdf}%
\caption{\label{fig:RCfull} 84~rotation curves and mass models for individual galaxies. Points and lines represent the same as those in Fig.~\ref{fig:RCex}. All galaxies except DDO~154 show the observed rotation velocity obtained from SPARC database~\cite{Lelli:2016zqa}; only DDO~154 shows the observed rotation velocity obtained from THINGS~\cite{deBlok:2008wp,Walter:2008psa} (The H {\scriptsize I} Nearby Galaxy Survey).}
\end{figure*}

In Cotton gravity, both large galaxies (UGC 2885 and UGC 2487 in Fig.~\ref{fig:RCex}) and small galaxies (NGC 3109 and NGC 100 in Fig.~\ref{fig:RCex}) can be explained. 
Figure~\ref{fig:RCfull} shows the rotation curves of 84 galaxies. 
The 84~galaxies are very different in morphologies, luminosities, masses, sizes, and gas fractions.
Nevertheless, 84~galaxy rotation curves can be explained by the single field equation~\eqref{eq:EFE}. 

\addtocounter{figure}{-1}
\begin{figure*}[b]
\includegraphics[width=60mm]{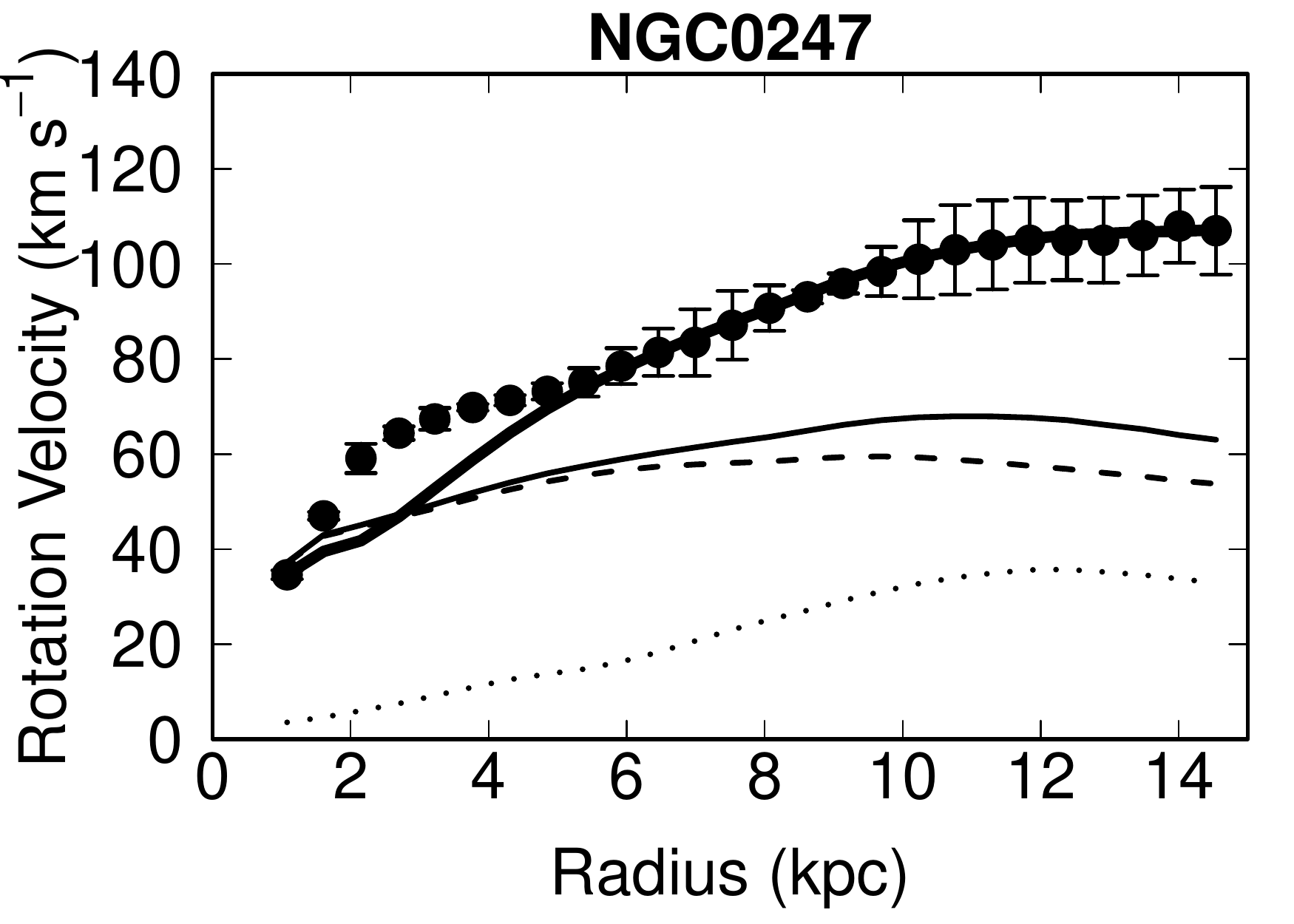}%
\includegraphics[width=60mm]{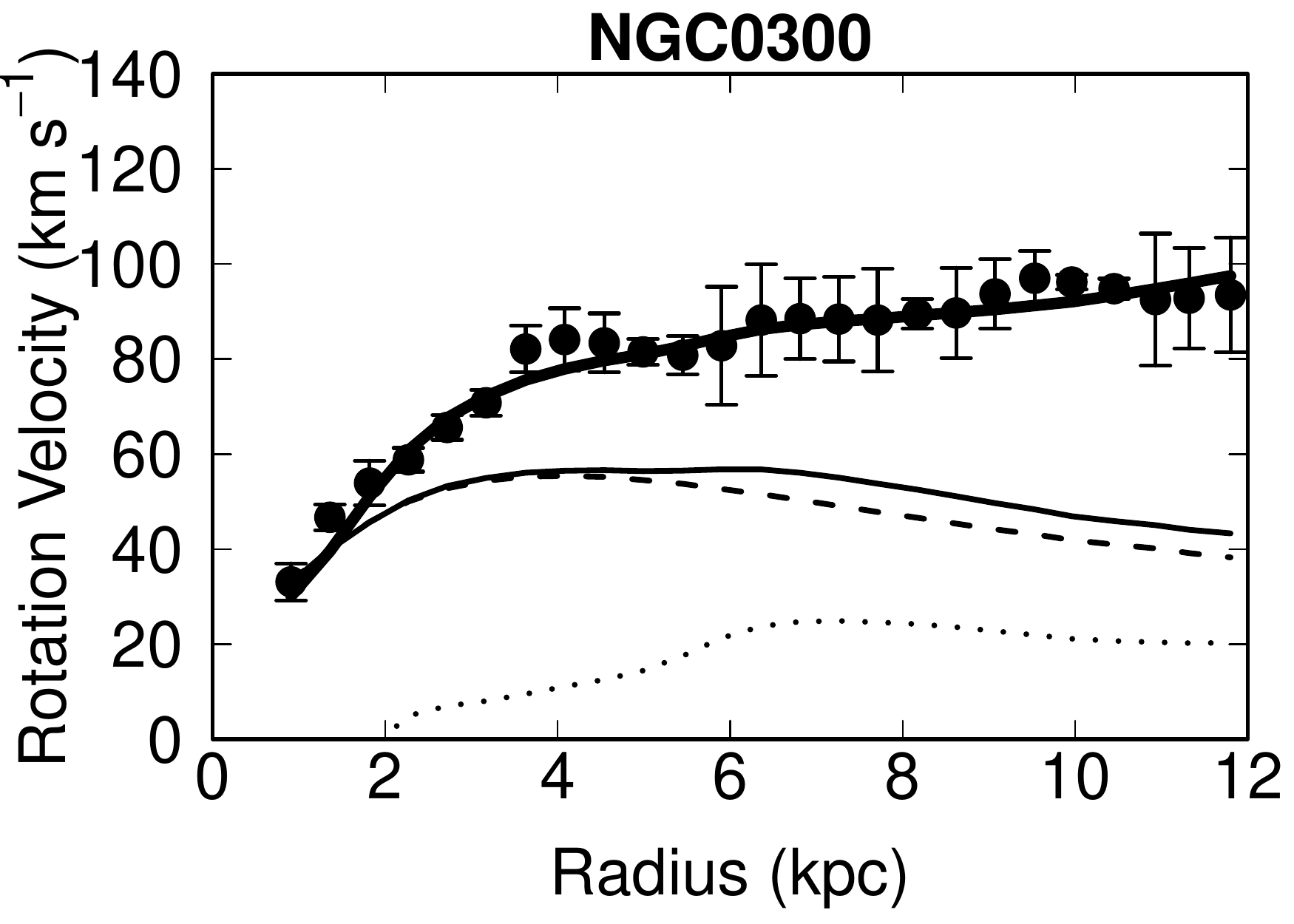}%
\includegraphics[width=60mm]{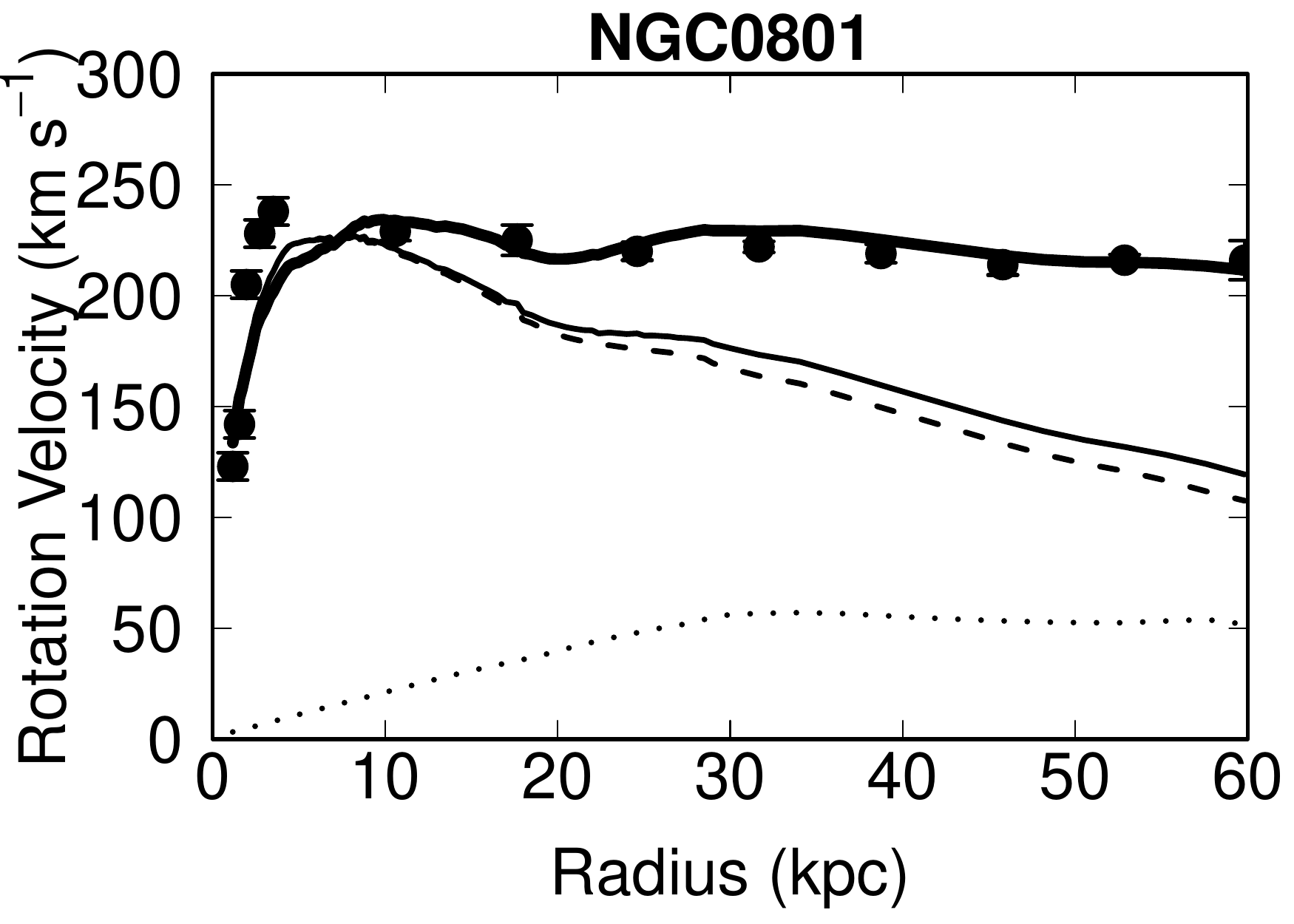}%
\\ \ \\
\includegraphics[width=60mm]{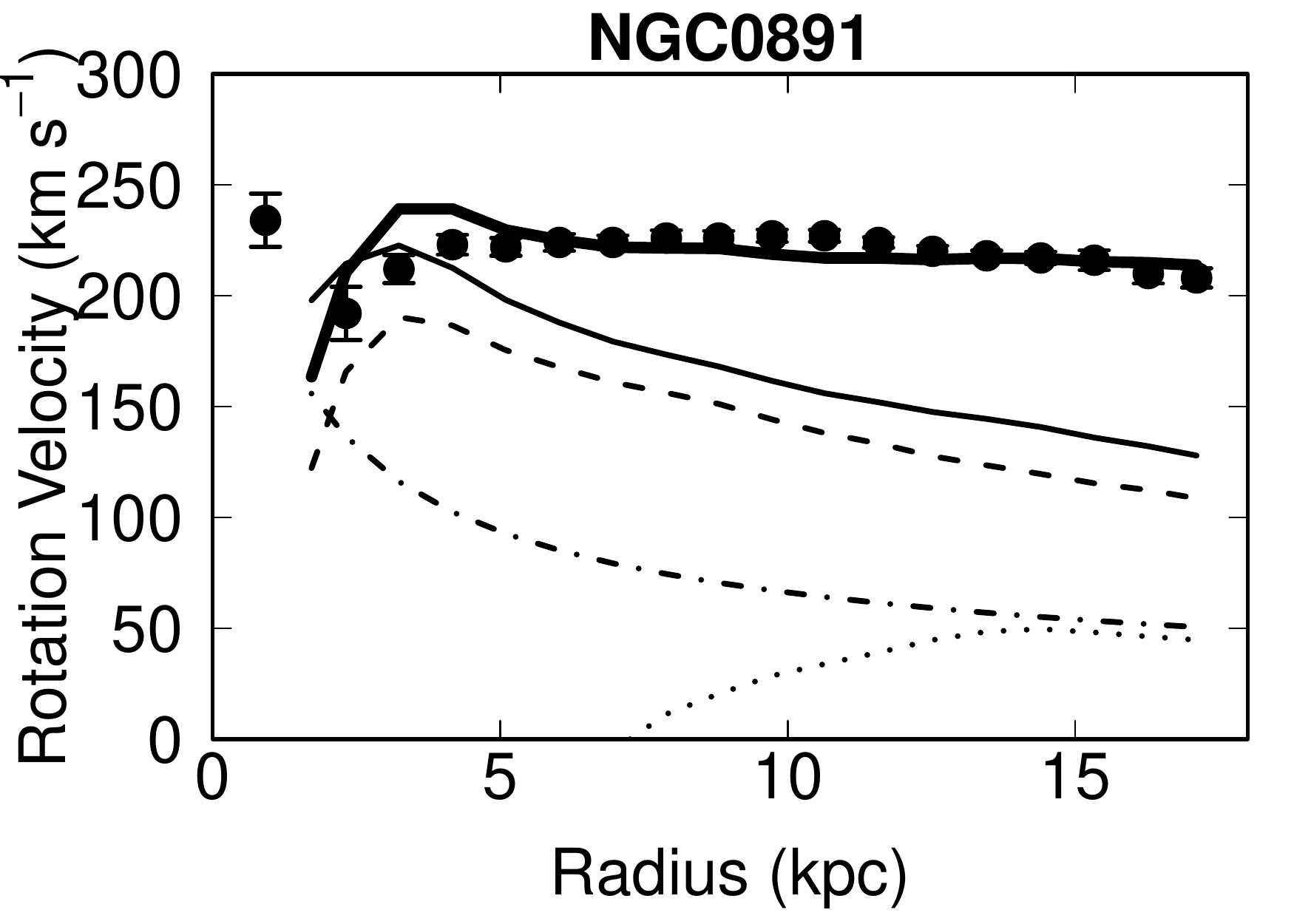}%
\includegraphics[width=60mm]{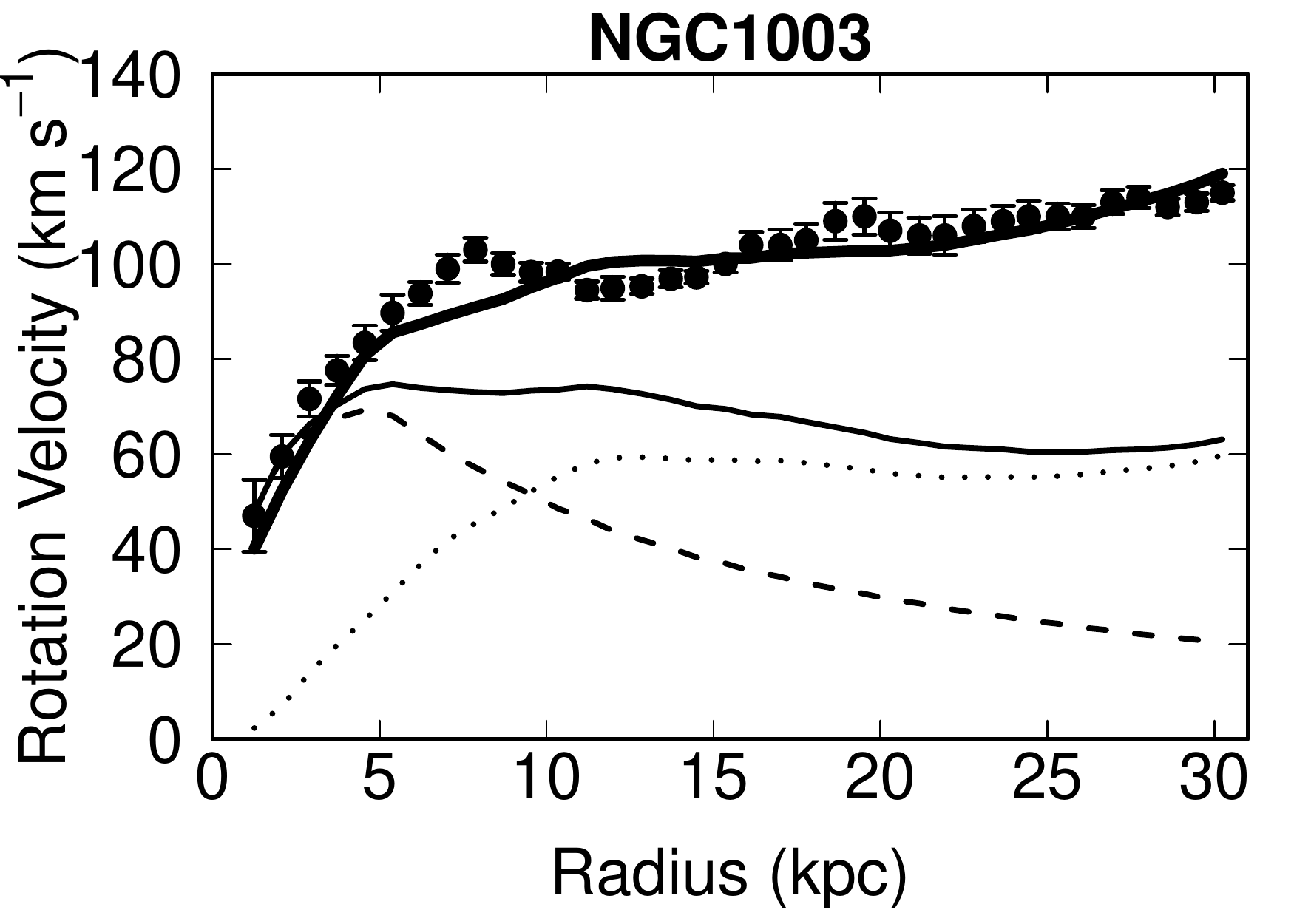}%
\includegraphics[width=60mm]{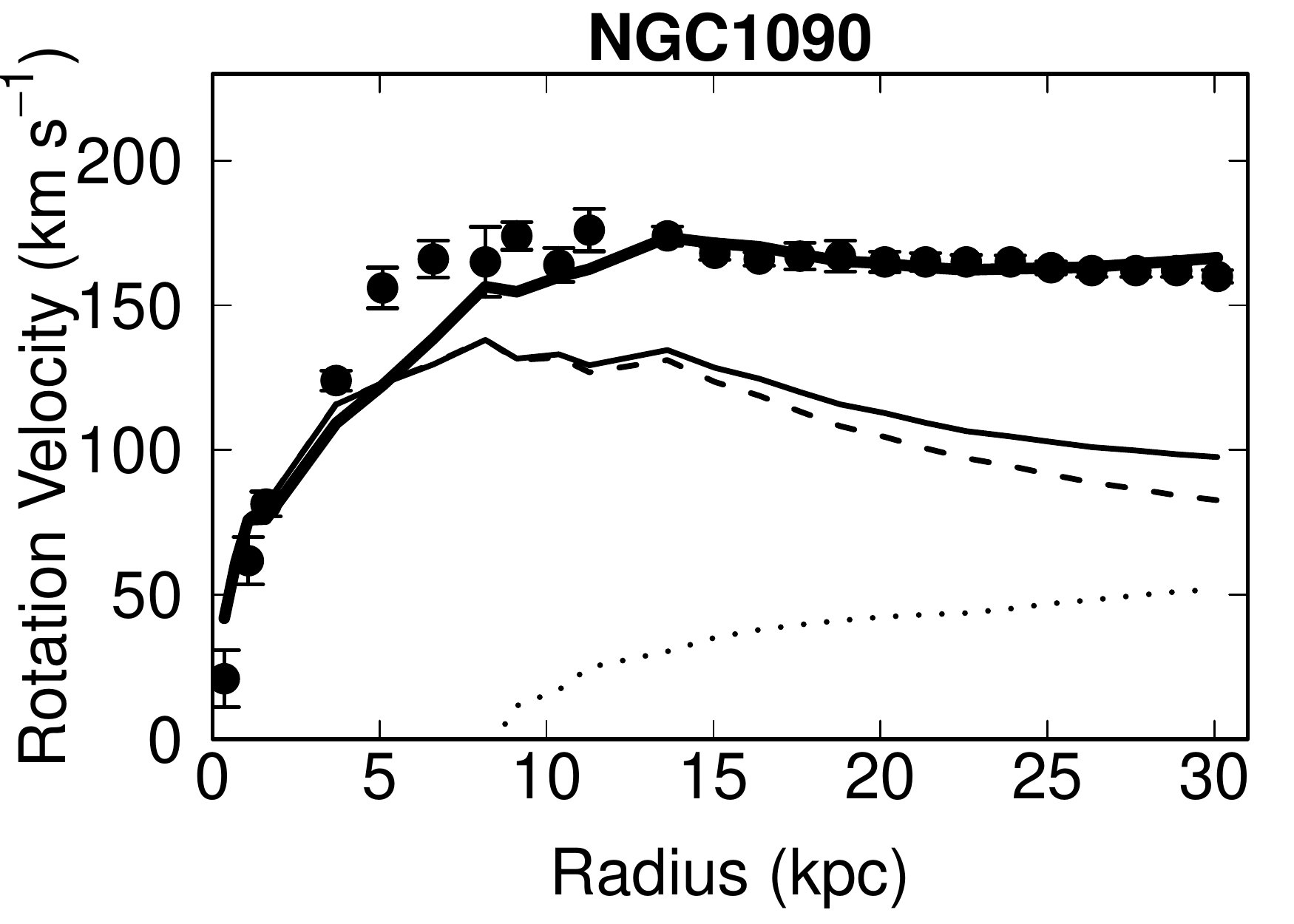}%
\\ \ \\
\includegraphics[width=60mm]{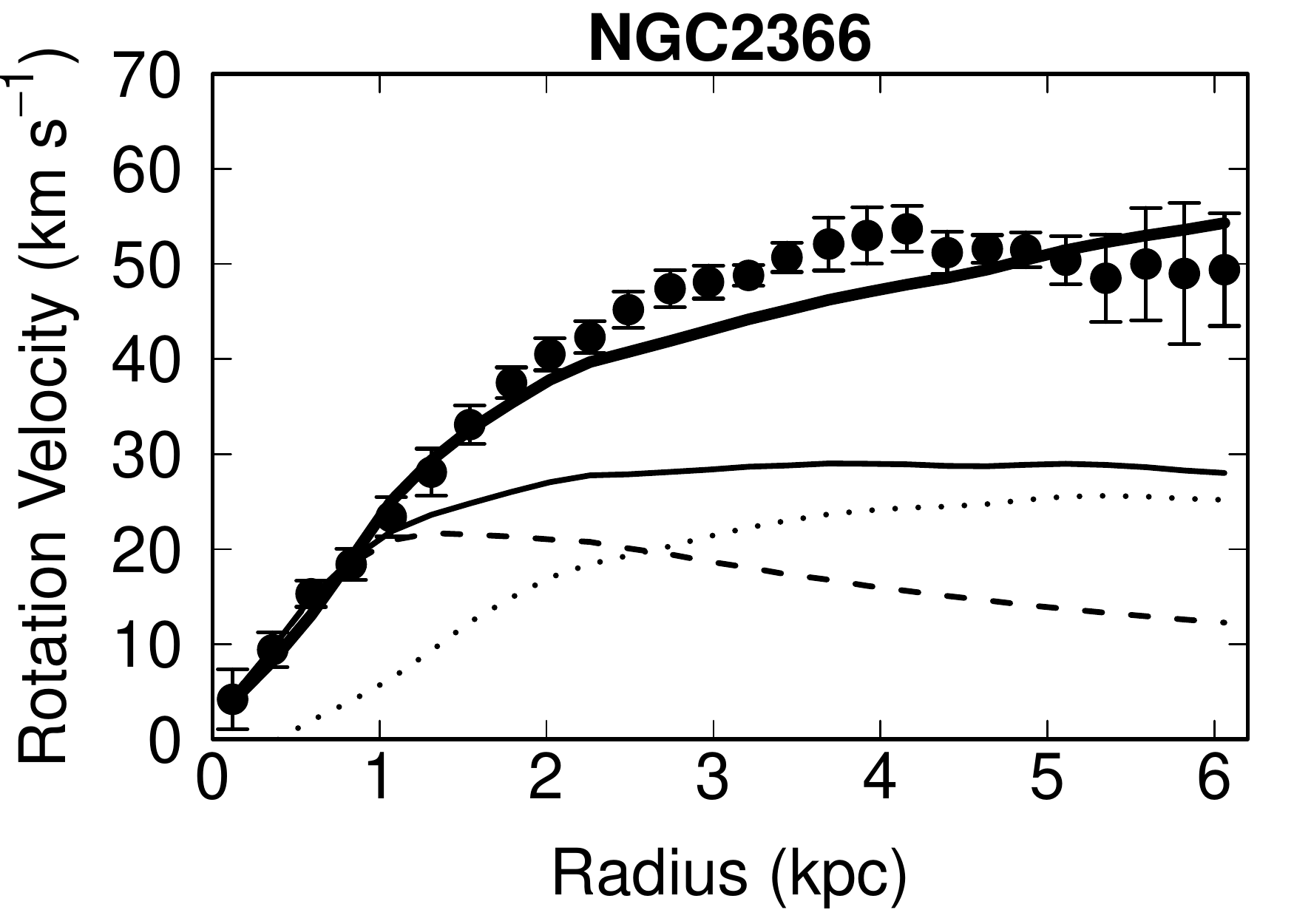}%
\includegraphics[width=60mm]{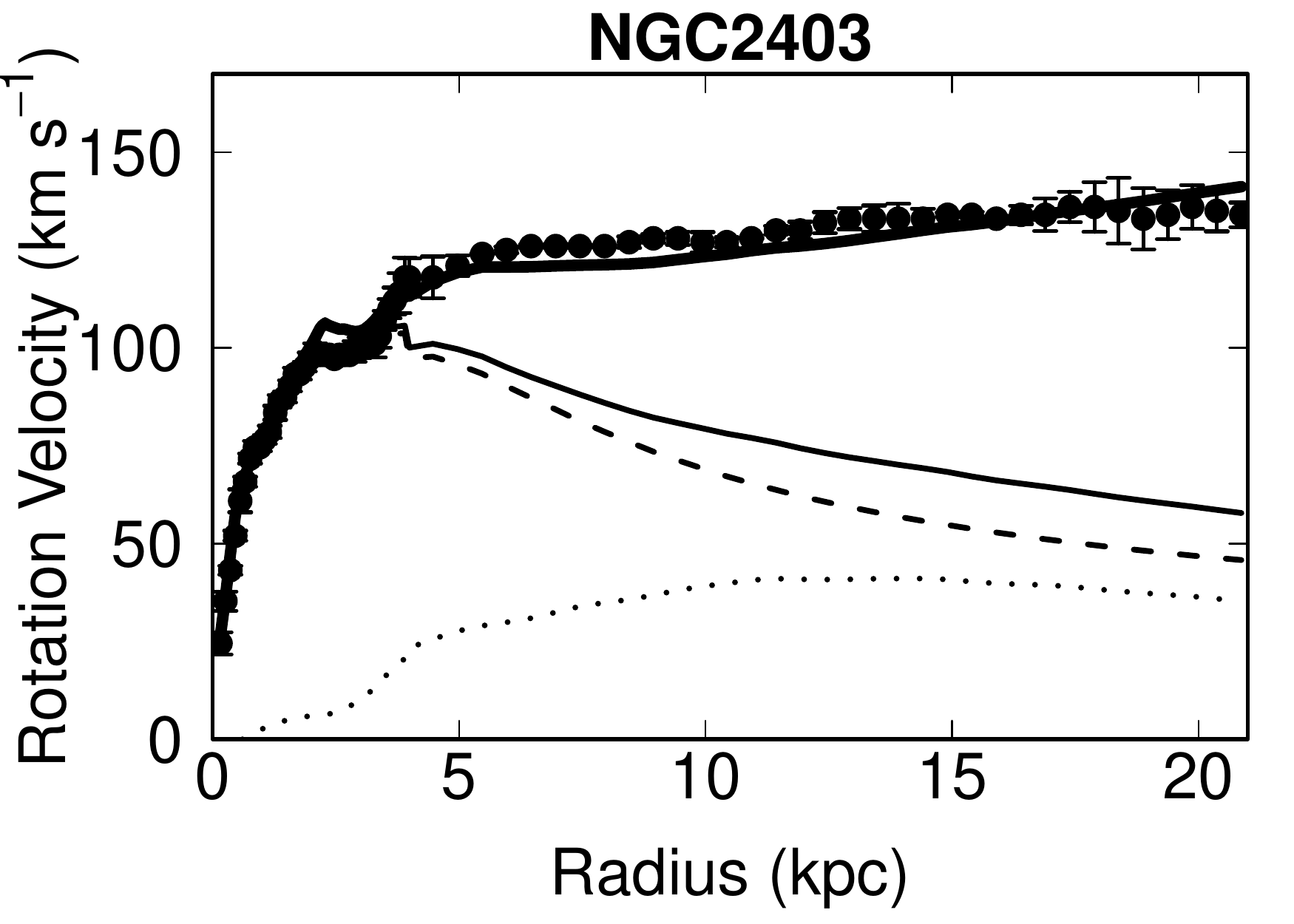}%
\includegraphics[width=60mm]{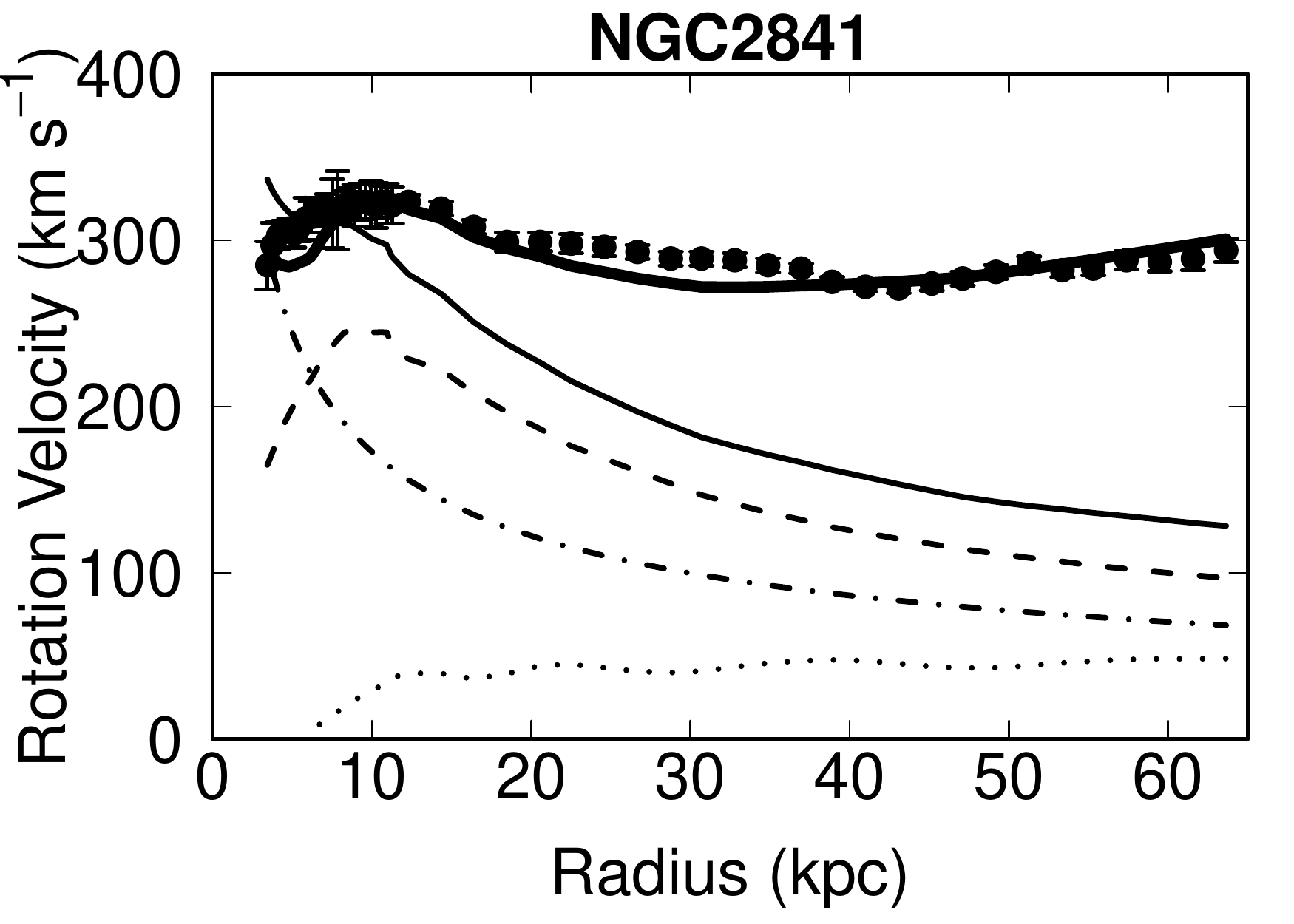}%
\\ \ \\
\includegraphics[width=60mm]{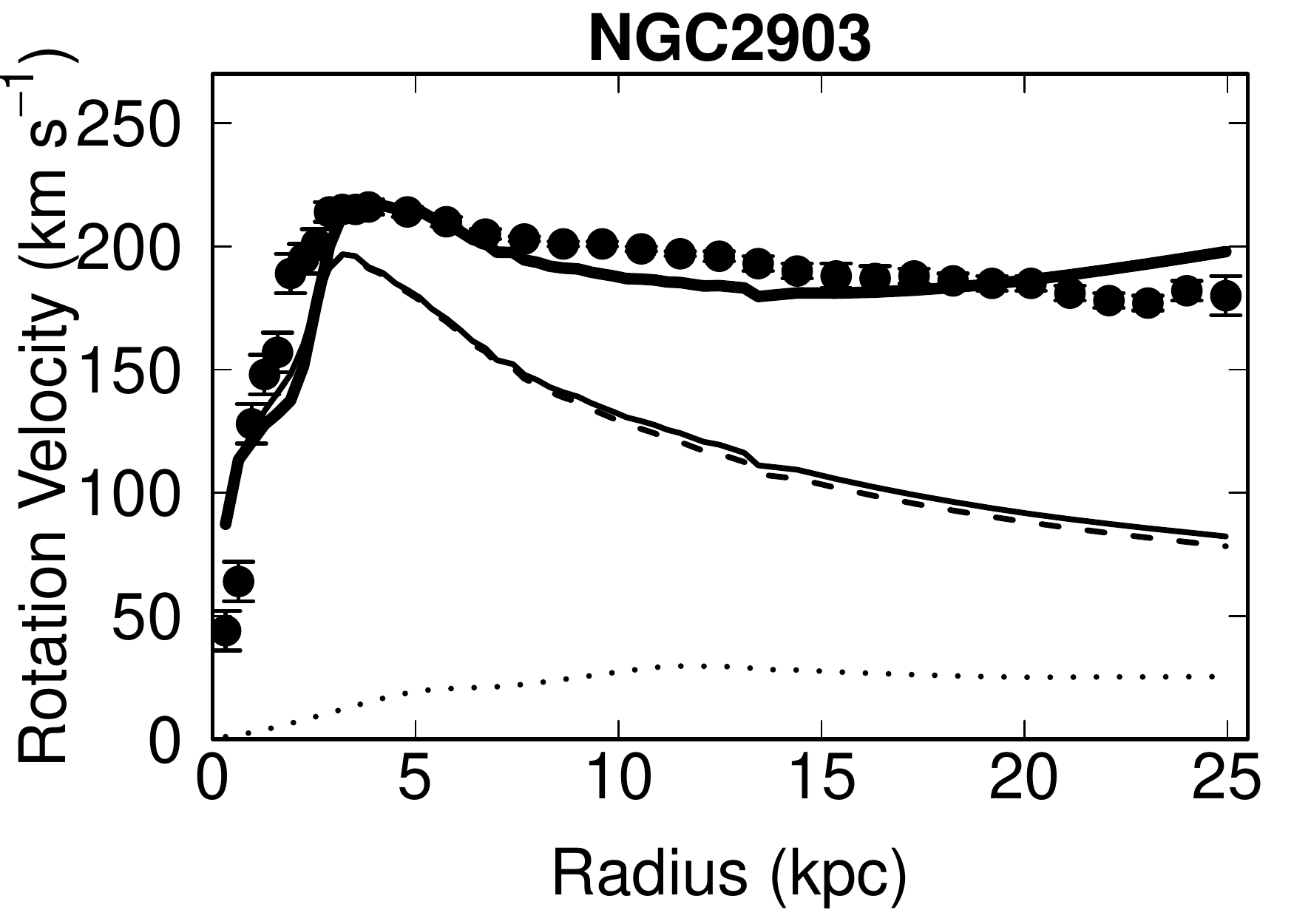}%
\includegraphics[width=60mm]{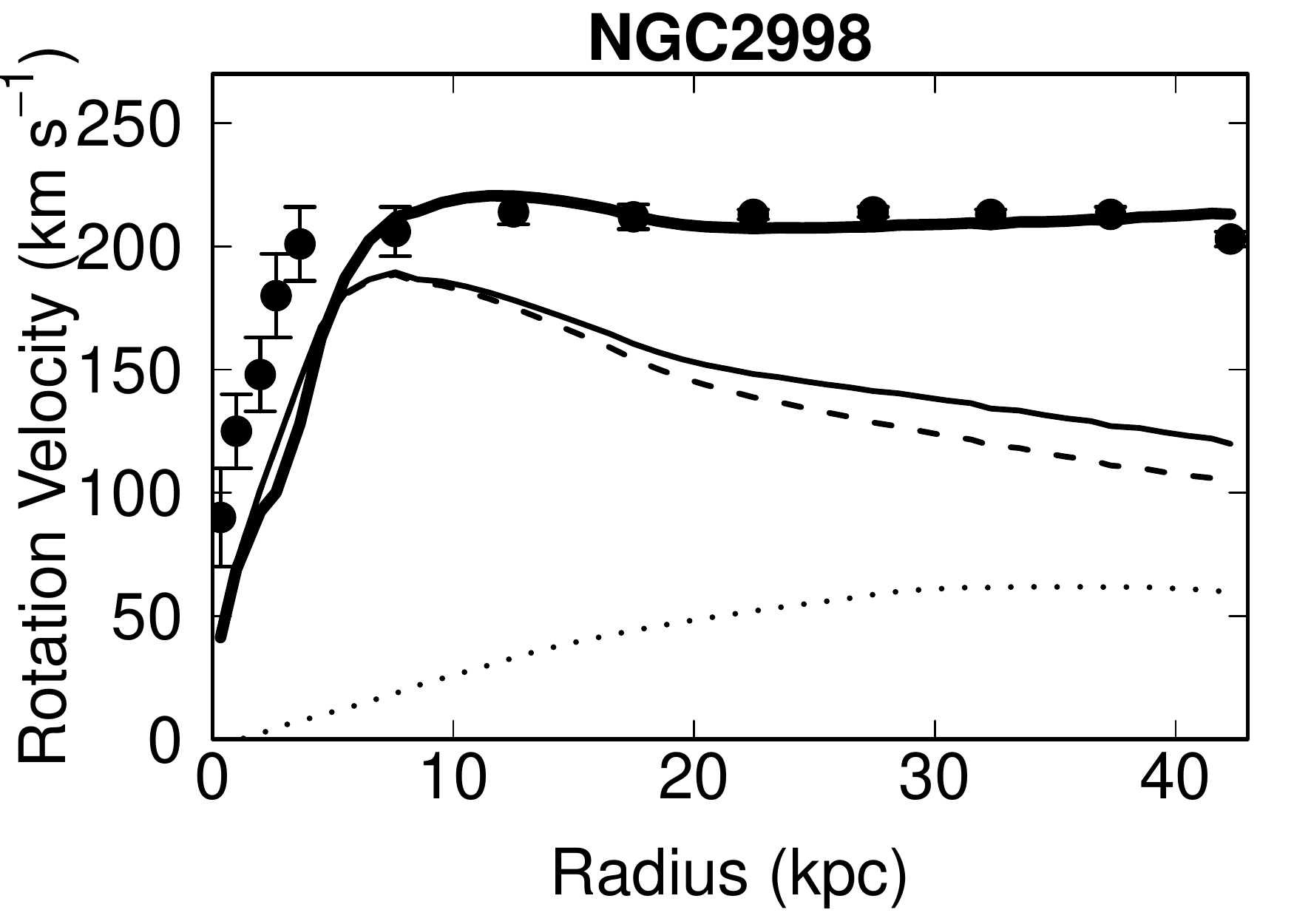}%
\includegraphics[width=60mm]{NGC3109_RC_V5.pdf}%
\caption{\label{fig:RCfull2} \textit{(continued)}.}
\end{figure*}

\addtocounter{figure}{-1}
\begin{figure*}[b]
\includegraphics[width=60mm]{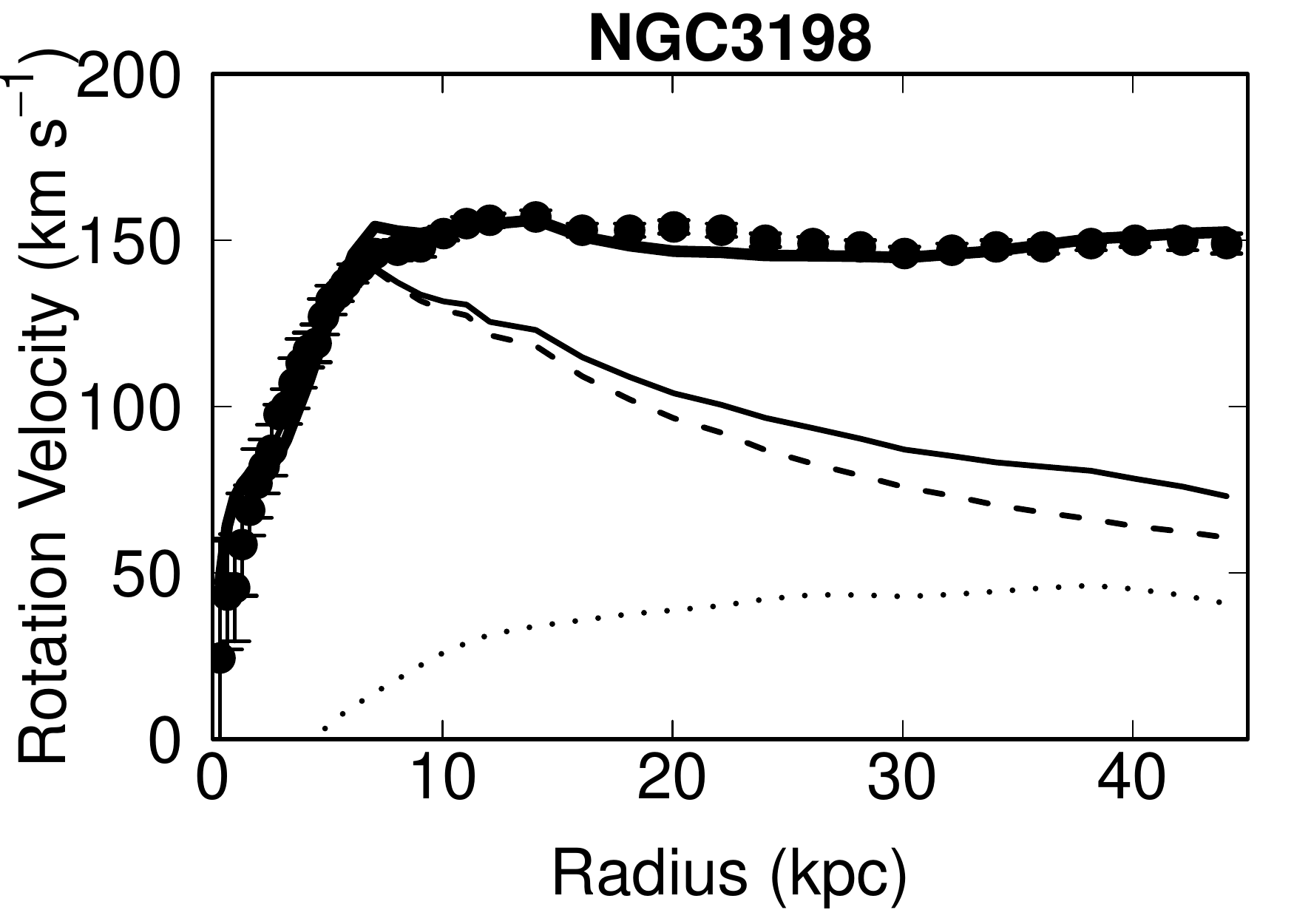}%
\includegraphics[width=60mm]{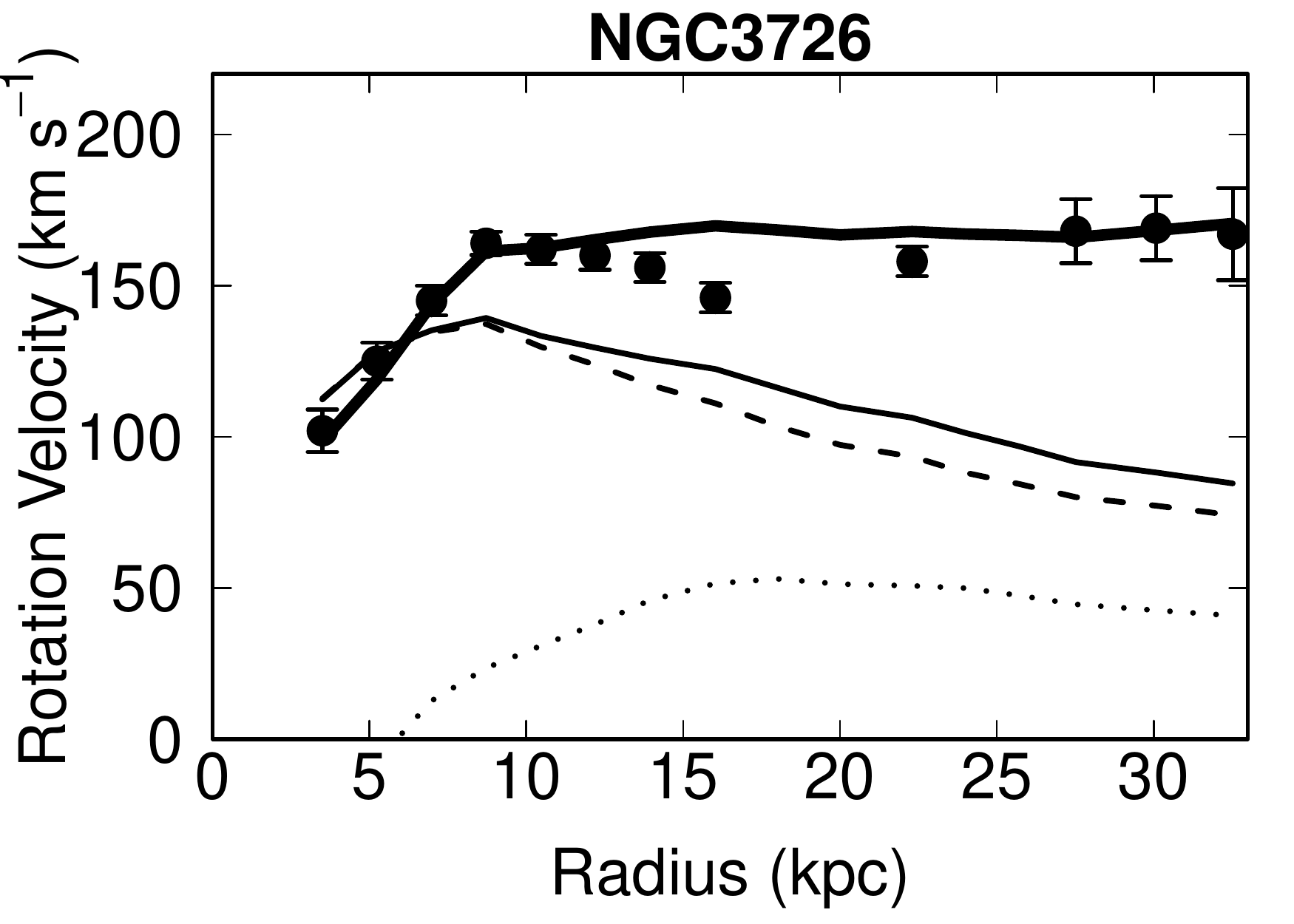}%
\includegraphics[width=60mm]{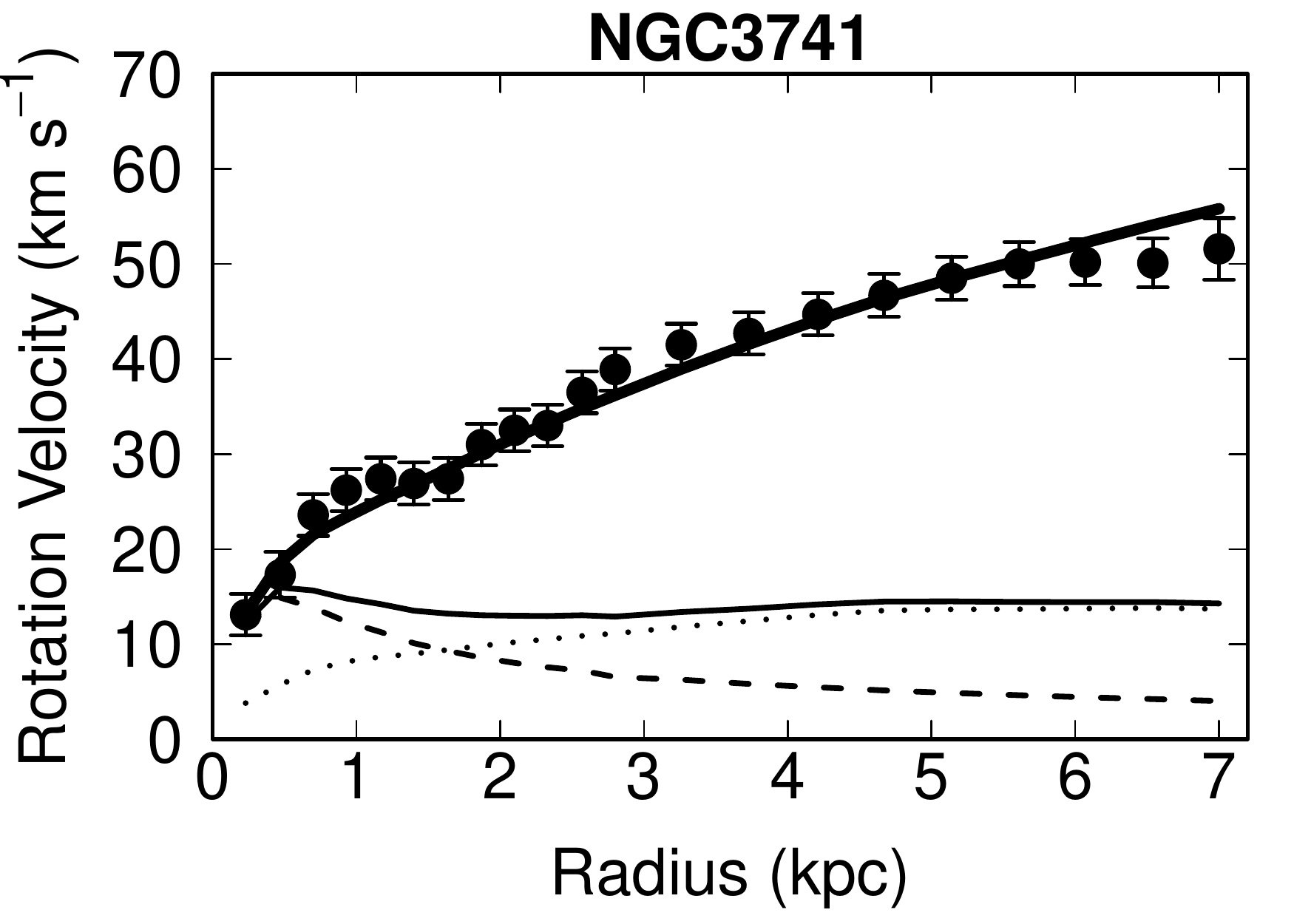}%
\\ \ \\
\includegraphics[width=60mm]{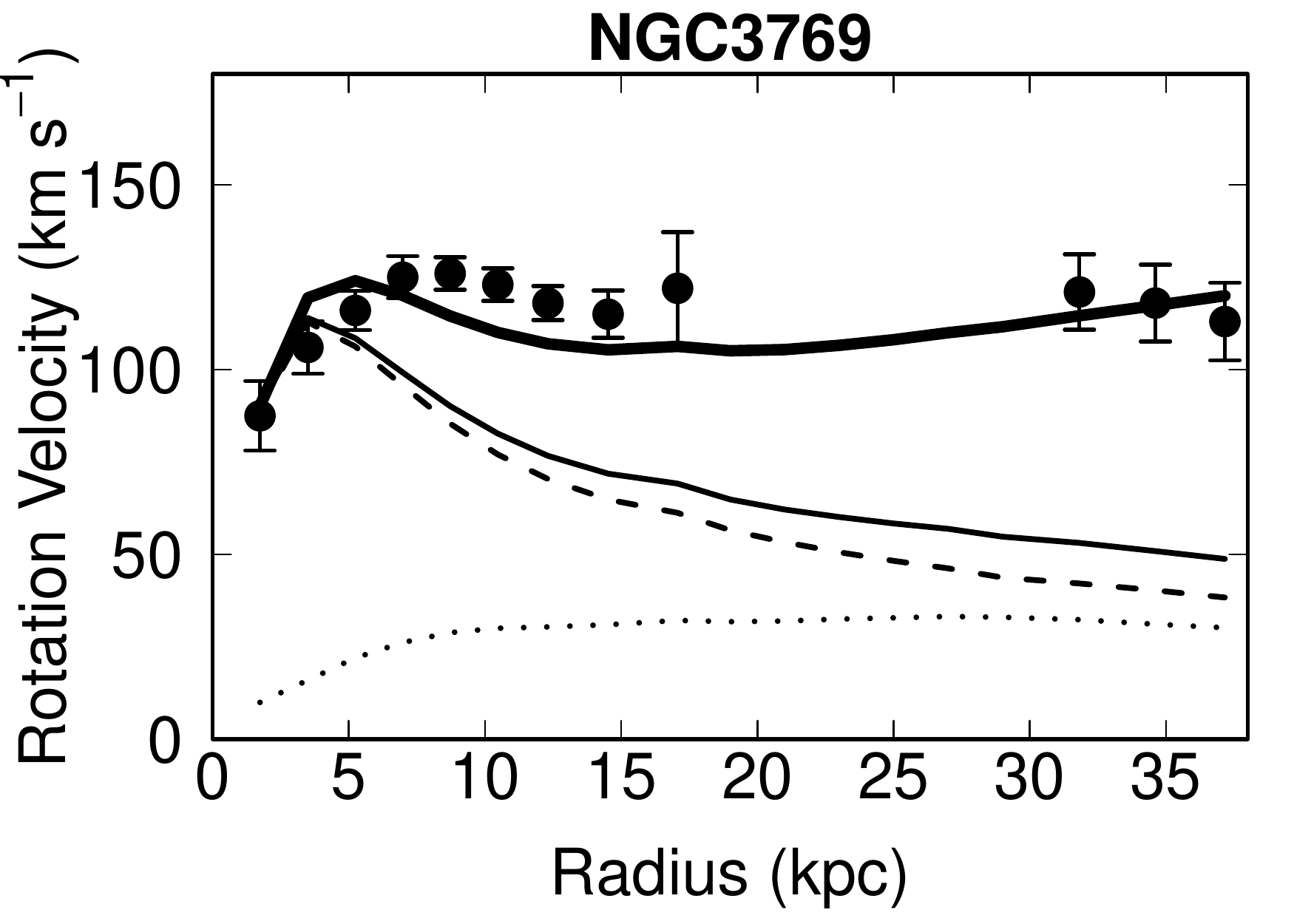}%
\includegraphics[width=60mm]{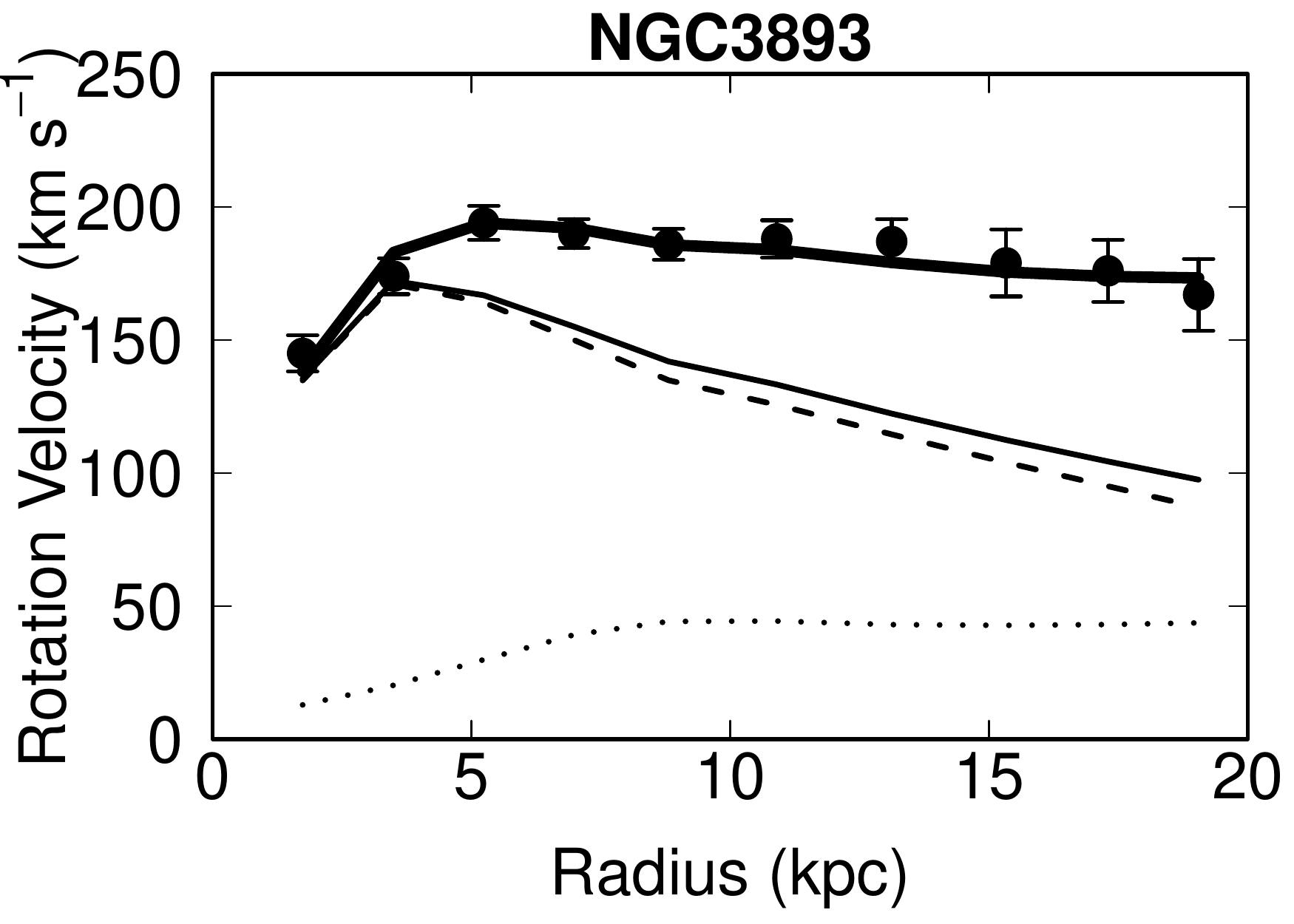}%
\includegraphics[width=60mm]{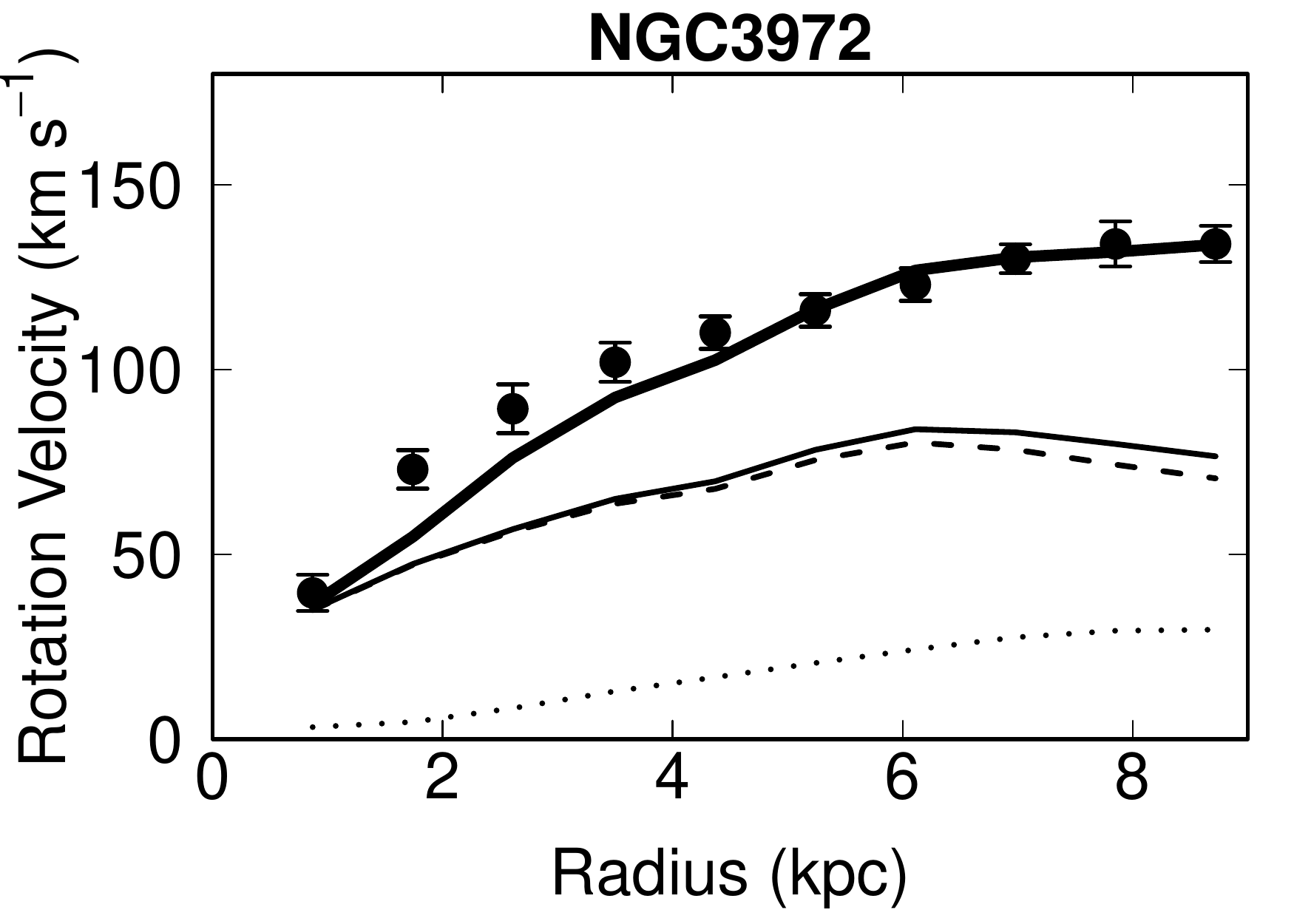}%
\\ \ \\
\includegraphics[width=60mm]{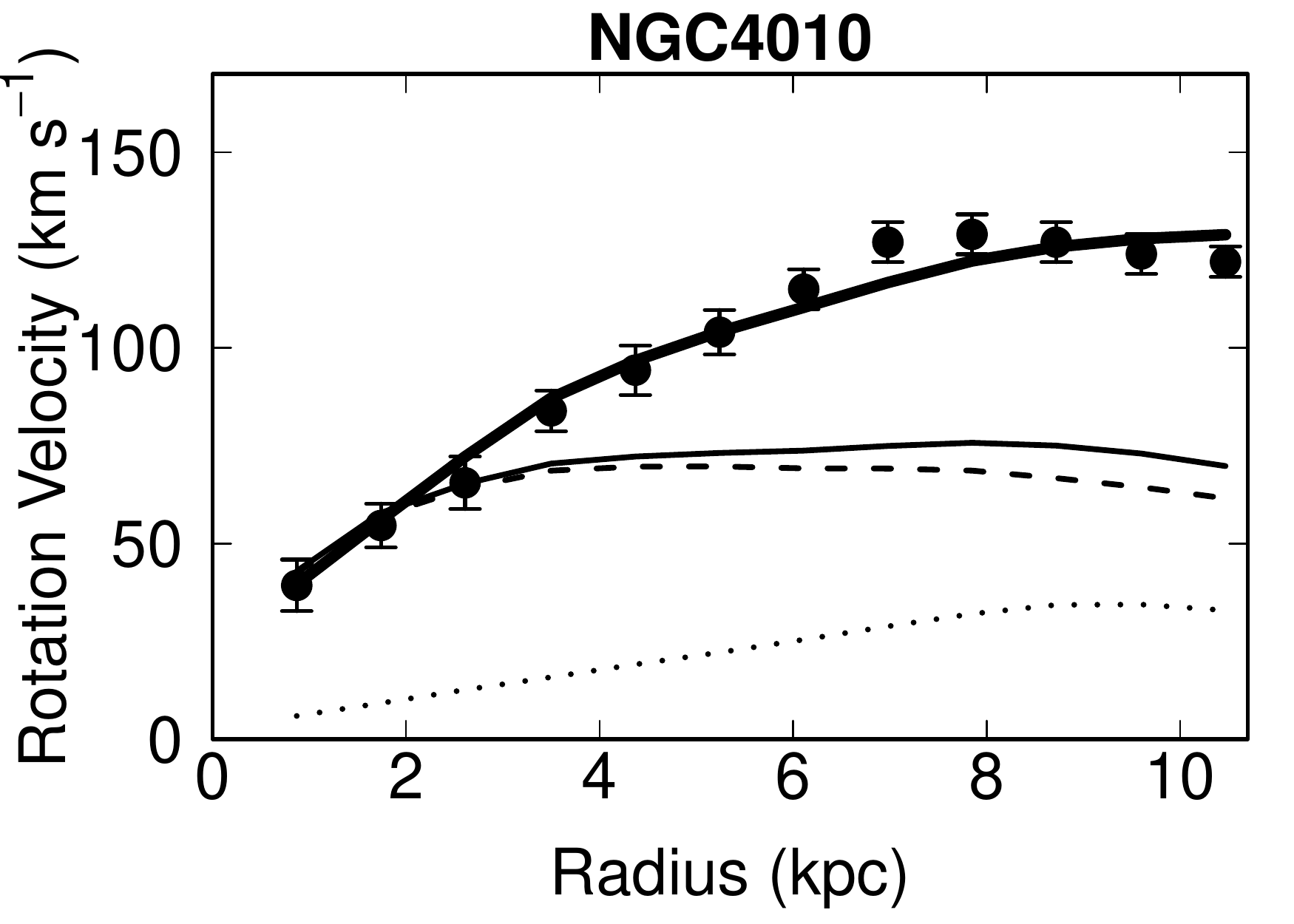}%
\includegraphics[width=60mm]{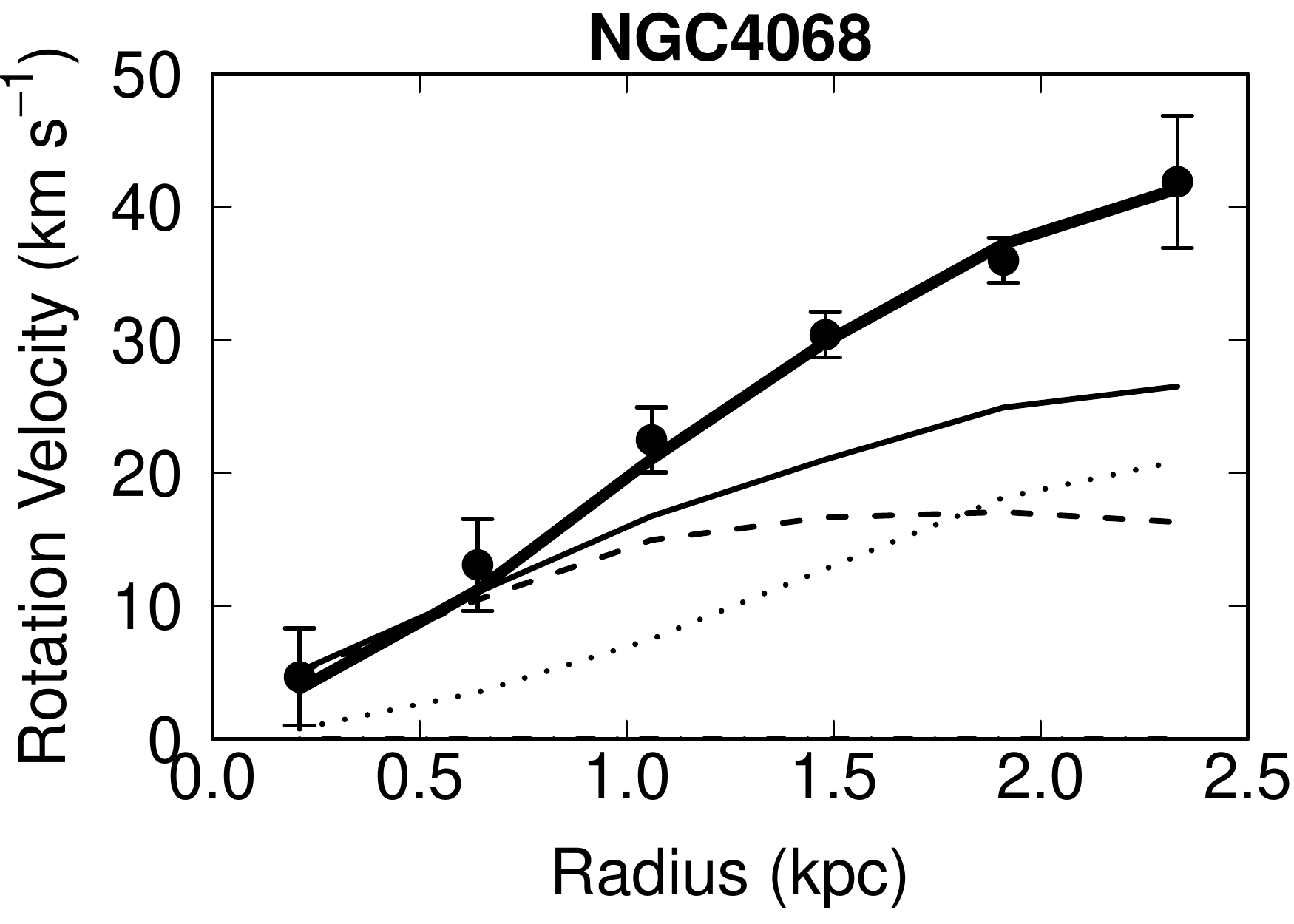}%
\includegraphics[width=60mm]{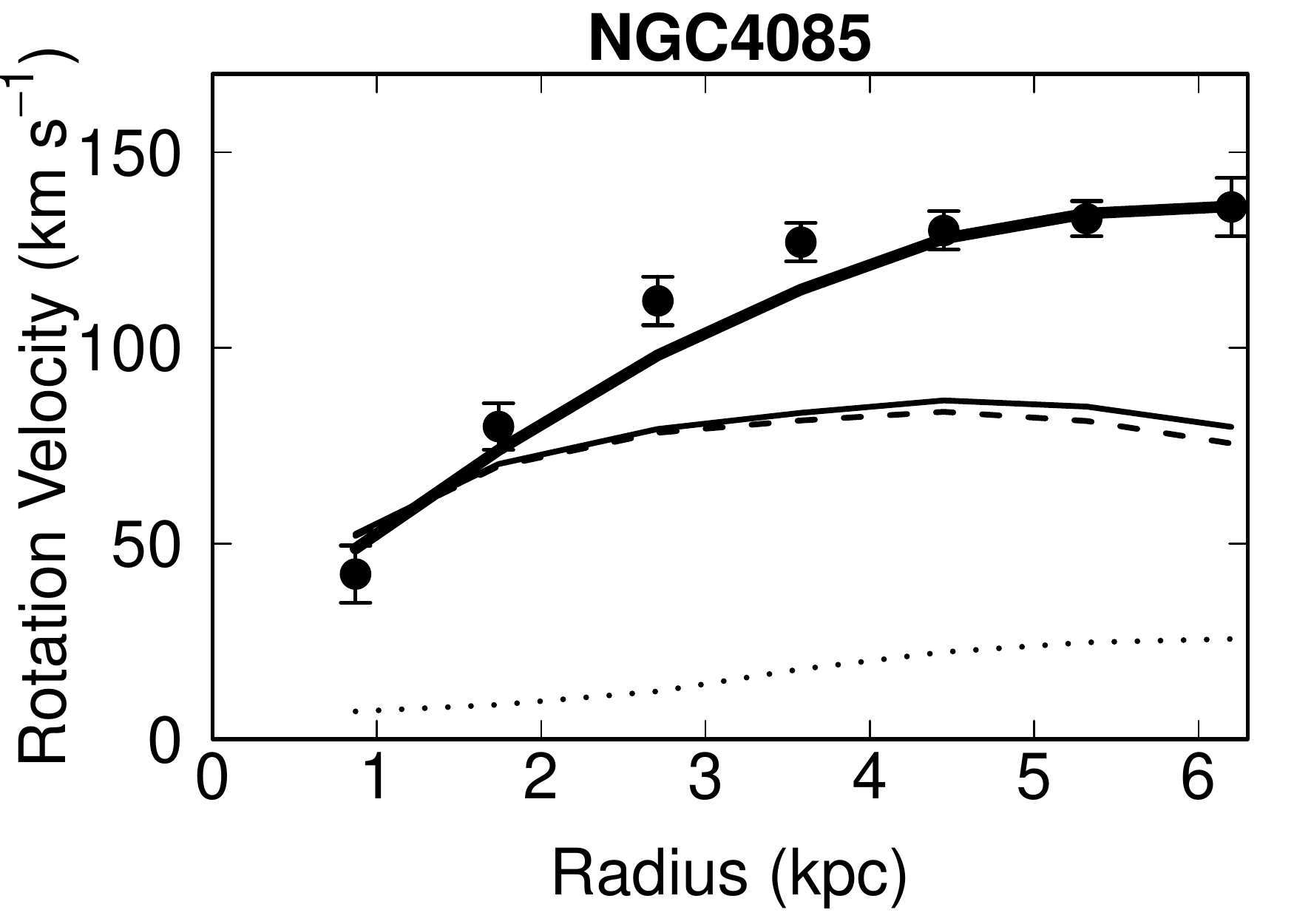}%
\\ \ \\
\includegraphics[width=60mm]{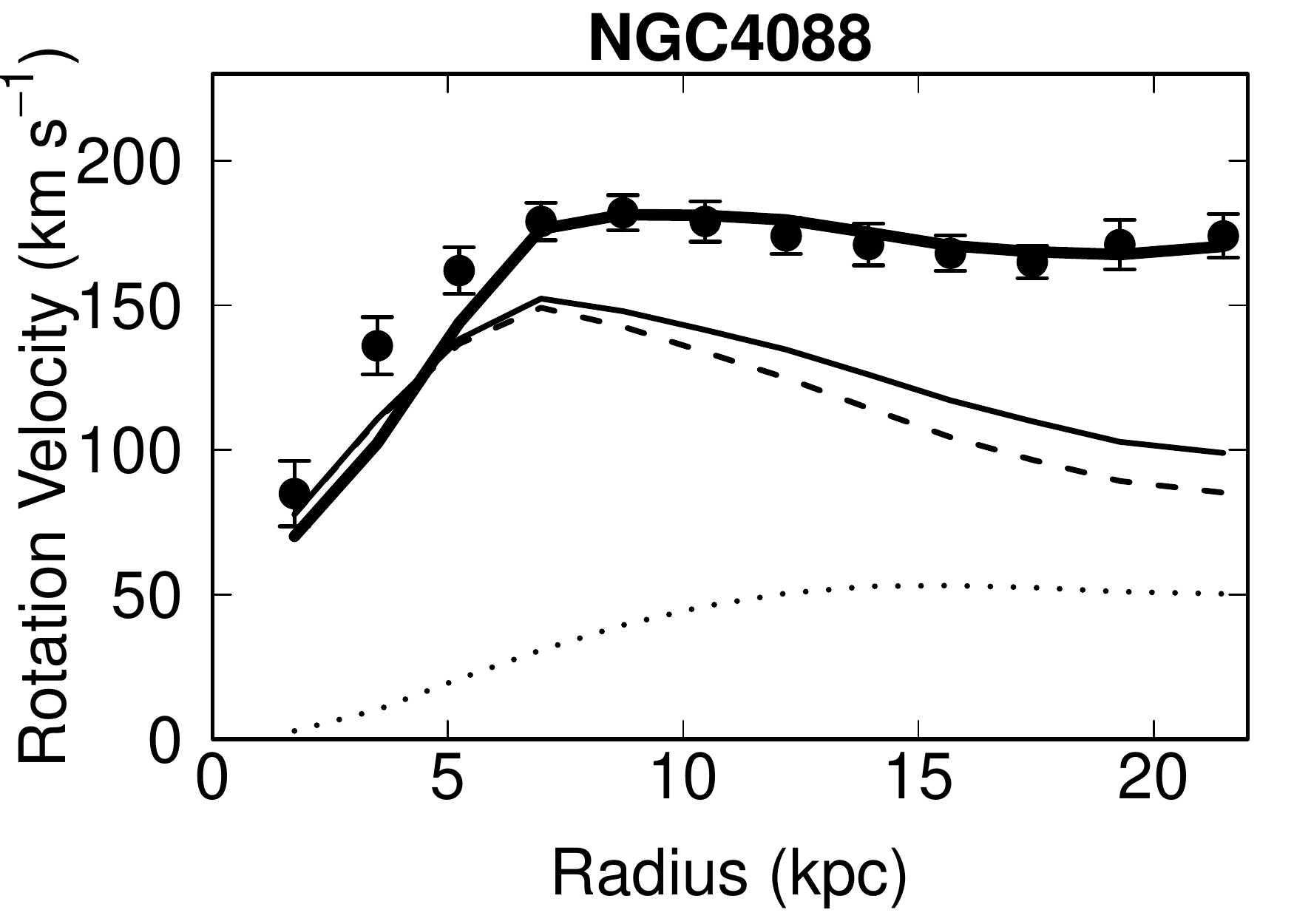}%
\includegraphics[width=60mm]{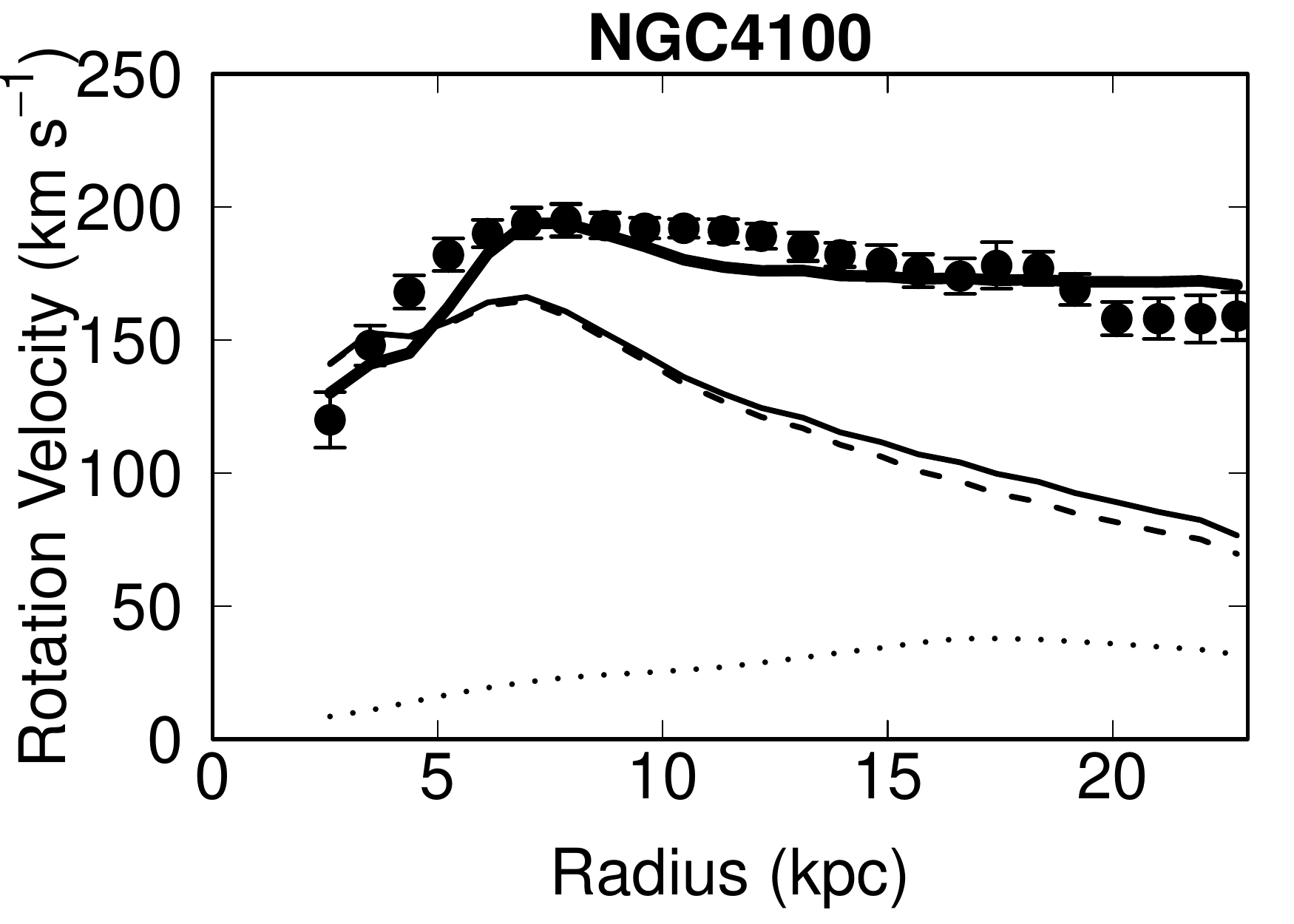}%
\includegraphics[width=60mm]{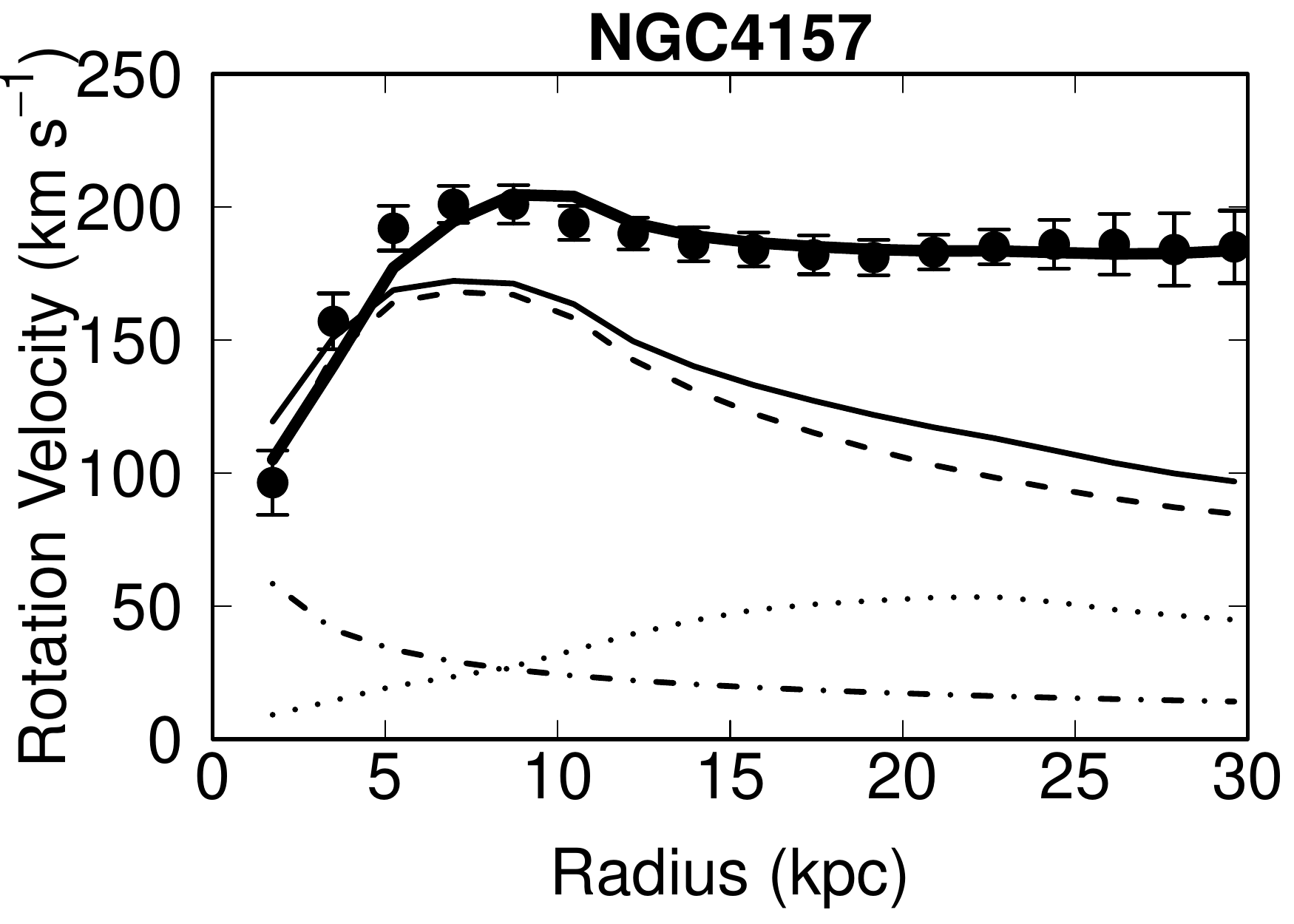}%
\caption{\label{fig:RCfull3} \textit{(continued)}.}
\end{figure*}

\addtocounter{figure}{-1}
\begin{figure*}[b]
\includegraphics[width=60mm]{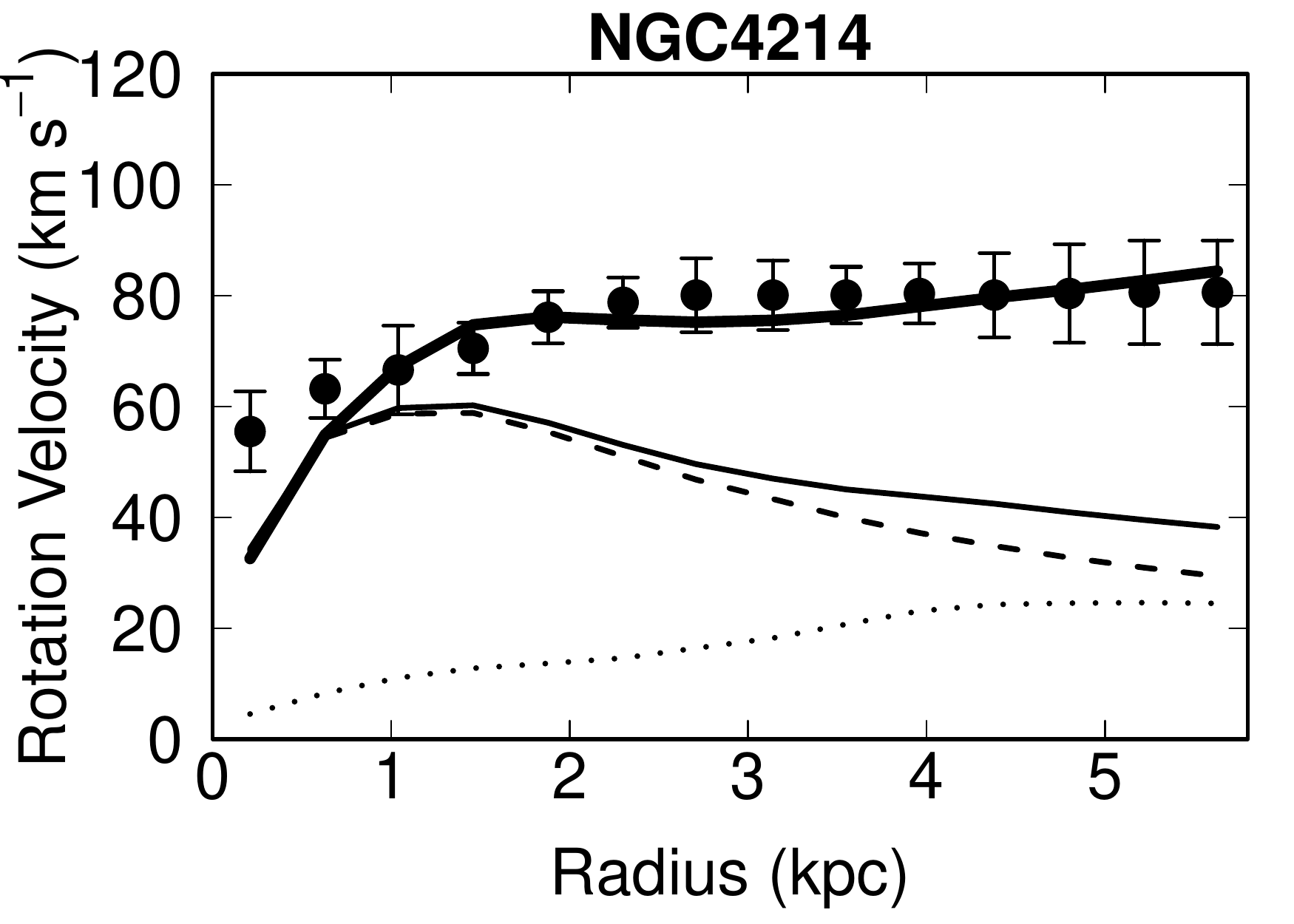}%
\includegraphics[width=60mm]{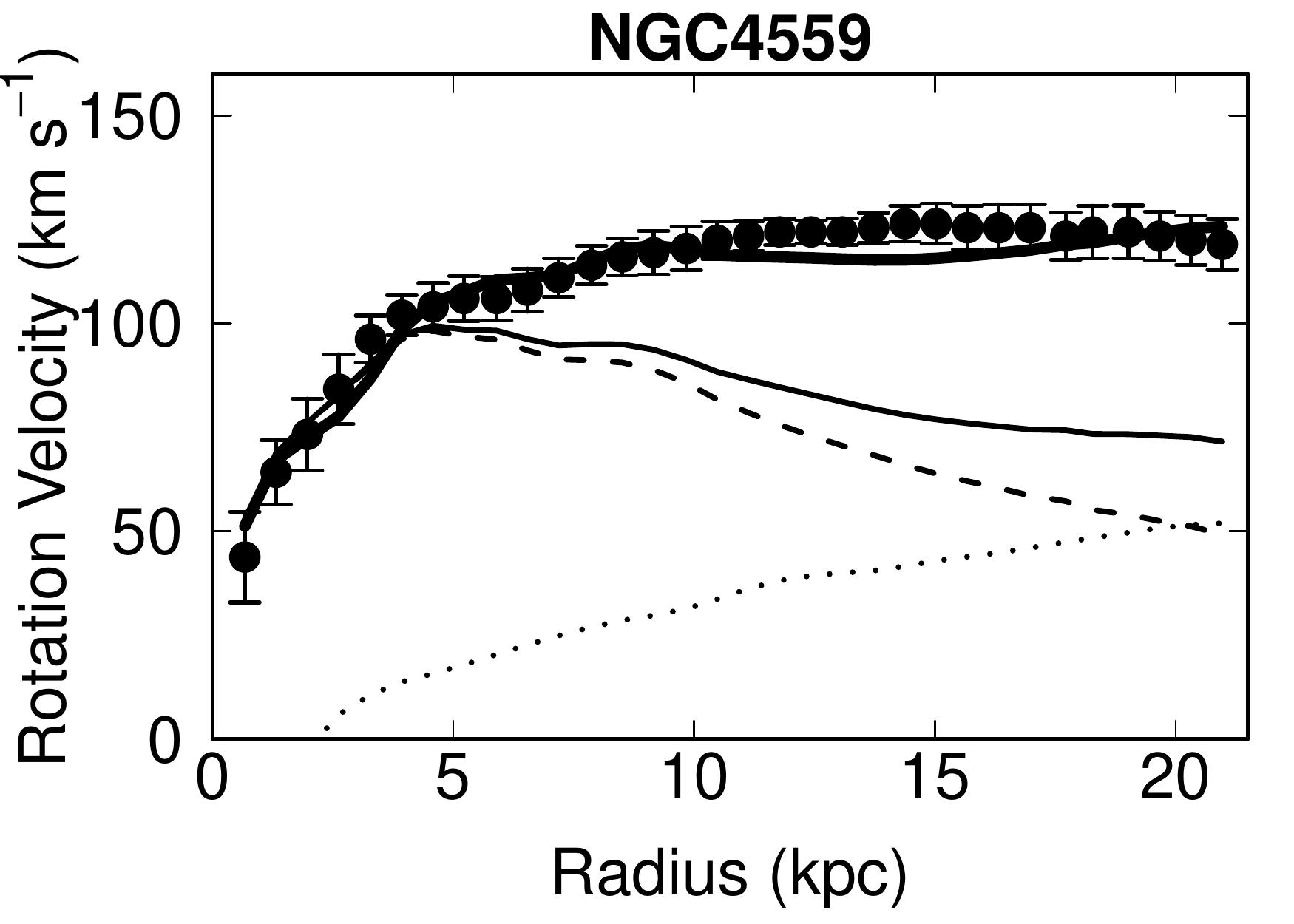}%
\includegraphics[width=60mm]{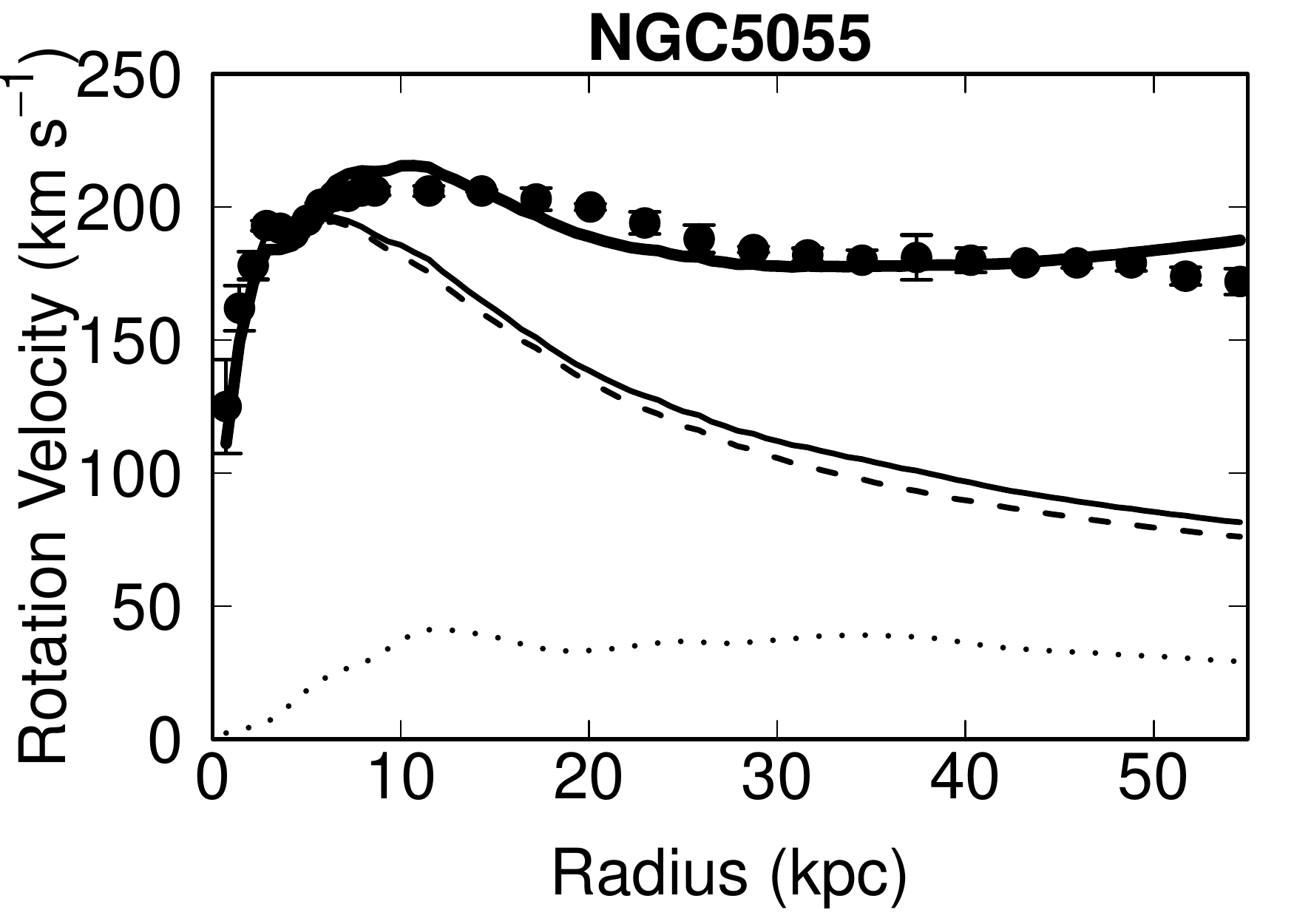}%
\\ \ \\
\includegraphics[width=60mm]{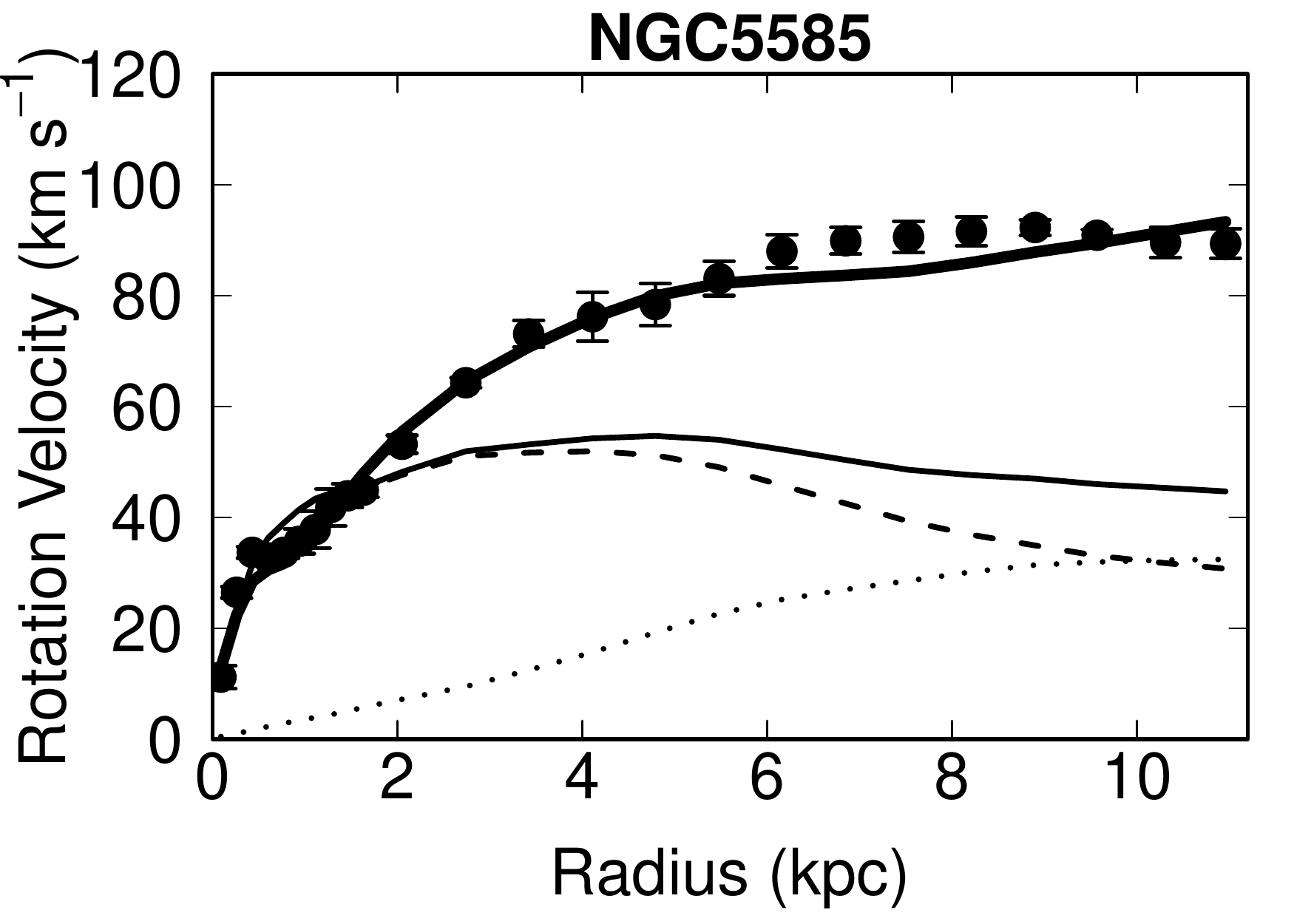}%
\includegraphics[width=60mm]{NGC5907_RC_V5.pdf}%
\includegraphics[width=60mm]{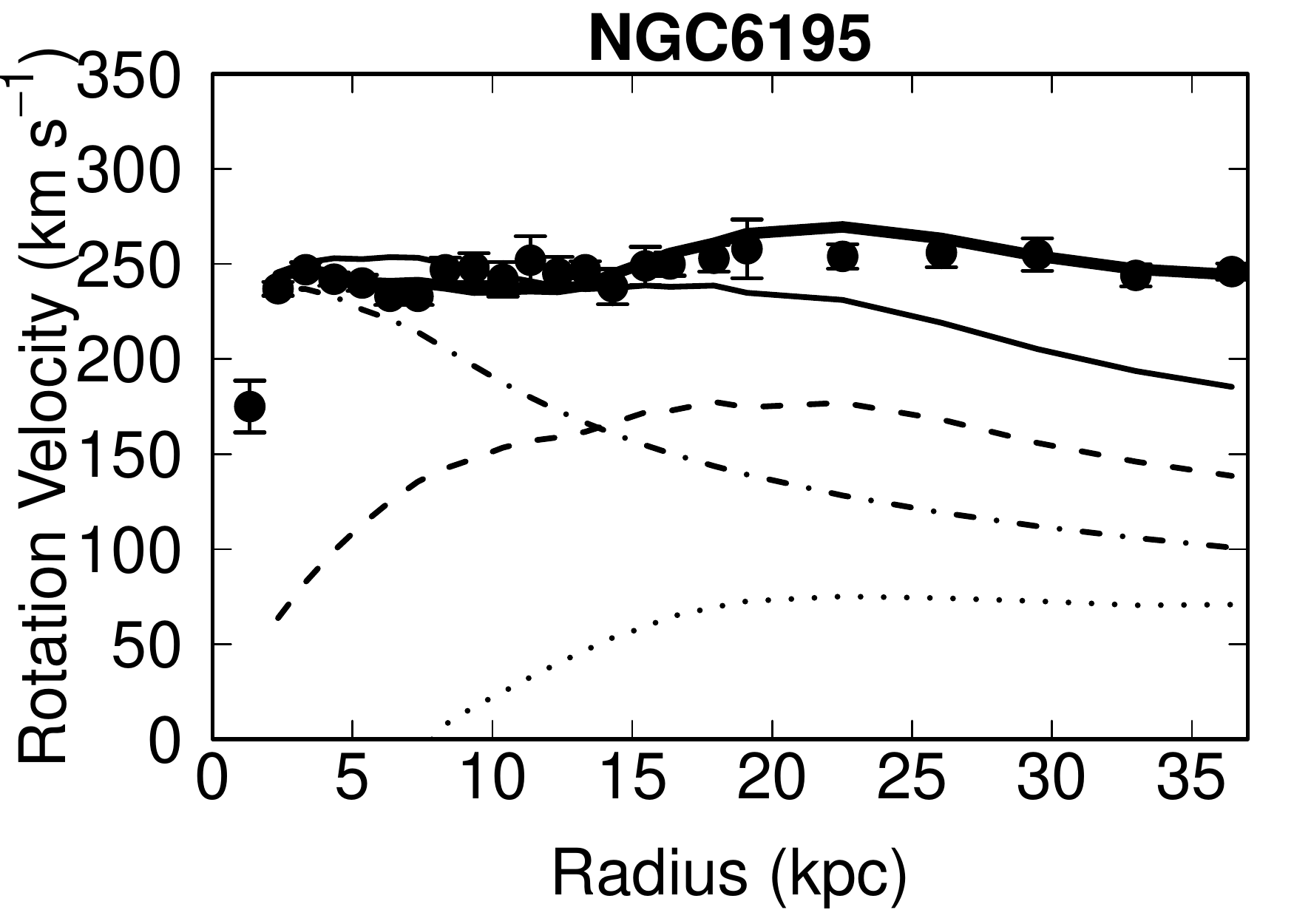}%
\\ \ \\
\includegraphics[width=60mm]{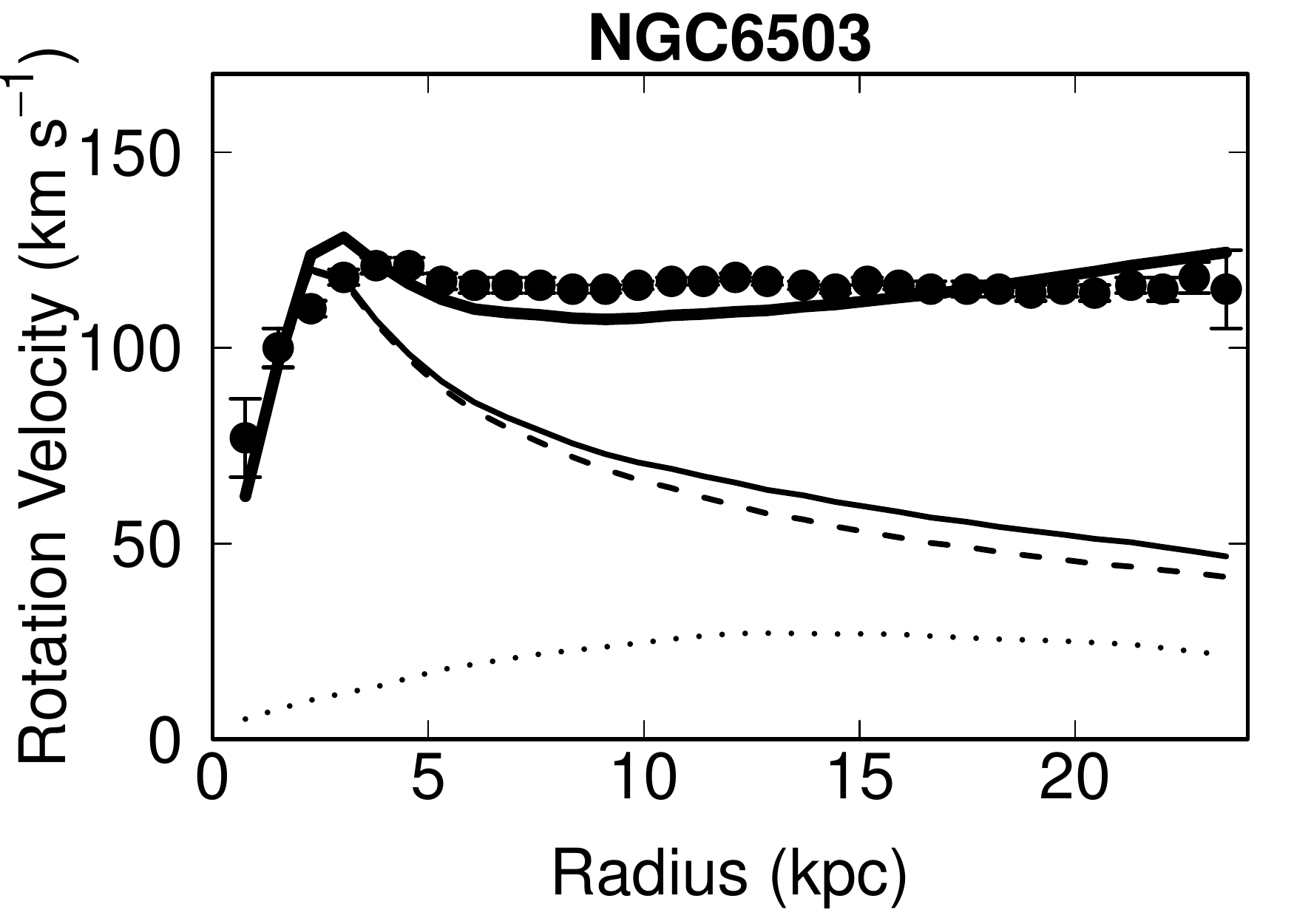}%
\includegraphics[width=60mm]{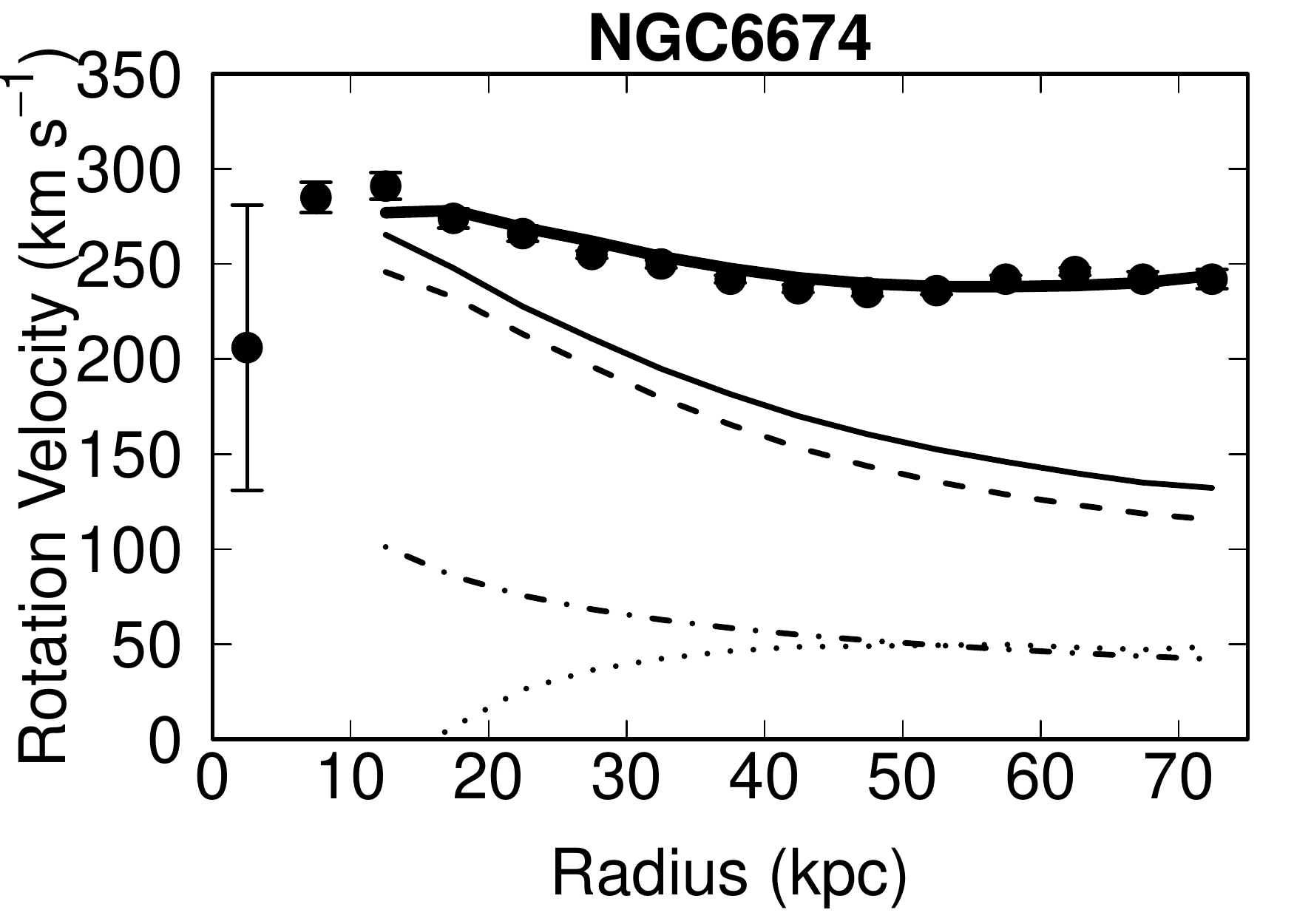}%
\includegraphics[width=60mm]{NGC7331_RC_V5.pdf}%
\\ \ \\
\includegraphics[width=60mm]{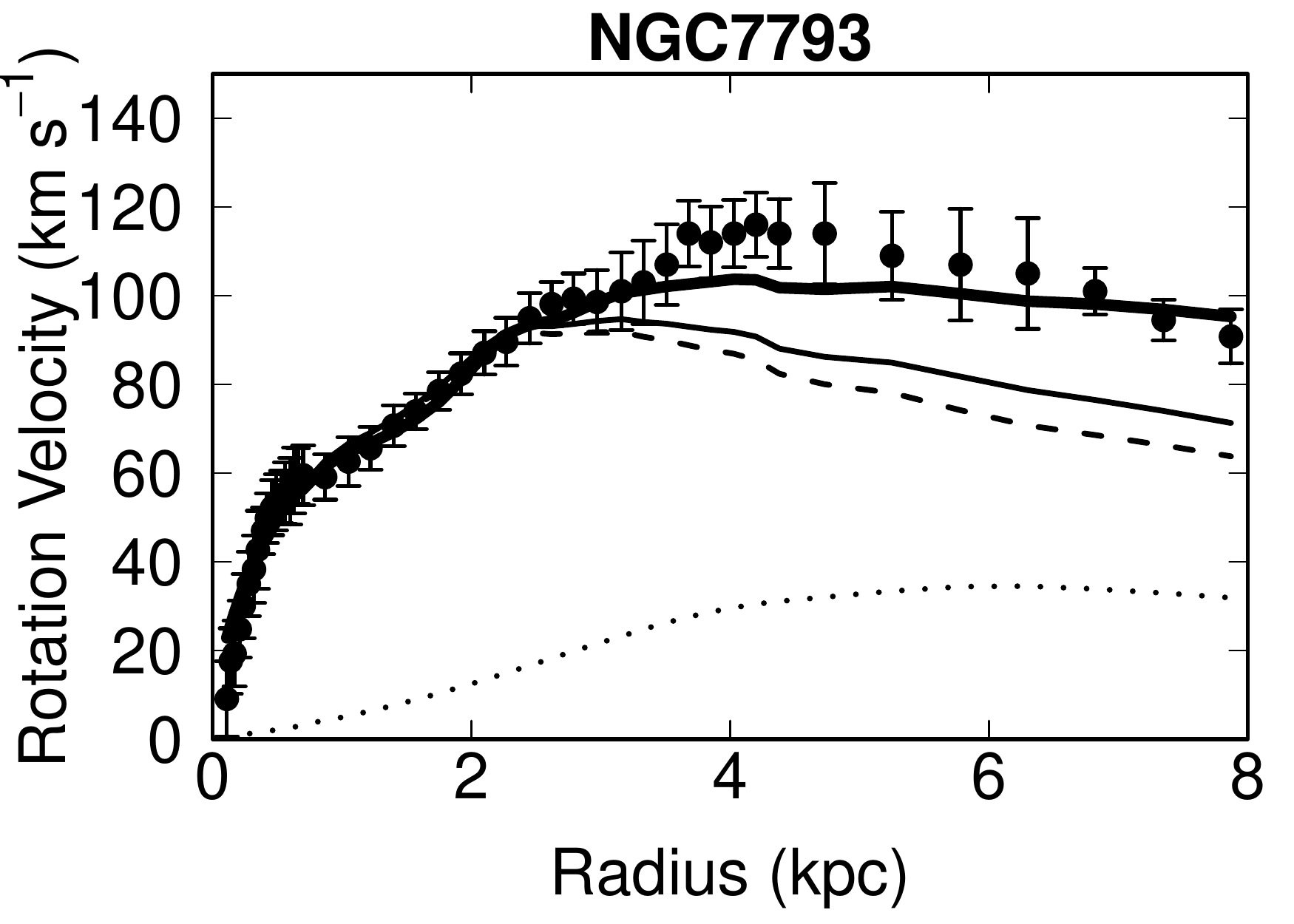}%
\includegraphics[width=60mm]{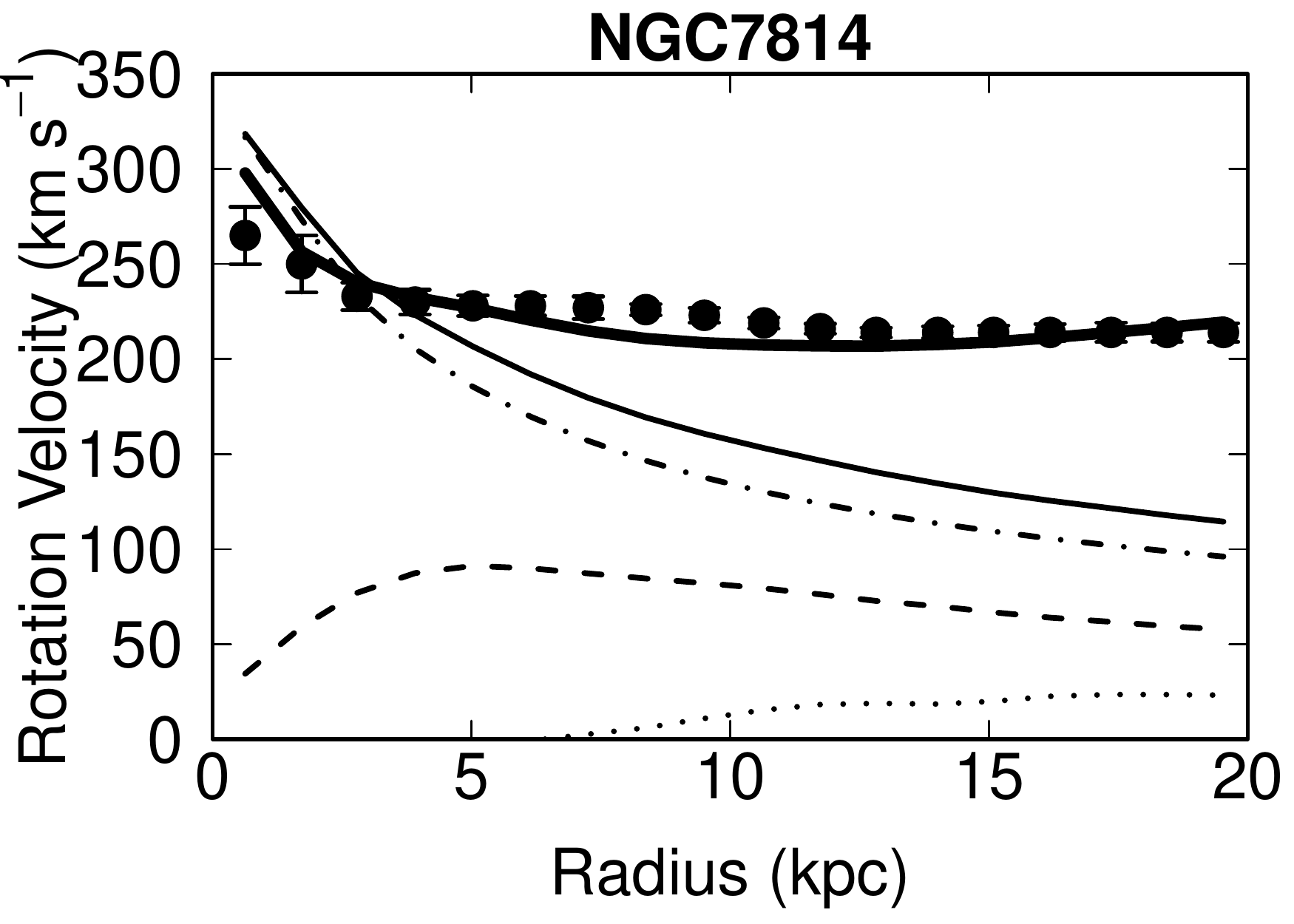}%
\includegraphics[width=60mm]{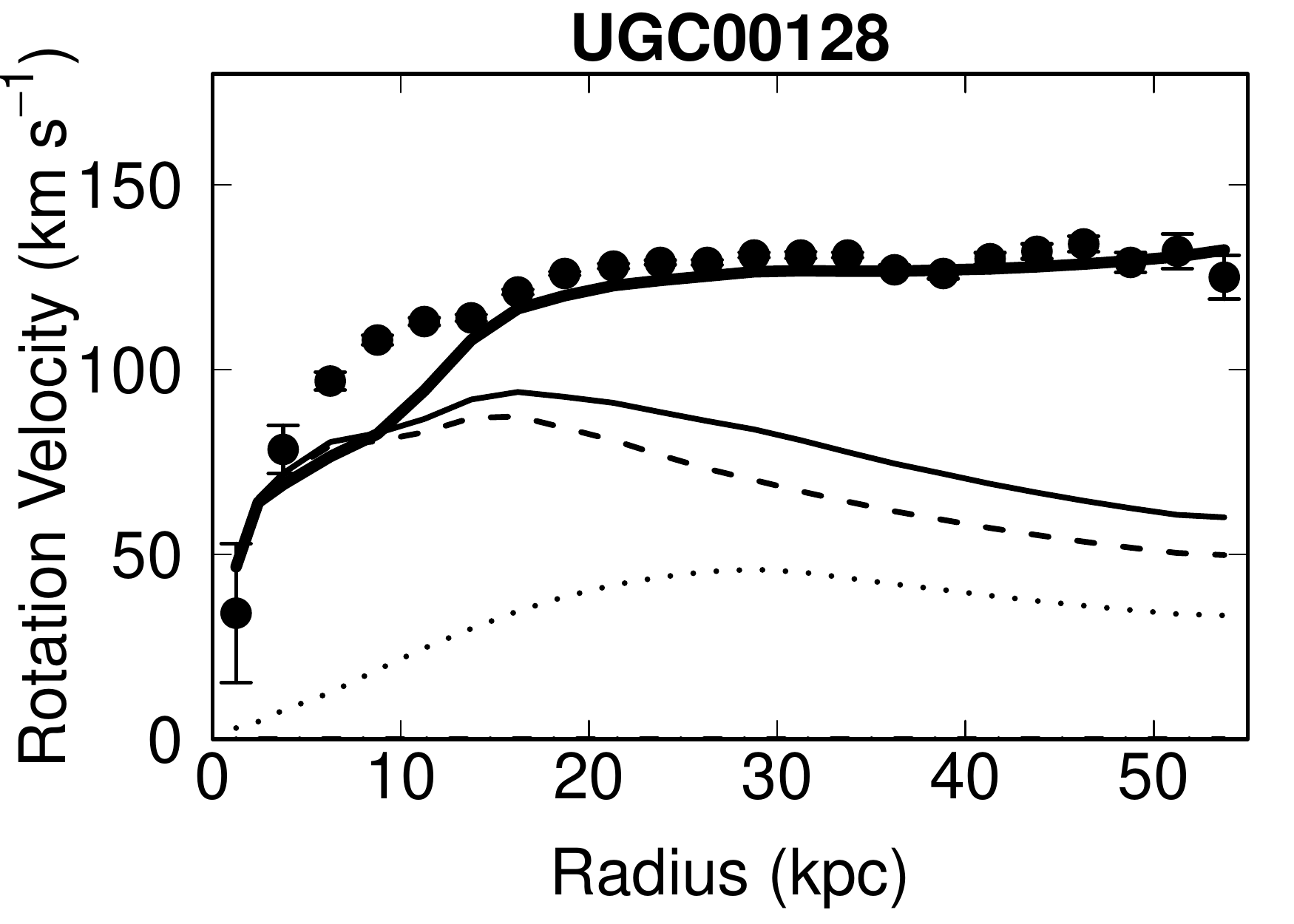}%
\caption{\label{fig:RCfull4} \textit{(continued)}.}
\end{figure*}

\addtocounter{figure}{-1}
\begin{figure*}[b]
\includegraphics[width=60mm]{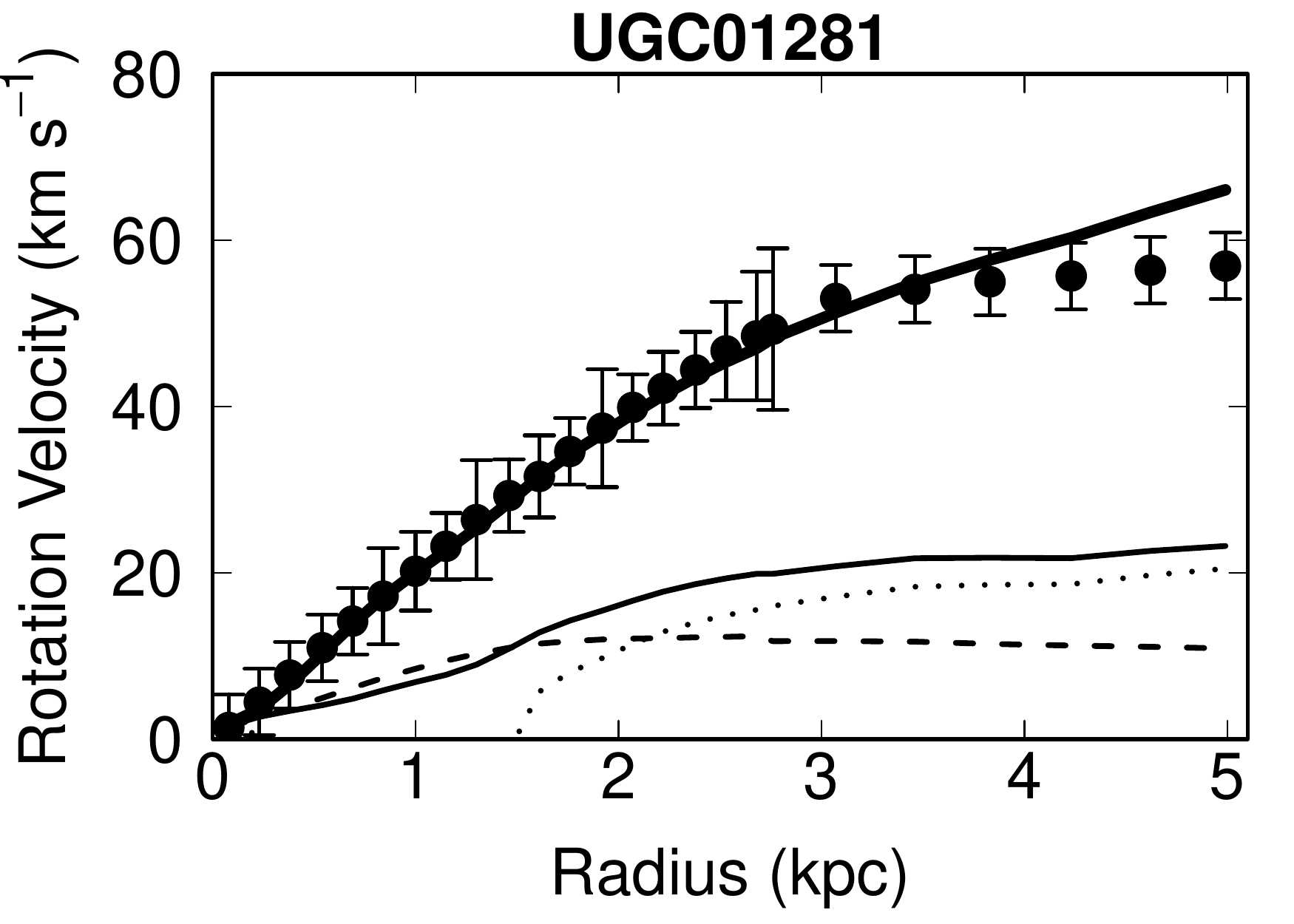}%
\includegraphics[width=60mm]{UGC02487_RC_V5.pdf}%
\includegraphics[width=60mm]{UGC02885_RC_V5.pdf}%
\\ \ \\
\includegraphics[width=60mm]{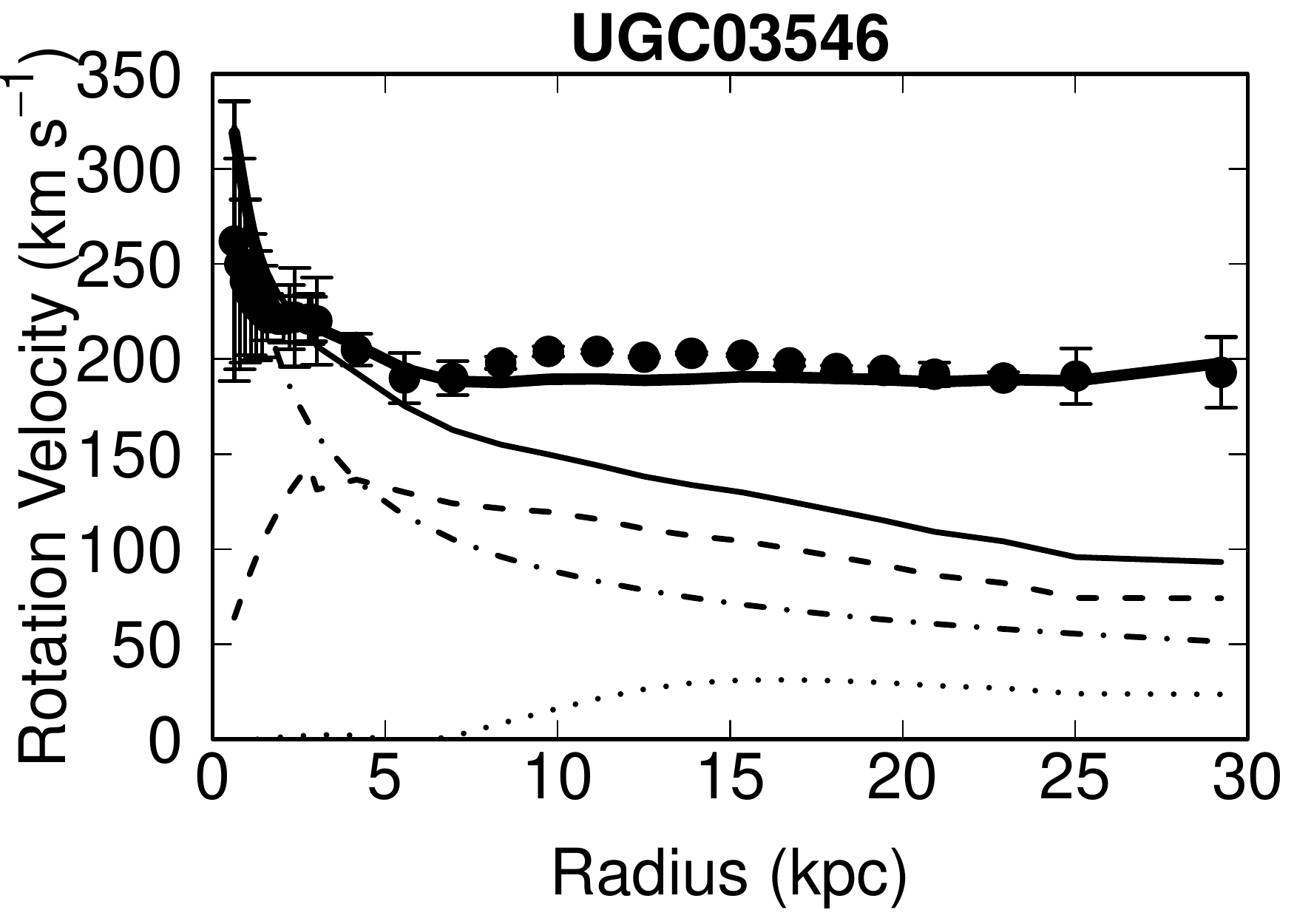}%
\includegraphics[width=60mm]{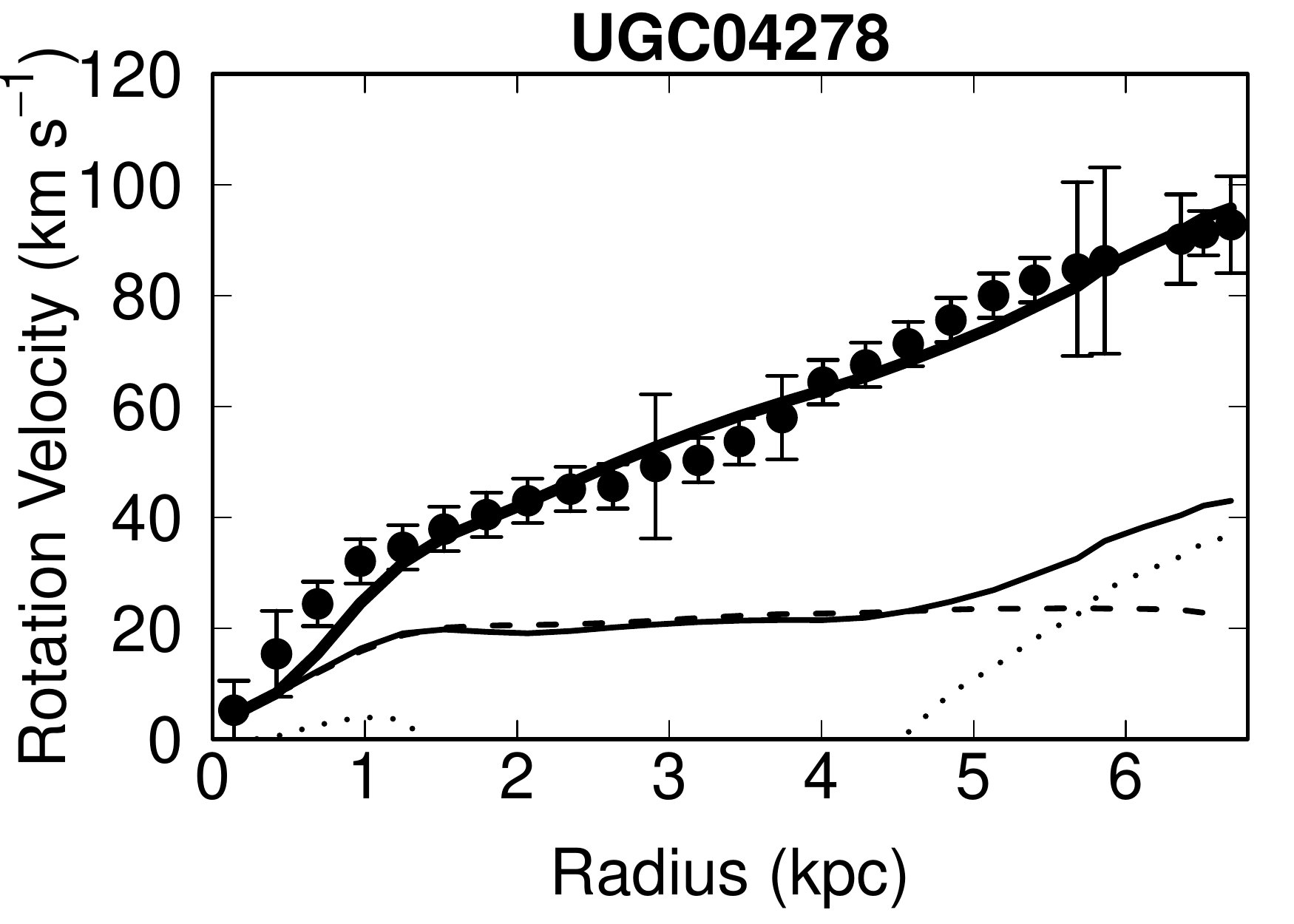}%
\includegraphics[width=60mm]{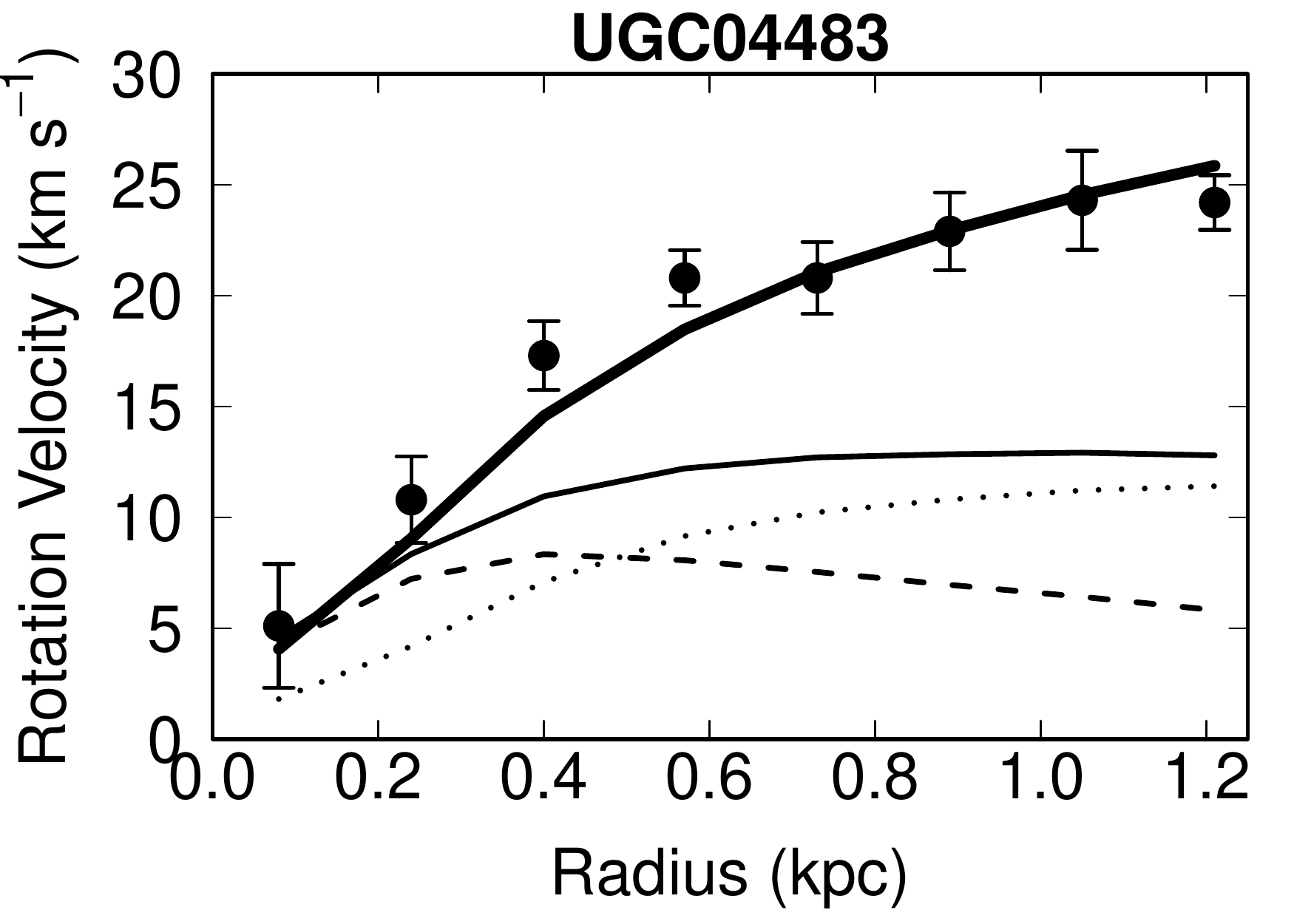}%
\\ \ \\
\includegraphics[width=60mm]{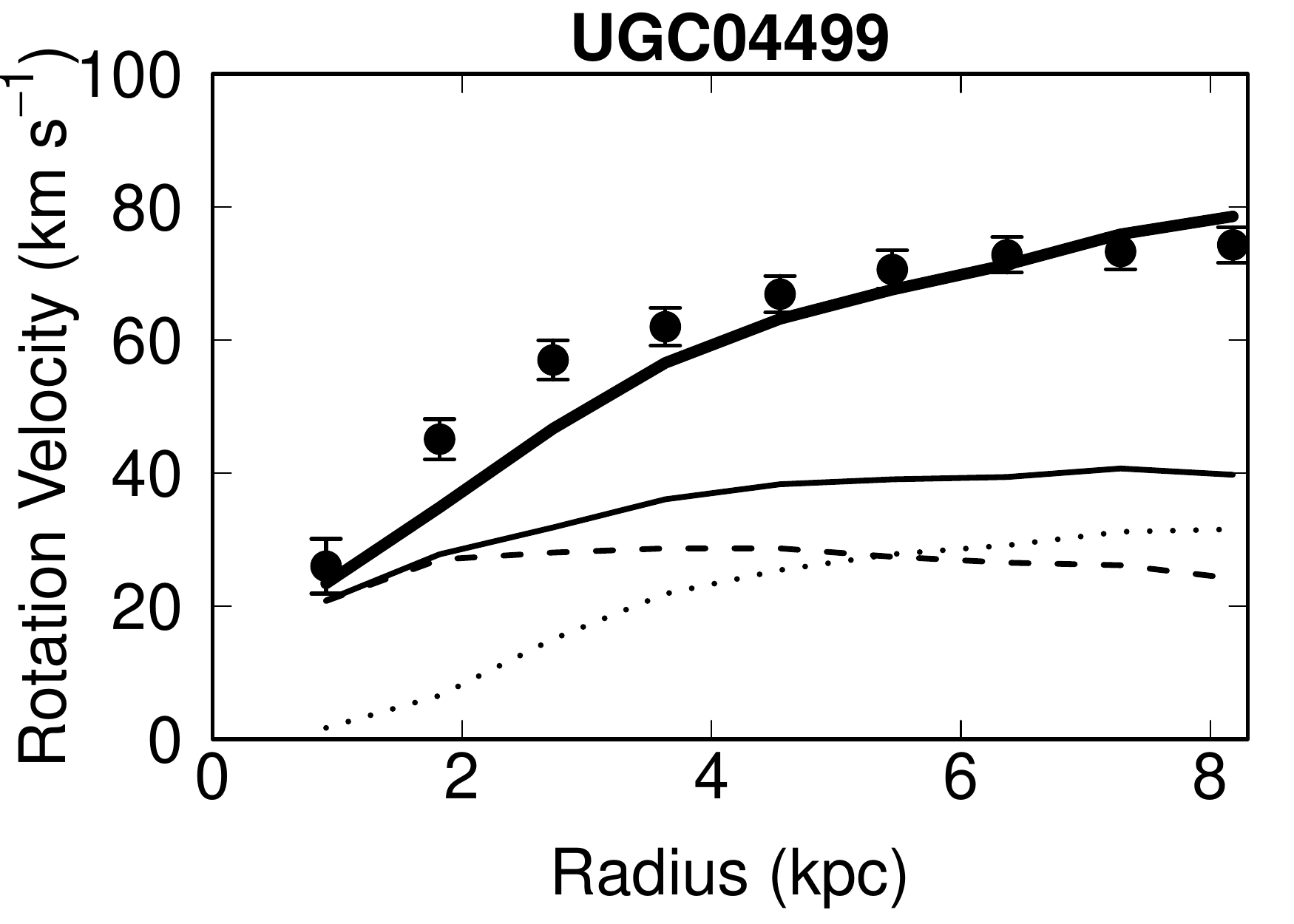}%
\includegraphics[width=60mm]{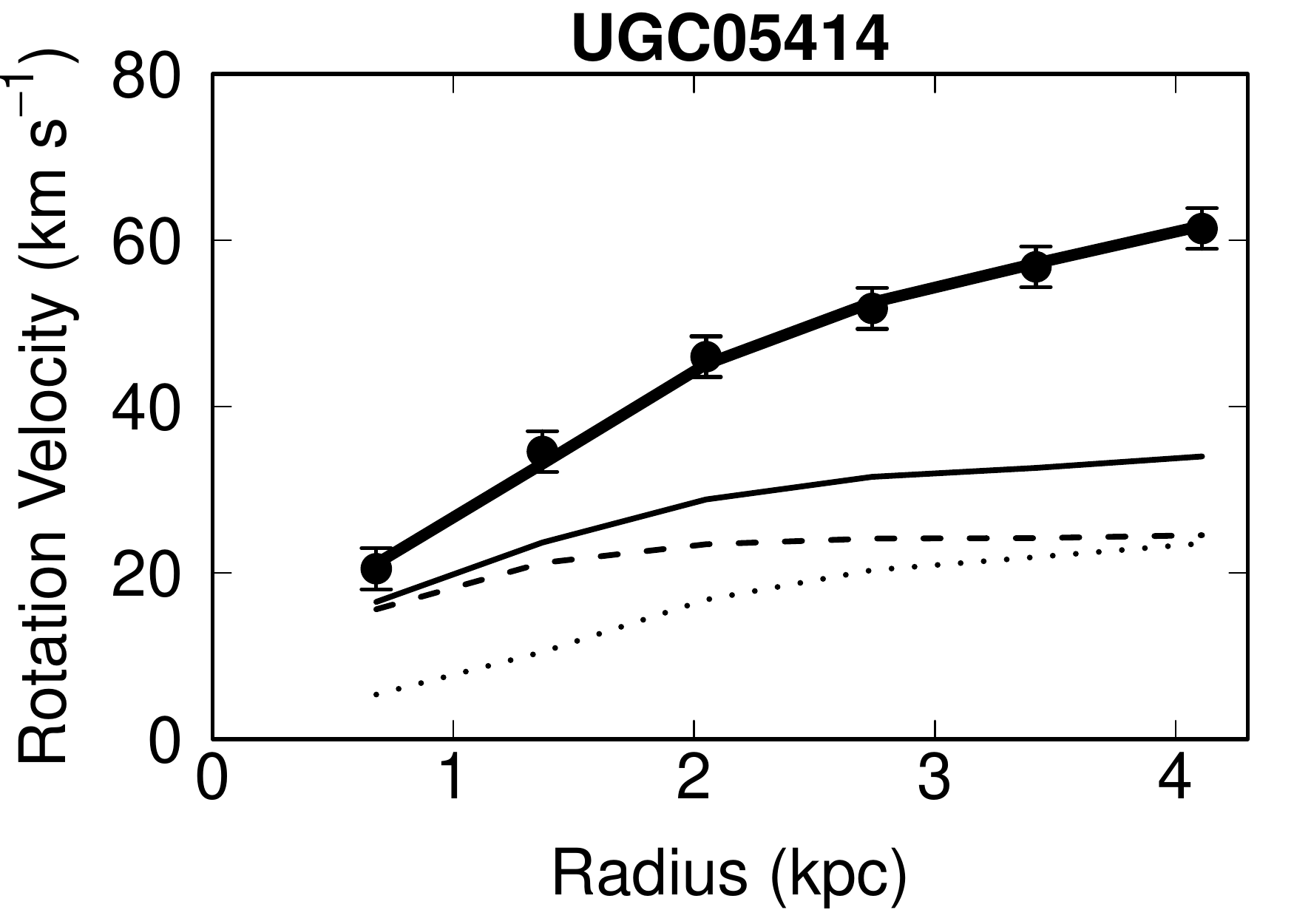}%
\includegraphics[width=60mm]{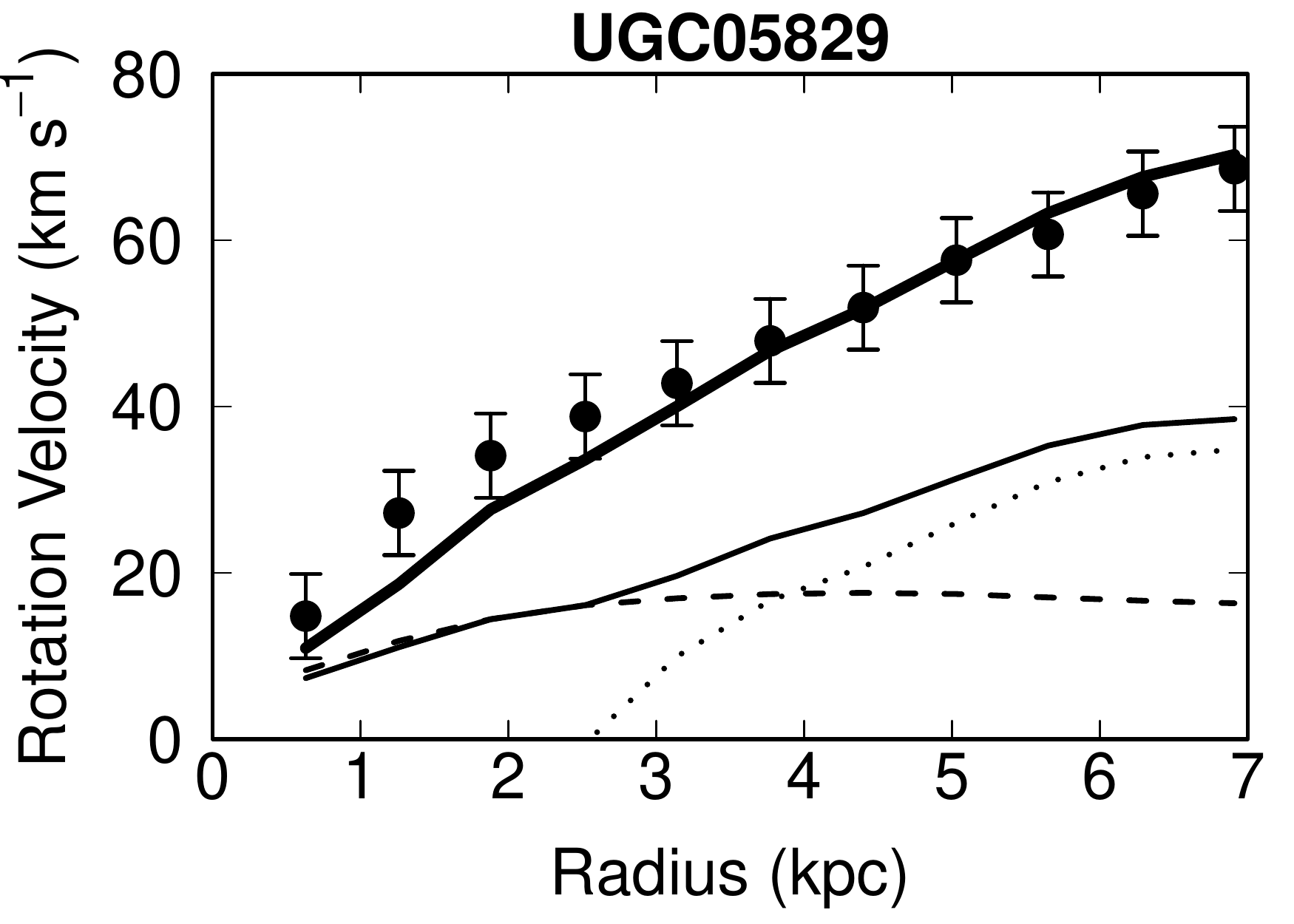}%
\\ \ \\
\includegraphics[width=60mm]{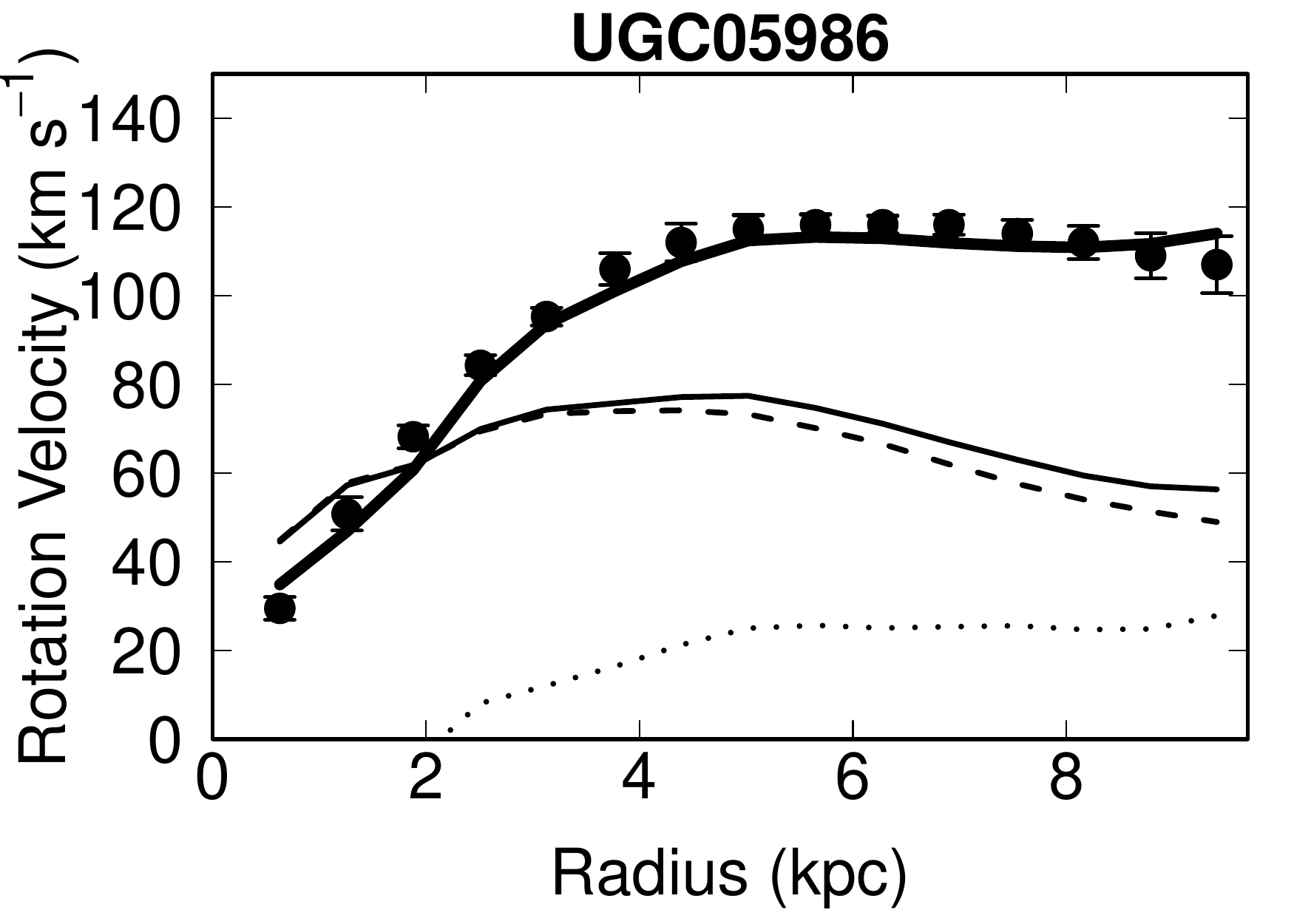}%
\includegraphics[width=60mm]{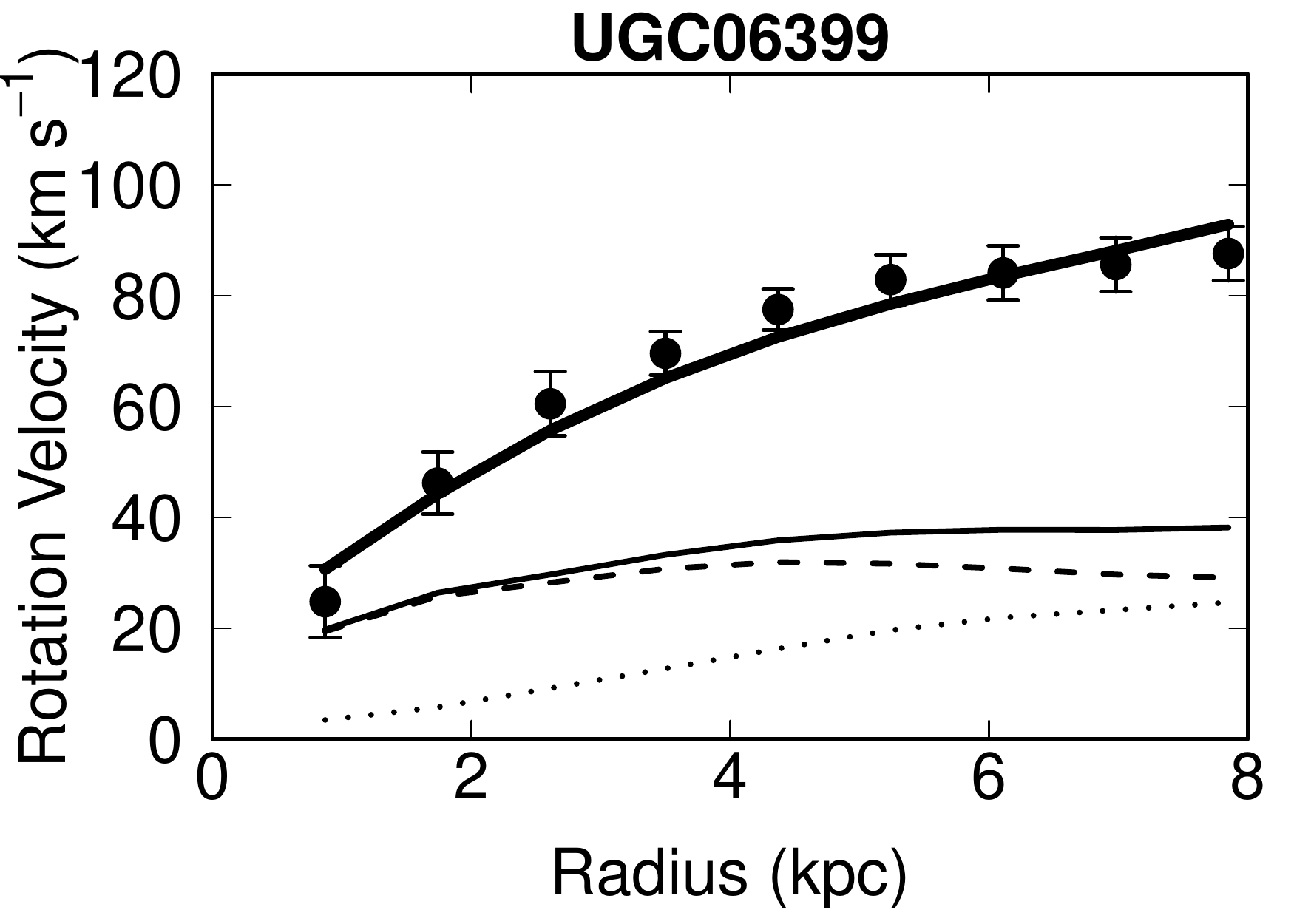}%
\includegraphics[width=60mm]{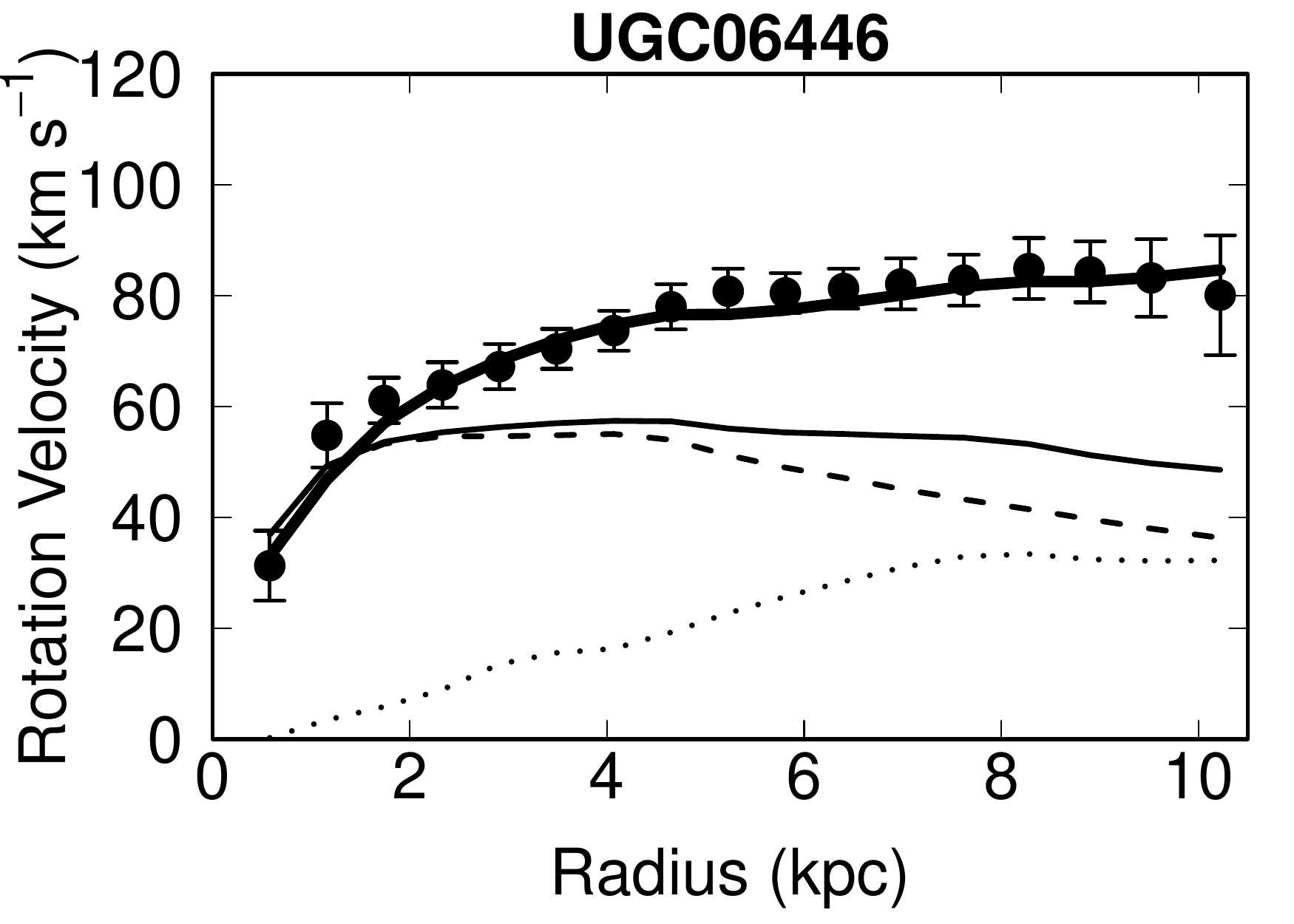}%
\caption{\label{fig:RCfull5} \textit{(continued)}.}
\end{figure*}

\addtocounter{figure}{-1}
\begin{figure*}[b]
\includegraphics[width=60mm]{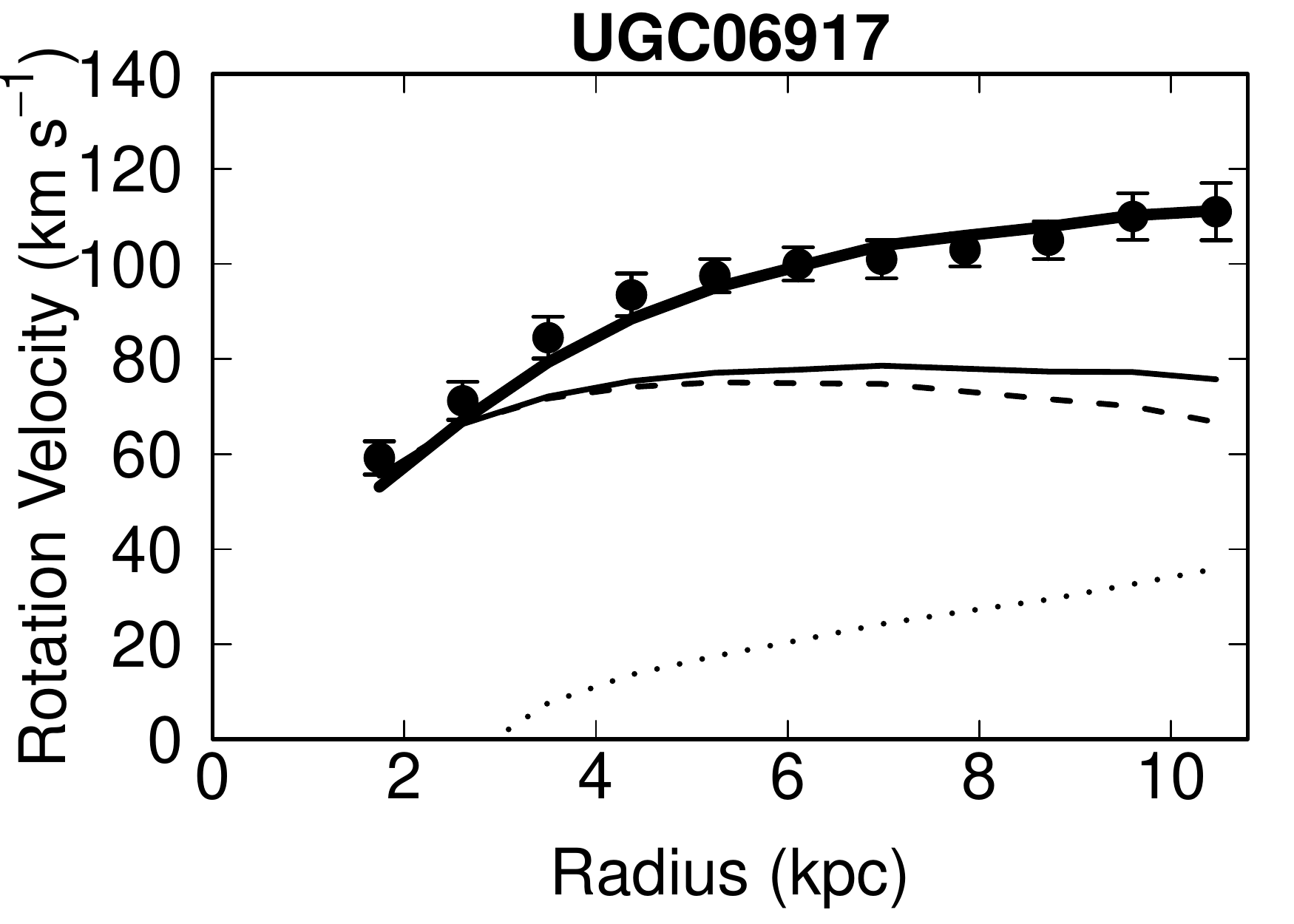}%
\includegraphics[width=60mm]{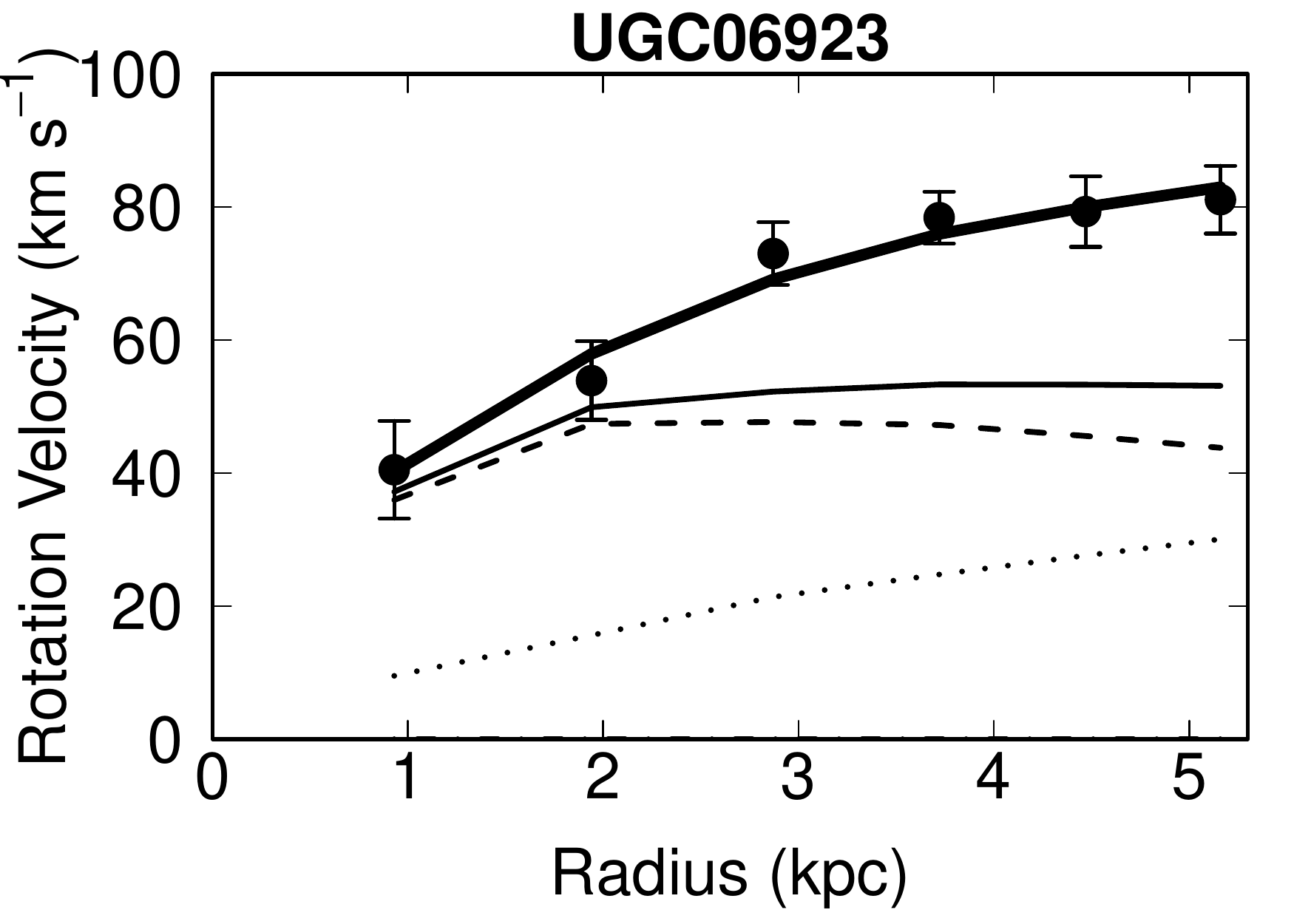}%
\includegraphics[width=60mm]{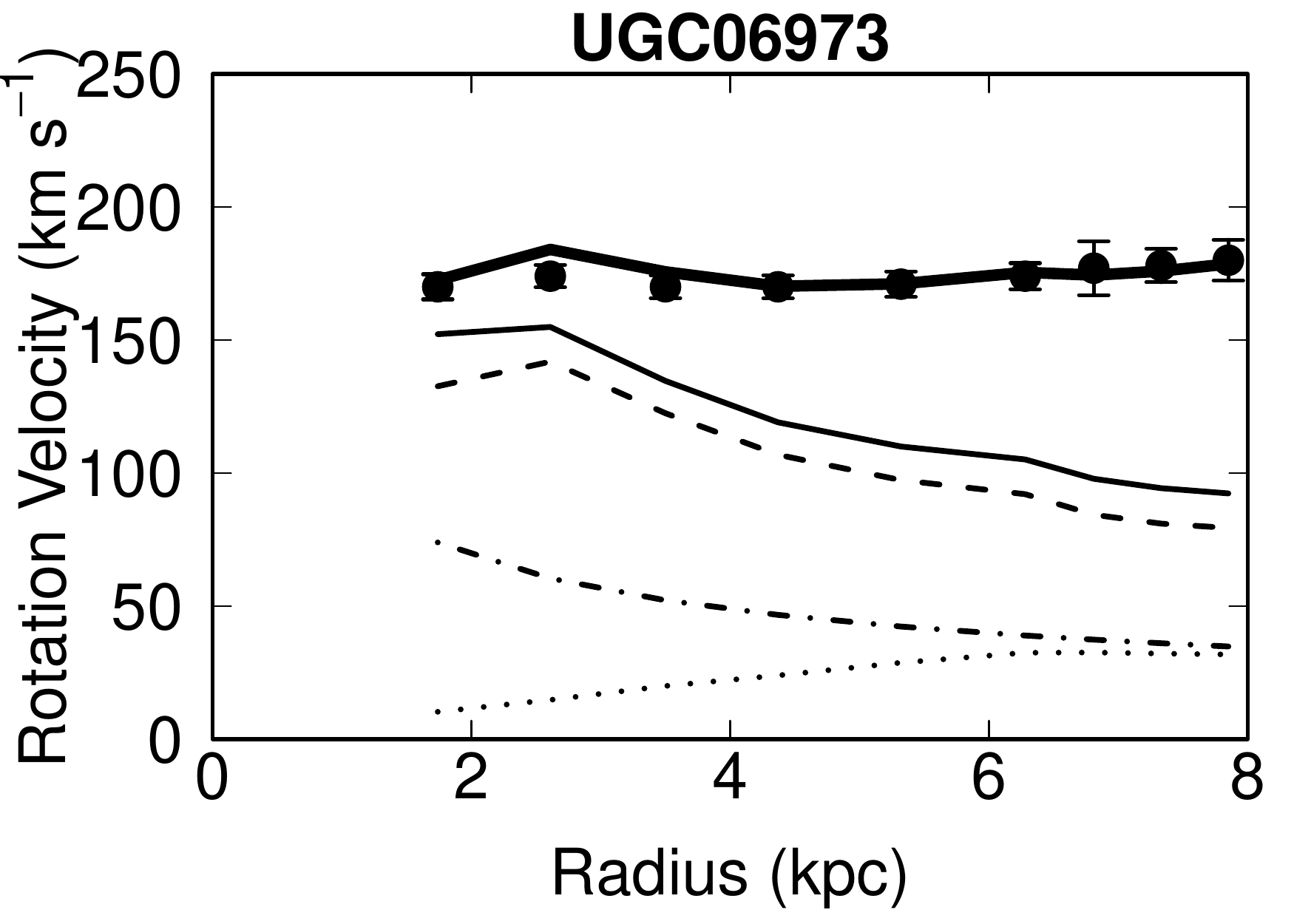}%
\\ \ \\
\includegraphics[width=60mm]{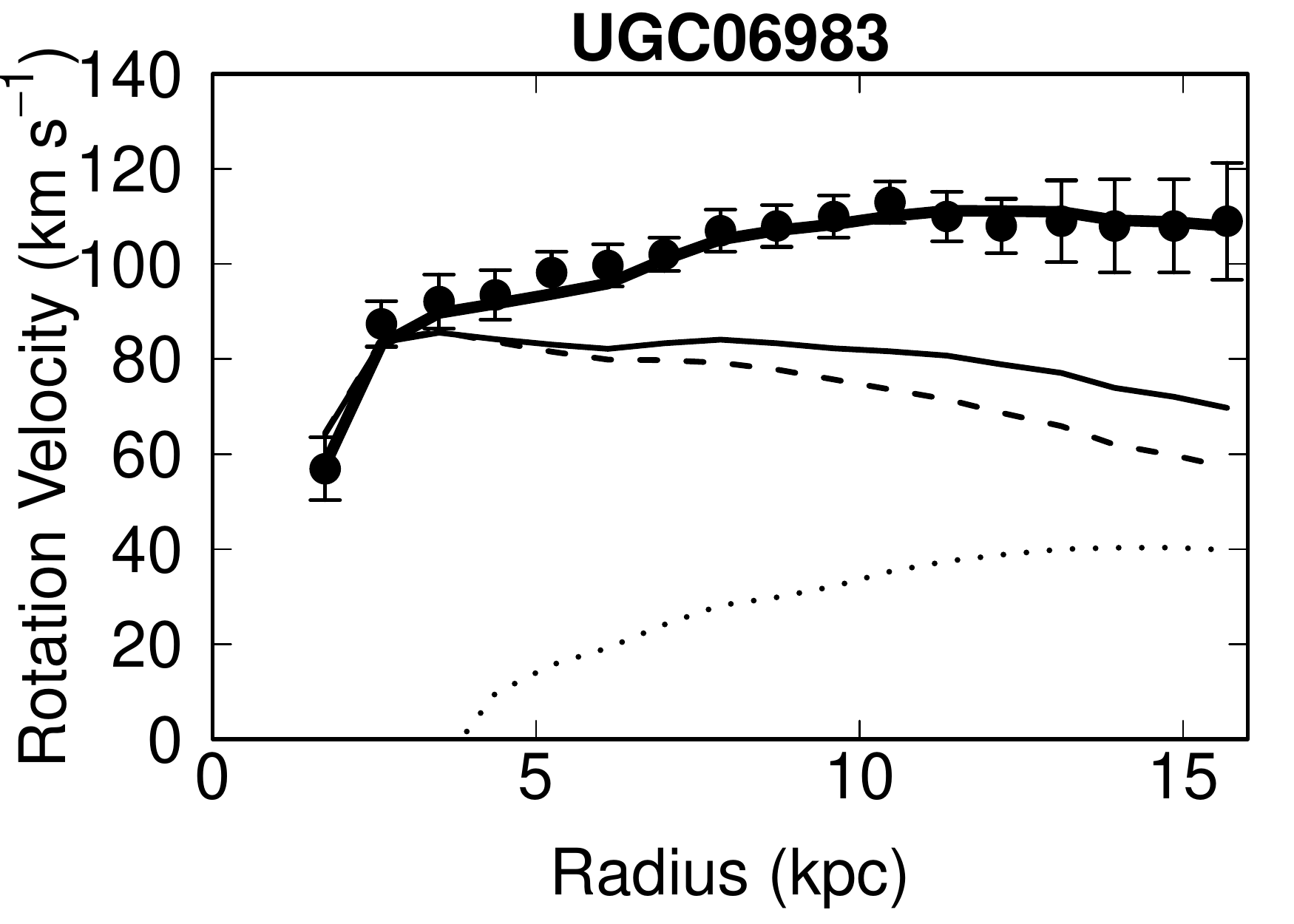}%
\includegraphics[width=60mm]{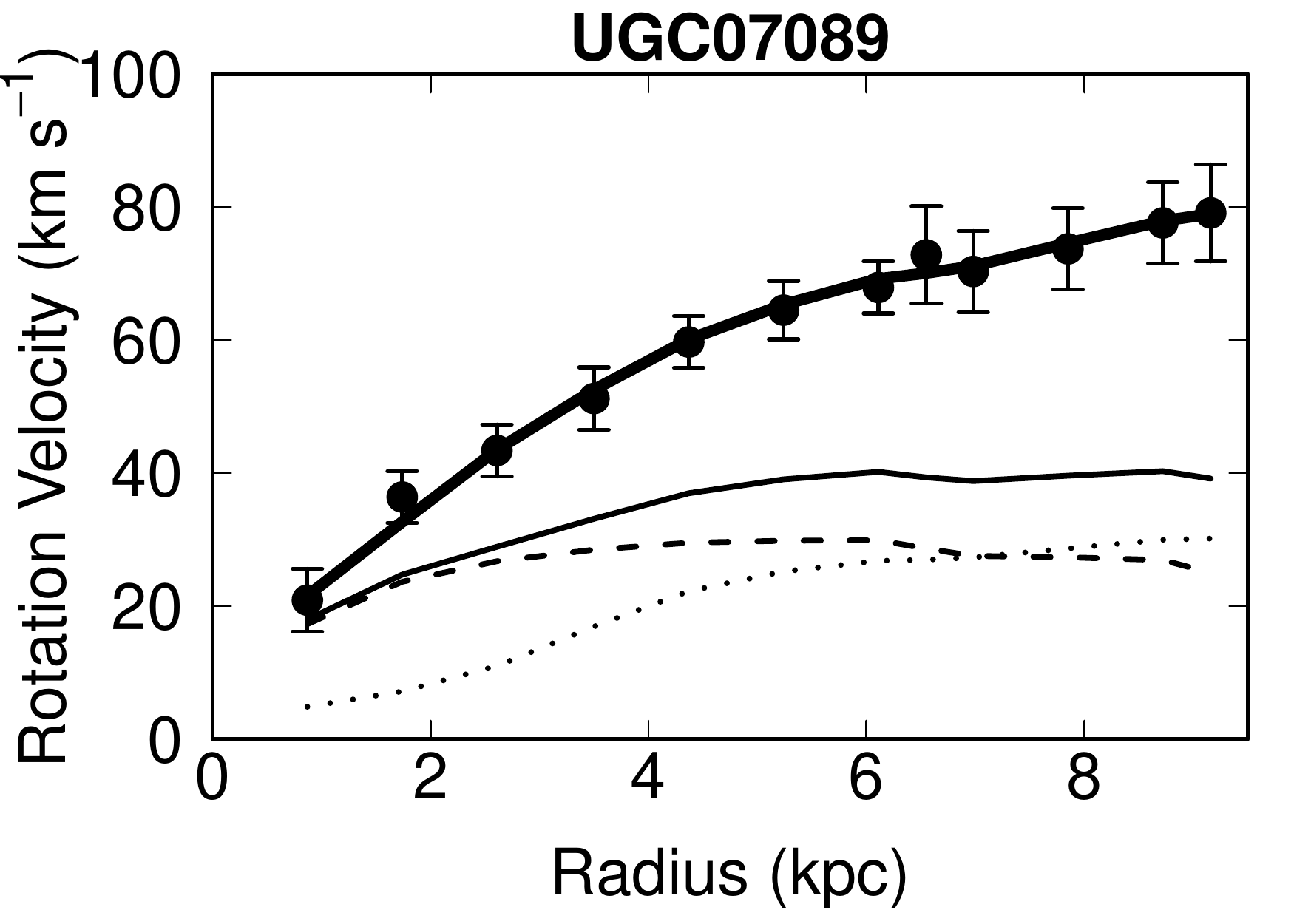}%
\includegraphics[width=60mm]{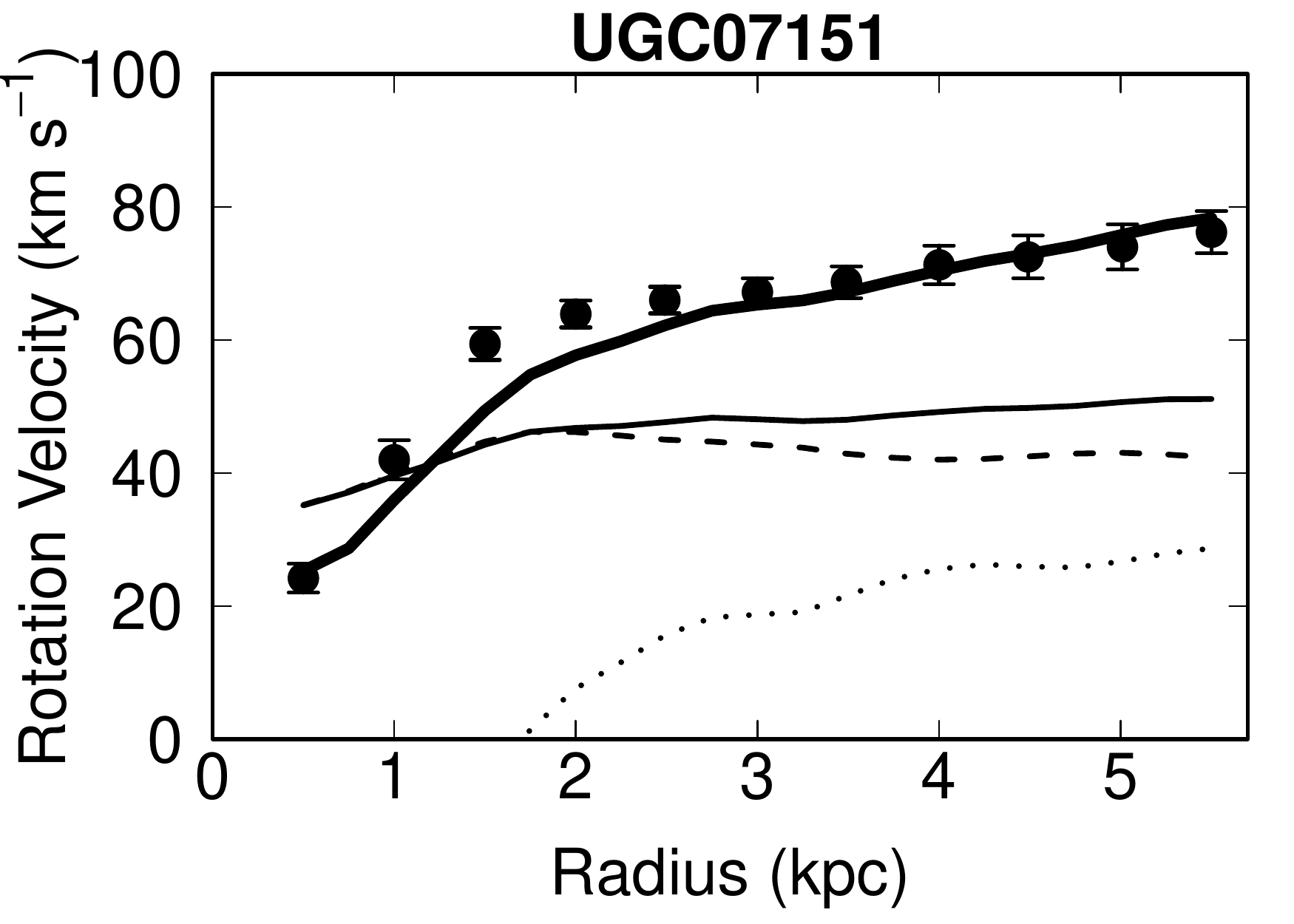}%
\\ \ \\
\includegraphics[width=60mm]{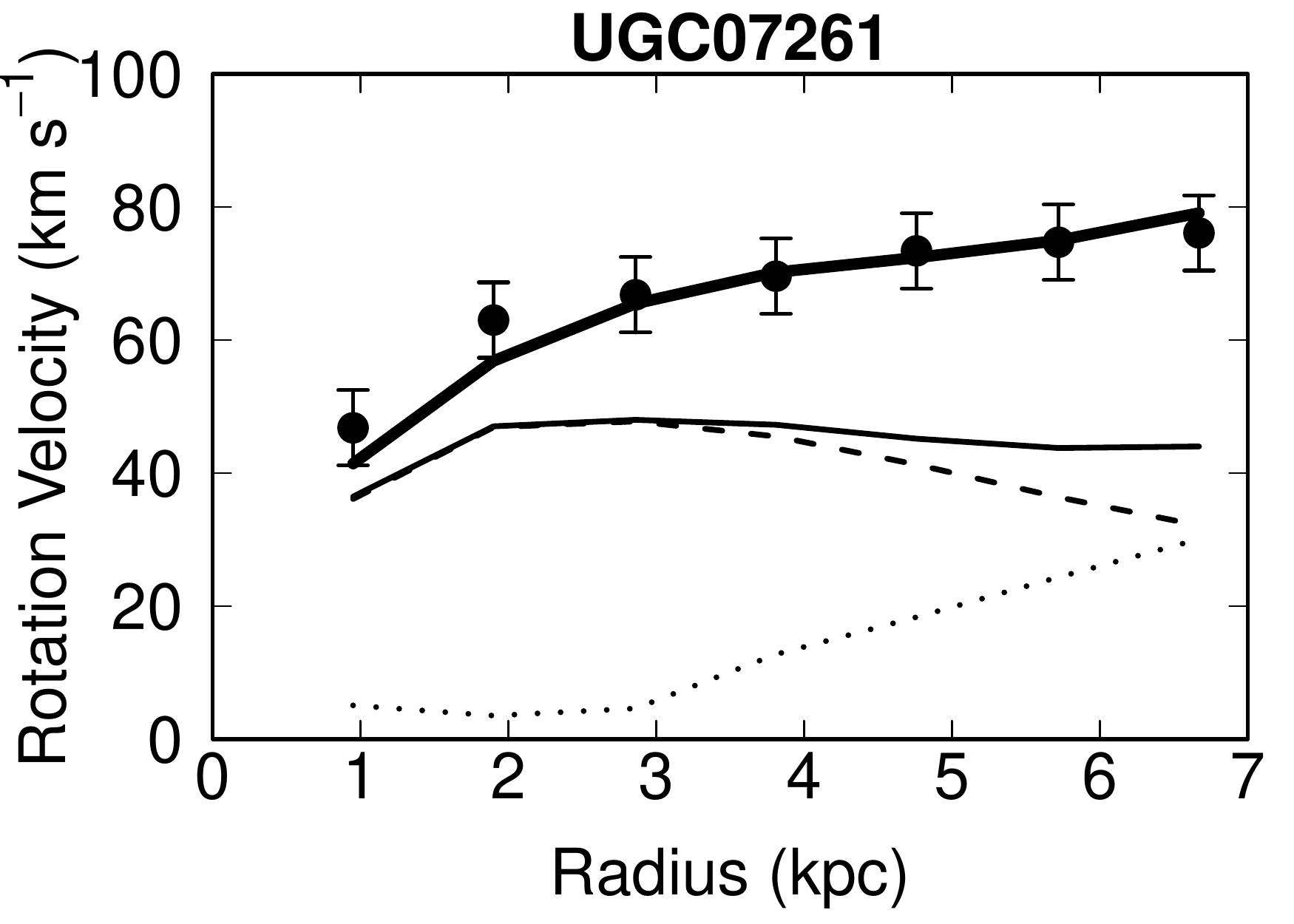}%
\includegraphics[width=60mm]{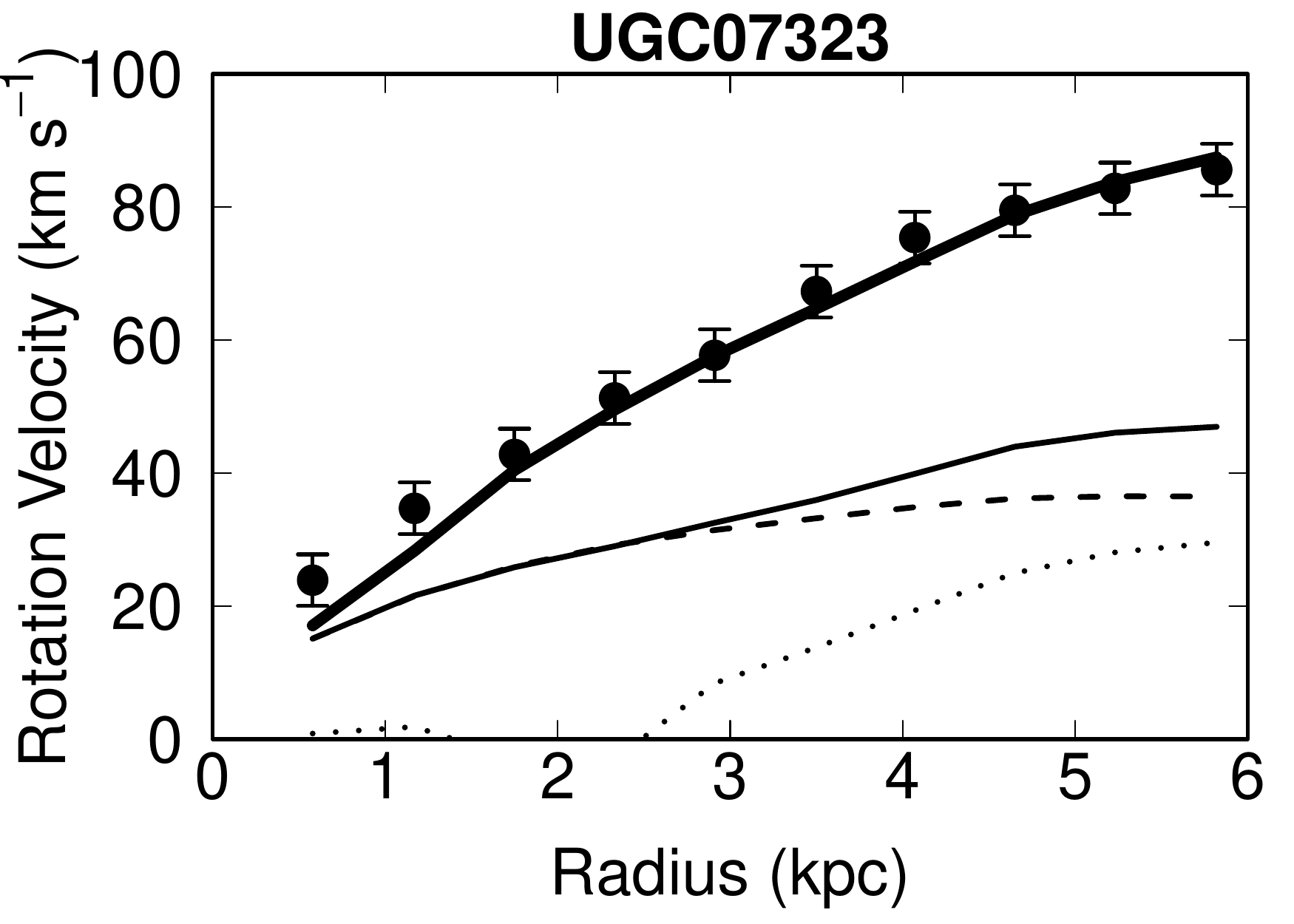}%
\includegraphics[width=60mm]{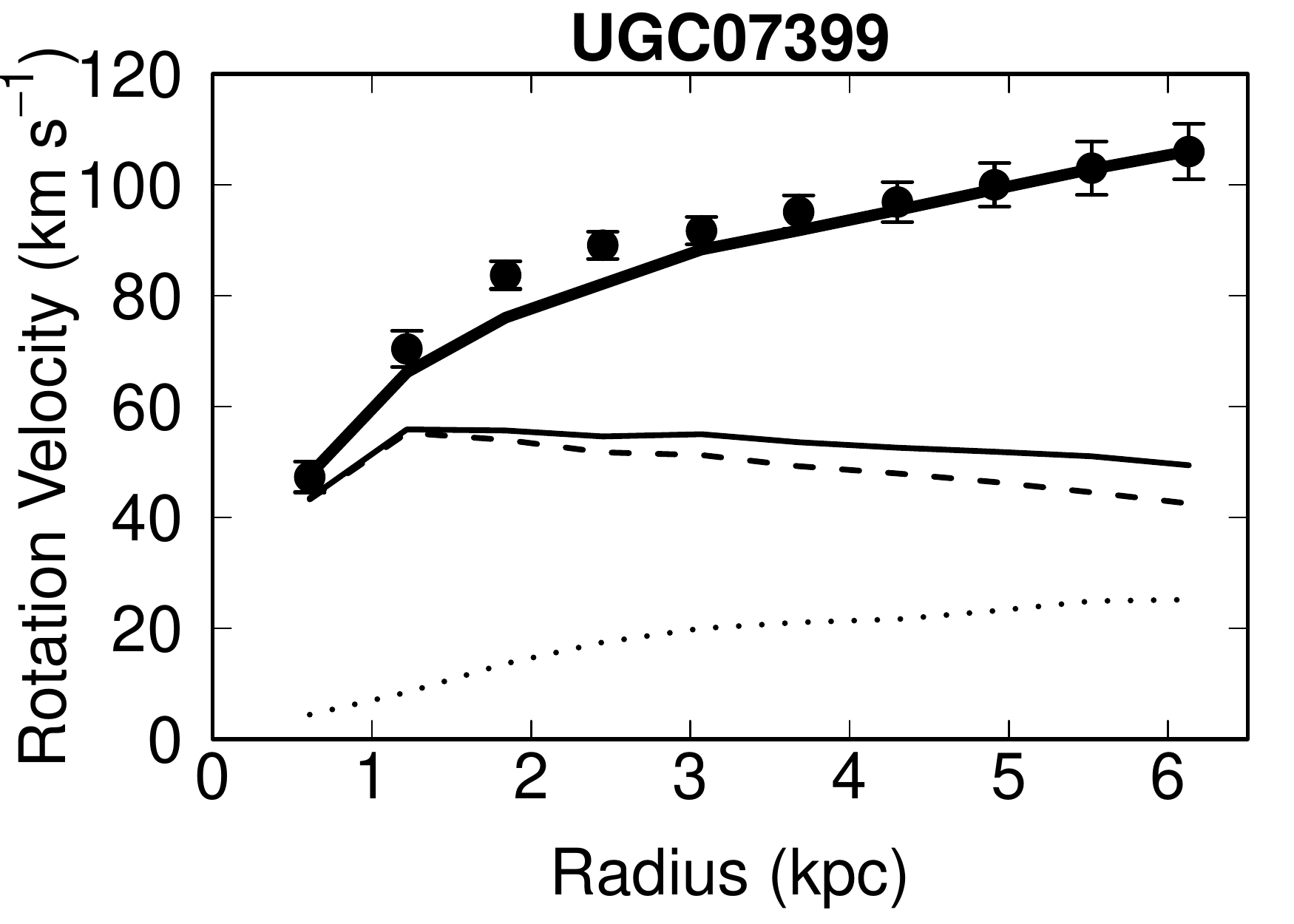}%
\\ \ \\
\includegraphics[width=60mm]{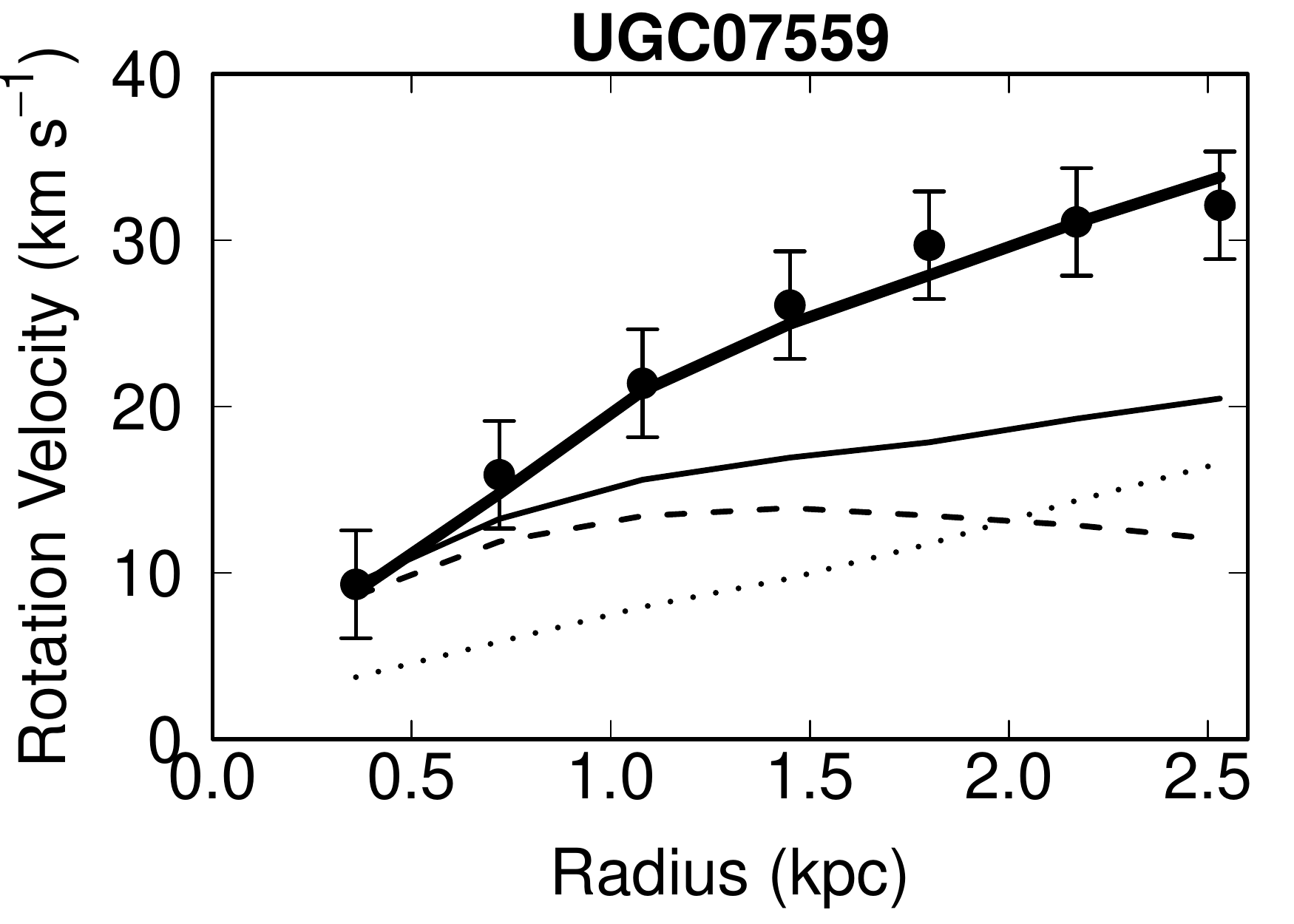}%
\includegraphics[width=60mm]{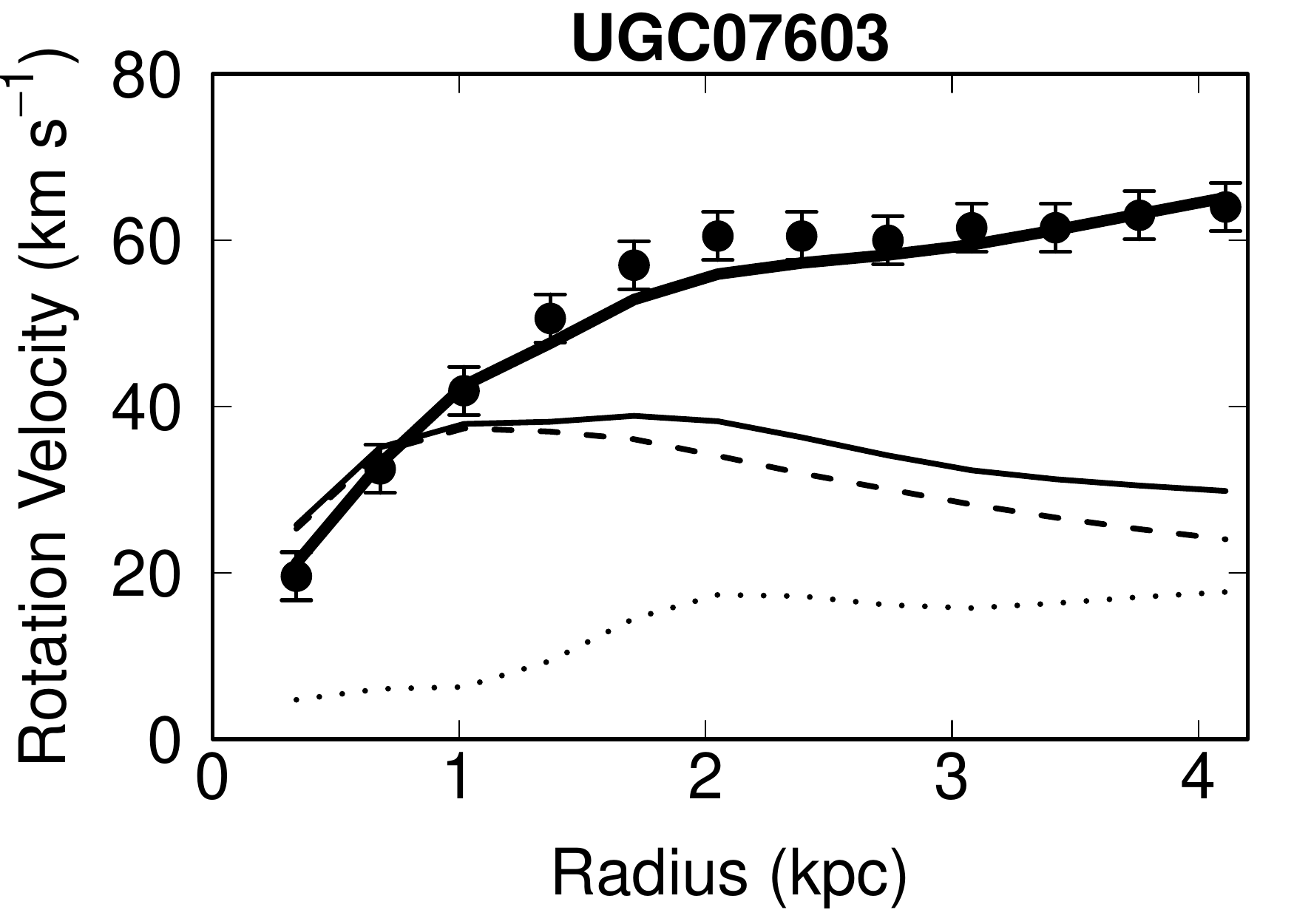}%
\includegraphics[width=60mm]{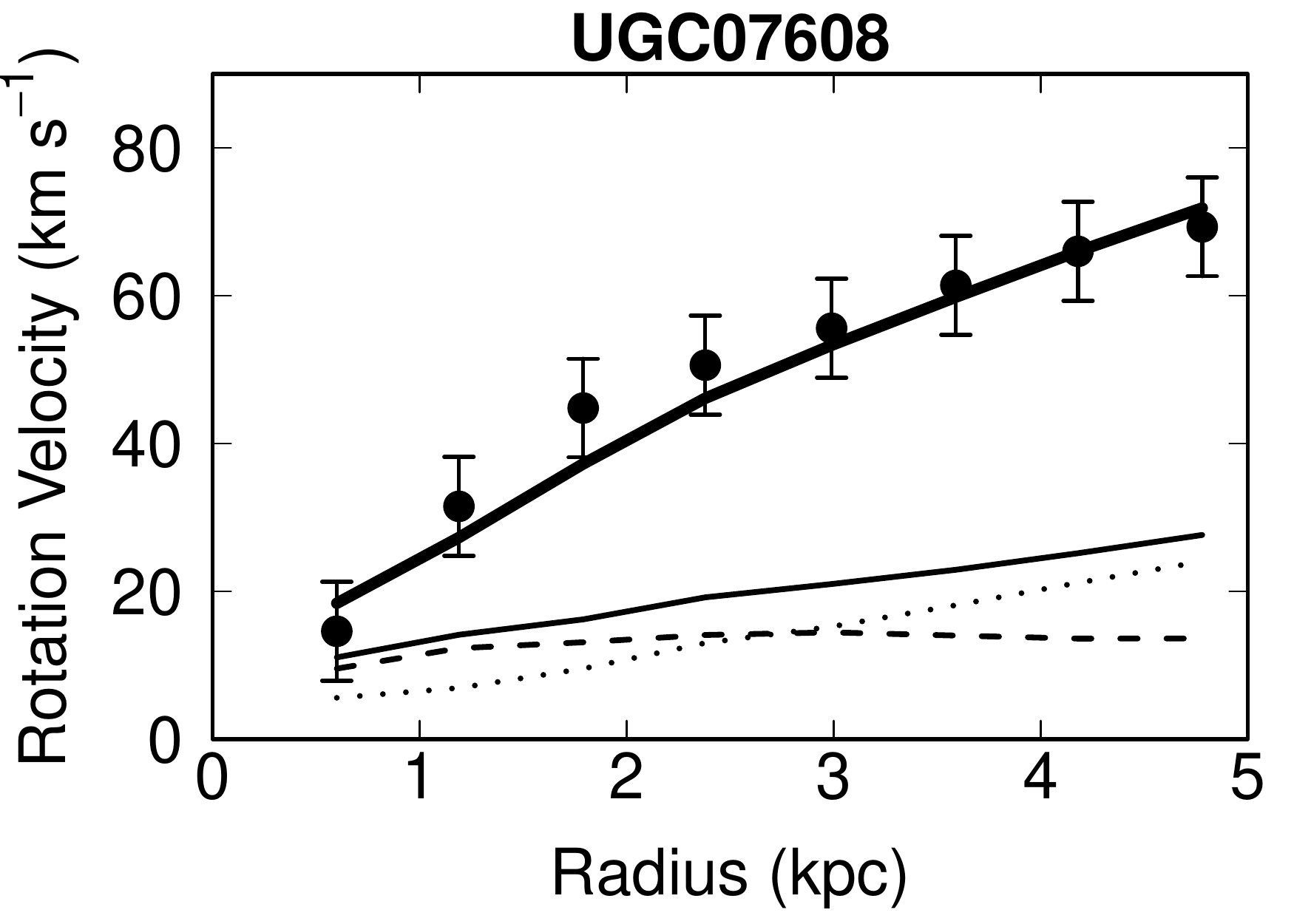}%
\caption{\label{fig:RCfull6} \textit{(continued)}.}
\end{figure*}

\addtocounter{figure}{-1}
\begin{figure*}[b]
\includegraphics[width=60mm]{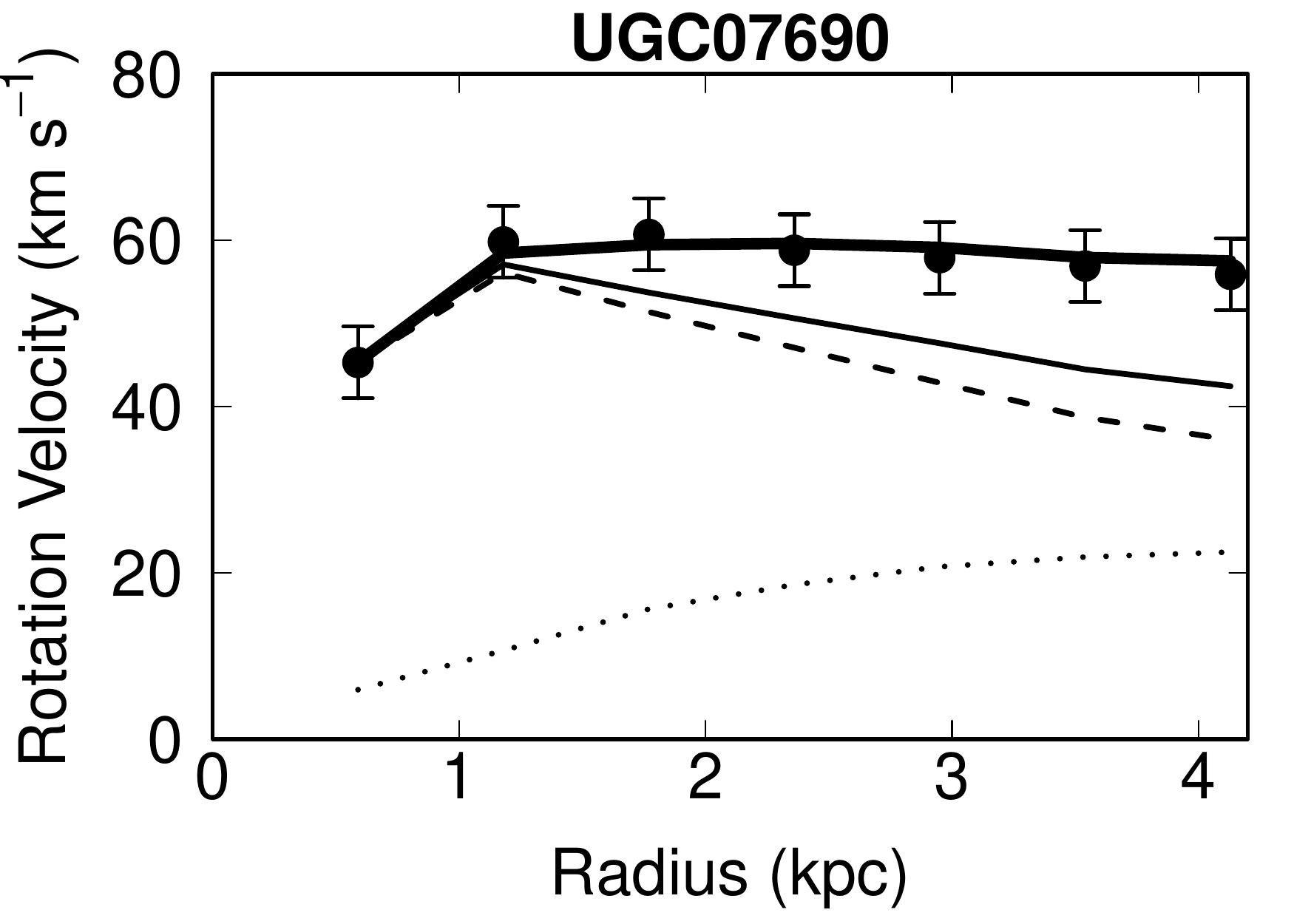}%
\includegraphics[width=60mm]{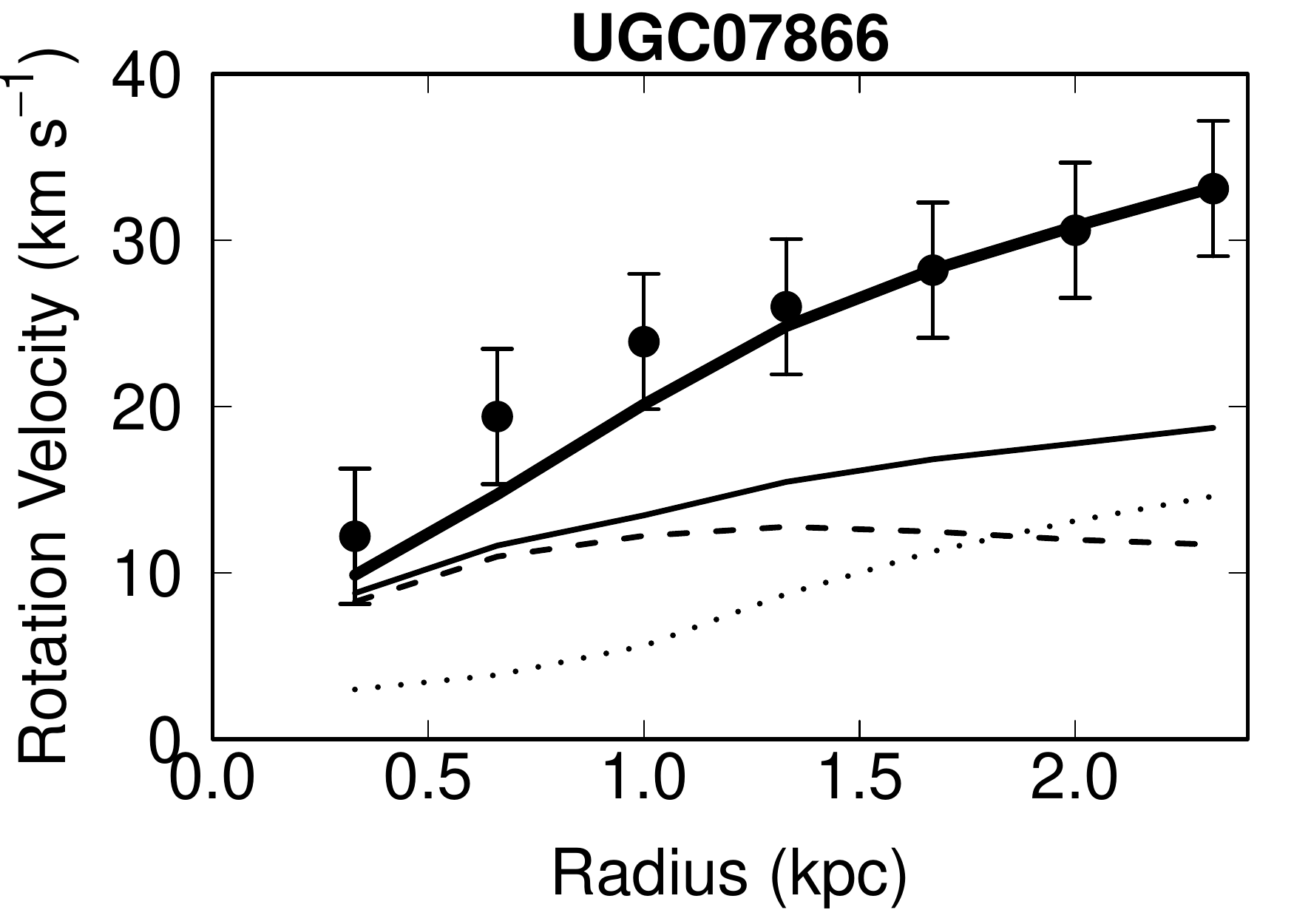}%
\includegraphics[width=60mm]{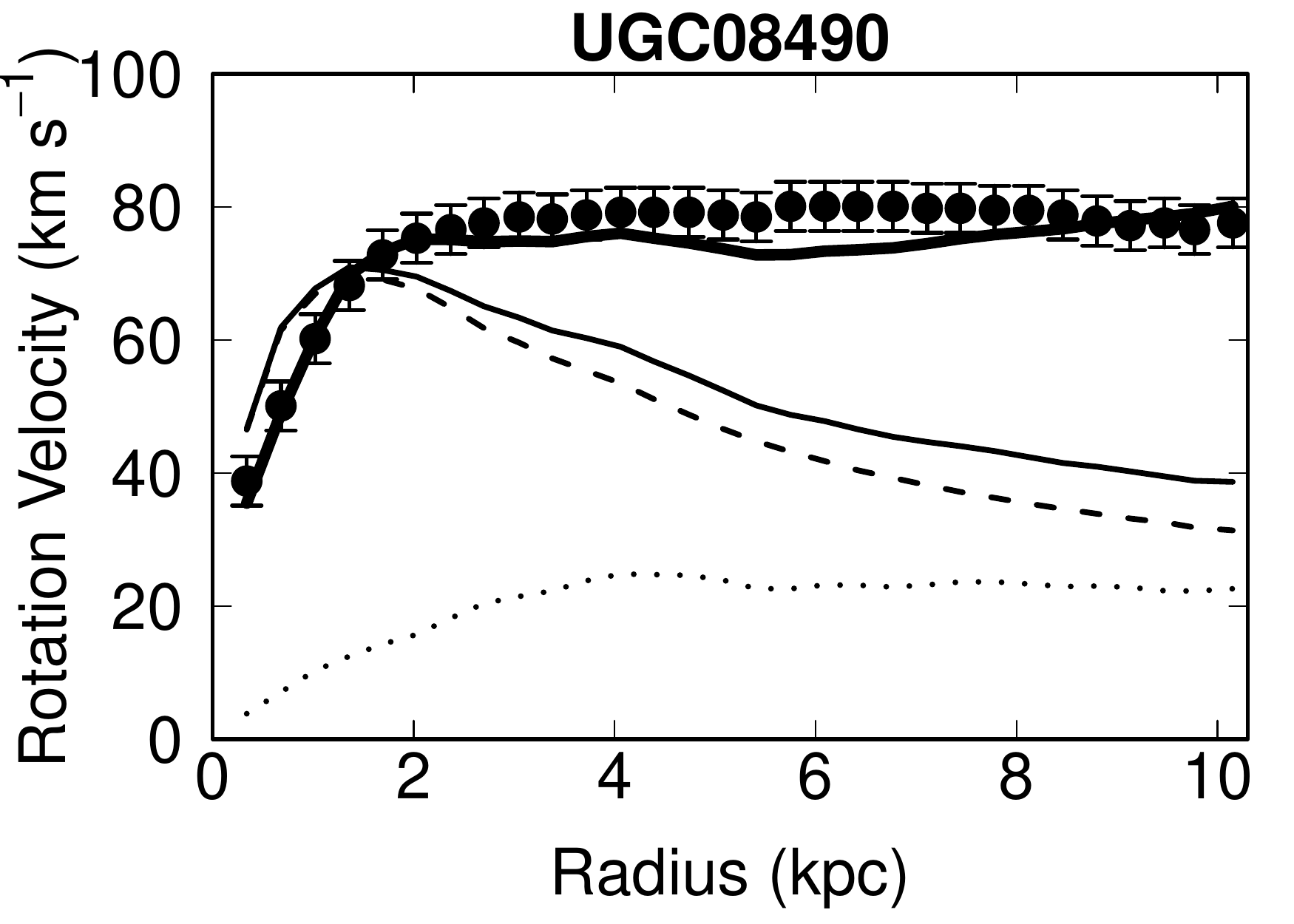}%
\\ \ \\
\includegraphics[width=60mm]{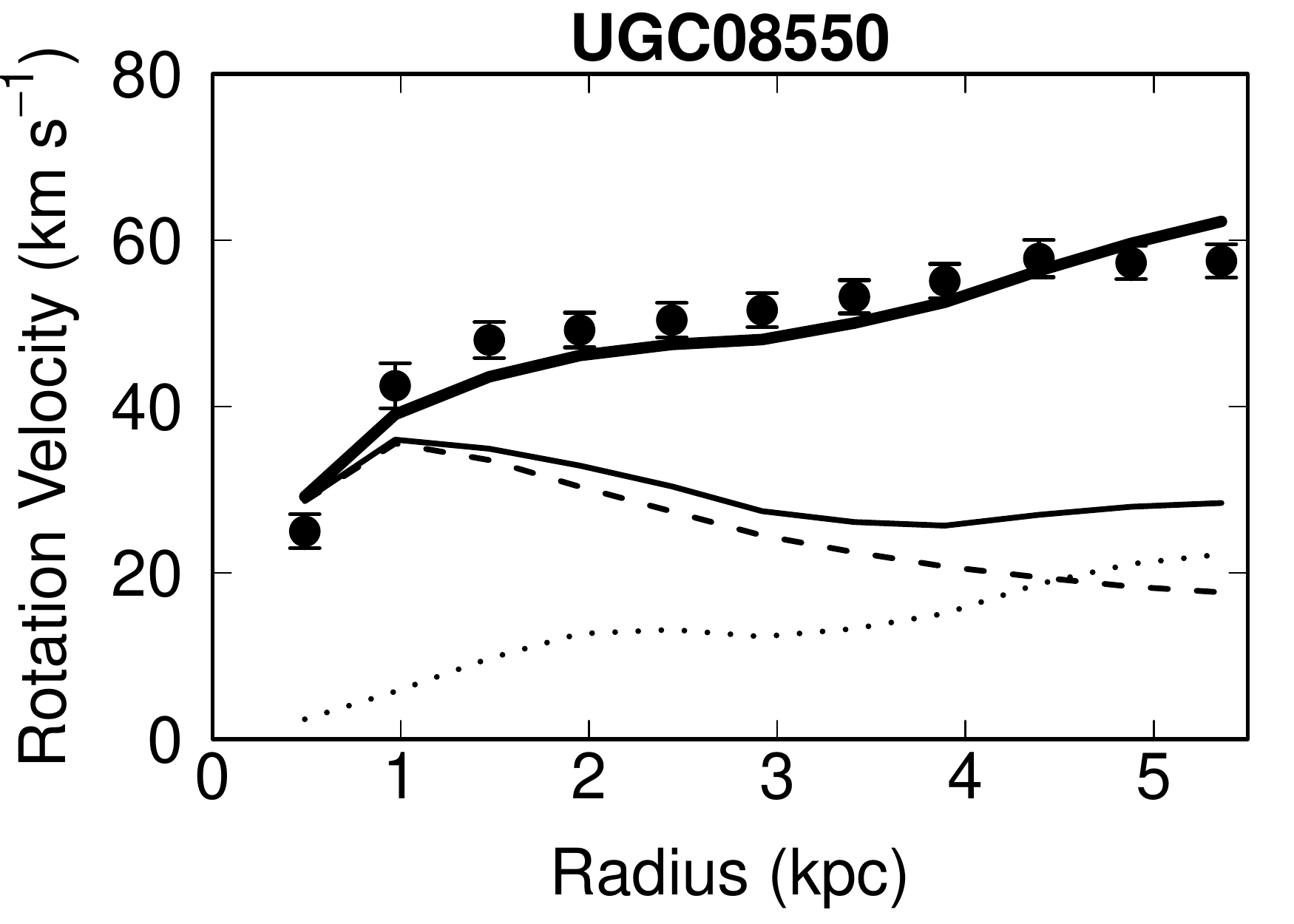}%
\includegraphics[width=60mm]{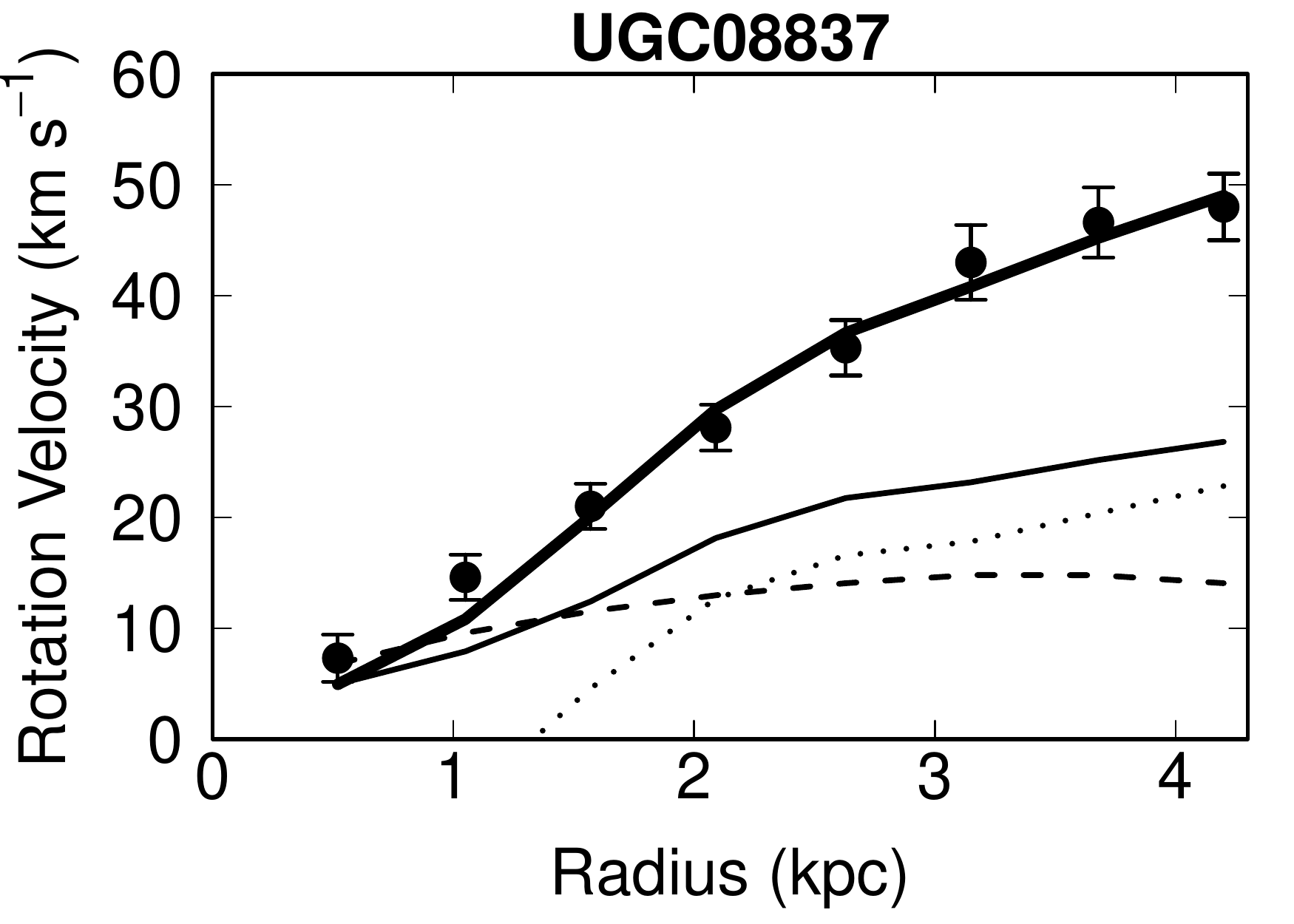}%
\includegraphics[width=60mm]{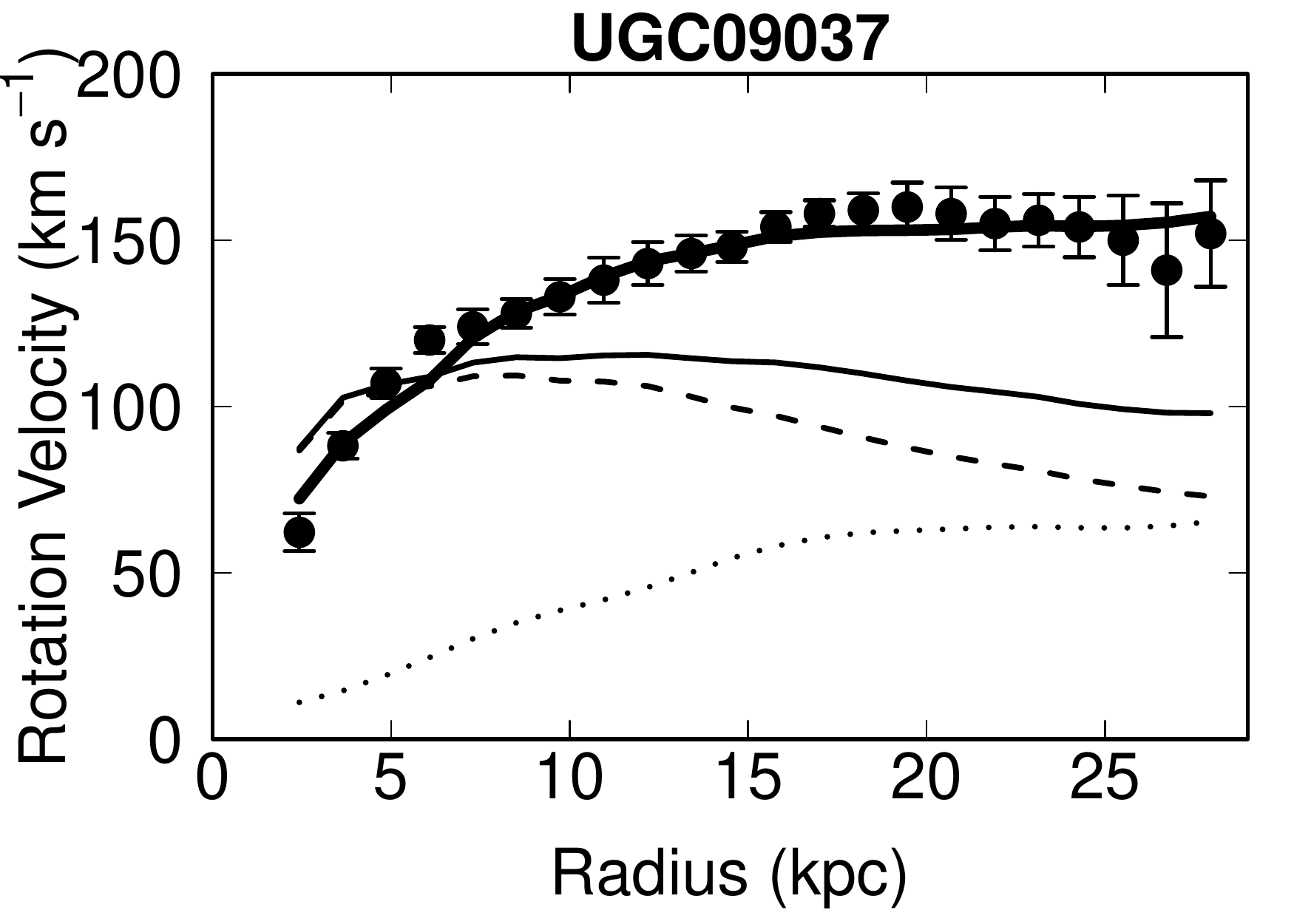}%
\\ \ \\
\includegraphics[width=60mm]{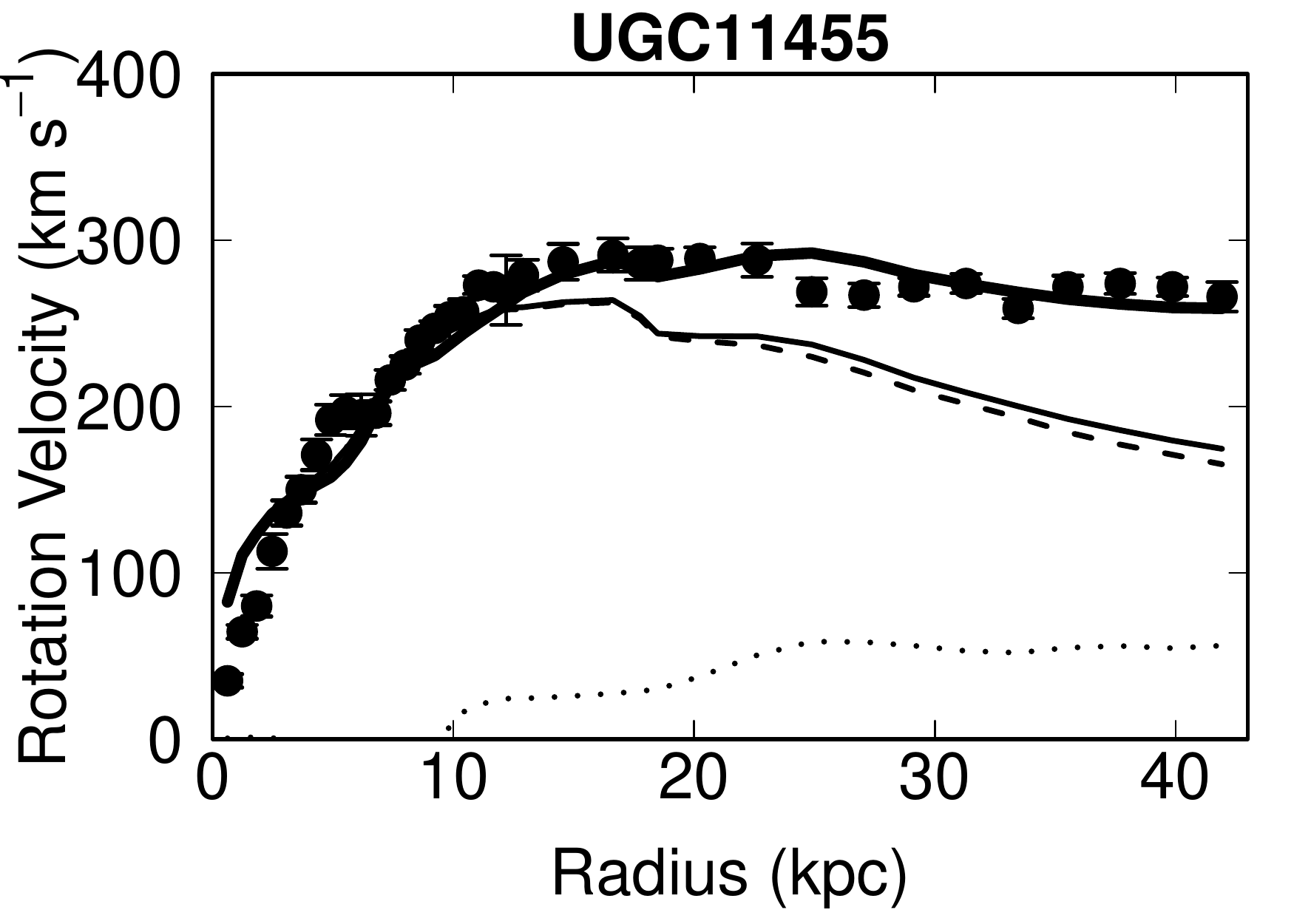}%
\includegraphics[width=60mm]{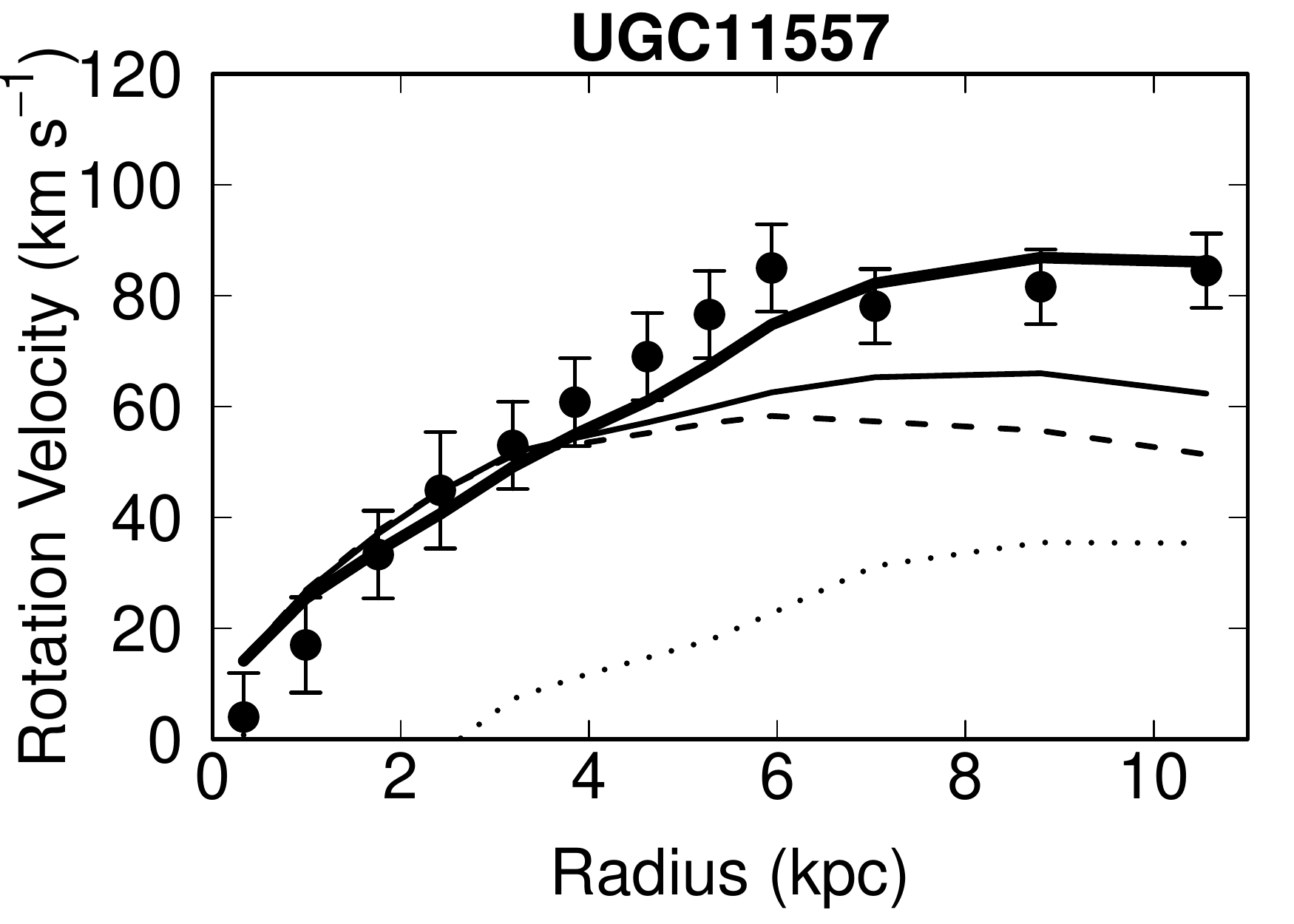}%
\includegraphics[width=60mm]{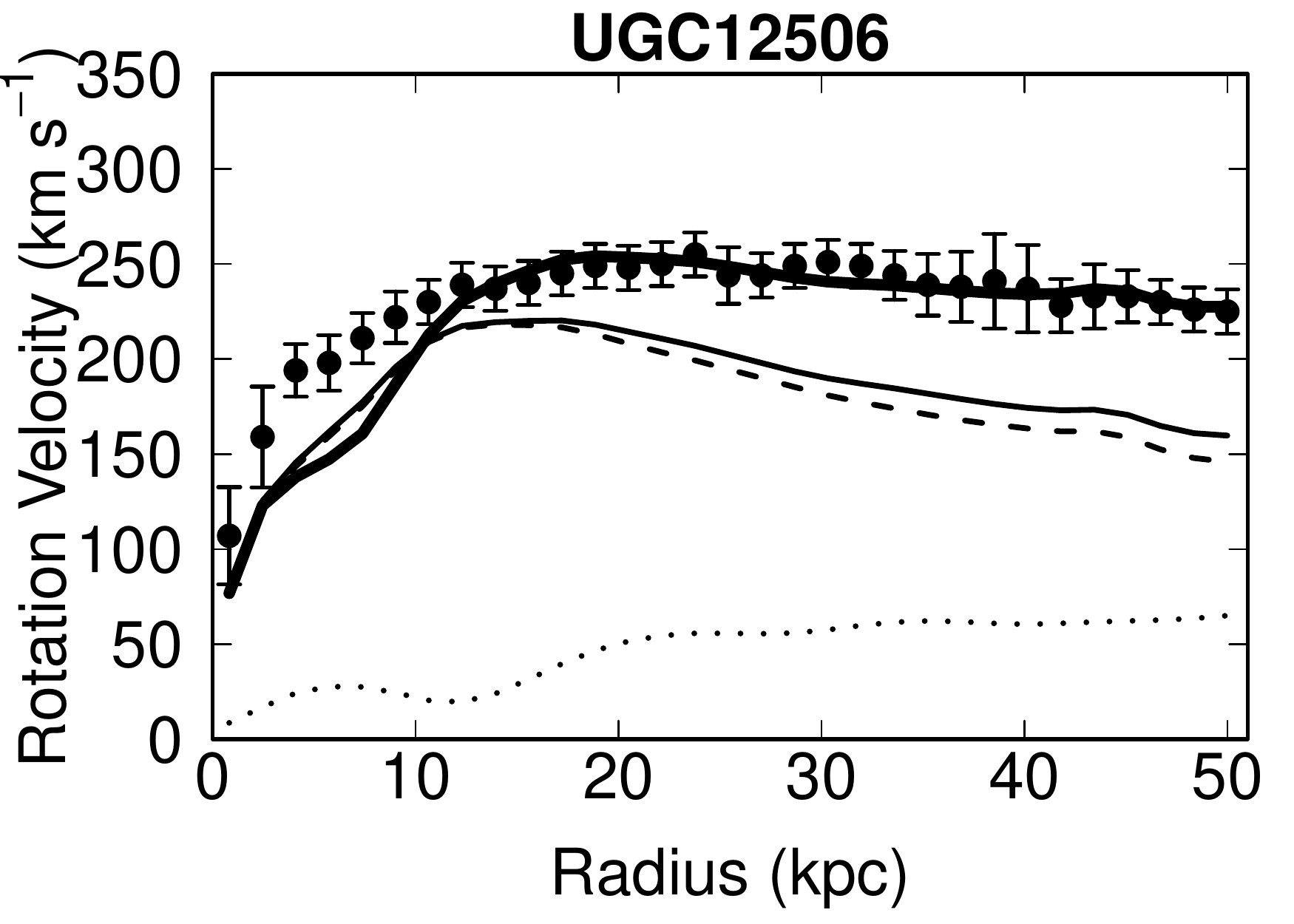}%
\\ \ \\
\includegraphics[width=60mm]{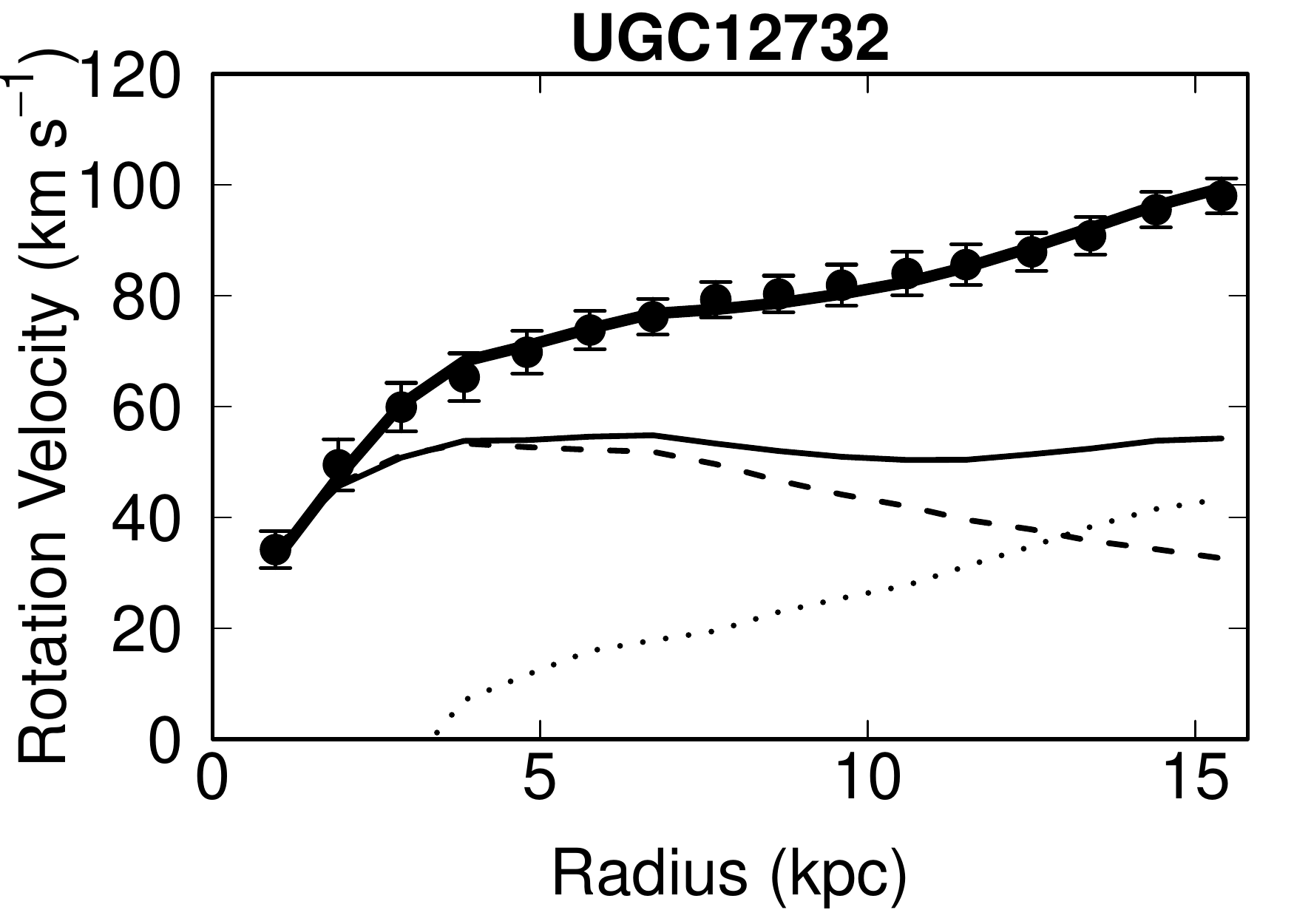}%
\includegraphics[width=60mm]{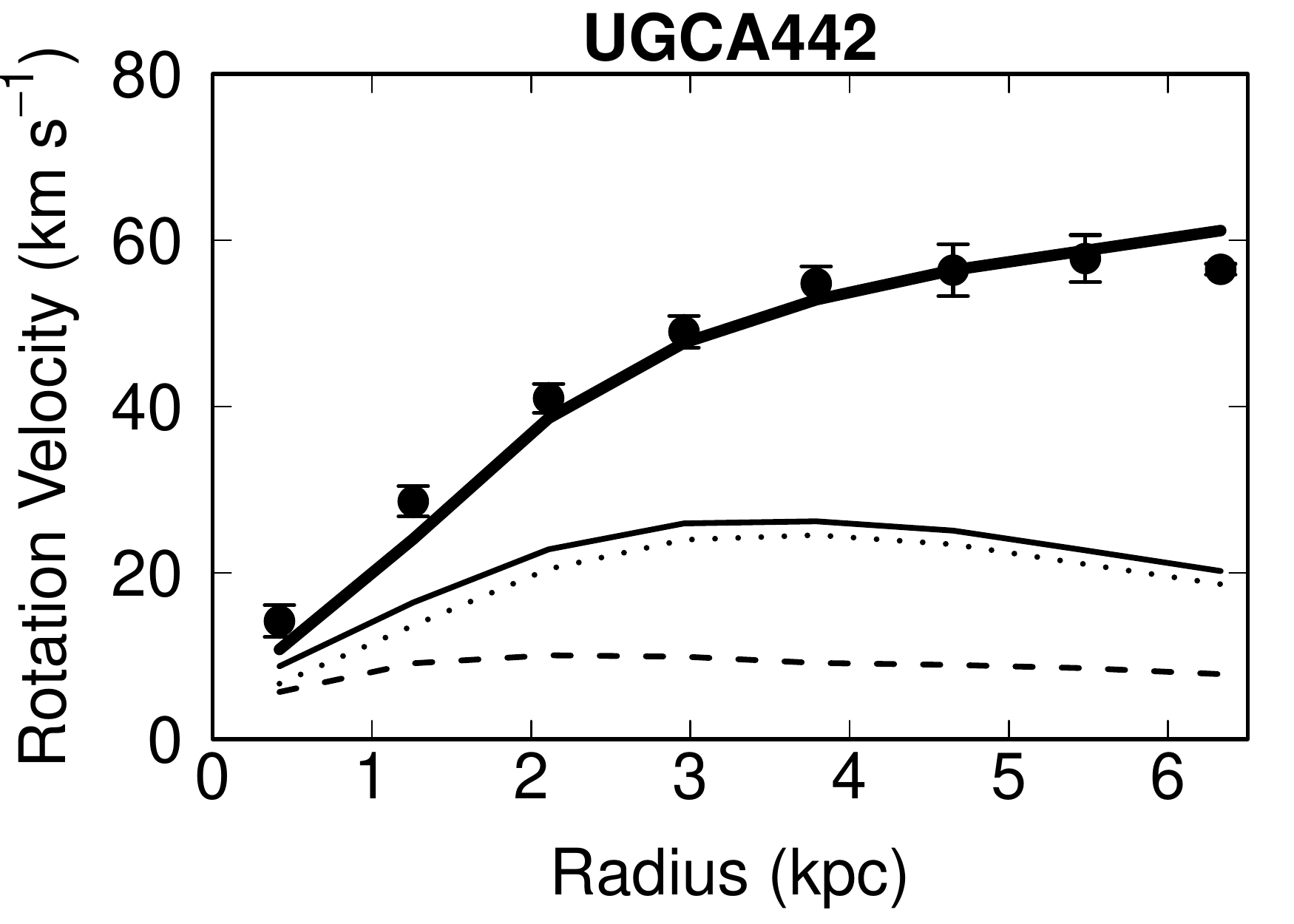}%
\includegraphics[width=60mm]{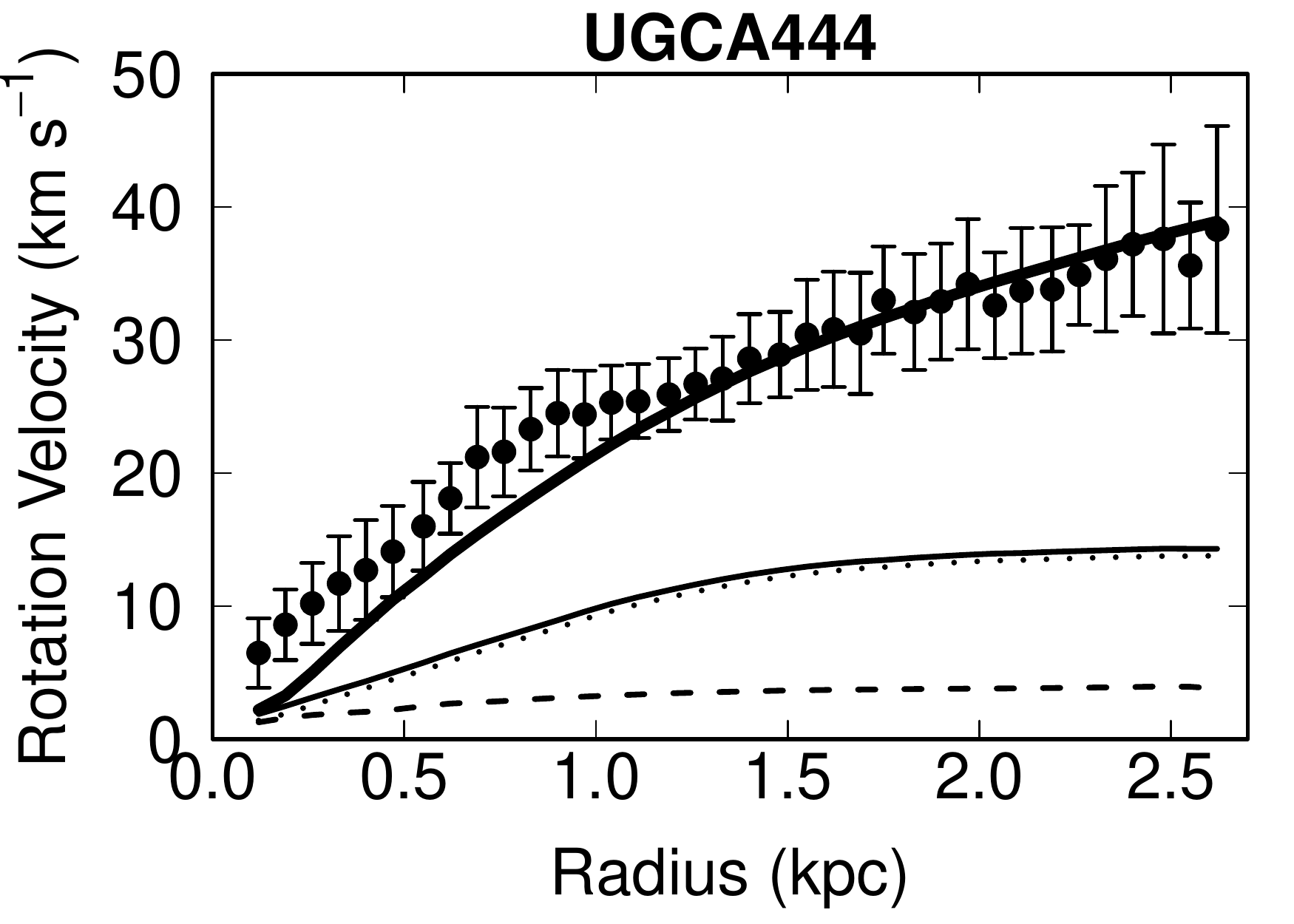}%
\caption{\label{fig:RCfull7} \textit{(continued)}.}
\end{figure*}

\subsection{Radial acceleration relation}

\begin{figure*}[t]
\includegraphics[width=90mm]{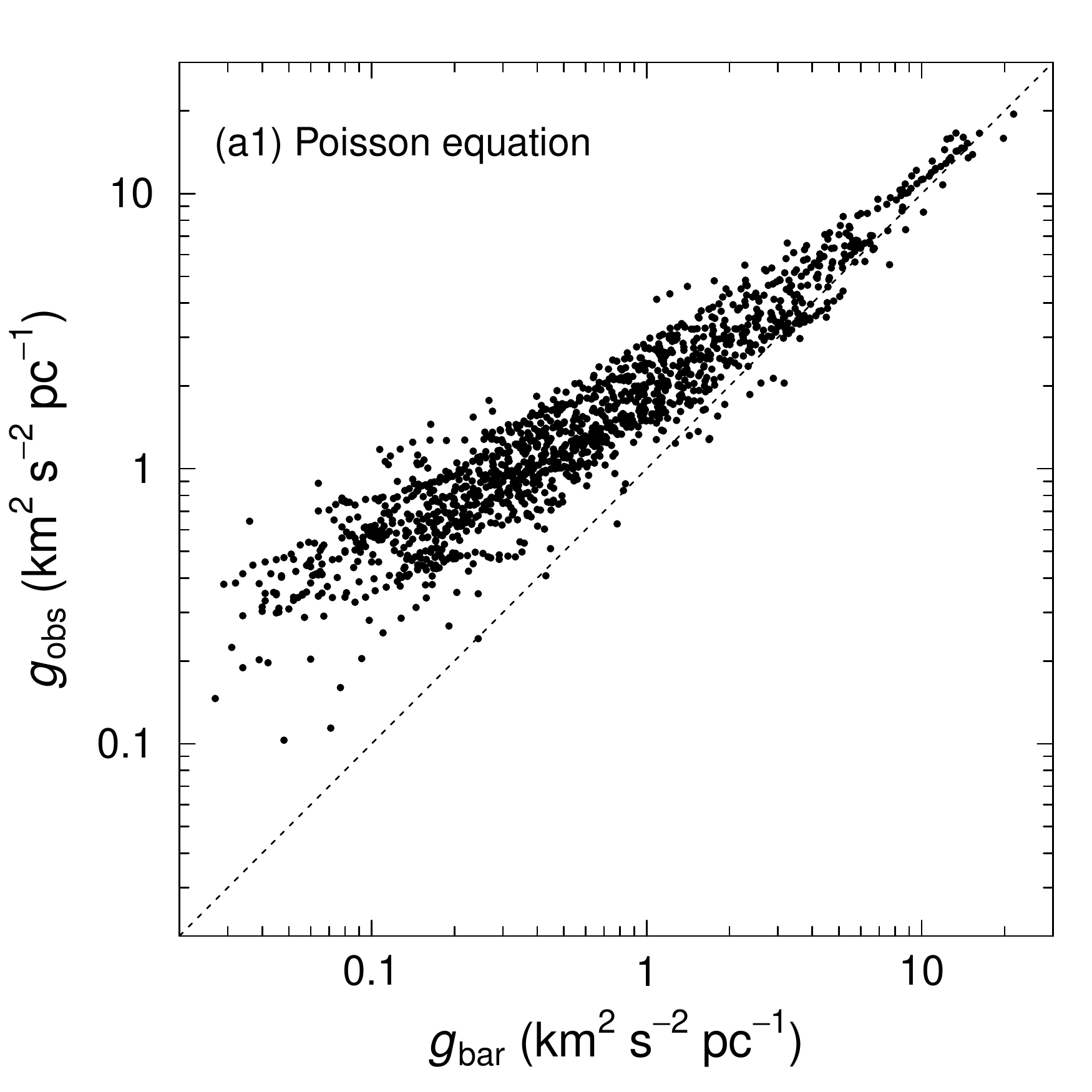}%
\includegraphics[width=90mm]{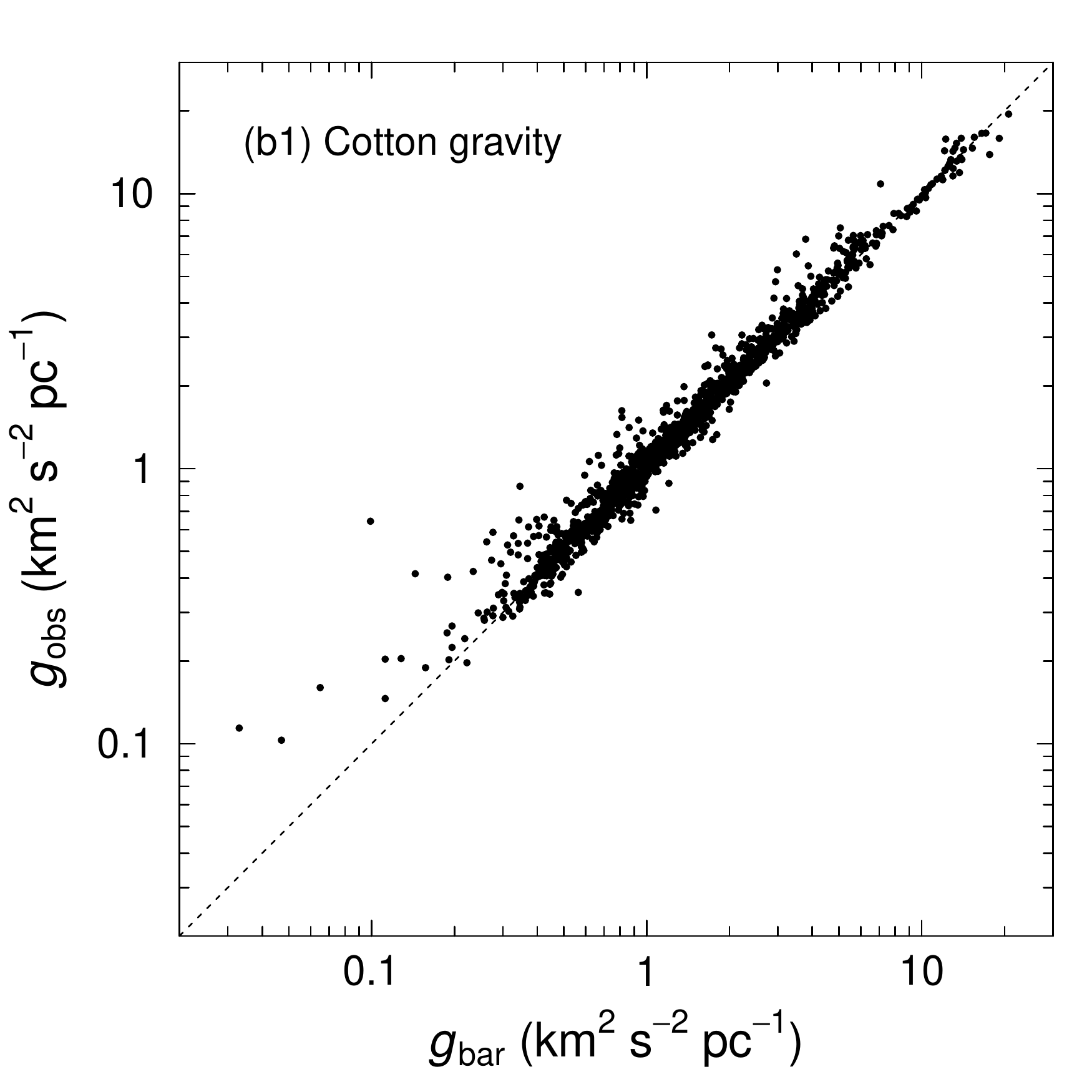}%
\\
\includegraphics[width=90mm]{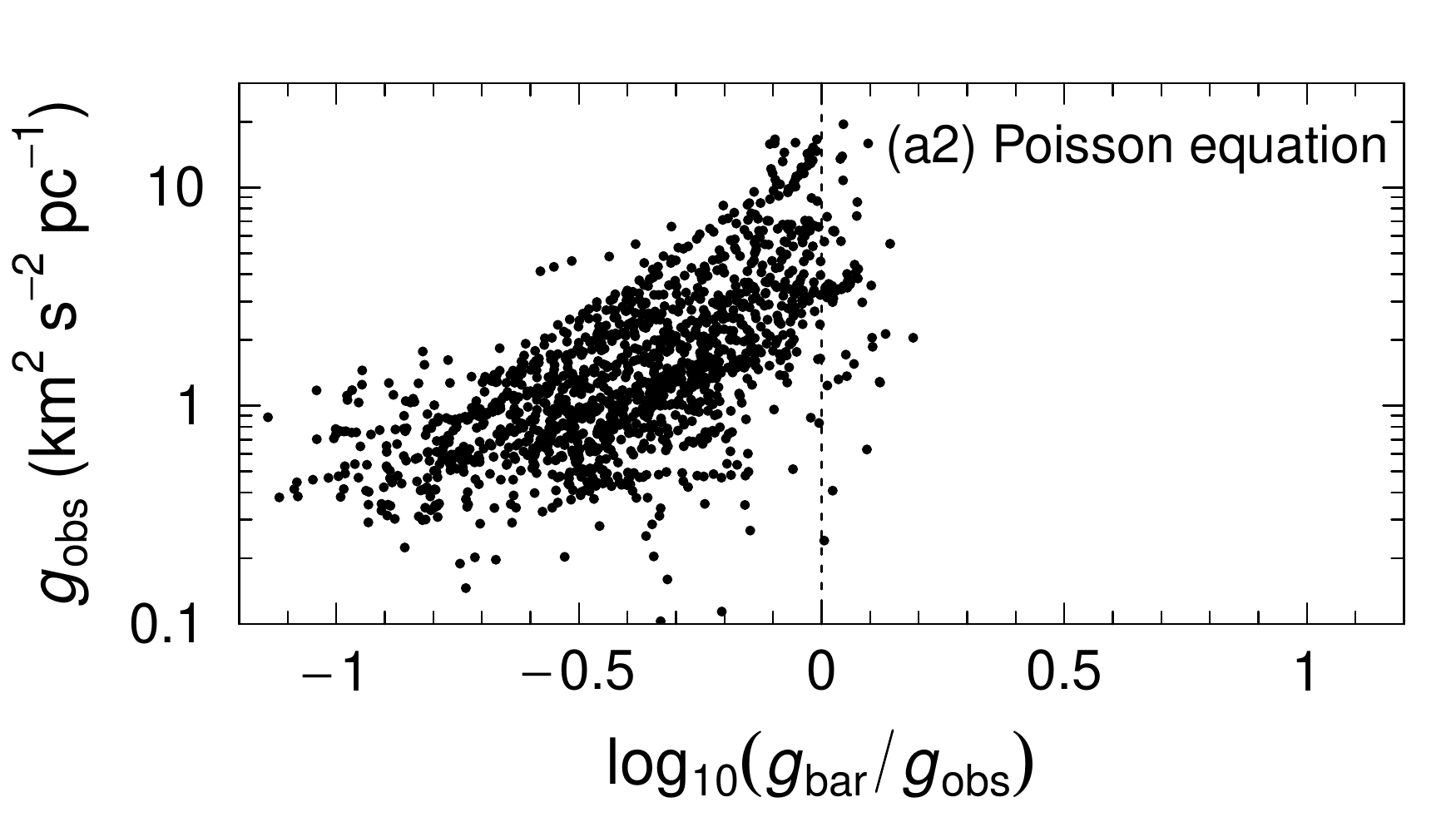}%
\includegraphics[width=90mm]{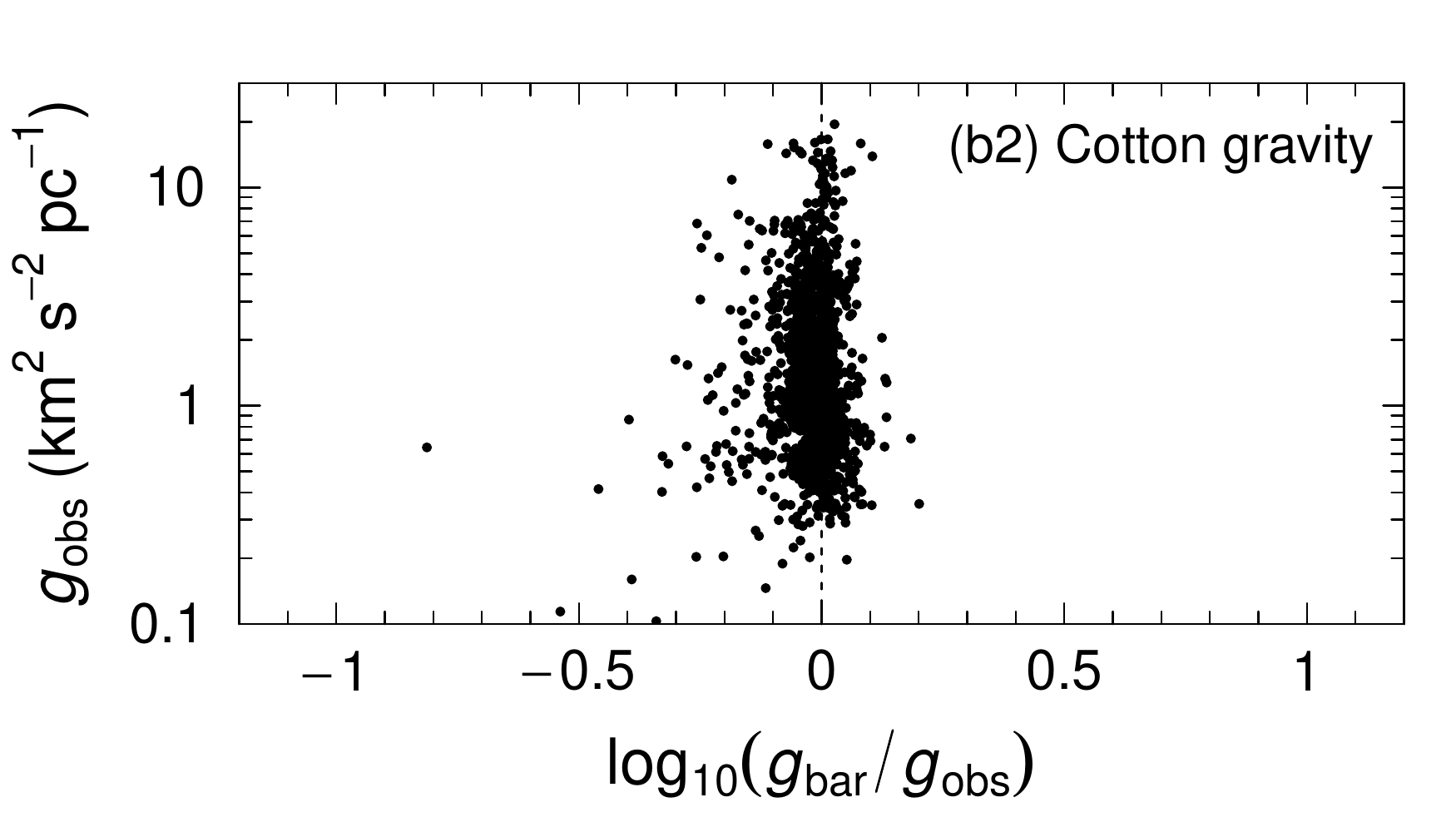}%
\\
\includegraphics[width=90mm]{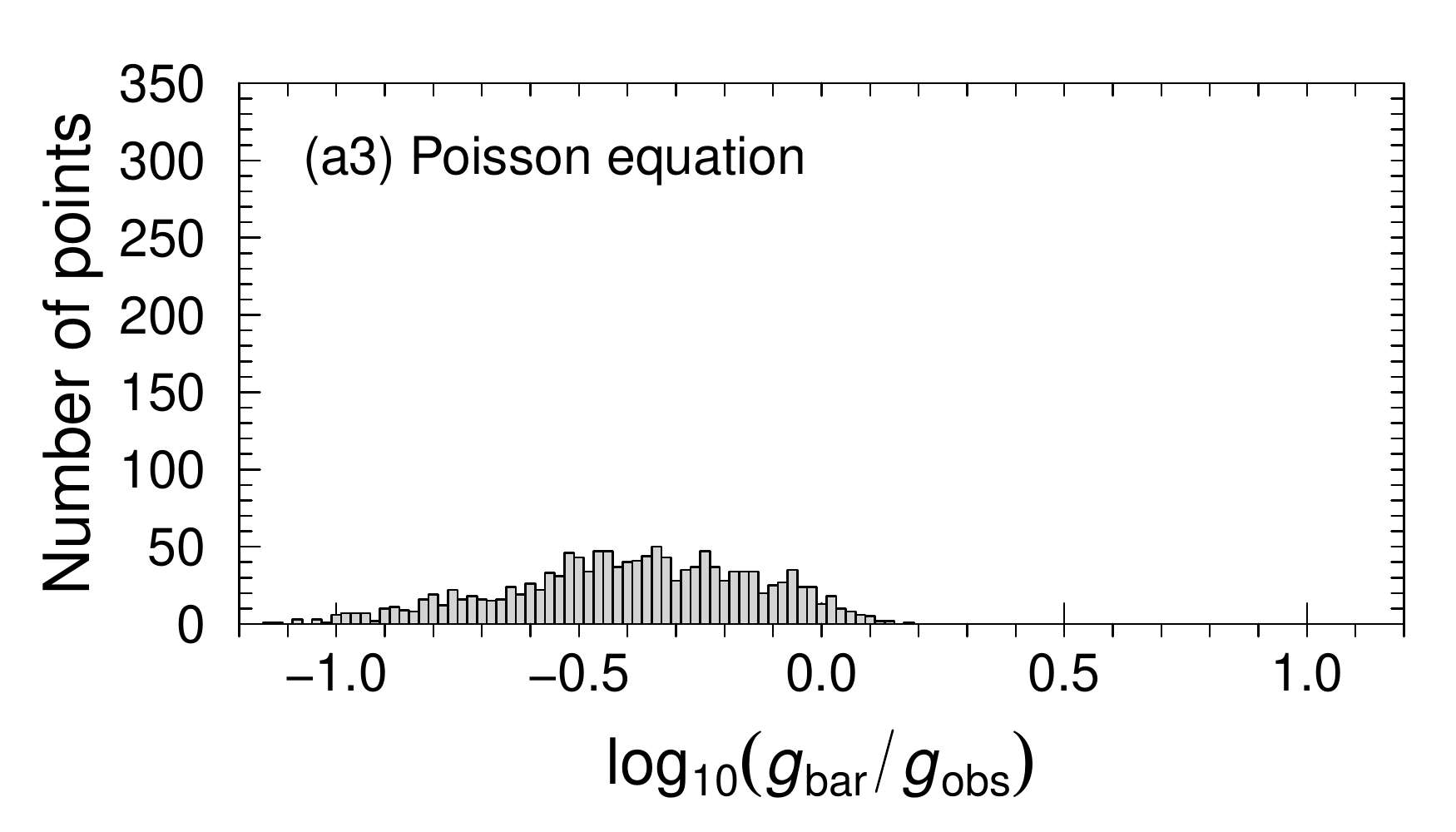}%
\includegraphics[width=90mm]{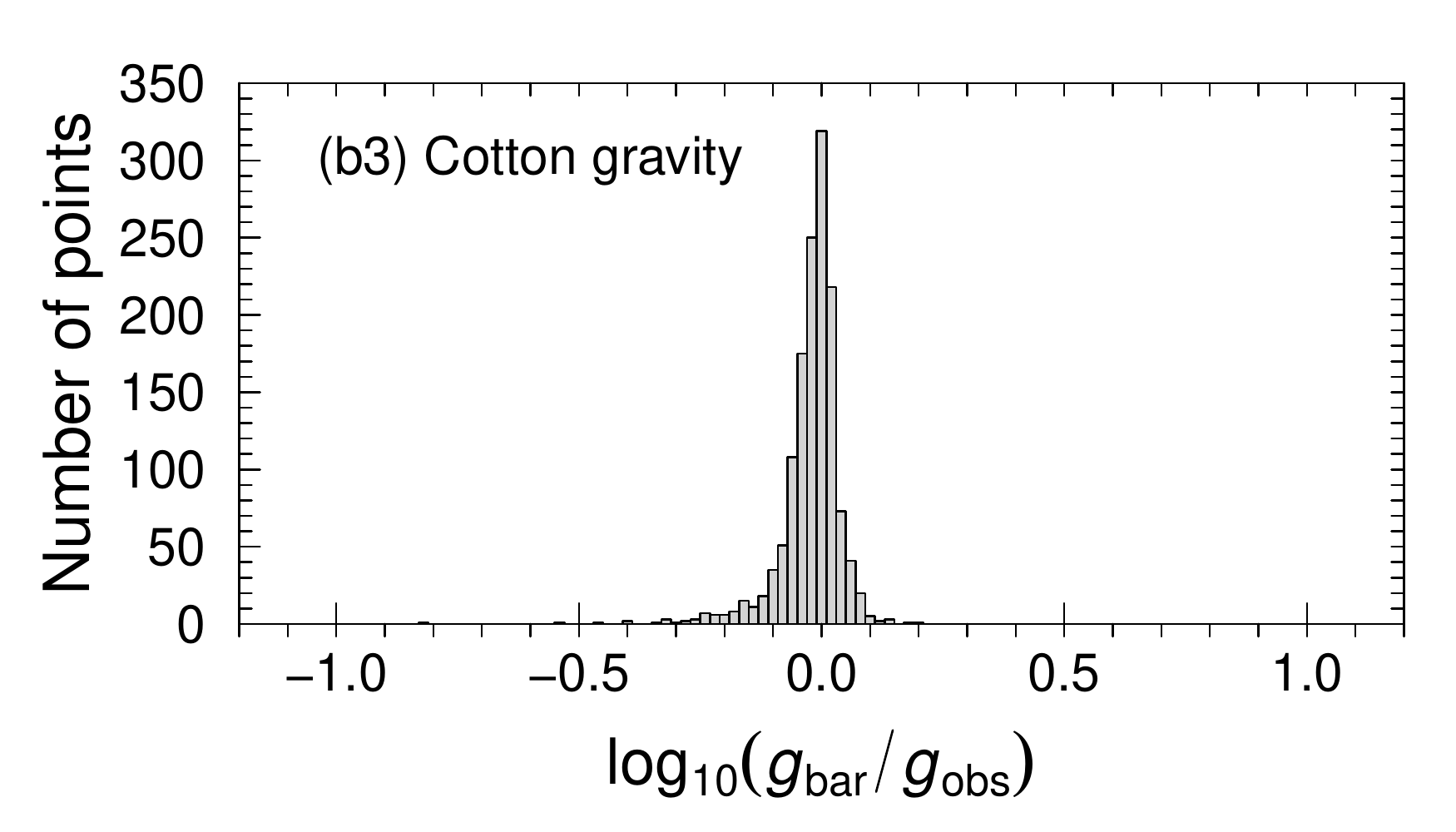}%
\caption{\label{fig:RAR} Radial acceleration relation~\cite{McGaugh:2016leg}: the observed gravitational acceleration, $g_{\rm obs}=V_{\rm obs}^2(R)/R$, is plotted against that calculated for the observed distributions of baryons, $g_{\rm bar}=\partial \Phi/\partial R$, where the gravitational potential $\Phi$ is determined by solving ({\bf a1}) the Poisson equation~\eqref{eq:Poisson}, and ({\bf b1}) the effective field equation~\eqref{eq:EFE} in Cotton gravity. 
1388 data points for 84 galaxies are shown in both panels. 
The data points in the most inner regions are rejected to minimize the affects of regularization in numerical computations (see, the appendix). 
The dashed line is the line of unity ({\bf a1} and {\bf b1}).
The middle panels ({\bf a2}) and ({\bf b2}) show the same as those of the top panels, but the horizontal axis is replaced by $\log_{10}(g_{\rm bar}/g_{\rm obs})$.
The bottom panels ({\bf a3}) and ({\bf b3}) show the distribution of $\log_{10} (g_{\rm bar}/g_{\rm obs})$, where the horizontal axis is the same as that of the middle panel.
The ({\bf a3}) shows that 1239 data points ($89\%$) are distributed in the range of $\log_{10} (g_{\rm bar}/g_{\rm obs}) < -0.07$ ($g_{\rm bar}/g_{\rm obs} < 0.85$). It indicates the need for dark matter. The ({\bf b3}) shows that 1179 data points ($85\%$) are distributed in the range of $-0.07 < \log_{10} (g_{\rm bar}/g_{\rm obs}) < 0.07$  ($0.85 < g_{\rm bar}/g_{\rm obs} < 1.17$). The distribution of ({\bf b3}) is very tight---indeed, it is not even the Gaussian (it is described by the Cauchy distribution with the location $-0.008$ and the scale $0.024$). It indicates that dark matter is unnecessary.}
\end{figure*}

Figure~\ref{fig:RAR} shows the radial acceleration relation~\cite{McGaugh:2016leg} for the present study. 
Here, the radial acceleration relation is a relation between the observed acceleration ($g_{\rm obs}$) and the calculated acceleration ($g_{\rm bar}$) for the baryons, where
\begin{equation}
	g_{\rm obs} = \frac{V_{\rm obs}^2(R)}{R},\quad \mbox{and} \quad
	g_{\rm bar} = \frac{\partial \Phi}{\partial R}. 
\end{equation}
The potential $\Phi$ is determined by solving the Poisson equation and the field equation~\eqref{eq:EFE} of Cotton gravity.
Figure~\ref{fig:RAR} shows two cases: ({\bf a}) for the Poisson equation, and ({\bf b}) for Cotton gravity. 

For the case of the Poisson equation, the calculated acceleration $g_{\rm bar}$ is smaller than the observed acceleration $g_{\rm obs}$ for $g_{\rm obs} \sim 1 \ {\rm km^2 \ s^{-2} \ pc^{-1}}$ or less [({\bf a}) in Fig~\ref{fig:RAR}]. It indicates the need for dark matter.

For the case of Cotton gravity, the calculated acceleration $g_{\rm bar}$ is in agreement with the observed acceleration $g_{\rm obs}$ even for $g_{\rm obs} \sim 1 \ {\rm km^2 \ s^{-2} \ pc^{-1}}$ or less [({\bf b}) in Fig~\ref{fig:RAR}]. Thus, in Cotton gravity, the $g_{\rm obs}$ can be explained by the baryons without the need for dark matter. 

\clearpage

\begin{figure}[t]
\includegraphics[width=86mm]{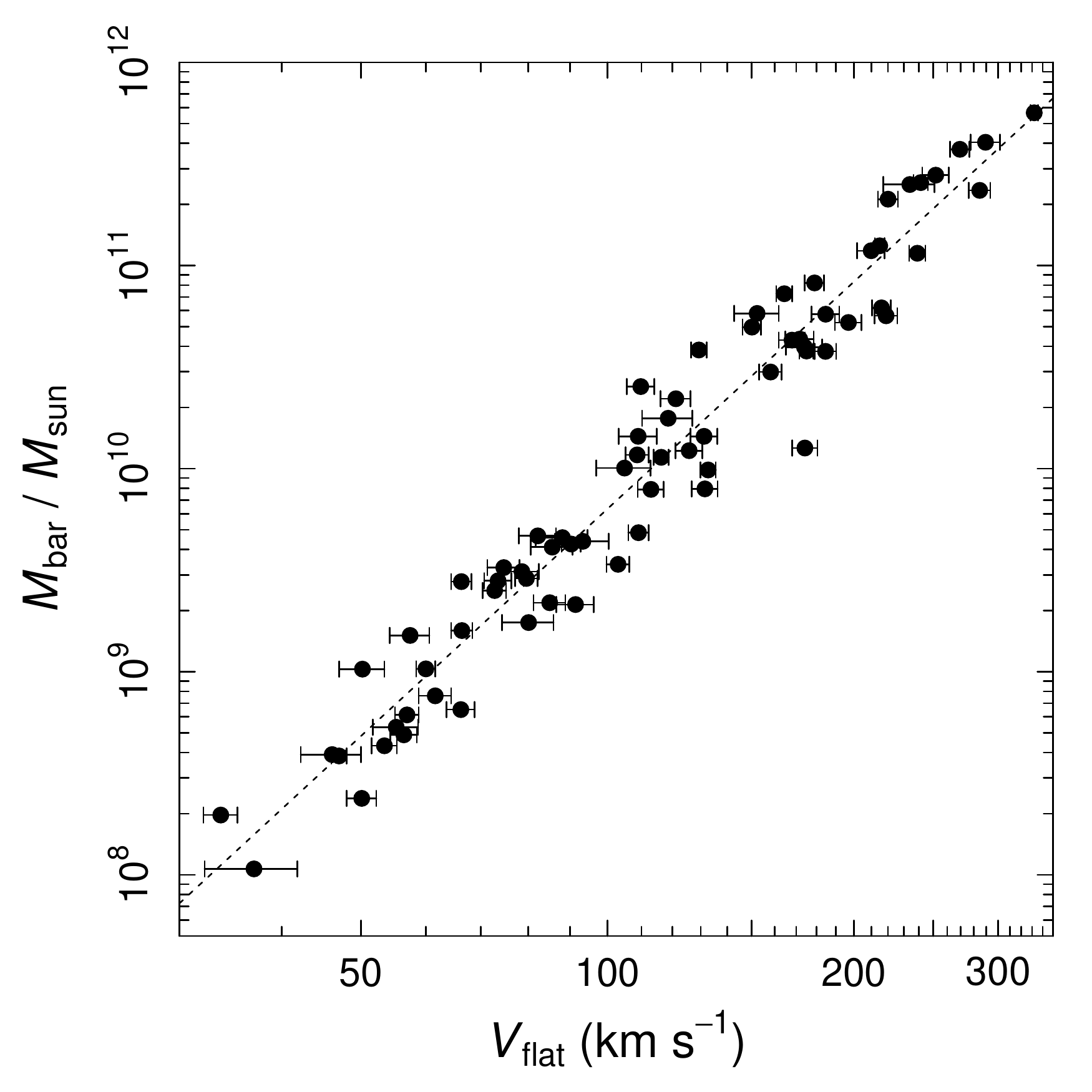}%
\caption{\label{fig:BTFR} Baryonic Tully-Fisher relation~\cite{McGaugh:2000sr}: the  baryonic mass $M_{\rm bar}$ as a function of the flat rotation speed~$V_{\rm flat}$. The baryonic mass $M_{\rm bar}$, a sum of the stellar mass ($M_\star=M_{\rm disk}+M_{\rm bulge}$) and the mass of gas ($M_{\rm gas}$), is listed in column~(7) of Table~\ref{tab:table1}. The flat rotation speed $V_{\rm flat}$ are available from the SPARC database~\cite{Lelli:2016zqa}. The dashed line represents a linear fit in the log-log plot: the slope is 3.72.
}
\end{figure}

\subsection{Baryonic Tully-Fisher relation}
The baryonic Tully-Fisher relation is an empirical relation between the total baryonic mass $M_{\rm bar}$ and the flat rotation speed $V_{\rm flat}$~\cite{McGaugh:2000sr}. 
Figure~\ref{fig:BTFR} shows the baryonic Tully-Fisher relation for the present study.
From the data, we find that
\begin{equation}
	M_{\rm bar}/M_\odot = 10^{2.37} \left(\frac{V_{\rm flat}}{{\rm km/s}}\right)^{3.72},
\end{equation}
or equivalently
\begin{equation}
	M_{\rm bar}/(10^9 M_\odot) = 6.5 \left(\frac{V_{\rm flat}}{100 \ {\rm km/s}}\right)^{3.72},
	\label{eq:BTFR}
\end{equation}
for $1 \times 10^ 8 < M_{\rm bar}/M_\odot < 6 \times 10^{11}$. 
Equation~\eqref{eq:BTFR} is a convenient form, because the coefficient is dimensionless and has a value of ${\cal O}(1)$.
In Cotton gravity, this relation is obtained as a consequence of the field equations.
Here, it should be noted that $V_{\rm flat}$ is just an effective concept---the rotation velocities should not be necessarily flat in Cotton gravity (Fig.~\ref{fig:RCfull}).
In general, the exponent is not exactly equals to 4 in the context of Cotton gravity.

\begin{figure}[t]
\includegraphics[width=86mm]{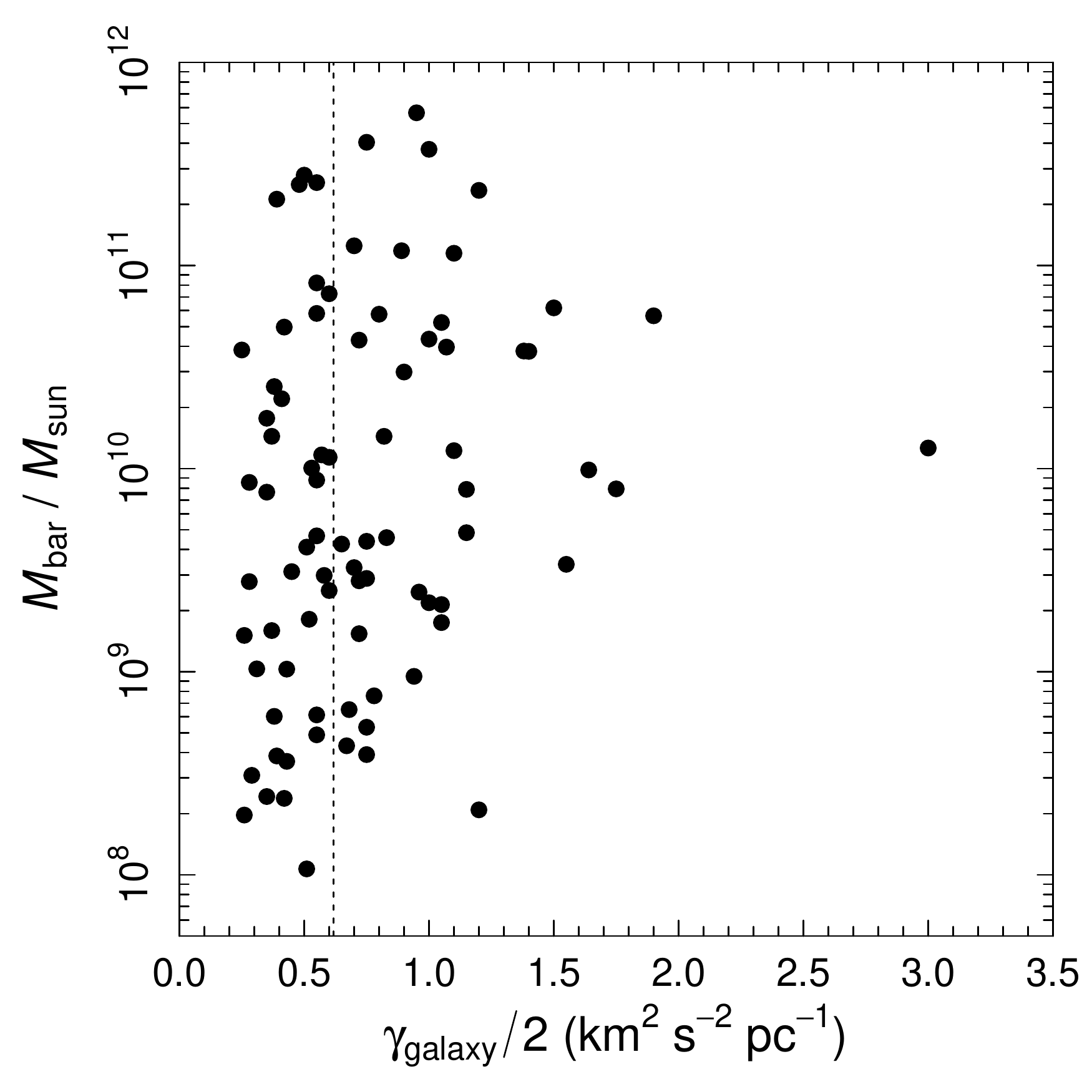}%
\\
\includegraphics[width=86mm]{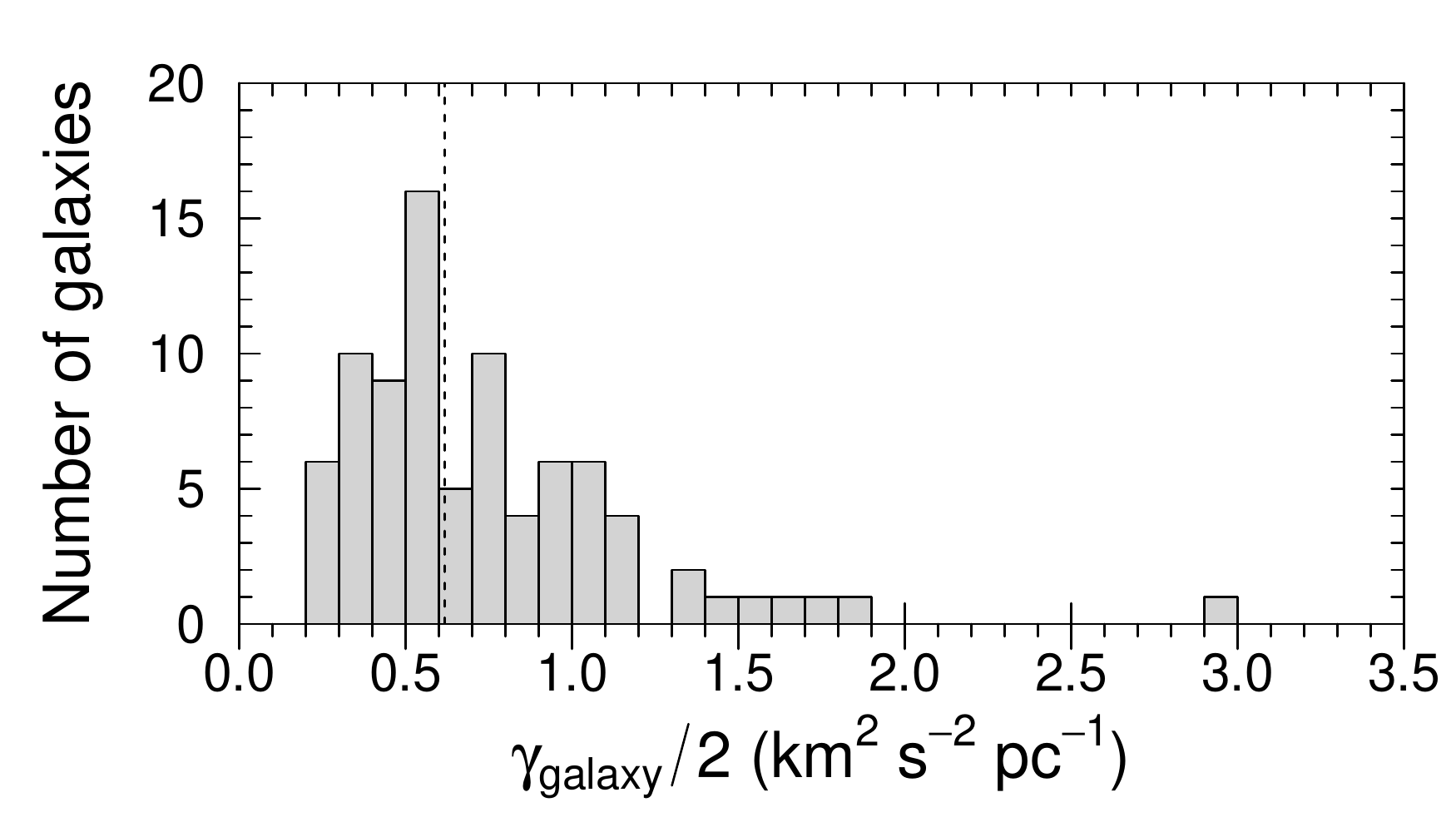}%
\caption{\label{fig:gamma}The distribution of the best fit values of $\gamma_{\rm galaxy}$.
{\it Upper}: the vertical axis is the same as that of Fig.~\ref{fig:BTFR}. The horizontal axis represents the values of $\gamma_{\rm galaxy}/2$ ($1/2$ is a convention). The vertical dashed line represents the value of $a_0/6 = 0.62 \ {\rm km^2 \ s^{-2} \ pc^{-1}}$, where $a_0=1.2\times 10^{-10} \ {\rm m \ s^{-2}} = 3.7  \ {\rm km^2 \ s^{-2} \ pc^{-1}}$ is a fundamental parameter of  MOND~\cite{Milgrom:1983pn}. 
{\it Lower}: a histogram of $\gamma_{\rm galaxy}/2$ is shown. The horizontal axis and the vertical dashed line are the same as those of the upper panel, respectively. The galaxy far away is UGC 6973.
}
\end{figure}

\subsection{Distribution of $\gamma_{\rm galaxy}$}
The parameter $\gamma_{\rm galaxy}$ should be determined for each galaxy. 
In Cotton gravity, it is theoretically unnatural to assume that $\gamma_{\rm galaxy}$ takes a single value for all galaxies. 
The particular value of $\gamma_{\rm galaxy}$ per galaxy is allowed, and it is a significant advantage of Cotton gravity.

There is no correlation between the baryonic mass $M_{\rm bar}$ and the value of $\gamma_{\rm galaxy}$ (Fig.~\ref{fig:gamma}). We find that 76 of 84 galaxies have the values in the range $ 0.2 < \gamma_{\rm galaxy}/2 < 1.3 \ {\rm km^2 \ s^{-2} \ pc^{-1}}$(Fig.~\ref{fig:gamma}). We also find that UGC 6973 has an exceptionally large value, $\gamma_{\rm galaxy}/2=3.0 \ {\rm km^2 \ s^{-2} \ pc^{-1}}$.

\begin{figure*}[t]
\includegraphics[width=60mm]{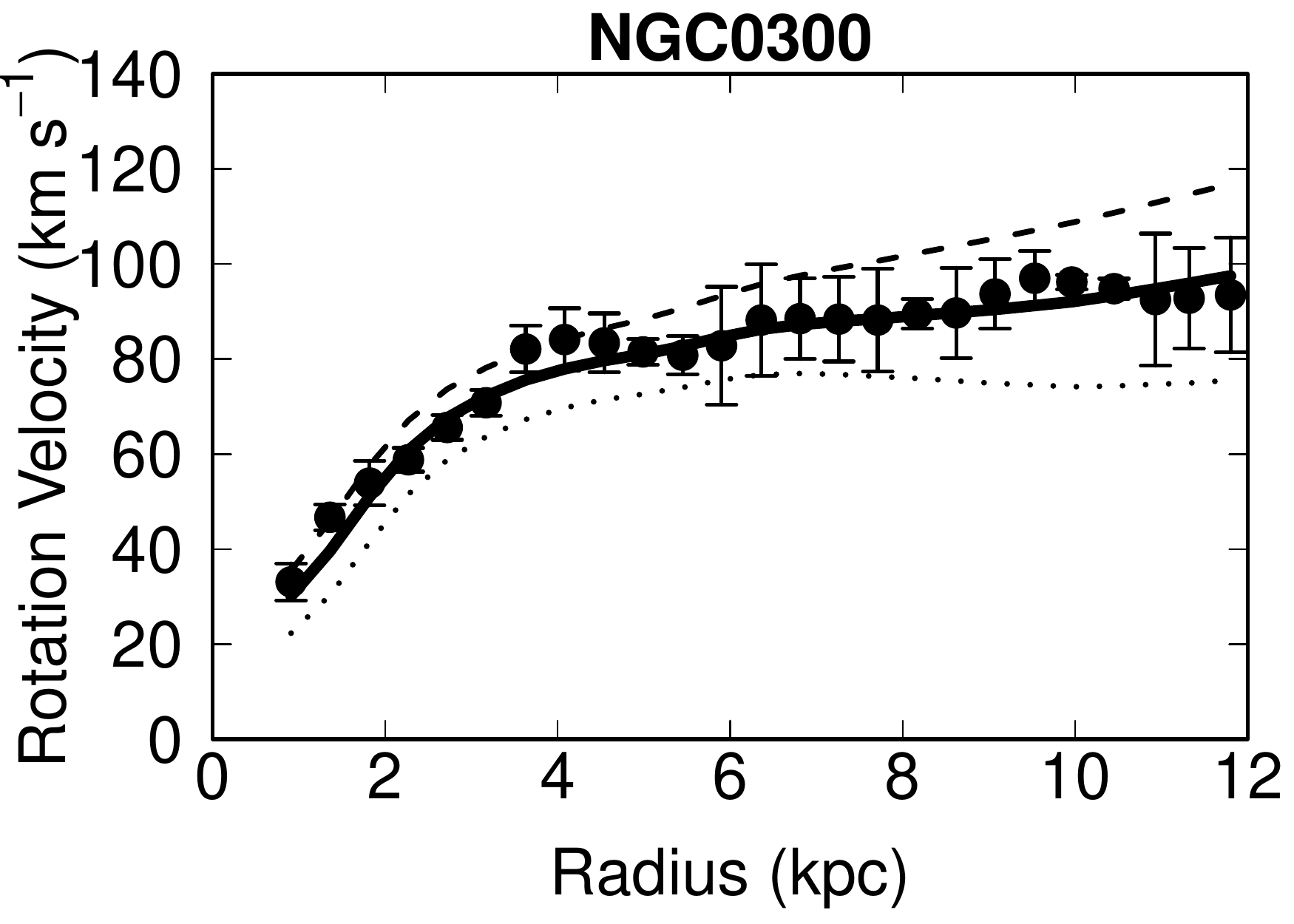}%
\includegraphics[width=60mm]{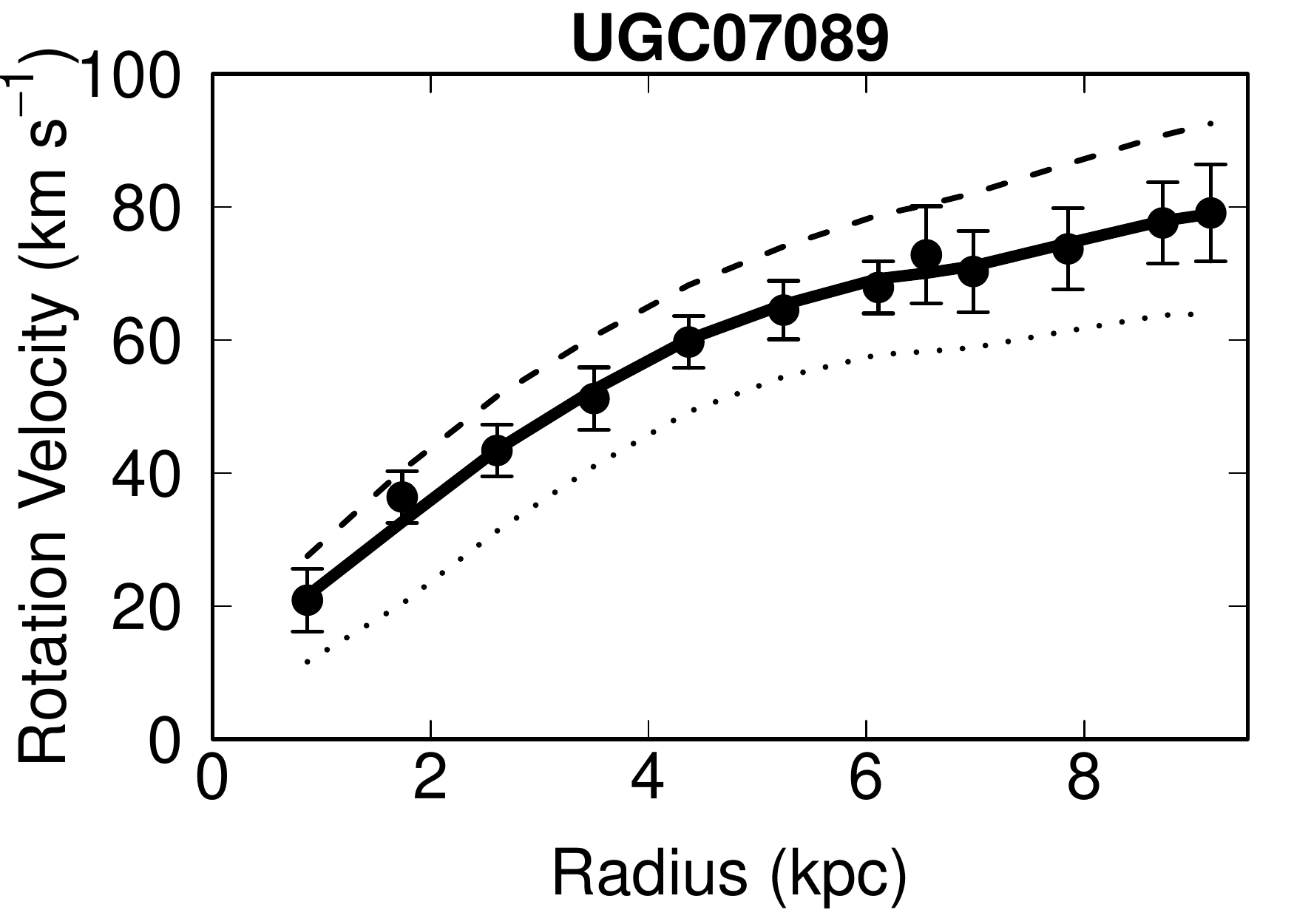}%
\includegraphics[width=60mm]{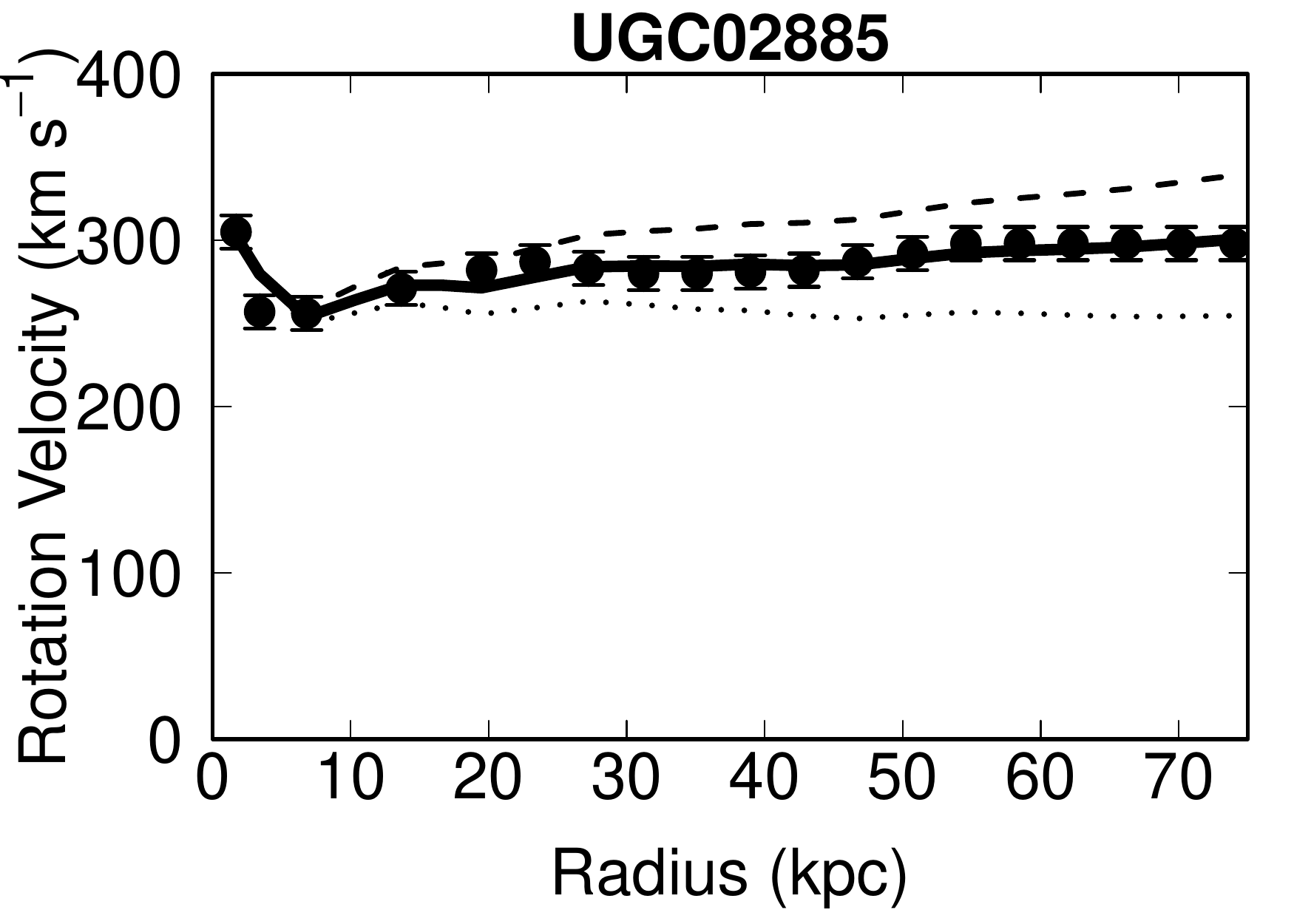}%
\caption{\label{fig:RC_gamma_depend}Examples of rotation curves with $50\%$ changes of $\gamma_{\rm galaxy}/2$. The points with error bars and the thick solid lines represent the same as those in Fig.~\ref{fig:RCex}. The calculated rotation velocities in Cotton gravity are shown when changing the values of $\gamma_{\rm galaxy}/2$ by 50\% from the best fit values; dashed lines for the 50\% larger case, and dotted lines for the 50\% smaller. Changing the values of $\gamma_{\rm galaxy}$ does not affect the whole shape of the curves.}
\end{figure*}

\subsection{Changing $\gamma_{\rm galaxy}$}

Figure~\ref{fig:RC_gamma_depend} shows the rotation velocities when changing the values of $\gamma_{\rm galaxy}/2$ by 50\% from the best fit values.
We find that changing the values of $\gamma_{\rm galaxy}/2$ does not affect the whole shape of the curves. 

Cotton gravity is flexible in the sense that it allows an additional parameter $\gamma_{\rm galaxy}$ for each galaxy. 
Therefore, one may think that the observed rotation curves can be easily explained by adjusting arbitrarily $\gamma_{\rm galaxy}$.
Figure~\ref{fig:RC_gamma_depend} indicates that it is not true; the changes of $\gamma_{\rm galaxy}$ affect only details, and does not affect the whole shape. In fact, it is nontrivial whether Cotton gravity can explain the observed rotation curves, even though the parameter $\gamma_{\rm galaxy}$ is allowed for each galaxy.

\section{\label{sec:discussion}Discussion and conclusions}
We found that Cotton gravity can explain the rotation curves of 84 galaxies without the need for dark matter. In Cotton gravity, the flat rotation curves and the nonflat rising shape of rotation curves can be explained in the same context without special assumptions (Fig.~\ref{fig:RCfull}). All we need is the data on the distribution of stars and gas, and the gravitational field equation~\eqref{eq:EFE}. This is satisfactory because the observed data (surface density) explains the observed data (rotation velocity).

Cotton gravity allows an extra linear term in the gravitational potential of a point mass.
The coefficient of the linear term $\gamma_{\rm galaxy}$ is a constant of integration. Therefore, it (and the mass $M$) should be determined separately for each galaxy. Various values of $\gamma_{\rm galaxy}$ are completely consistent with Eq.~\eqref{eq:FE}.
According to this flexibility of Cotton gravity, galaxy rotation curves can be explained.

From this work, we have observed the following: 
the observed potential $\Phi$ and the observed mass densities of 84 galaxies does not satisfy the Poisson equation, but (at least approximately) satisfy the effective field equation~\eqref{eq:EFE}. This is empirical. That is, what we found is that the effective field equation~\eqref{eq:EFE} has the particular solution that is consistent with the observed gravitational potential $\Phi$ and the baryonic masses. 

Here it should be also mentioned the following. Cotton gravity was proposed with the pure theoretical motivation as a generalization of  general relativity~\cite{Harada:2021bte}. It was not intended to solve a particular problem in astrophysics. In particular, the galaxy rotation curve was not the motivation for Cotton gravity. Nevertheless, Cotton gravity can explain the rotation curves---it is just a consequence of the field equations of Cotton gravity. 

The results of this paper suggest the following:
\begin{enumerate}
	\item When gravity is extremely weak, the law of gravity may deviate from general relativity.
	\item Galaxies are typical systems in which such deviations from general relativity cannot be ignored.
	\item Galaxies have no dark matter halos. 
\end{enumerate}

These may be alternative to the dark matter paradigm. 
For understanding gravity and matter in the universe, it is necessary to test the laws of gravity in the extremely weak regimes.
Galaxies are unique physical systems in which the effects of very weak gravity can be precisely observed. Investigating galaxies would help to test the law of gravity applicable to extremely weak regimes.

This work is the first quantitative test of Cotton gravity. 
The first test suggests that Cotton gravity is a possible candidate as a generalization of general relativity. Further investigations would be expected.

\begin{acknowledgments}
This work would not be possible without SPARC~\cite{Lelli:2016zqa} (Spitzer Photometry and Accurate Rotation Curves). 
This work was supported by JSPS KAKENHI Grant No. JP22K03599.
\end{acknowledgments}

\appendix*
\section{NUMERICAL CALCULATIONS}

\subsection{Poisson equation}

In the axially symmetric system---we assume that the rotating galaxies have the axial symmetry---, the Poisson equation~\eqref{eq:Poisson} can be written in a cylindrical coordinate,
\begin{equation}
	\left[\frac{\partial^2}{\partial R^2} + \frac{1}{R}\frac{\partial}{\partial R}	+\frac{\partial^2}{\partial z^2}\right] \Phi(R, |z|)
	= 4 \pi G \rho(R,|z|),
	\label{eq:Poisson_Axial}
\end{equation}
where $R$ is the radial distance and $z$ is the height. 

The mass density of baryons has two components,
\begin{equation}
	\rho_{\rm bar} = \rho_{\rm disk} + \rho_{\rm gas}.
\end{equation}
For galaxies with the central bulge, $\rho_{\rm bulge}$ should be added (only 11 galaxies in this work). 
We assume that the stellar disk and gas have a small but finite thickness---though, we find that the results are not sensitive to finite thickness. We assume that the bulge is spherical. 

For the stellar disk, we adopt the following profile,
\begin{equation}
	\rho_{\rm disk} (R, |z|) =\Upsilon_\star \Sigma_{\rm disk} (R) \frac{e^{-|z|/z_{\rm d}}}{2z_{\rm d}},
\end{equation}
where $\Upsilon_\star$ is the mass-to-light ratio (listed in column~(6) of Table~\ref{tab:table1}), $\Upsilon_\star \Sigma_{\rm disk} (R)$ is the surface density, and $z_{\rm d}$ is the scale height.
We use $z_{\rm d} = 0.196 (R_{\rm d}/{\rm kpc})^{0.633} \ {\rm kpc}$~\cite{Lelli:2016zqa,Bershady:2010}, where $R_{\rm d}$ is the scale length. Consequently, we obtain
\begin{eqnarray}
	&&4\pi G \rho_{\rm disk} \nonumber\\
	&&=	\frac{137.9e^{-|z|/z_{\rm d}}}{(R_{\rm d}/{\rm kpc})^{0.633}} 
		\frac{\Upsilon_{\star}}{M_\odot/L_\odot}
		\frac{\Sigma_{\rm disk}(R)}{L_\odot/{\rm pc^2}}
		\quad {\rm km^2 \ s^{-2} \ kpc^{-2}}. \nonumber\\
		\label{eq:rho_disk}
\end{eqnarray}
The values of the surface brightness $\Sigma_{\rm disk}(R)$ and the scale length $R_{\rm d}$ are available from SPARC~\cite{Lelli:2016zqa}. 

For the gas, we adopt the following profile,
\begin{equation}
	\rho_{\rm gas} (R, |z|) =1.33 \Sigma_{\rm HI}(R) \frac{{\rm sech^2}\left(z/z_{\rm g}\right)}{2z_{\rm g}},
\end{equation}
where 1.33 is a factor to account for the cosmic abundance of helium, $\Sigma_{\rm HI}(R)$ is the radial H {\scriptsize I} surface density, and $z_{\rm g}$ is the scale height. 
We assume that $z_{\rm g}=z_{\rm d}$ for simplicity.
We collect the data of $\Sigma_{\rm HI}(R)$ from the references in column~(11) of Table~\ref{tab:table1}. 
Consequently, we have
\begin{eqnarray}
	&&4\pi G \rho_{\rm gas}\nonumber\\
	&&=	\frac{183.4{\rm sech^2}\left(z/z_{\rm g}\right)}{(R_{\rm d}/{\rm kpc})^{0.633}} 
		\frac{\Sigma_{\rm HI} (R)}{M_\odot/{\rm pc^2}}
		\quad
		{\rm km^2 \ s^{-2} \ kpc^{-2}}. \nonumber\\
		\label{eq:rho_gas}
\end{eqnarray}

For the bulge, we adopt the mass-to-light ratio $1.4 \Upsilon_{\star}$ for all galaxies. 
Therefore, the bulge density is given by
\begin{eqnarray}
	\rho_{\rm bulge} (r) 
	= -\frac{1.4 \Upsilon_{\star}}{\pi} \int_r^\infty \frac{d\Sigma_{\rm bulge} (R)}{dR} \frac{dR}{\sqrt{R^2-r^2}}.
	\nonumber\\
\end{eqnarray}
The values of the surface brightness $\Sigma_{\rm bulge} (R)$ are available from SPARC~\cite{Lelli:2016zqa}.

We consider the following discrete coordinates,
\begin{equation}
	R = R_i \geq 0, \quad
	|z| = z_j \geq 0,\quad
	i,j=0,1,2,\cdots
\end{equation}
where we set $R_0=0$ and $z_0=0$.
These coordinates are not necessarily homogeneous as
$R_{i+2} - R_{i+1} \not= R_{i+1} - R_{i}$ and $z_{j+2} - z_{j+1} \not= z_{j+1} - z_{j}$.

The central finite differences are given by 
\begin{subequations}
\begin{eqnarray}
	\frac{\partial }{\partial R} \Phi(R, |z|)
	&=& \frac{\Phi_{i+1,j} - \Phi_{i-1,j}}{R_{i+1} - R_{i-1}},
	\label{eq:fist_finite_difR}
	\\
	\frac{\partial }{\partial z} \Phi(R, |z|)
	&=& \frac{\Phi_{i,j+1} - \Phi_{i,j-1}}{z_{j+1} - z_{j-1}},
	\label{eq:fist_finite_difz}	
\end{eqnarray}
\end{subequations}	
where $\Phi (R_i,z_j) = \Phi_{i,j}$. 

The second order finite difference is given by
\begin{eqnarray}
	&&\frac{\partial^2}{\partial R^2}\Phi(R, |z|) 
	=  \frac{4\left( \Phi_{i+1,j} + \Phi_{i-1,j} - 2 \Phi_{i,j} \right)}{(R_{i+1}-R_{i-1})^2} \nonumber\\
	&& \quad
	- \frac{4(R_{i+1}+R_{i-1}-2R_i)(\Phi_{i+1,j} - \Phi_{i-1,j})}{(R_{i+1}-R_{i-1})^3},  \nonumber\\
	\label{eq:second_finite_difR}
\end{eqnarray}
where the second term is necessary when the coordinate $R_i$ is not homogeneous---if the coordinate $R_i$ is homogeneous, then the second term vanishes. 
Similarly we have
\begin{eqnarray}
	&&\frac{\partial^2}{\partial z^2}\Phi(R, |z|)
	= \frac{4\left( \Phi_{i,j+1} + \Phi_{i,j-1} - 2 \Phi_{i,j} \right)}{(z_{j+1}-z_{j-1})^2} \nonumber\\
	&&\quad
	- \frac{4(z_{j+1}+z_{j-1}-2z_j)( \Phi_{i,j+1} - \Phi_{i,j-1})}{(z_{j+1}-z_{j-1})^3}.
	\label{eq:second_finite_difz}	
\end{eqnarray}
The following comes from the operator $(\bm{r}\cdot\nabla)^2$ in Eq.~\eqref{eq:EFE},
\begin{eqnarray}
	&&\frac{\partial^2}{\partial R \partial z}\Phi (R, |z|) \nonumber\\
	&&= \frac{\Phi_{i+1,j+1} - \Phi_{i+1,j-1} - \Phi_{i-1,j+1} + \Phi_{i-1,j-1}}{(R_{i+1}-R_{i-1})(z_{j+1}-z_{j-1})}.
	\qquad
	\label{eq:second_finite_difRz}
\end{eqnarray}
Equation~\eqref{eq:second_finite_difRz} holds even if the coordinates $R_i$ and $z_j$ are not homogeneous.

\begin{widetext}
For $i, j >0$, from Eqs.~\eqref{eq:fist_finite_difR},~\eqref{eq:second_finite_difR}, and~\eqref{eq:second_finite_difz}, the Poisson equation~\eqref{eq:Poisson_Axial} is written by
\begin{eqnarray}
	&& \frac{4\left( \Phi_{i+1,j} + \Phi_{i-1,j} - 2\Phi_{i,j} \right)}{(R_{i+1} - R_{i-1})^2}
	-\frac{4(R_{i+1}+R_{i-1}-2R_i)(\Phi_{i+1,j} - \Phi_{i-1,j})}{(R_{i+1}-R_{i-1})^3} 
	+ \frac{\Phi_{i+1,j} - \Phi_{i-1,j}}{R_i(R_{i+1}-R_{i-1})}	
			\nonumber\\
	&& \quad + \frac{4\left( \Phi_{i,j+1} + \Phi_{i, j-1} - 2\Phi_{i,j} \right)}{(z_{j+1}-z_{j-1})^2}	
	-  \frac{4(z_{j+1}+z_{j-1}-2z_j)( \Phi_{i,j+1} - \Phi_{i, j-1} )}{(z_{j+1}-z_{j-1})^3}	
	=4\pi G \rho_{i,j}
	\qquad \mbox{for $i, j >0$,}
	\label{eq:Poisson_Axial2}
\end{eqnarray}
where $\rho (R_i, z_j) = \rho_{i,j}$.

Solving Eq.~\eqref{eq:Poisson_Axial2} with respect to $\Phi_{i,j}$, we find the formula for $i, j>0$,
\begin{eqnarray}
	&&\left[\frac{2}{(R_{i+1}-R_{i-1})^2} + \frac{2}{(z_{j+1}-z_{j-1})^2}\right]\Phi_{i,j} 
	 =\frac{ \Phi_{i+1,j} + \Phi_{i-1,j}}{(R_{i+1}-R_{i-1})^2}
	-\frac{(R_{i+1}+R_{i-1}-2R_i)(\Phi_{i+1,j} - \Phi_{i-1,j})}{(R_{i+1}-R_{i-1})^3} \nonumber\\
	&& \quad 
	+\frac{\Phi_{i+1,j} - \Phi_{i-1,j}}{4R_i(R_{i+1}-R_{i-1})} 
	+ \frac{\Phi_{i,j+1} + \Phi_{i, j-1} }{(z_{j+1}-z_{j-1})^2}	
	- \frac{(z_{j+1}+z_{j-1}-2z_j)( \Phi_{i,j+1} - \Phi_{i, j-1} )}{(z_{j+1}-z_{j-1})^3}	
	-\pi G \rho_{i,j}
	\quad \mbox{for $i, j>0$.}
	\label{eq:Poisson_ij>0}
\end{eqnarray}

For $i>0$ and $j=0$, we obtain the formula by replacing as $z_{j-1}=-z_{j+1}$ and $\Phi_{i,j-1}=\Phi_{i,j+1}$ in Eq.~\eqref{eq:Poisson_ij>0}, 
\begin{eqnarray}
	&&\left[\frac{2}{(R_{i+1}-R_{i-1})^2} + \frac{1}{2z_{j+1}^2}\right]\Phi_{i,j}
	=
	\frac{\Phi_{i+1,j} + \Phi_{i-1,j} }{(R_{i+1}-R_{i-1})^2}
	-\frac{(R_{i+1}+R_{i-1}-2R_i)(\Phi_{i+1,j} - \Phi_{i-1,j})}{(R_{i+1}-R_{i-1})^3} \nonumber\\
	&&\qquad
	+\frac{\Phi_{i+1,j} - \Phi_{i-1,j}}{4R_i(R_{i+1}-R_{i-1})} 
	+ \frac{\Phi_{i,j+1}}{2z_{j+1}^2}	
	-\pi G \rho_{i,j}
	\quad \mbox{for $i>0$ and $j=0$.}
	\label{eq:Poisson_i>0j=0}
\end{eqnarray}

For $i=0$, the second term in Eq.~\eqref{eq:Poisson_Axial} is singular. We can avoid it as follows. 
Consider the two dimensional case, as example. 
Transforming the cylindrical coordinates $(R,\varphi)$ to the Cartesian coordinates $(x, y)$ in two dimensions, the central finite Laplacian at the origin is given by
\begin{equation}
	\nabla^2 \Phi = \frac{\Phi (\Delta x, 0) + \Phi (-\Delta x, 0) - 2 \Phi (0,0)}{(\Delta x)^2} 
	+ \frac{\Phi (0, \Delta y) + \Phi (0, -\Delta y) - 2 \Phi (0,0)}{(\Delta y)^2}.
	\label{eq:finite_Laplace_origin1}
\end{equation}
When $\Delta x = \Delta y = \delta$, Eq.~\eqref{eq:finite_Laplace_origin1} is given by
\begin{eqnarray}
	\nabla^2 \Phi 
	= \frac{1}{\delta^2}\left( \Phi (\delta, 0) + \Phi (-\delta, 0) + \Phi (0, \delta) + \Phi (0, -\delta) - 4 \Phi (0,0) \right)
	= \frac{4}{\delta^2}\left( \Phi (\delta, 0) - \Phi (0,0) \right),
	\label{eq:finite_Laplace_origin2}
\end{eqnarray}
where $\Phi (\delta, 0) = \Phi (-\delta, 0) = \Phi (0, \delta) = \Phi (0, -\delta)$ has been used (this holds in the axially symmetric system).

Applying Eq.~\eqref{eq:finite_Laplace_origin2} to the present case, we find that 
\begin{equation}
	\left[ \frac{\partial^2}{\partial R^2} + \frac{1}{R}\frac{\partial}{\partial R} \right] \Phi (R, |z|)
	= \frac{4\left( \Phi_{i+1,j} - \Phi_{i,j} \right)}{R_{i+1}^2} 
	\quad \mbox{for $i=0$.}
\end{equation}
Therefore, for $i=0$ and $j>0$, the Poisson equation~\eqref{eq:Poisson_Axial} is written by
\begin{eqnarray}
	\frac{4\left( \Phi_{i+1,j} - \Phi_{i,j} \right)}{R_{i+1}^2} 
	+\frac{4\left( \Phi_{i,j+1} + \Phi_{i,j-1} - 2 \Phi_{i,j} \right)}{(z_{j+1}-z_{j-1})^2}
	-\frac{4(z_{j+1}+z_{j-1}-2z_j)(\Phi_{i,j+1} - \Phi_{i, j-1})}{(z_{j+1}-z_{j-1})^3}	
	=4 \pi G \rho_{i,j}
	\nonumber\\
	\quad \mbox{for $i=0$ and $j>0$.}	
	\label{eq:Poisson_Axial3}
\end{eqnarray}
Solving Eq.~\eqref{eq:Poisson_Axial3} with respect to $\Phi_{i,j}$, we obtain the formula for $i=0$ and $j>0$,
\begin{eqnarray}
	\left[ \frac{1}{R_{i+1}^2} + \frac{2}{(z_{j+1}-z_{j-1})^2} \right]\Phi_{i,j} 
	=
	\frac{\Phi_{i+1,j}}{R_{i+1}^2} 
	 + \frac{\Phi_{i,j+1} + \Phi_{i, j-1}}{(z_{j+1}-z_{j-1})^2}
	-  \frac{(z_{j+1}+z_{j-1}-2z_j)(\Phi_{i,j+1} - \Phi_{i, j-1})}{(z_{j+1}-z_{j-1})^3}	
	-\pi G \rho_{i,j} \nonumber\\
	\mbox{for $i=0$ and $j>0$.\qquad}
	\label{eq:Poisson_i=0j>0}
\end{eqnarray}

For $i=j=0$, we obtain the formula  by replacing as $z_{j-1}=-z_{j+1}$ and $\Phi_{i,j-1}=\Phi_{i,j+1}$ in Eq.~\eqref{eq:Poisson_i=0j>0},
\begin{eqnarray}
	\left[ \frac{1}{R_{i+1}^2} + \frac{1}{2z_{j+1}^2} \right]\Phi_{i,j}
	= 
	\frac{\Phi_{i+1,j}}{R_{i+1}^2}  
	 + \frac{\Phi_{i,j+1}}{2z_{j+1}^2} 
	-\pi G \rho_{i,j}
	\qquad \mbox{for $i=j=0$.}
	\label{eq:Poisson_origin} 
\end{eqnarray}

\begin{figure*}[t]
\includegraphics[width=60mm]{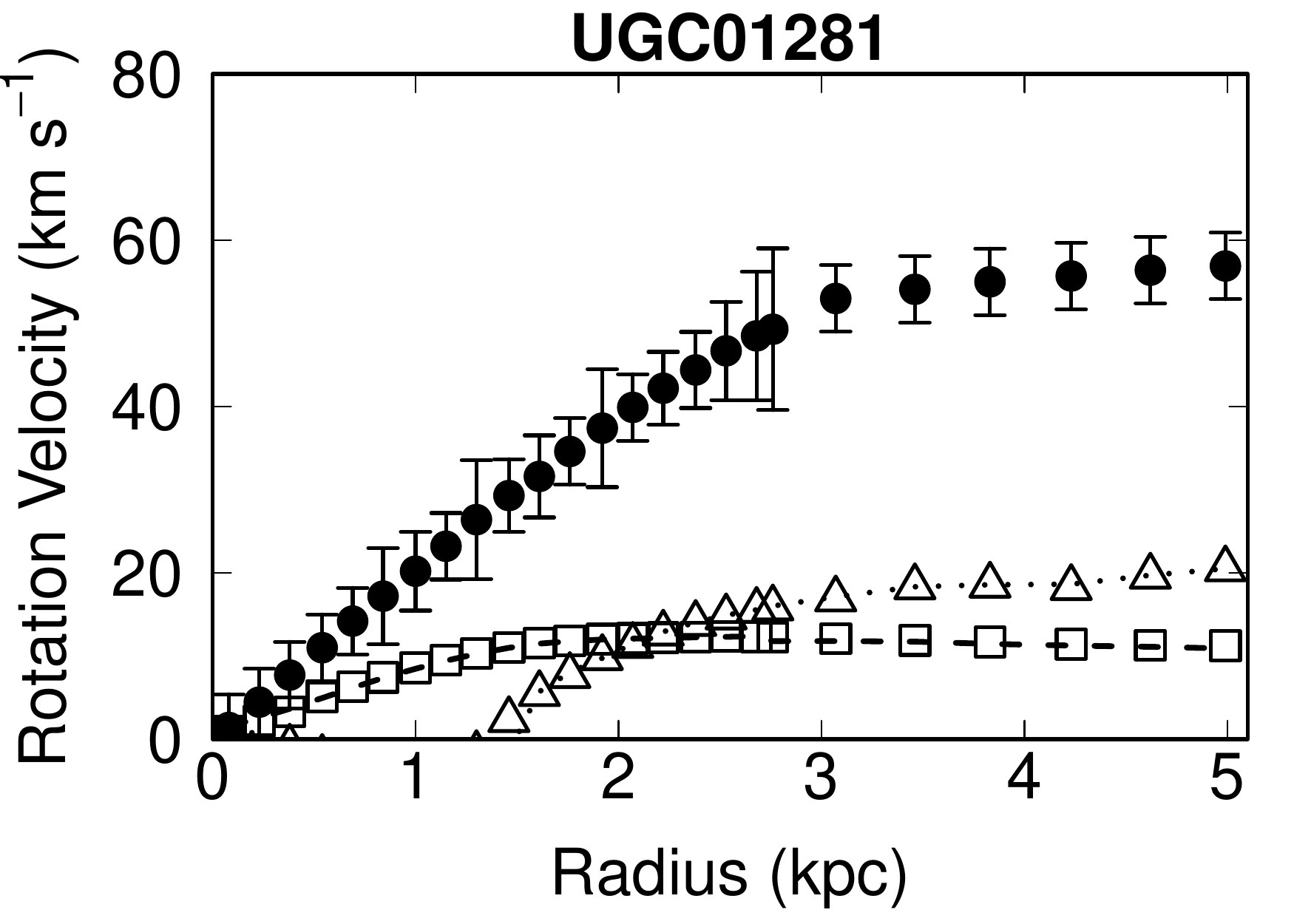}%
\includegraphics[width=60mm]{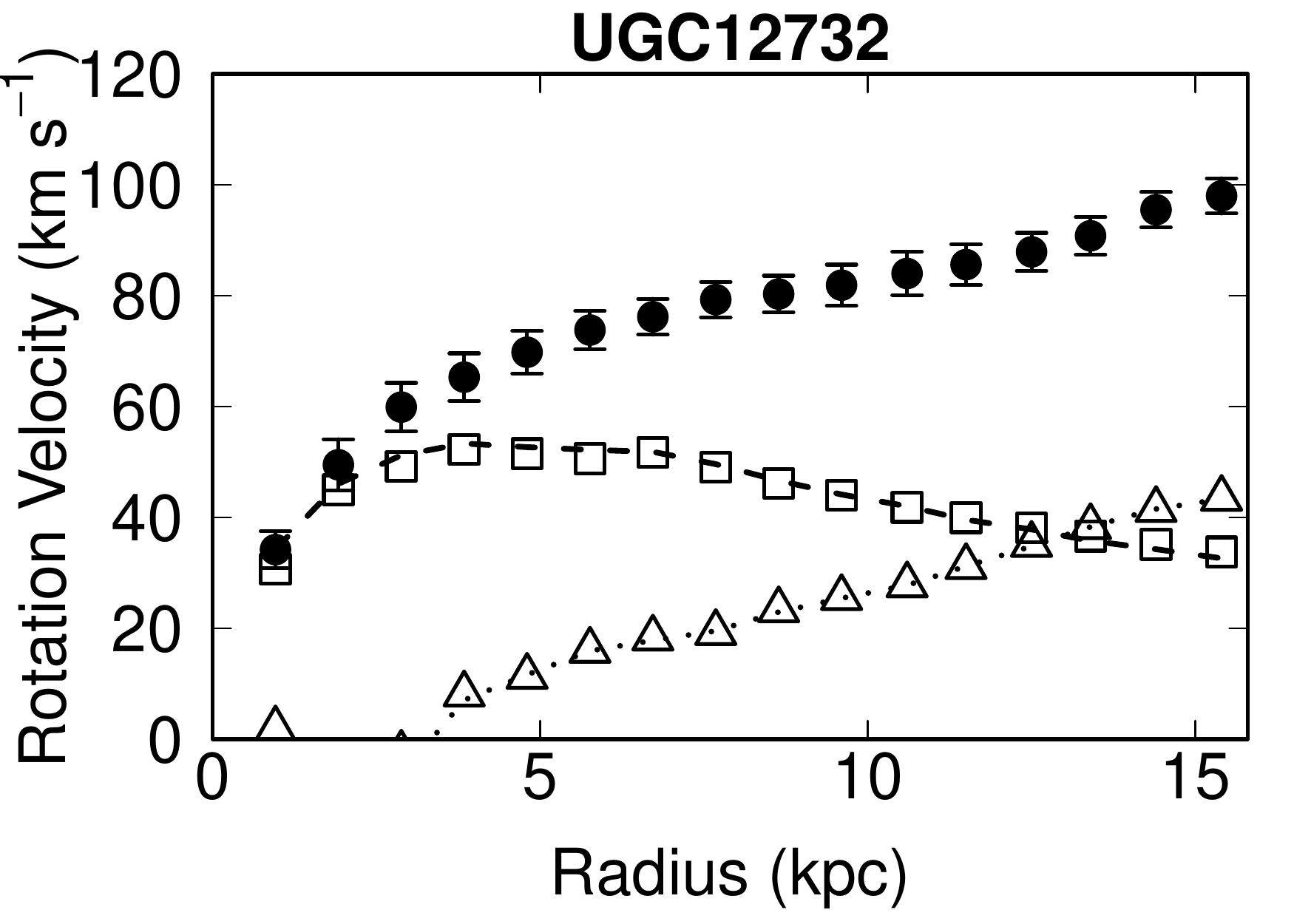}%
\includegraphics[width=60mm]{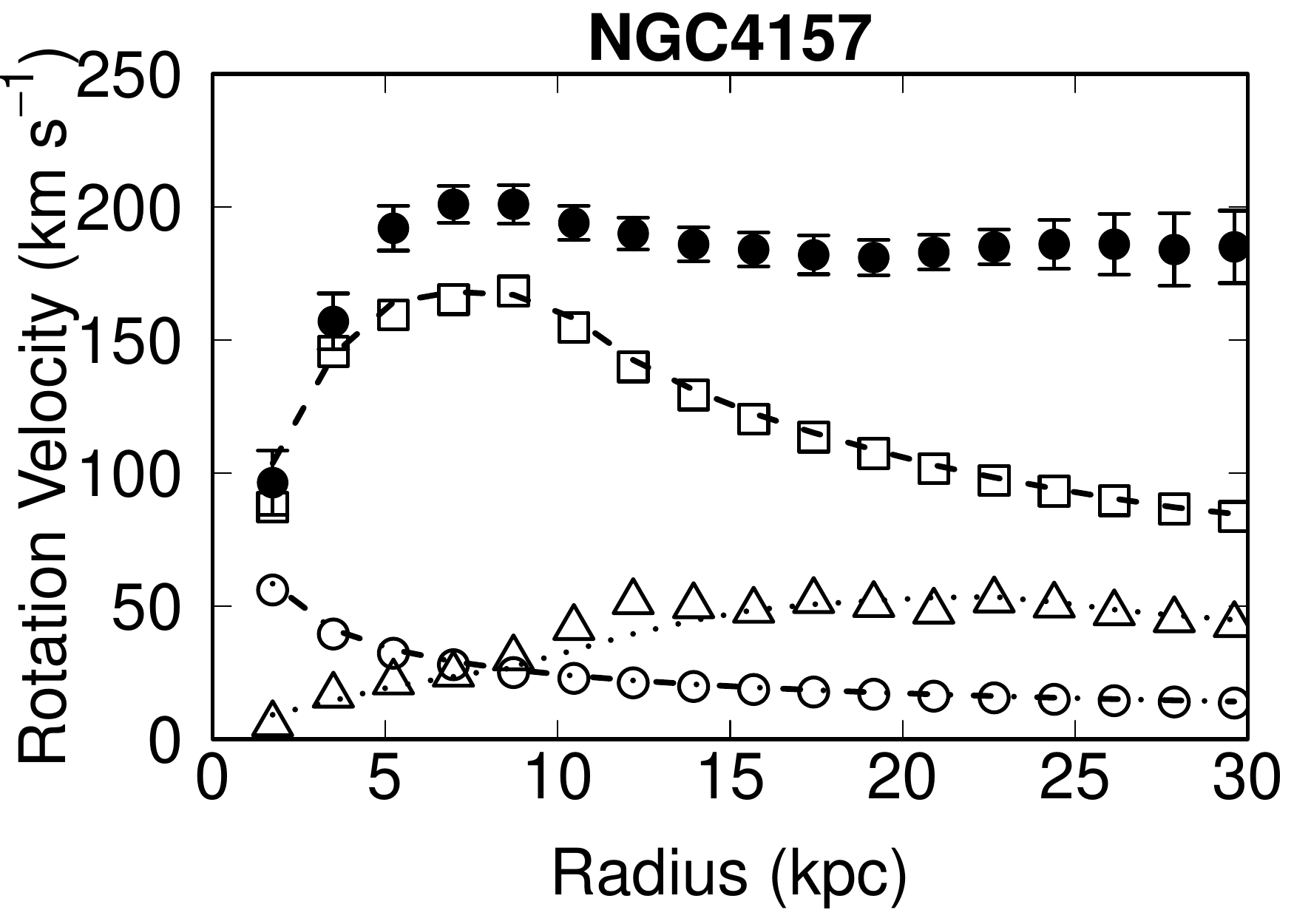}%
\caption{\label{fig:PoissonTest} Examples of rotation curves of galaxies.
Three lines (dashed for the stellar disk; dotted for the gas; dash-dotted for the bulge when present) are computed by using the formulas for the Poisson equation given in the appendix. Three open symbols (square for the stellar disk; triangle for the gas; circle for the bulge when present) represent the previous results~\cite{Lelli:2016zqa}. In all cases, lines and symbols are shown for the mass-to-light ratio $\Upsilon_\star$ given in Table~\ref{tab:table1}. Our results are completely consistent with the previous ones~\cite{Lelli:2016zqa}, though different computational methods are used. For NGC 4157, a tiny deviation for the gas component (dotted line and open triangle) is found, but it would be due to the difference of the input data for the H {\scriptsize I} surface density.}
\end{figure*}

Using the formulas~\eqref{eq:Poisson_ij>0}, \eqref{eq:Poisson_i>0j=0}, \eqref{eq:Poisson_i=0j>0}, and \eqref{eq:Poisson_origin}, we can compute the gravitational potential $\Phi$ via the Poisson equation.
Following the $1/r$ solution of the Poisson equation, we set the boundary condition $\Phi=0$ at far from the center of galaxies. 
This boundary condition enables us to compute $\Phi$ for each baryonic component: stellar disk, bulge, and gas. 
(In Cotton gravity, following the solution~\eqref{eq:sol_Cotton}, we adopt the boundary condition~\eqref{eq:bc}.
 So we can compute $\Phi$ only for the sum of the baryonic components, rather than the each component.)

Our results are completely consistent with the previous ones (Fig.~\ref{fig:PoissonTest}) for each baryonic component. In the previous study~\cite{Lelli:2016zqa}, the potential $\Phi$ was computed by using the different method. Therefore, this is a good check for our numerical calculations.


\subsection{Cotton gravity}

For Cotton gravity, we numerically solve the effective field equation~\eqref{eq:EFE}, rather than the field equations~\eqref{eq:FE}.
In the effective field equation~\eqref{eq:EFE}, $1/|\bm{r}|^2$ term is singular at the origin. 
To avoid it in the numerical computations, 
we introduce the functions $F_1$ and $F_2$ to regularize the effective field equation~\eqref{eq:EFE} as follows, 
\begin{eqnarray}
	\left[\nabla^2 - \frac{F_1}{3|\bm{r}|^2} \left( (\bm{r}\cdot \nabla)^2 + 4(\bm{r}\cdot\nabla)+2\right) \right]\Phi 
	= \frac{16\pi G}{3}\rho \left(1 - \frac{F_2}{4}\right),
	\label{eq:EFE2}
\end{eqnarray}
where $F_1$ and $F_2$ satisfy
\begin{subequations}
\begin{eqnarray}
	&&F_1 (\bm{r}) \rightarrow 1, \ F_2 (\bm{r}) \rightarrow 0 \quad \mbox{for $|\bm{r}| \rightarrow \infty$},\quad\\
	&&F_1 (\bm{r}) \simeq 1, \ F_2 (\bm{r}) \simeq 0 \quad \mbox{except for $|\bm{r}| \simeq 0$},\quad\\
	&&F_1 (\bm{r})/|{\bm r}|^2 \rightarrow 0, \ F_2 (\bm{r}) \rightarrow 1 \quad \mbox{for $|\bm{r}| \rightarrow 0$}.\quad
\end{eqnarray}
\end{subequations}
This guarantees that Eq.~\eqref{eq:EFE2} reduces to Eq.~\eqref{eq:EFE} except near the origin; it also guarantees that Eq.~\eqref{eq:EFE2} reduces to Eq.~\eqref{eq:Poisson} at the origin as an approximation.

Here, we adopt the following functions,
\begin{eqnarray}
	F_1 (R, |z|) &=& \tanh^2 \left(\frac{R^2+z^2}{r_{\rm c}^2} \right),
\end{eqnarray}
and $F_2(R, |z|)=1-F_1(R, |z|)$, where we set $r_{\rm c}=R_{\rm d}/3$ (we set $r_{\rm c}=R_{\rm d}$ for some larger galaxies). 
The particular choice of $F_1$ and $F_2$ is not sensitive to the results except near the origin.
To minimize the affects of regularization, the points for $R<0.1R_{\rm last}$ (where $R_{\rm last}$ is the radius at the most outer data) are rejected in Fig.~\ref{fig:RAR}.

We write the functions $F_1$ and $F_2$ as
\begin{eqnarray}
	F_1 (R_i, z_j) = F_{1 i,j},\quad
	F_2 (R_i, z_j) = F_{2 i,j}.
\end{eqnarray}

In the axially symmetric system, the regularized effective field equation~\eqref{eq:EFE2} is written by
\begin{eqnarray}
	&&\left[\frac{\partial^2}{\partial R^2}+\frac{1}{R}\frac{\partial}{\partial R} + \frac{\partial^2}{\partial z^2}
	 - \frac{F_1(R,|z|)}{3(R^2+z^2)} \left( R^2\frac{\partial^2}{\partial R^2} + z^2 \frac{\partial^2}{\partial z^2} + 2Rz\frac{\partial^2}{\partial R \partial z}
	 + 4 R \frac{\partial}{\partial R} + 4 z \frac{\partial}{\partial z}
	 +2\right) \right]\Phi(R,|z|) \nonumber\\
	&&= \frac{16\pi G}{3}\rho(R,|z|) \left( 1 - \frac{F_2 (R, |z|)}{4}\right).
	\label{eq:EFE_Axial}
\end{eqnarray}

For $i, j >0$, from Eqs.~\eqref{eq:fist_finite_difR},~\eqref{eq:fist_finite_difz},~\eqref{eq:second_finite_difR},~\eqref{eq:second_finite_difz}, and~\eqref{eq:second_finite_difRz}, the field equation~\eqref{eq:EFE_Axial} is written by
\begin{eqnarray}
	&& \frac{4\left( \Phi_{i+1,j} + \Phi_{i-1,j} - 2\Phi_{i,j} \right)}{(R_{i+1} - R_{i-1})^2}
	-\frac{4(R_{i+1}+R_{i-1}-2R_i)(\Phi_{i+1,j} - \Phi_{i-1,j})}{(R_{i+1}-R_{i-1})^3}
	 + \frac{\Phi_{i+1,j} - \Phi_{i-1,j}}{R_i(R_{i+1}-R_{i-1})}			
	\nonumber\\
	&& + \frac{4\left( \Phi_{i,j+1} + \Phi_{i, j-1} - 2\Phi_{i,j} \right)}{(z_{j+1}-z_{j-1})^2}	
	-  \frac{4(z_{j+1}+z_{j-1}-2z_j)(\Phi_{i,j+1} - \Phi_{i, j-1})}{(z_{j+1}-z_{j-1})^3}	 \nonumber\\
	&& -\frac{F_{1 i,j}}{3(R_i^2 + z_j^2)}
		\left[ \frac{4R_i^2 \left( \Phi_{i+1,j} + \Phi_{i-1,j} -2 \Phi_{i,j}\right)}{(R_{i+1}-R_{i-1})^2}
		- \frac{4R_i^2(R_{i+1}+R_{i-1}-2R_i)(\Phi_{i+1,j} - \Phi_{i-1,j})}{(R_{i+1} - R_{i-1})^3} 
		+\frac{4R_i(\Phi_{i+1,j}-\Phi_{i-1,j})}{R_{i+1}-R_{i-1}}	\right]	\nonumber\\
	&& - \frac{F_{1 i,j}}{3(R_i^2 + z_j^2)}
		\left[ \frac{4z_j^2\left( \Phi_{i,j+1} + \Phi_{i,j-1} - 2 \Phi_{i,j} \right)}{(z_{j+1}-z_{j-1})^2}
		- \frac{4z_j^2(z_{j+1}+z_{j-1}-2z_j)(\Phi_{i,j+1} - \Phi_{i,j-1})}{(z_{j+1}-z_{j-1})^3} 
		+\frac{4z_j(\Phi_{i,j+1}-\Phi_{i,j-1})}{z_{j+1}-z_{j-1}}\right]		\nonumber\\
	&&- \frac{F_{1 i,j}}{3(R_i^2 + z_j^2)}
		\left[
		\frac{2R_i z_j(\Phi_{i+1,j+1} - \Phi_{i+1,j-1} - \Phi_{i-1,j+1} + \Phi_{i-1,j-1})}{(R_{i+1}-R_{i-1})(z_{j+1}-z_{j-1})}
		+2\Phi_{i,j}
		\right] 
	=\frac{16 \pi G}{3} \rho_{i,j} \left( 1 - \frac{F_{2 i,j}}{4}\right)
		\ \mbox{for $i, j >0$.}\nonumber\\
		\label{eq:EFE_Axial_ij>0}
\end{eqnarray}
Solving Eq.~\eqref{eq:EFE_Axial_ij>0} with respect to $\Phi_{i,j}$, we obtain the formula for $i,j>0$,
\begin{eqnarray}
	&&\left[
		\frac{2}{(R_{i+1}-R_{i-1})^2}+\frac{2}{(z_{j+1}-z_{j-1})^2}
		-\frac{F_{1 i,j}}{3(R_i^2+z_j^2)}
		\left(\frac{2 R_i^2}{(R_{i+1}-R_{i-1})^2}+\frac{2 z_j^2}{(z_{j+1}-z_{j-1})^2} - \frac{1}{2}\right)
	     \right] \Phi_{i,j} \nonumber\\
	&=&
		\frac{\Phi_{i+1,j} + \Phi_{i-1,j}}{(R_{i+1}-R_{i-1})^2} 
			-\frac{(R_{i+1}+R_{i-1}-2R_i)(\Phi_{i+1,j} - \Phi_{i-1,j})}{(R_{i+1}-R_{i-1})^3} 	
			+ \frac{\Phi_{i+1,j} - \Phi_{i-1,j}}{4R_i(R_{i+1}-R_{i-1})}	
			   \nonumber\\
	&&	 + \frac{\Phi_{i,j+1} + \Phi_{i, j-1}}{(z_{j+1}-z_{j-1})^2}	 
		- \frac{(z_{j+1}+z_{j-1}-2z_j)(\Phi_{i,j+1} - \Phi_{i, j-1})}{(z_{j+1}-z_{j-1})^3}		\nonumber\\
	&&-\frac{F_{1 i,j}}{3(R_i^2 + z_j^2)}
		\left[ \frac{R_i^2\left( \Phi_{i+1,j} + \Phi_{i-1,j} \right)}{(R_{i+1}-R_{i-1})^2} 
		- \frac{R_i^2(R_{i+1}+R_{i-1}-2R_i)(\Phi_{i+1,j} - \Phi_{i-1,j})}{(R_{i+1} - R_{i-1})^3} 
		 +\frac{R_i \left( \Phi_{i+1,j} - \Phi_{i-1,j} \right)}{R_{i+1}-R_{i-1}}	\right]	\nonumber\\
	&&-\frac{F_{1 i,j}}{3(R_i^2 + z_j^2)}
		\left[\frac{z_j^2\left( \Phi_{i,j+1} + \Phi_{i, j-1} \right)}{(z_{j+1}-z_{j-1})^2}
		-  \frac{z_j^2(z_{j+1}+z_{j-1}-2z_j)(\Phi_{i,j+1} - \Phi_{i, j-1})}{(z_{j+1}-z_{j-1})^3}	
		+ \frac{z_j\left( \Phi_{i,j+1} - \Phi_{i,j-1} \right)}{z_{j+1}-z_{j-1}}
		\right]	\nonumber\\
	&&-\frac{F_{1 i,j}}{3(R_i^2 + z_j^2)}
	\left[
	\frac{R_i z_j\left( \Phi_{i+1,j+1} - \Phi_{i+1,j-1} - \Phi_{i-1,j+1} + \Phi_{i-1,j-1} \right)}{2(R_{i+1}-R_{i-1})(z_{j+1}-z_{j-1})}
	\right]
	-\frac{4\pi G}{3}\rho_{i,j}\left( 1 - \frac{F_{2 i,j}}{4}\right)
	\quad \mbox{for $i, j>0$.}\nonumber\\
	\label{eq:EFE_ij>0}
\end{eqnarray}

For $i>0$ and $j=0$, we obtain the formula by replacing as $z_{j-1}=-z_{j+1}$, $\Phi_{i,j-1}=\Phi_{i,j+1}$, and $\Phi_{i\pm1,j-1}=\Phi_{i\pm1,j+1}$ in Eq.~\eqref{eq:EFE_ij>0} and by using $z_0 = 0$, 
\begin{eqnarray}
	&&\left[
		\frac{2}{(R_{i+1}-R_{i-1})^2}+\frac{1}{2z_{j+1}^2}
	-\frac{F_{1 i,j}}{3R_i^2}
		\left(\frac{2 R_i^2}{(R_{i+1}-R_{i-1})^2} - \frac{1}{2}\right)
	     \right] \Phi_{i,j} \nonumber\\
	&=&
		\frac{\Phi_{i+1,j} + \Phi_{i-1,j}}{(R_{i+1}-R_{i-1})^2} 
			-\frac{(R_{i+1}+R_{i-1}-2R_i)(\Phi_{i+1,j} - \Phi_{i-1,j})}{(R_{i+1}-R_{i-1})^3} 
			+ \frac{\Phi_{i+1,j} - \Phi_{i-1,j}}{4R_i(R_{i+1}-R_{i-1})}	
		 + \frac{\Phi_{i,j+1} }{2z_{j+1}^2}	
			\nonumber\\
	&& -\frac{F_{1 i,j}}{3}
	\left[
	 \frac{\Phi_{i+1,j} + \Phi_{i-1,j}}{(R_{i+1}-R_{i-1})^2} 
		- \frac{(R_{i+1}+R_{i-1}-2R_i)(\Phi_{i+1,j} - \Phi_{i-1,j}) }{(R_{i+1} - R_{i-1})^3} 
		 +\frac{\Phi_{i+1,j} - \Phi_{i-1,j}}{R_i(R_{i+1}-R_{i-1})} \right]  \nonumber\\
	 &&-\frac{4\pi G}{3}\rho_{i,j}\left( 1 - \frac{F_{2 i,j}}{4}\right) 
		\qquad \mbox{for $i>0$ and $j=0$.} 
	\label{eq:EFE_i>0j=0}
\end{eqnarray}

For $i=0$ and $j>0$, the field equation~\eqref{eq:EFE_Axial} can be written by
\begin{eqnarray}
	&&\frac{4\left( \Phi_{i+1,j} - \Phi_{i,j} \right)}{R_{i+1}^2}
	+\frac{4\left( \Phi_{i,j+1} + \Phi_{i,j-1} - 2 \Phi_{i,j} \right)}{(z_{j+1}-z_{j-1})^2}
	-\frac{4(z_{j+1}+z_{j-1}-2z_j)(\Phi_{i,j+1} - \Phi_{i, j-1})}{(z_{j+1}-z_{j-1})^3}		\nonumber\\
	&&
	 -\frac{F_{1 i,j}}{3}
	\left[
	\frac{4(\Phi_{i,j+1}+\Phi_{i,j-1}-2\Phi_{i,j})}{(z_{j+1}-z_{j-1})^2}
	-\frac{4(z_{j+1}+z_{j-1}-2z_j)(\Phi_{i,j+1} - \Phi_{i, j-1})}{(z_{j+1}-z_{j-1})^3}	
	+\frac{4(\Phi_{i,j+1}-\Phi_{i,j-1})}{z_j(z_{j+1}-z_{j-1})}
	+\frac{2\Phi_{i,j}}{z_j^2}  \right]
	\nonumber\\
	&=&\frac{16\pi G}{3}\rho_{i,j} \left( 1 - \frac{F_{2 i,j}}{4}\right)
		\qquad  \mbox{for $i=0$ and $j>0$.} 
\end{eqnarray}
Solving this equation with respect to $\Phi_{i,j}$, we obtain the formula for $i=0$ and $j>0$,
\begin{eqnarray}
	&&\left[\frac{1}{R_{i+1}^2} 
	+ \frac{2}{(z_{j+1}-z_{j-1})^2} 
	-\frac{F_{1 i,j}}{3}
	\left(\frac{2}{(z_{j+1}-z_{j-1})^2} - \frac{1}{2z_j^2}\right)
	 \right]\Phi_{i,j} \nonumber\\
	&=& \frac{\Phi_{i+1,j}}{R_{i+1}^2} 
	+ \frac{\Phi_{i,j+1}+\Phi_{i,j-1}}{(z_{j+1}-z_{j-1})^2}
	-\frac{(z_{j+1}+z_{j-1}-2z_j)(\Phi_{i,j+1} - \Phi_{i, j-1}) }{(z_{j+1}-z_{j-1})^3} \nonumber\\
	&&
	-\frac{F_{1 i,j}}{3}
	\left[
	 \frac{\Phi_{i,j+1}+\Phi_{i,j-1}}{(z_{j+1}-z_{j-1})^2}	
	-\frac{(z_{j+1}+z_{j-1}-2z_j)(\Phi_{i,j+1} - \Phi_{i, j-1}) }{(z_{j+1}-z_{j-1})^3}	
	+\frac{\Phi_{i,j+1}-\Phi_{i,j-1}}{z_j(z_{j+1}-z_{j-1})}	
	\right]
	-\frac{4\pi G}{3}\rho_{i,j} \left( 1 - \frac{F_{2 i,j}}{4}\right) \nonumber\\
	&&\hspace{12cm} \mbox{for $i=0$ and $j>0$.}
	\label{eq:EFE_i=0j>0}
\end{eqnarray}

For $i=j=0$, we confirm that the field equation reduces to Eq.~\eqref{eq:Poisson_origin} by replacing as $z_{j-1}=-z_{j+1}$, $\Phi_{i,j-1}=\Phi_{i,j+1}$ in Eq.~\eqref{eq:EFE_i=0j>0} and by using $z_0 = 0$, $F_{1 i,j}=0$, and $F_{2 i,j}=1$. This is just a convenient approximation for $i=j=0$. 

Using the formulas~\eqref{eq:EFE_ij>0}, \eqref{eq:EFE_i>0j=0}, \eqref{eq:EFE_i=0j>0},~\eqref{eq:Poisson_origin} and the boundary condition~\eqref{eq:bc}, we can numerically compute the potential $\Phi$ via the effective field equation in Cotton gravity. 
The results are shown in Fig.~\ref{fig:RCfull}.
\end{widetext}

\clearpage

\bibliography{references}
\end{document}